\documentclass[aip,reprint]{revtex4-1}

\usepackage{graphicx}
\usepackage{amsmath}
\usepackage{amsfonts}
\newcounter{chem}
\newcounter{temp}
\usepackage{caption}
\usepackage{subcaption}
\usepackage[version=4]{mhchem}
\usepackage{dcolumn}
\usepackage{bm}
\usepackage{etoolbox}
\usepackage[colorlinks = true, urlcolor = blue, citecolor=blue,linkcolor=blue]{hyperref}
\usepackage{multirow}
\usepackage{xcolor}

\usepackage[normalem]{ulem}

\usepackage{listings}

\usepackage{xr}
\IfFileExists{SupplementaryMaterial.aux}{%
 \externaldocument[S-]{SupplementaryMaterial}%
}{%
\newlabel{S-FirstPage}{{}{1}{}{Doc-Start}{}}
\newlabel{S-eq:ME_O}{{S\mbox  {-}1}{1}{}{equation.1.1}{}}
\newlabel{S-eq:ME_N4S}{{S\mbox  {-}2}{1}{}{equation.1.2}{}}
\newlabel{S-eq:ME_N2D}{{S\mbox  {-}3}{1}{}{equation.1.3}{}}
\newlabel{S-eq:ME_C}{{S\mbox  {-}4}{1}{}{equation.1.4}{}}
\newlabel{S-eq:ME_NOi}{{S\mbox  {-}5}{1}{}{equation.1.5}{}}
\newlabel{S-eq:ME_CNl}{{S\mbox  {-}6}{1}{}{equation.1.6}{}}
\newlabel{S-eq:ME_COm}{{S\mbox  {-}7}{1}{}{equation.1.7}{}}
\newlabel{S-eq:omega_I_NO}{{S\mbox  {-}8}{2}{}{equation.1.8}{}}
\newlabel{S-eq:omega_I_CN}{{S\mbox  {-}9}{2}{}{equation.1.9}{}}
\newlabel{S-eq:omega_I_CO}{{S\mbox  {-}10}{2}{}{equation.1.10}{}}
\newlabel{S-eq:omega_E_NO_CN}{{S\mbox  {-}11}{2}{}{equation.1.11}{}}
\newlabel{S-eq:omega_E_NO_CO}{{S\mbox  {-}12}{2}{}{equation.1.12}{}}
\newlabel{S-eq:omega_E_CN_CO}{{S\mbox  {-}13}{2}{}{equation.1.13}{}}
\newlabel{S-eq:omega_D_NO}{{S\mbox  {-}14}{2}{}{equation.1.14}{}}
\newlabel{S-eq:omega_D_CN}{{S\mbox  {-}15}{2}{}{equation.1.15}{}}
\newlabel{S-eq:omega_D_CO}{{S\mbox  {-}16}{2}{}{equation.1.16}{}}
\newlabel{S-eq:K_I}{{S\mbox  {-}17}{2}{}{equation.1.17}{}}
\newlabel{S-eq:K_I_CN}{{S\mbox  {-}18}{2}{}{equation.1.18}{}}
\newlabel{S-eq:K_I_CO}{{S\mbox  {-}19}{3}{}{equation.1.19}{}}
\newlabel{S-eq:K_E_NO_CN}{{S\mbox  {-}20}{3}{}{equation.1.20}{}}
\newlabel{S-eq:K_E_NO_CO}{{S\mbox  {-}21}{3}{}{equation.1.21}{}}
\newlabel{S-eq:K_E_CN_CO}{{S\mbox  {-}22}{3}{}{equation.1.22}{}}
\newlabel{S-eq:K_D_NO}{{S\mbox  {-}23}{3}{}{equation.1.23}{}}
\newlabel{S-eq:K_D_CN}{{S\mbox  {-}24}{3}{}{equation.1.24}{}}
\newlabel{S-eq:K_D_CO}{{S\mbox  {-}25}{3}{}{equation.1.25}{}}
\newlabel{S-eq:g_NO}{{S\mbox  {-}26}{3}{}{equation.1.26}{}}
\newlabel{S-eq:g_CN}{{S\mbox  {-}27}{3}{}{equation.1.27}{}}
\newlabel{S-eq:g_CO}{{S\mbox  {-}28}{3}{}{equation.1.28}{}}
\newlabel{S-eq:k_glob_E_NO_CN}{{S\mbox  {-}29}{3}{}{equation.1.29}{}}
\newlabel{S-eq:k_glob_E_NO_CO}{{S\mbox  {-}30}{3}{}{equation.1.30}{}}
\newlabel{S-eq:k_glob_E_CN_CO}{{S\mbox  {-}31}{3}{}{equation.1.31}{}}
\newlabel{S-eq:k_glob_E_NO_CO_p}{{S\mbox  {-}32}{4}{}{equation.1.32}{}}
\newlabel{S-eq:k_glob_E_CN_CO_p}{{S\mbox  {-}33}{4}{}{equation.1.33}{}}
\newlabel{S-eq:k_glob_D_NO}{{S\mbox  {-}34}{4}{}{equation.1.34}{}}
\newlabel{S-eq:k_glob_D_CN}{{S\mbox  {-}35}{4}{}{equation.1.35}{}}
\newlabel{S-eq:k_glob_D_CO}{{S\mbox  {-}36}{4}{}{equation.1.36}{}}
\newlabel{S-BranchingRatioNOC}{{S\mbox  {-}1}{5}{Branching ratio for the \ce {NO(X^2\Pi ) + C(^3P)} asymptote overlaid with the rovibrational state distribution of \ce {NO(X^2\Pi )} at five temperatures}{figure.caption.7}{}}
\newlabel{S-tab:FittingParametersTauRTauV}{{S\mbox  {-}I}{5}{Curve fit parameters for the rotational and vibrational relaxation times for the \ce {NO(X^2 \Pi ) + C(^3P)}, \ce {CN(X^2 \Sigma ^{+}) + O(^3P)} and \ce {CO(X^1 \Sigma ^{+}) + N(^4S,^2D)} collisional systems}{table.caption.15}{}}
\newlabel{S-RotEnergyLoss}{{S\mbox  {-}II}{5}{Fitting parameters for the rotational energy loss coefficient for the \ce {NO(X^2 \Pi ) + C(^3P)}, \ce {CN(X^2 \Sigma ^{+}) + O(^3P)} and \ce {CO(X^1 \Sigma ^{+}) + N(^4S,^2D)} collisional systems}{table.caption.22}{}}
\newlabel{S-BranchingRatioCON4S}{{S\mbox  {-}2}{6}{Branching ratio for the \ce {CO(X^1\Sigma ^+) + N(^4S)} asymptote overlaid with the rovibrational state distribution of \ce {CO(X^1\Sigma ^+)} at five temperatures}{figure.caption.8}{}}
\newlabel{S-BranchingRatioCON2D}{{S\mbox  {-}3}{6}{Branching ratio for the \ce {CO(X^1\Sigma ^+) + N(^2D)} asymptote overlaid with the rovibrational state distribution at five temperatures}{figure.caption.9}{}}
\newlabel{S-VibEnergyLoss}{{S\mbox  {-}III}{6}{Fitting parameters for the vibrational energy loss coefficient for the \ce {NO(X^2 \Pi ) + C(^3P)}, \ce {CN(X^2 \Sigma ^{+}) + O(^3P)} and \ce {CO(X^1 \Sigma ^{+}) + N(^4S,^2D)} collisional systems}{table.caption.23}{}}
\newlabel{S-LastBibItem}{{4}{6}{}{table.caption.33}{}}
\newlabel{S-PES_Specific_NO_C_2000K}{{S\mbox  {-}4}{7}{Potential energy surface-specific rovibrational distribution of the \ce {NO(X^2\Pi ) + C(^3P)} collisional system at $T_t$~=~10,000~K and $T_v$~=~2000~K (dissociation and exchange processes excluded). The distributions are coloured by the vibrational quantum number}{figure.caption.12}{}}
\newlabel{S-PES_Specific_CN_O_2000K}{{S\mbox  {-}5}{8}{Potential energy surface-specific rovibrational distribution of the \ce {CN(X^2\Sigma ^{+}) + O(^3P)} collisional system at $T_t$~=~10,000~K and $T_v$~=~2000~K (dissociation and exchange processes excluded). The distributions are coloured by the vibrational quantum number}{figure.caption.13}{}}
\newlabel{S-PESSPecific_TauRTauV}{{S\mbox  {-}6}{9}{Potential energy surfaces specific relaxation times for \ce {NO(X^2\Pi ) + C(^3P)} and \ce {CN(X^2\Sigma ^{+}) + O(^3P)}}{figure.caption.14}{}}
\newlabel{S-DissEnergyMolar_NOC}{{S\mbox  {-}7}{9}{Temporal evolution of the average rotational and vibrational energies of \ce {NO(X^2\Pi )} and of its mole fraction, under the action of the total dissociation, at $T = 7500$, 10,000 and 20,000~K (heterogeneous exchange reactions excluded)}{figure.caption.17}{}}
\newlabel{S-DissEnergyMolar_CON4S}{{S\mbox  {-}8}{10}{Temporal evolution of the average rotational and vibrational energies of \ce {CO(X^1 \Sigma ^{+})} when colliding with \ce {N(^4S)} and of its mole fraction, under the action of the total dissociation, at $T = 7500$, 10,000 and 20,000~K (heterogeneous exchange reactions excluded)}{figure.caption.18}{}}
\newlabel{S-DissEnergyMolar_CON2D}{{S\mbox  {-}9}{10}{Temporal evolution of the average rotational and vibrational energies of \ce {CO(X^1 \Sigma ^{+})} when colliding with \ce {N(^2D)} and of its mole fraction, under the action of the total dissociation, at $T = 7500$, 10,000 and 20,000~K (heterogeneous exchange reactions excluded)}{figure.caption.19}{}}
\newlabel{S-CumulativeProduction}{{S\mbox  {-}10}{11}{Time-cumulative depletion of the diatomic species and global dissociation rate coefficient of the \ce {NO(X^2\Pi ) + C(^3P)}, \ce {CO(X^1 \Sigma ^{+}) + N(^4S)} and \ce {CO(X^1 \Sigma ^{+}) + N(^2D)} collisional systems at $T = 10{,}000$~K (heterogeneous exchange reactions excluded)}{figure.caption.20}{}}
\newlabel{S-MidQSS_Others}{{S\mbox  {-}11}{12}{Rovibrational distributions of the \ce {NO(X^2\Pi ) + C(^3P)}, \ce {CO(X^1 \Sigma ^{+}) + N(^4S)} and \ce {CO(X^1 \Sigma ^{+}) + N(^2D)} collisional systems at the midpoint of the QSS regime at $T = 10{,}000$~K (heterogeneous exchange processes excluded). The left panels present the population distributions coloured by the vibrational quantum number, and the right panels a zoom of the rovibrational states near the rotationless dissociation limit, coloured by the state-specific energy deficit}{figure.caption.21}{}}
\newlabel{S-ProdDep_CNO_CON4S}{{S\mbox  {-}12}{13}{Distribution of the exchange production and depletion rates for \ce {CN(X^2\Sigma ^{+}) + O(^3P) <=> CO(X^1\Sigma ^{+}) + N(^4S)} at peak production time for five temperatures. The grey dots indicate that the production and depletion rates are below or above a threshold}{figure.caption.25}{}}
\newlabel{S-ProdDep_CNO_CON2D}{{S\mbox  {-}13}{13}{Distribution of the exchange production and depletion rates for \ce {CN(X^2\Sigma ^{+}) + O(^3P) <=> CO(X^1\Sigma ^{+}) + N(^2D)} at peak production time for five temperatures. The grey dots indicate that the production and depletion rates are below or above a threshold}{figure.caption.26}{}}
\newlabel{S-ProdDep_NOC_CON4S}{{S\mbox  {-}14}{14}{Distribution of the exchange production and depletion rates for \ce {NO(X^2\Pi ) + C(^3P) <=> CO(X^1\Sigma ^{+}) + N(^4S)} at peak production time for five temperatures. The grey dots indicate that the production and depletion rates are below or above a threshold}{figure.caption.27}{}}
\newlabel{S-ProdDep_NOC_CON2D}{{S\mbox  {-}15}{14}{Distribution of the exchange production and depletion rates for \ce {NO(X^2\Pi ) + C(^3P) <=> CO(X^1\Sigma ^{+}) + N(^2D)} at peak production time for five temperatures. The grey dots indicate that the production and depletion rates are below or above a threshold}{figure.caption.28}{}}
\newlabel{S-GlobalRateExchange_NOC}{{S\mbox  {-}16}{15}{Comparison of the global rate coefficients of the heterogeneous exchange reactions and their corresponding rovibrational distributions of \ce {NO(X^2\Pi )} with and without dissociation kinetics at the midpoint of the QSS regime at $T = 10{,}000$~K. The initial mole fraction was $\chi _{\ce {NO}} = \chi _{\ce {C}} = 0.5$}{figure.caption.29}{}}
\newlabel{S-GlobalRateExchange_CON4S}{{S\mbox  {-}17}{16}{Comparison of the global rate coefficients of the heterogeneous exchange reactions and their corresponding rovibrational distributions of \ce {CO(X^1\Sigma ^{+})} with and without dissociation kinetics at the midpoint of the QSS regime at $T = 10{,}000$~K. The initial mole fraction was $\chi _{\ce {CO}} = \chi _{\ce {N(^4S)}} = 0.5$}{figure.caption.30}{}}
\newlabel{S-GlobalRateExchange_CON2D}{{S\mbox  {-}18}{17}{Comparison of the global rate coefficients of the heterogeneous exchange reactions and their corresponding rovibrational distributions of \ce {CO(X^1\Sigma ^{+})} with and without dissociation kinetics at the midpoint of the QSS regime at $T = 10{,}000$~K. The initial mole fraction was $\chi _{\ce {CO}} = \chi _{\ce {N(^2D)}} = 0.5$}{figure.caption.31}{}}
\newlabel{S-ArrheniusCoefficientFit}{{S\mbox  {-}IV}{18}{Recommended modified Arrhenius fitting parameters for the \ce {NO(X^2 \Pi ) + C(^3P)}, \ce {CN(X^2 \Sigma ^{+}) + O(^3P)} and \ce {CO(X^1 \Sigma ^{+}) + N(^4S,^2D)} collisional systems}{table.caption.33}{}}
\newlabel{S-LastPage}{{}{18}{}{}{}}
}

\newenvironment{chequation}{%
 \setcounter{temp}{\value{equation}}%
 \setcounter{equation}{\value{chem}}%
}{%
 \setcounter{chem}{\value{equation}}%
 \setcounter{equation}{\value{temp}}%
}

\colorlet{mygray}{black!30}
\colorlet{mygreen}{green!60!blue}
\colorlet{mymauve}{red!60!blue}

\usepackage{tikz}
\usetikzlibrary{arrows.meta,calc}
\usetikzlibrary{fit}
\usetikzlibrary{shapes.geometric, arrows}

\tikzstyle{io} = [trapezium, trapezium left angle=70, trapezium right angle=110, minimum width=3cm, minimum height=1cm, text centered, draw=black, fill=gray, fill opacity = 0.25,text opacity=1]
\tikzstyle{startstop} = [rectangle, rounded corners, minimum width=3cm, minimum height=1cm,text centered, draw=black]
\tikzstyle{process} = [rectangle, aspect=1,minimum width=1cm, minimum height=1cm, text centered, text width=3cm, draw=black, text opacity=1]
\tikzstyle{decision} = [diamond, aspect=2,minimum width=1cm, minimum height=1cm, text centered, draw=black, text opacity=1]
\tikzstyle{box} = [draw,blue,inner sep=20pt]
\tikzstyle{arrow} = [thick,->,>=stealth]
\tikzset{font={\fontsize{8pt}{12}\selectfont}}

\makeatletter
\def\@email#1#2{%
 \endgroup
 \patchcmd{\titleblock@produce}
 {\frontmatter@RRAPformat}
 {\frontmatter@RRAPformat{\produce@RRAP{*#1\href{mailto:#2}{#2}}}\frontmatter@RRAPformat}
 {}{}
}%
\makeatother

\begin{document}

\title{Rovibrational state-specific quasi-classical trajectory calculation and master equation analysis of the CNO system in high-energy collisions}

\author{C. H. B. Civrais}
\email{civraisc@mst.edu (Corresponding Author)}
\affiliation{Department of Mechanical and Aerospace Engineering, University of California, Irvine, CA 92697, USA}
\affiliation{Present institution: Mechanical and Aerospace Engineering Department, Missouri University of Science and Technology, Rolla, MO, 65401}

\author{Breno R. L. Galvão}
\affiliation{Centro Federal de Educação Tecnológica de Minas Gerais, CEFET-MG, Av. Amazonas 5253, (30421-169), Belo Horizonte, Minas Gerais, Brazil}

\author{Sung Min Jo}
\affiliation{UCF Center of Excellence in Hypersonics and Space Propulsion, University of Central Florida, Orlando, FL 32816, USA}

\author{Marco Panesi}
\email{mpanesi@uci.edu (Corresponding Author)}
\affiliation{Department of Mechanical and Aerospace Engineering, University of California, Irvine, CA 92697, USA}

\date{\today}

\begin{abstract}
The \ce{CNO} system governs the formation and depletion of the \ce{CN}, \ce{CO}, and \ce{NO} radiative species relevant to combustion, astrophysical, and hypersonic applications. The kinetic processes of the \ce{NO(X^2\Pi) + C(^3P)}, \ce{CN(X^2\Sigma^{+}) + O(^3P)} and \ce{CO(X^1\Sigma^+) + N(^4S,^2D)} collisional systems are herein investigated using rovibrational state-specific quasi-classical trajectory calculations and master equation analysis for thermal conditions relevant to hypersonic applications. The inelastic processes are found to proceed primarily through rotational excitation of the low-lying states and through rovibrational excitation of the high-lying states. For most collisional systems, the rotational and vibrational relaxation times are shown to exhibit comparable scales and unconventional temperature dependences, and the assumption of decoupled energy modes is found to be invalid. The dissociation processes are found to be governed by the high-lying rovibrational states and are strongly correlated to the state-specific energy deficit rather than to vibrational-specific dynamics.
At $T$ = 10,000~K, between 72 \% and 93 \% of the depletion of the diatomic species occurs within a quasi-steady-state (QSS) regime for all collisional systems. However, for decreasing temperatures, it is observed that a large portion of the depletion occurs outside the QSS regime, which indicates that the QSS assumption holds for moderate to high thermal conditions for the CNO system. In contrast, it was found that no such regime is established for exchange reactions, which occur outside quasi-steady-state conditions, demonstrating that the QSS assumption is invalid for modelling exchange reactions. The \ce{NO(X^2\Pi) + C(^3P)} system is found to proceed predominantly towards \ce{CN(X^2\Sigma^{+}) + O(^3P)} at low temperatures and towards the \ce{CO(X^1\Sigma^+) + N(^4S,^2D)} channels at high temperatures with a preference for the formation of ground-state atomic nitrogen ( \ce{N(^4S)}  ). Additionally, it is shown that the fraction of the reaction proceeding towards the \ce{CO(X^1\Sigma^+) + N(^4S,^2D)} channels remains lower than in the available shock-tube measurements. The thermal exchange rate coefficients are found to  reproduce the  available shock-tube measurements and high-fidelity calculations. Finally, consistent similarities are identified across the inelastic, exchange, and dissociation processes in collisional systems, offering promising physical foundations for reduced-order model development.
\end{abstract}

\maketitle

\section{Introduction}

The \ce{CNO} system exhibits a rich chemistry that governs the formation and depletion of three major radiative species, i.e.~ \ce{CN}, \ce{CO}, and \ce{NO}. The \ce{CNO} system incorporates the kinetic processes of three collisional systems, i.e.~ \ce{NO(X^2\Pi) + C(^3P)}, \ce{CN(X^2\Sigma^{+}) + O(^3P)} and \ce{CO(X^1\Sigma^+) + N(^4S,^2D)}, that has attracted significant interest across diverse scientific communities, ranging from combustion~\cite{Smoot1998Reburning,Lamoureux2016Modeling}, astrophysics~\cite{Herbst1973formation,Boger2005dense}, and hypersonic applications~\cite{park1989nonequilibrium,park1994review,Park1999Stagnation,Park2001Chemical}. In combustion applications, the first two collisional systems have received considerable attention as they provide pathways for the \ce{NO$_x$} reburning mechanism, allowing for pollutant reduction~\cite{Smoot1998Reburning,Lamoureux2016Modeling}. In the context of interstellar applications, these collisional systems also received an important investigation due to \ce{CN} radicals acting as the main precursors to complex molecules forming, required to understand the chemical evolution of interstellar clouds~\cite{Boger2005dense}. Finally, in the context of high-enthalpy and high-velocity applications, the \ce{CNO} system presents a great interest to allow for mission planning and vehicle design for the next generation of NASA missions towards Mars~\cite{TibereInglesse2024Reconstruction,Fawley2025Magnetohydrodynamic,Johnston2014Modeling,Jaffe2017Mars} and Venus~\cite{Fawley2025Magnetohydrodynamic,Borner2026Noble,Cabrera2026Pioneer,meurisse2026venusaerosol}, and gain physical insight in the formation of radiative species from the ablation by-products in the shock-layer for Earth's reentry~\cite{martin2015kinetic,prata2022aircarbon,Rao2024OSPREE,Rao2026OSPREE}. Additionally, beyond planetary entry applications, the accurate modelling of this system is of significant importance for the interpretation of ground-test flow diagnostics, e.g.~inductively-coupled plasma (ICP) torches, within which ablation processes are investigated for flow conditions representative of hypersonic flights~\cite{Meyers2026PlasmatronX,Kumar2027Coils}.

Among the possible atom-diatom combinations, the \ce{NO(X^2\Pi) + C(^3P)} and \ce{CN(X^2\Sigma^{+}) + O(^3P)} collisional systems received most attention from both experimental and theoretical investigations. Indeed, most of the past studies have focused on these collisional systems due their adiabatic, barrier-less, and exothermic pathways leading to the formation of both ground \ce{N(^4S)} and electronic excited \ce{N(^2D)} atomic nitrogen, allowing for the reactions to be probed under low thermal conditions. Specifically, the \ce{NO(X^2\Pi) + C(^3P)} collisional system was first investigated by Dubrin~\textit{et al.}~\cite{Dubrin1964Reactions} to investigate the branching ratio between the two reactive asymptote, and concluded that the reaction proceeds predominantly through the formation of \ce{CN(X^2\Sigma^{+}) + O(^3P)}. This conclusion was subsequently consolidated by a series of experimental work spanning the late 1960s to the early 1990s, combining flash-photolysis with atomic resonance and ultraviolet absorption~\cite{Braun1969Flash,donovan1970recent,Husain1971Study,Husain1971Kinetic,Husain1975Kinetic,Husain1982Collisional}, velocity-selected molecular beams~\cite{Krause1981Velocity}, and crossed pulsed supersonic beam apparatuses coupled with laser-induced fluorescence detection of the \ce{CN(X^2\Sigma^{+})} product~\cite{Dorthe1985Reactive,Sekiya1986Vibrational,Naulin1988Kinematic,Costes1989pulsed,Costes1990Reactive,Costes1990dissociation,Dorthe1991Fast,Naulin1991ReactionDynamics,Costes1993Translational}. The overall rate coefficient for the removal of \ce{C(^3P)} by \ce{NO} has also been measured from flame~\cite{Becker1988Spectroscopic} to fast-flow~\cite{Bergeat1999FastFlow} conditions, and for temperatures as low as 15~K by means of the CRESU technique~\cite{Chastaing2000Direct}, confirming the barrier-less character of the reaction. However, the predominance of the \ce{CN(X^2\Sigma^{+}) + O(^3P)} channel was questioned by Dean~\textit{et al.}~\cite{Dean1991ShockTube}, who used shock-tube measurements of the ground-state \ce{N(^4S)} concentration between 2430 and 4040~K and demonstrated that the exchange reaction proceeds predominantly through the formation of \ce{CO(X^1\Sigma^{+}) + N} at approximately 60 \%, independently of the temperature. These observations were later supported by theoretical calculations including nonadiabatic transitions~\cite{koner2018cno}. The \ce{CN(X^2\Sigma^{+}) + O(^3P)} collisional system has also been characterised over a broader temperature range by means of flash-photolysis, laser-induced fluorescence, and shock-tube experiments~\cite{Boden1968Kinetics,Schmatjko1978Reaktionen,Louge1984Cyanogen,Davidson1991ShockCN,Titarchuk1995Kinetics}, complemented by high-temperature shock-tube measurements of related \ce{CN} reactions~\cite{Lindackers1990High}. These studies demonstrated that the exchange rate coefficients can be considered as quasi-constant, to a good level of approximation, over the temperature range probed experimentally. In contrast to its counterparts, the \ce{CO(X^1\Sigma^+) + N(^4S,^2D)} collisional systems have been studied exclusively as the product channel of the other two, with no direct experimental measurements provided for these reactive mechanisms.

\begin{figure*}
 \centering
 \resizebox{\textwidth}{!}{\begin{tikzpicture}[
 level/.style={thick}, 
 term/.style={font=\small}, 
 connect/.style={-,thick,gray}
 ]


 \node[inner sep=0pt] (img2A1ONC) at (3, -13) {\includegraphics[width=6.5cm]{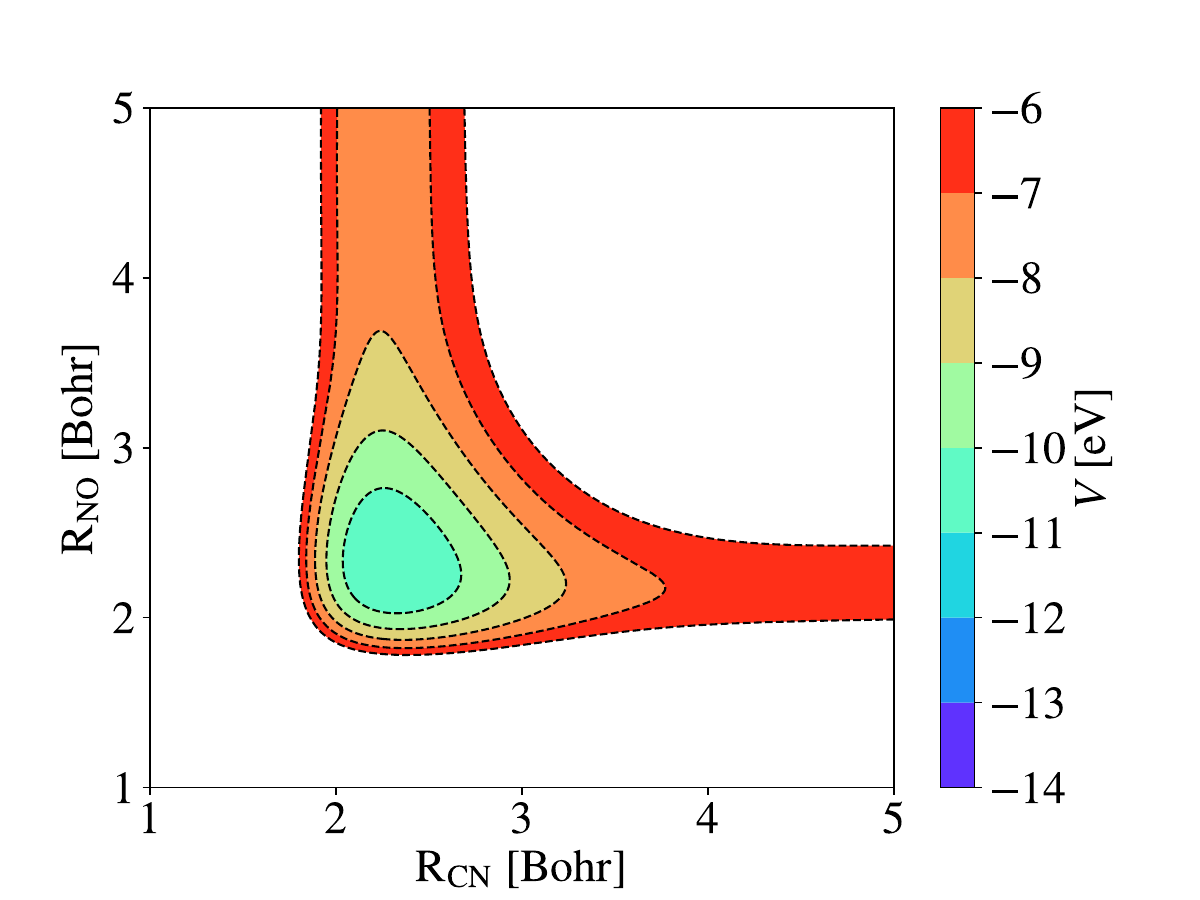}};
 \node[anchor=south west, font=\small] at ([xshift=1cm, yshift=0.6cm]img2A1ONC.south west) {\textbf{\ce{CNO($^2A^{\prime},^2A^{\prime\prime}$)}}};
 \node[inner sep=0pt] (img2A1NOC) at (22, -4) {\includegraphics[width=6.5cm]{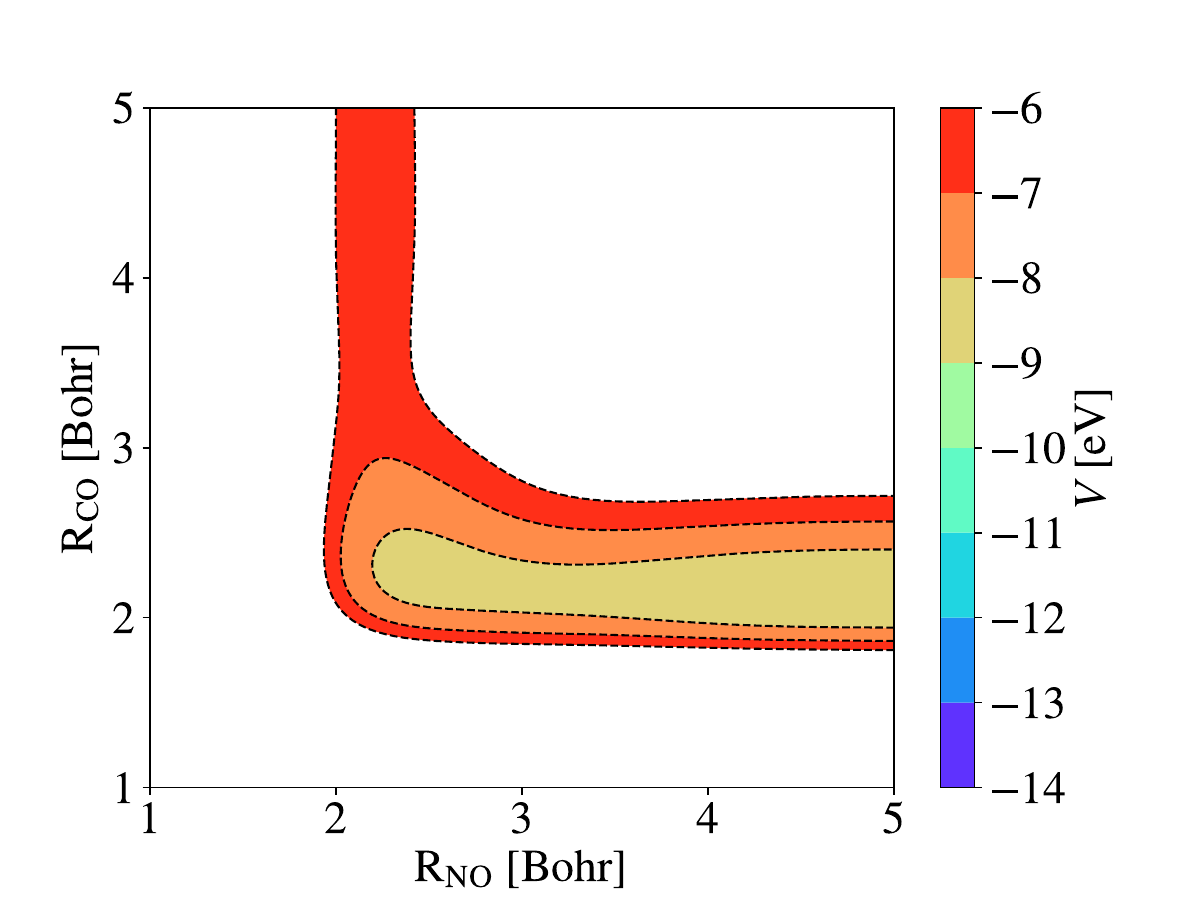}};
 \node[anchor=south west, font=\small] at ([xshift=1cm, yshift=0.6cm]img2A1NOC.south west) {\textbf{\ce{CON($^2A^{\prime},^2A^{\prime\prime}$)}}};
 \node[inner sep=0pt] (img2A1OCN) at (22, -15) {\includegraphics[width=6.5cm]{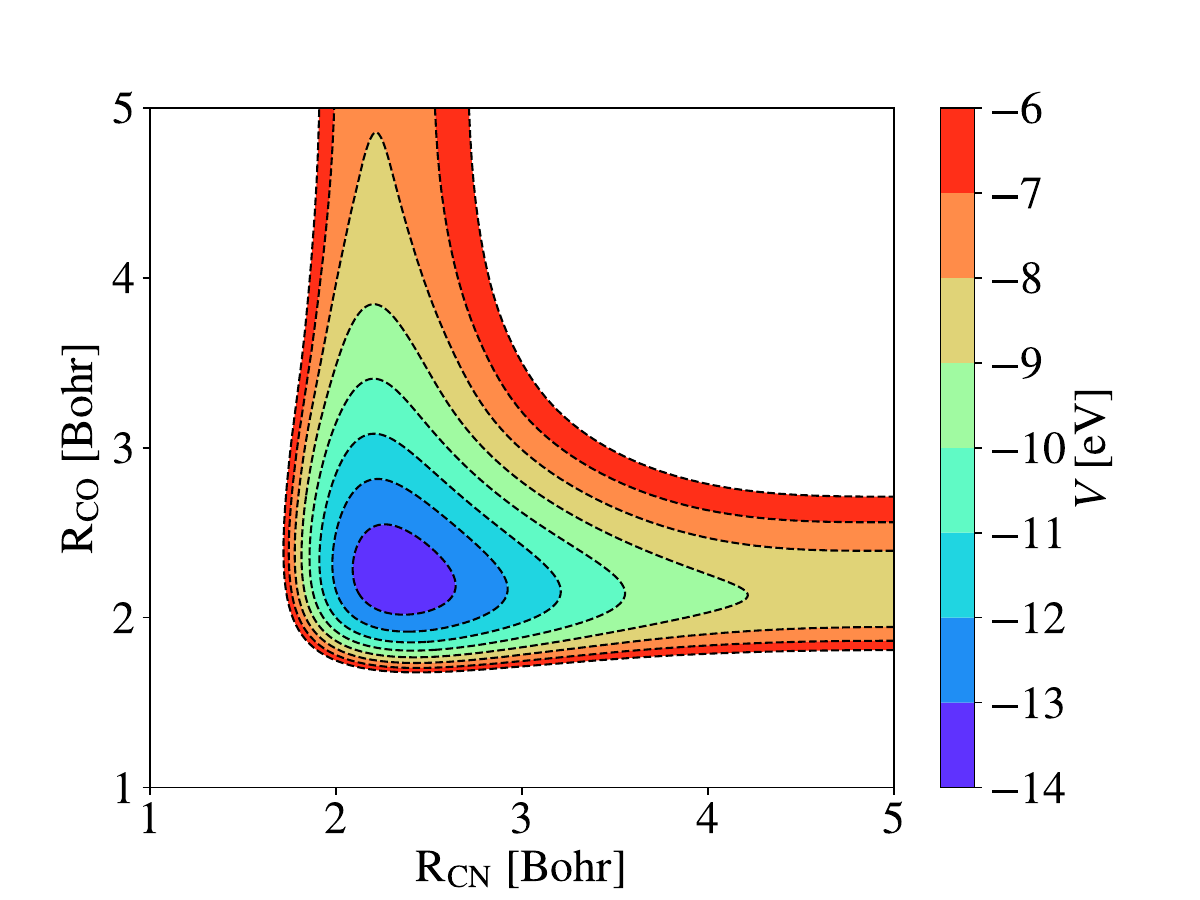}};
 \node[anchor=south west, font=\small] at ([xshift=1cm, yshift=0.6cm]img2A1OCN.south west) {\textbf{\ce{NCO($^2A^{\prime},^2A^{\prime\prime}$)}}};
 \node[inner sep=0pt] (img4A2OCN) at (9.5, -16) {\includegraphics[width=6.5cm]{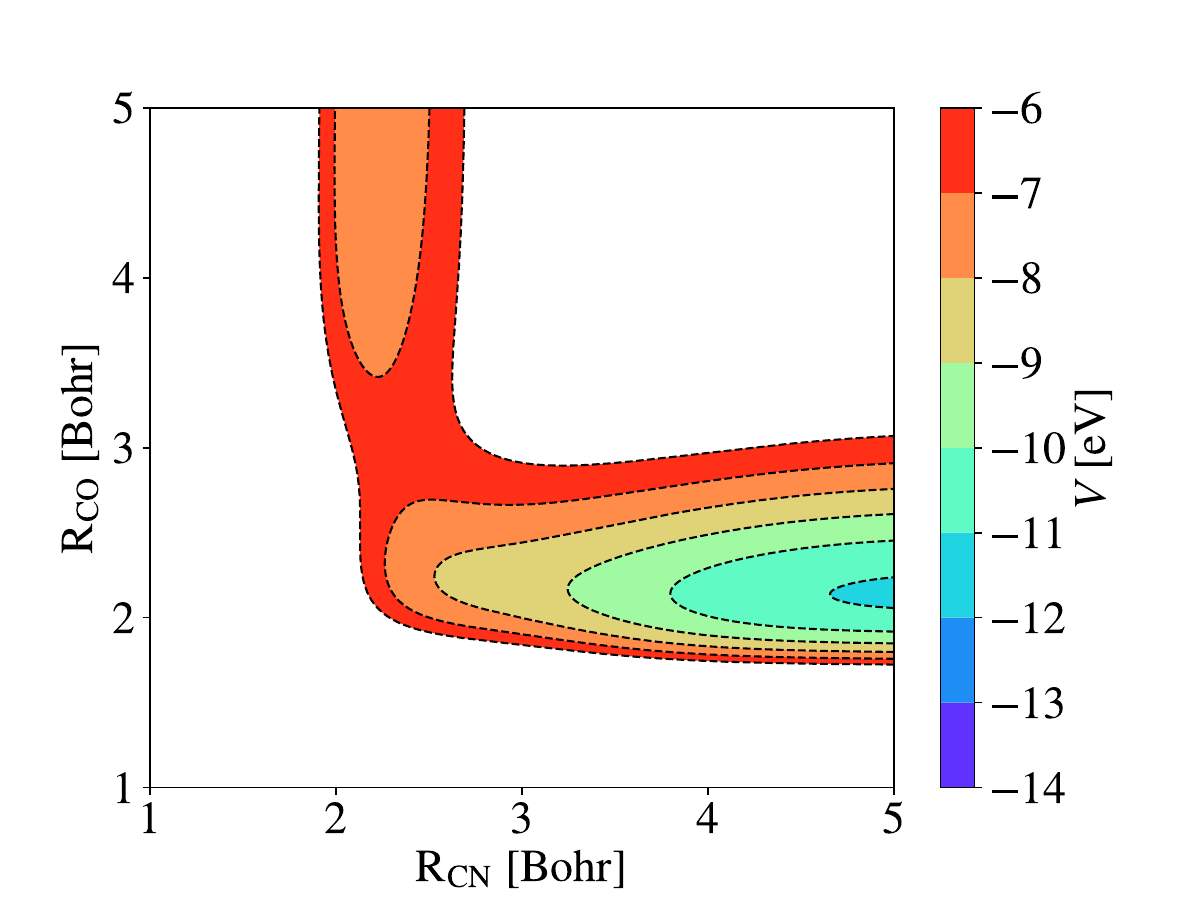}};
 \node[anchor=south west, font=\small] at ([xshift=1cm, yshift=0.6cm]img4A2OCN.south west) {\textbf{\ce{NCO($^4A^{\prime\prime}$)}}};
 \node[inner sep=0pt] (img4A2ONC) at (3, -2) {\includegraphics[width=6.5cm]{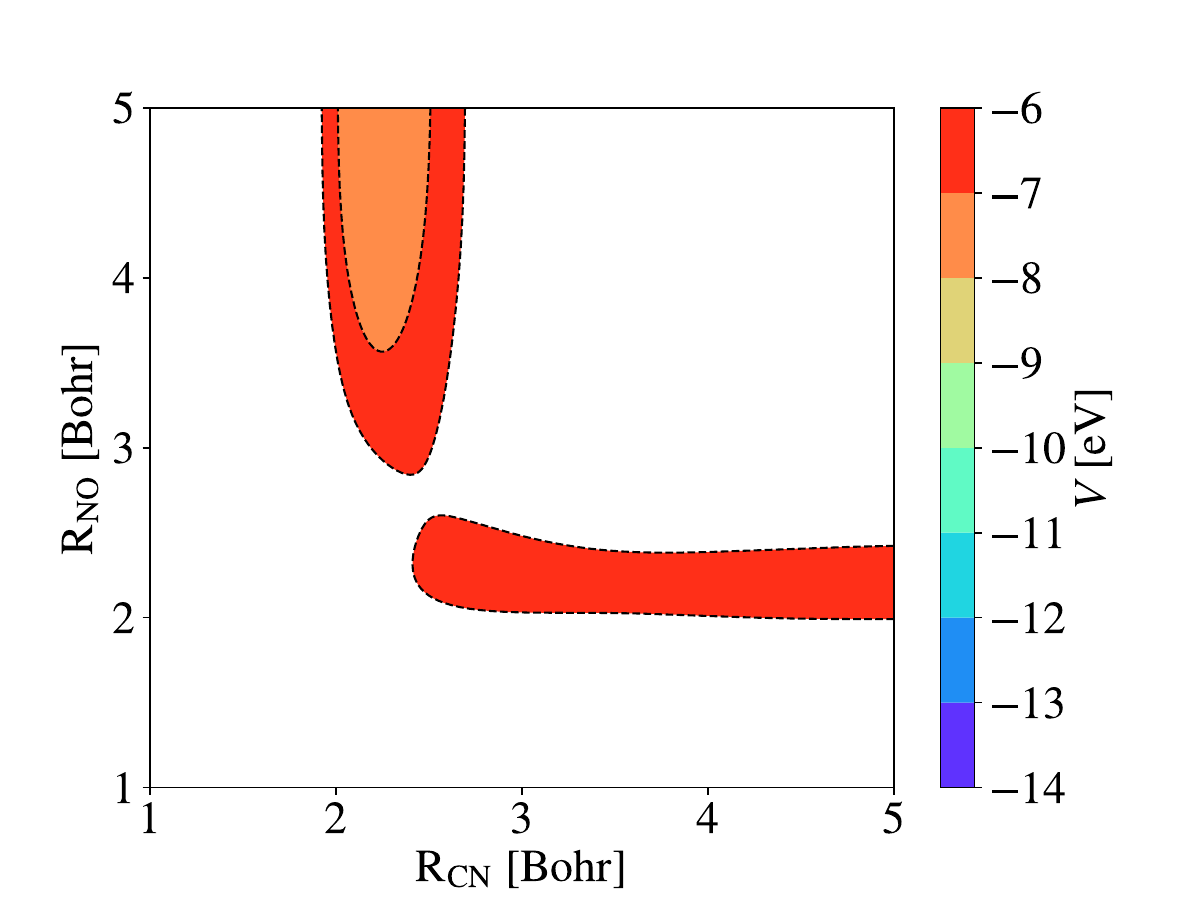}};
 \node[anchor=south west, font=\small] at ([xshift=1cm, yshift=0.6cm]img4A2ONC.south west) {\textbf{\ce{CNO($^4A^{\prime\prime}$)}}};
 
 \draw[->] (-1.5,-14.5) -- (-1.5,0.5) node[midway,left,rotate=90,yshift = +1cm]{Energy (eV)};
 \foreach \y in {-14,-12,...,0} {
 \draw (-1.6,\y) -- (-1.4,\y) node[left=5pt]{\y};
 }
 
 
 \draw[level] (9.5,-6.546) -- (12.5,-6.546) node[midway,above, xshift = 0.cm]{\textbf{\ce{NO(X^2\Pi) + C(^3P)}}};
 
 \draw[level] (-1,-7.817) -- (2,-7.817) node[midway,above, xshift = 0cm]{\textbf{\ce{CN(X^2\Sigma^{+}) + O(^3P)}}};
 
 \draw[level] (20,-11.155) -- (23,-11.155) node[midway,above, xshift = 0cm]{\textbf{\ce{CO(X^1\Sigma^+) + N(^4S)}}};
 \draw[level] (20,-8.776) -- (23,-8.776) node[midway,above, xshift = 0cm]{\textbf{\ce{CO(X^1\Sigma^+) + N(^2D)}}};
 
 \draw[level] (14,0) -- (17,0) node[midway,above]{\textbf{\ce{C(^3P)} + \ce{N(^4S)} + \ce{O(^3P)}}};
 
 \draw[level] (4,-10.939) -- (7,-10.939) node[midway,above]{\ce{CNO($^2A^{\prime},^2A^{\prime\prime}$)}};
 \draw[level] (15.5,-13.643) -- (18.5,-13.643) node[midway,above]{\ce{NCO($^2A^{\prime},^2A^{\prime\prime}$)}};
 \draw[level] (12,-8.417) -- (15,-8.417) node[midway,above]{\ce{CON($^2A^{\prime},^2A^{\prime\prime}$)}};
 
 \draw[level] (11.5,-10.295) -- (14.5,-10.295) node[midway,above]{\ce{cyc($^2A^{\prime\prime}$)}};
 
 \draw[level] (8,-10.552) -- (11,-10.552) node[midway,above]{\ce{NCO($^4A^{\prime\prime}$)}};
 \draw[level] (5,-8.2852) -- (8,-8.2852) node[midway,above]{\ce{CNO($^4A^{\prime\prime}$)}};
 
 \draw[connect, line width=0.9pt, dashed] (15.5,-13.643) -- (7,-10.939) node[midway, below, yshift = -0.25cm] {\Large $\bf ^2A^{\prime},^2A^{\prime\prime}$}; 
 \draw[connect, line width=0.9pt, dashed] (20,-8.776) -- (15,-8.417); 
 \draw[connect, line width=0.9pt, dashed] (18.5,-13.643) -- (20,-8.776); 
 \draw[connect, line width=0.9pt, dashed] (7,-10.939) -- (9.5,-6.546); 
 \draw[connect, line width=0.9pt, dashed] (4,-10.939) -- (2,-7.817); 
 \draw[connect, line width=0.9pt, dashed] (12,-8.417) -- (9.5,-6.546); 
 
 \draw[connect, line width=0.9pt, dash pattern=on 8pt off 3pt,gray] (15.5,-13.643) -- (14.5,-10.295); 
 \draw[connect, line width=0.9pt, dash pattern=on 8pt off 3pt,gray] (11.5,-10.295) -- (9.5,-6.546) node[midway, left] {\Large $\bf ^2A^{\prime\prime}$}; 
 
 \draw[connect, line width=0.9pt, dash pattern=on 5pt off 2pt on 1pt off 2pt on 1pt off 2pt] (20,-11.155) -- (12.5,-6.546); 
 \draw[connect, line width=0.9pt, dash pattern=on 5pt off 2pt on 1pt off 2pt on 1pt off 2pt] (20,-11.155) -- (11,-10.552); 
 \draw[connect, line width=0.9pt, dash pattern=on 5pt off 2pt on 1pt off 2pt on 1pt off 2pt] (8,-10.552) -- (2,-7.817); 
 \draw[connect, line width=0.9pt, dash pattern=on 5pt off 2pt on 1pt off 2pt on 1pt off 2pt] (5,-8.2852) -- (2,-7.817); 
 \draw[connect, line width=0.9pt, dash pattern=on 5pt off 2pt on 1pt off 2pt on 1pt off 2pt] (8,-8.2852) -- (9.5,-6.546) node[midway, above, xshift = -0.15cm] {\Large $\bf ^4A^{\prime\prime}$}; 
 
 \draw[connect, line width=0.9pt](12.5,-6.546) -- (14,0); 
 \draw[connect, line width=0.9pt](2,-7.817) -- (14,0); 
 \draw[connect, line width=0.9pt](20,-11.155) -- (17,0); 
 \draw[connect, line width=0.9pt](20,-8.776) -- (17,0); 

 \node[anchor=north east] at (current bounding box.north east) {
 {\Large
 \begin{tabular}{ll}
 \tikz\draw[level,dashed,gray] (-0.5,0) -- (0.5,0); & $^2A^{\prime}$ \\
 \tikz\draw[level,dash pattern=on 8pt off 3pt,gray] (-0.5,0) -- (0.5,0); & $^2A^{\prime\prime}$ \\
 \tikz\draw[level,dash pattern=on 5pt off 2pt on 1pt off 2pt on 1pt off 2pt,gray] (-0.5,0) -- (0.5,0); & $^4A^{\prime\prime}$ \\
 \tikz\draw[level,solid,gray] (-0.5,0) -- (0.5,0); & Dissociation \\
 \end{tabular}
 }
 };

 \end{tikzpicture}}
 \caption{Correlation diagram for the \ce{CNO} system on the $^2A^{\prime}$, $^2A^{\prime\prime}$ and $^4A^{\prime\prime}$ PESs developed by Galvão and co-workers.~\cite{goncalves2018cno,alves2021dmbe,alves2023high,Alves2019Quasiclassical,Lu2024Theoretical,Galvao2025Electronic} The zero of energy is set to the atomisation energy of the \ce{CNO} system, i.e.~ \ce{C(^3P) + N(^4S) + O(^3P)}. The minima and asymptotes are represented as horizontal levels. Their correlations are indicated by connecting lines corresponding to a given symmetry state. The atomisation limit and the excited-atom asymptotes \ce{NO(X^2\Pi) + C(^1D)} and \ce{CN(X^2\Sigma^{+}) + O(^1D)} are also shown, whereas the transition states are not represented. The inset figures indicate the topology of the $^2A^{\prime},^2A^{\prime\prime}$ and $^4A^{\prime\prime}$ PESs for the \ce{CNO} isomers at an inter-bond angle fixed at $180^{\circ}$.}
 \label{fig:CorrelationDiagram}
\end{figure*}

To allow for an accurate interpretation of the experimental measurements, theoretical studies were also conducted to identify the electronic states and reactive pathways of the \ce{CNO} system. Specifically, until the late 2000s, the \ce{NO(X^2\Pi_r) + C(^3P)} collisional system was assumed to proceed only towards the \ce{CN(X^2\Sigma^+) + O(^3P)} or \ce{CO(X^1\Sigma^+) + N(^2D)} asymptotes. This assumption was based on the early study of Halvick~\textit{et al.}~\cite{Halvick1984Theoretical}, in which only the doublet surfaces were considered directly accessible, and the quartet surfaces were assumed to be reached through spin-orbit coupling. However, this assumption was not supported by the experimental measurements. Schmatjko and Wolfrum~\cite{Schmatjko1978Reaktionen} detected about 20 \% of \ce{N(^4S)} in the related \ce{CN(X^2\Sigma^{+}) + O(^3P)} reaction, and Dean~\textit{et al.}~\cite{Dean1991ShockTube} and Bergeat~\textit{et al.}~\cite{Bergeat1999FastFlow} subsequently reported a substantial formation of \ce{N(^4S)} in the \ce{NO(X^2\Pi) + C(^3P)} reaction. The formation of the ground-state nitrogen atom was subsequently addressed by Abrahamsson et al.~\cite{Abrahamsson2008New}, who demonstrated that the quartet surface provides a barrierless, adiabatic pathway to the \ce{CO(X^1\Sigma^+) + N(^4S)} asymptote, and closely reproduces the crossed-beam branching ratio reported by Naulin et al.~\cite{Naulin1991ReactionDynamics}, which was incorrectly reproduced when only the doublet surfaces were considered. Following the early analytical surfaces of Rayez, Halvick and co-workers~\cite{Rayez1986Three,Halvick1987Three} and these observations, three sets of \textit{ab initio} potential energy surfaces (PESs) were developed by different groups~\cite{Persson1995theoretical,Simonson1995Quasiclassical,Andersson2000QCT,Andersson2000Improved,Andersson2003Computational,Abrahamsson2008New,Abrahamsson2009Dynamics,Abrahamsson2006Classical,Frankcombe2012Adiabatic,koner2018cno,goncalves2018cno,alves2021dmbe,alves2023high,Alves2019Quasiclassical,Lu2024Theoretical,Galvao2025Electronic}. The three reactive channels of the \ce{CNO} system are now recognised to proceed through a shared manifold of three low-lying electronic states. The correlation diagram of the three attractive potential energy surfaces developed by Galvão and coworkers~\cite{goncalves2018cno,alves2021dmbe,alves2023high,Alves2019Quasiclassical,Lu2024Theoretical,Galvao2025Electronic} is depicted in Fig.~\ref{fig:CorrelationDiagram}. The three sets of surfaces served as the basis for quasi-classical trajectory (QCT)~\cite{Halvick1987Three,Rayez1986Three,Persson1995theoretical,Andersson2000QCT,Andersson2000Improved,Abrahamsson2008New,Andersson2003Computational,Abrahamsson2009Dynamics,Abrahamsson2006Classical,Simonson1995Quasiclassical,Frankcombe2012Adiabatic,koner2018cno,goncalves2018cno,alves2021dmbe,alves2023high,Alves2019Quasiclassical,Lu2024Theoretical,Galvao2025Electronic} and quantum wave-packet~\cite{Abrahamsson2006Classical,Lu2024Theoretical} calculations. These calculations provide the thermal rate coefficients, the product-state distributions of the three reactive channels, and the temperature-dependent branching ratio~\cite{koner2018cno,Lu2024Theoretical}, for thermal conditions up to 20,000~K for \ce{NO(X^2\Pi) + C(^3P)} ~\cite{koner2018cno}.

While these experimental measurements and dynamics calculations provide physical insights into the macroscopic properties of these collisional systems, in the context of high-enthalpy and high-velocity applications where insights on the microscopic scales are required to allow for model development, these studies provide little information. One approach to gaining an understanding of the kinetics of chemistry systems at the microscopic scale is to construct a state-specific kinetics database using QCT calculations on the aforementioned \textit{ab initio} PES and perform master equation analysis. This approach informs on the rate coefficients of each process occurring on the collisional system allowing for an accurate development of chemistry models, e.g. reduced order models~\cite{Venturi2020Data,Zanardi2023Adaptive,Jacobsen2024Information,Zanardi2025PetrovGalerkin} or QCT-informed chemistry models~\cite{Kim2021Modification,Torres2024High,gimelshein2019nonequilibrium,civrais2024qkpartI,civrais2024qkpartII,civrais2025VibronicQK}.

To this end, the present article aims to investigate the kinetic processes of the \ce{NO(X^2\Pi) + C(^3P)}, \ce{CN(X^2\Sigma^{+}) + O(^3P)}, and \ce{CO(X^1\Sigma^+) + N(^4S,^2D)} collisional systems for conditions representative of hypersonic applications. It consists of constructing a rovibrational state-to-state kinetics database using QCT calculation on the state-of-the-art \textit{ab initio} PESs. Subsequently, it involves the extraction of the moments of the distribution functions to obtain quantities of interests and propose a reduced order representation of the kinetic database of the collisional systems.

\section{Physical Modelling}
\subsection{Potential Energy Surfaces}

The \ce{CNO} system involves three reactive entrance channels that evolve adiabatically on a common manifold of two doublet, i.e.~ $^2A^{\prime}$, $^2A^{\prime\prime}$, and one quartet, i.e.~ $^4A^{\prime\prime}$, potential energy surfaces. The doublet surfaces are adiabatically accessible to the \ce{NO(X^2\Pi) + C(^3P)}, \ce{CN(X^2\Sigma^+) +O(^3P)} and \ce{CO(X^1\Sigma^{+}) + N(^2D)} entrance channels, whereas the quartet surface is adiabatically accessible to the \ce{NO(X^2\Pi) + C(^3P)}, \ce{CN(X^2\Sigma^+) + O(^3P)} and \ce{CO(X^1\Sigma^{+}) + N(^4S)} entrance channels. In the literature, three sets of \textit{ab initio} potential energy surfaces have been developed to support the experimental measurements conducted on the \ce{CNO} system.

The first set of potential energy surfaces was developed over a decade beginning in the mid-1990s by Andersson and co-workers.~\cite{Persson1995theoretical,Andersson2000QCT,Andersson2000Improved,Abrahamsson2008New,Andersson2003Computational,Abrahamsson2009Dynamics,Abrahamsson2006Classical,Simonson1995Quasiclassical,Frankcombe2012Adiabatic} This set of PESs originates from the work of Persson~\textit{et al.}~\cite{Persson1995theoretical} which was the first attempt to construct a global \textit{ab initio} PES for the $^2A^{\prime\prime}$ surface with the inclusion of intermediates. The \textit{ab initio} energies were calculated using single-state complete active space second-order perturbation theory (CASPT2) with the standard Fock matrix, and fitted onto a compact support functional form, i.e.~many-body expansion (MBE)~\cite{Simonson1995Quasiclassical}. However, the $^2A^{\prime\prime}$ surface of Persson~\textit{et al.}~\cite{Persson1995theoretical} presented several shortcomings. The long-range interactions were disregarded, and the surface was excessively attractive for non-linear approaches of \ce{C(^3P)} to \ce{NO(X^2\Pi)} due to a skewed fit towards the linear configuration, caused by 80 \% of the \textit{ab initio} energy points representing the linear configurations, under-resolving the rest of the surface. Therefore, Andersson~\textit{et al.}~\cite{Andersson2000Improved} further improved the $^2A^{\prime\prime}$ surface by adopting the many-body expansion approach but including long-range interactions, i.e.~electrostatic, dispersion forces, and induction, and calculating additional \textit{ab initio} energies for non-linear configurations. Simultaneously, Andersson~\textit{et al.}~\cite{Andersson2000QCT} developed an analytical fit of the $^2A^{\prime}$ surface following the same approach as for the $^2A^{\prime\prime}$. Following the same approach used for the doublet surfaces, Abrahamsson~\textit{et al.}~\cite{Abrahamsson2008New} developed the quartet surface with a total of 4947 \textit{ab initio} energy points, which was shown to improve significantly the reproduction of the experimentally measured branching ratio of the \ce{NO(X^2\Pi) + C(^3P)} system reported by Naulin~\textit{et al.} \cite{Naulin1991ReactionDynamics} 

 Subsequently, Koner~\textit{et al.} \cite{koner2018cno} constructed a global PES for the \ce{CNO} system following the reproducing kernel Hilbert space (RKHS) approach using over 15,000 \textit{ab initio} energy points per surface. These points were calculated with the multi-reference configuration interaction level of theory including the Davidson correction (MRCI+Q) and with the augmented Dunning-type correlation-consistent polarised valence triple-zeta (aug-cc-pVTZ) basis set. The reference wavefunction was computed using the complete active space self-consistent field (CASSCF) method.

Finally, the third set of PESs was recently developed by Galvão and co-workers.~\cite{goncalves2018cno,alves2021dmbe,alves2023high,Alves2019Quasiclassical,Lu2024Theoretical,Galvao2025Electronic} The construction of these PESs has been primarily motivated by the necessity to improve the long-range interactions to accurately describe the reactivity of barrierless reactions, especially at low temperatures. These PESs have been constructed by fitting a double many-body expansion (DMBE) onto a set of over 2000 \textit{ab initio} energy points per state. The DMBE allows a realistic description of the long-range multipolar interaction terms, i.e.~electrostatic and dispersion, which are known to contribute greatly to the rate coefficients of the \ce{CNO} system.~\cite{Alves2019Quasiclassical} The \textit{ab initio} energy points were calculated using explicitly correlated multi-reference configuration interaction with a quadruple-zeta basis set (MRCI-F12/cc-pVQZ-F12) and the Davidson (+Q) correction.

Considering that the latest set of PESs has been constructed using the state-of-the-art electron correlation method, this set is adopted in the present article. Considering the deep potential wells of the three diatomic molecules constituting the \ce{CNO} system, which imply a large number of rovibrational states, the computational cost of the database is considerable. The two doublet states are therefore assumed to exhibit similar dynamics, which allows the calculations to be restricted to the $^2A^{\prime}$ and $^4A^{\prime\prime}$ PESs. The validity of this assumption is examined in subsequent sections.

\subsection{Nonadiabatic Coupling} \label{Sec:Nonadiabatic}

The \ce{CNO} system exhibits true Renner--Teller effects. Specifically, the linear intermediates of the \ce{CNO} system, i.e.~ \ce{CNO}, \ce{CON} and \ce{NCO}, are degenerate and correlate to a $^2\Pi$ symmetry, imposing that the electronic and vibrational wavefunctions cannot be separated.~\cite{Persson1995theoretical} Halvick~\textit{et al.} \cite{Halvick1984Theoretical} have shown that the \ce{CNO} system presents several crossings between surfaces, of which the \ce{CNO(^4\Pi)} -- \ce{CNO(^2\Sigma^{+})} crossing is of greatest interest, allowing for nonadiabatic transitions between the doublet and quartet states, which are coupled by spin-orbit interactions.~\cite{Minaev1989MINDO} However, Halvick~\textit{et al.}~\cite{Halvick1984Theoretical} have also demonstrated that if nonadiabatic transitions were to occur between the $^4\Pi$ and $^2\Sigma^{+}$ surfaces, it would favour the formation of \ce{CNO($^2\Sigma^{+}$)} resulting in a preferred formation of \ce{CN(X^2\Sigma^{+}) + O(^3P)} over \ce{CN(A^2\Pi) + O(^3P)}. Past studies have included nonadiabatic transitions between the doublet and quartet states in the \ce{CNO} system to assess their importance for the \ce{CN(X^2\Sigma^{+}) + O(^3P) -> CO(X^1\Sigma^{+}) + N(^4S,^2D)},~\cite{Abrahamsson2009Dynamics} \ce{NO(X^2\Pi) + C(^3P) -> CO(X^1\Sigma^+) + N(^4S,^2D)} \cite{koner2018cno} and \ce{CO(X^1\Sigma^{+}) + N(^2D) -> CO(X^1\Sigma^{+}) + N(^4S)} \cite{Galvao2025Electronic} reactions. For these collisional systems, nonadiabatic transitions appear to have a negligible contribution to the heterogeneous exchange processes for temperatures representative of hypersonic applications. Therefore, in the present article, nonadiabatic transitions are neglected.

\subsection{State-to-state Kinetic Database}
\label{Sec:StSKineticDatabase}

The collisional systems of the \ce{CNO} system involve three types of elementary kinetic processes:
\begin{itemize}
 \item rovibrational excitation and de-excitation through inelastic reactions,
 \begin{chequation}
 \begin{align}
 \ce{NO($i$) + C <=>[k^{\text{I}, \ce{NO}}_{i\rightarrow j}][k^{\text{I}, \ce{NO}}_{j\rightarrow i}] NO($j$) + C}
 \label{NOC_Inelastic}
 \end{align}
 \end{chequation}

 \begin{chequation}
 \begin{align}
 \ce{CN($l$) + O <=>[k^{\text{I}, \ce{CN}}_{l\rightarrow m}][k^{\text{I}, \ce{CN}}_{m\rightarrow l}] CN($m$) + O}
 \label{CNO_Inelastic}
 \end{align}
 \end{chequation}

 \begin{chequation}
 \begin{align}
 \ce{CO($n$) + N($p$) <=>[k^{\text{I}, \ce{CO}}_{n,p\rightarrow o,p}][k^{\text{I}, \ce{CO}}_{o,p\rightarrow n,p}] CO($o$) + N($p$)}
 \label{CON_Inelastic}
 \end{align}
 \end{chequation}

 \item rovibrational energy transfer through the heterogeneous exchange reactions,
 \begin{chequation}
 \begin{align}
 \ce{NO($i$) + C <=>[k^{\text{E}, \ce{NO}}_{i\rightarrow l}][k^{\text{E}, \ce{CN}}_{l\rightarrow i}] CN($l$) + O}
 \label{NOC_Ex1}
 \end{align}
 \end{chequation}
 
 \begin{chequation}
 \begin{align}
 \ce{NO($i$) + C <=>[k^{\text{E}, \ce{NO}}_{i\rightarrow n,p}][k^{\text{E}, \ce{CO}}_{n,p\rightarrow i}] CO($n$) + N($p$)}
 \label{NOC_Ex2}
 \end{align}
 \end{chequation}

 \begin{chequation}
 \begin{align}
 \ce{CN($l$) + O <=>[k^{\text{E}, \ce{CN}}_{l\rightarrow n,p}][k^{\text{E}, \ce{CO}}_{n,p\rightarrow l}] CO($n$) + N($p$)}
 \label{CNO_Ex1}
 \end{align}
 \end{chequation}

 \item rovibrational dissociation and recombination reactions of bound and quasi-bound states,
 \begin{chequation}
 \begin{align}
 \ce{NO($i$) + C <=>[k^{\text{D}, \ce{NO}}_{i\rightarrow c}][k^{\text{R}, \ce{NO}}_{c \rightarrow i}] N + O + C}
 \label{NOC_Dissociation}
 \end{align}
 \end{chequation}

 \begin{chequation}
 \begin{align}
 \ce{CN($l$) + O <=>[k^{\text{D}, \ce{CN}}_{l\rightarrow c}][k^{\text{R}, \ce{CN}}_{c \rightarrow l}] C + N + O}
 \label{CNO_Dissociation}
 \end{align}
 \end{chequation}

 \begin{chequation}
 \begin{align}
 \ce{CO($n$) + N($p$) <=>[k^{\text{D}, \ce{CO}}_{n,p\rightarrow c}][k^{\text{R}, \ce{CO}}_{c \rightarrow n,p}] C + O + N}
 \label{CON_Dissociation}
 \end{align}
 \end{chequation}
 
\end{itemize}

 Note that \ref{NOC_Dissociation} -- \ref{CON_Dissociation} represent the total dissociation kinetic processes.

In these processes, $k$ is the rate coefficient and the superscripts $()^{\text{I}}$, $()^{\text{E}}$, $()^{\text{D}}$ and $()^{\text{R}}$ denote the process considered. These processes, i.e.~\ref{NOC_Inelastic} -- \ref{CON_Dissociation}, are herein presented in a compact form in which the indices $()_i$, $()_l$, and $()_n$ refer to the pre-collision and $()_j$, $()_m$, and $()_o$ to the post-collision rovibrational states of \ce{NO}, \ce{CN}, and \ce{CO}, stored in subsets $\mathcal{I_{\ce{NO}}}$, $\mathcal{I_{\ce{CN}}}$, and $\mathcal{I_{\ce{CO}}}$, respectively, by increasing energies. The index $()_p$ denotes the electronic state of atomic nitrogen, stored in $\mathcal{I_{\ce{N}}}$, with $p=0$ referring to \ce{N(^4S)} and $p=1$ to \ce{N(^2D)}; since nonadiabatic transitions are neglected, the electronic state of the nitrogen collision partner is preserved in the inelastic processes. The subscript $()_{\text{c}}$ refers to the continuum limit representing the asymptotic atomisation limit. The relation between the global indices, e.g.~ $i$, and the rovibrational states, e.g.~($v,J$), is expressed as $i = i(v,J),\ i\ \in \ \mathcal{I}$ with $v=0,\dots,v_{\text{max}}$ and $J = 0,\dots,J_{\text{max}}(v)$. The rovibrational states of each molecular species are obtained by integrating the radial Schr{ \" o}dinger equation with the Numerov scheme, with the Wentzel--Kramers--Brillouin (WKB) semi-classical approximation \cite{bernstein1979AtomMolecule} as a complementary step to compute the vibrational period and to identify the quasi-bound states. The diatomic potentials are the two-body terms of the DMBE PESs, and the procedure yields 6838, 7293 and 13,617 bound and quasi-bound levels for \ce{NO(X^2\Pi)}, \ce{CN(X^2\Sigma^{+})} and \ce{CO(X^1\Sigma^+)}, respectively.

The collisional trajectories are performed using \texttt{CoarseAIR},~\cite{Venturi2020Data,Venturi2020Bayesian} a modernised version of the original Vectorized Variable timestep Trajectory Code (\texttt{VVTC}) code developed at NASA Ames Research Center by Schwenke.~\cite{Schwenke1988Calculations} This numerical code has been extensively applied to the construction of similar rovibrational state-specific kinetic databases.~\cite{jo2023rovibrational,Macdonald2018constructionI,Macdonald2018constructionII,jo2022rovibrational,civrais2026NOC,Sharma2022HCN} 

Considering that the doublet and quartet states correlate with different reactive asymptotes, trajectories are propagated separately on each of the two surfaces, i.e.~$^2A^{\prime}$ and $^4A^{\prime\prime}$.

The electronic degeneracy is defined as the ratio between the partition function of the electronic states of the compound triatomic system spanned by the PES on which the trajectories are run, and the product of the electronic partition functions of the two reactants. Because the fine-structure levels of the compound correlate adiabatically to specific fine-structure levels of the reactants, the numerator is evaluated for the asymptote to which it correlates. For \ce{NO + C}, \ce{CN + O} and \ce{CO + N(^4S,^2D)} it reads, respectively,
\begin{equation}
 g^{\ce{NO}}_e
 = \frac{q_{\ce{CNO}}}
 {q_{\ce{C}(^3P_{0,1,2})}\, q_{\ce{NO}(X^2\Pi_{1/2,3/2})}},
 \label{ge_NO}
\end{equation}

\begin{equation}
 g^{\ce{CN}}_e
 = \frac{q_{\ce{CNO}}}
 {q_{\ce{O}(^3P_{2,1,0})}\, q_{\ce{CN}(X^2\Sigma^{+})}},
 \label{ge_CN}
\end{equation}

\begin{equation}
 g^{\ce{CO},\,^2A}_e
 = \frac{q_{\ce{CNO}(^2A)}}{q_{\ce{N}(^2D)}\, q_{\ce{CO}(X^1\Sigma^+)}},
 \label{ge_CO2A}
\end{equation}
\begin{equation}
 g^{\ce{CO},\,^4A}_e
 = \frac{q_{\ce{CNO}(^4A)}}{q_{\ce{N}(^4S)}\, q_{\ce{CO}(X^1\Sigma^+)}}.
 \label{ge_CO4A}
\end{equation}

The partition functions of the atomic species are defined as follows. Considering that atomic carbon and oxygen exhibit three fine-structure levels, their partition functions read $q_{\ce{C}(^3P_{0,1,2})} = 1 + 3 e^{-23.6/T} + 5 e^{-62.4/T}$ and $q_{\ce{O}(^3P_{2,1,0})} = 5 + 3 e^{-228 / T} + e^{-326 / T}$. Similarly, \ce{N(^2D)} exhibits spin-orbit splitting resulting in a partition function defined as $q_{\ce{N(^2D)}} = 6 + 4 e^{-12.53 / T}$. In contrast, the ground electronic state of atomic nitrogen does not present any fine structure resulting in a partition function of $q_{\ce{N(^4S)}} = 4$. 

 Among the diatomic species, only nitric oxide exhibits spin-orbit splitting, i.e.~two spin-orbit states. Their partition functions therefore read $q_{\ce{NO (X^2\Pi)}} = 2 + 2 e^{-172.4/T}$, $q_{\ce{CN(X^2\Sigma^{+})}} = 2$ and $q_{\ce{CO(X^1\Sigma^+)}} = 1$.

The partition function of the compound system refers to the electronic state of the PES considered, that is $q^{\ce{CN+O}}_{\ce{CNO}(^2A)} = q^{\ce{CO+N}}_{\ce{CNO}(^2A)} = 2$ and $q^{\ce{CN+O}}_{\ce{CNO}(^4A)} = q^{\ce{CO+N}}_{\ce{CNO}(^4A)} = 4$. For the \ce{CN(X^2\Sigma^{+}) + O(^3P)} asymptote, the eight states of the three surfaces correlate to the lowest fine-structure level, i.e.~\ce{O(^3P_2)}, and for the \ce{CO(X^1\Sigma^+) + N(^2D)} asymptote the four states of the doublet surfaces correlate to \ce{N(^2D_{5/2})}. For the \ce{NO(X^2\Pi) + C(^3P)} asymptote, the most recent theoretical studies\cite{Abrahamsson2008New,goncalves2018cno,koner2018cno,alves2021dmbe,alves2023high} demonstrate that the surfaces correlate to the lowest asymptotic fine-structure states. Following the adiabatic correlation of the fine-structure levels, the $^2A^{\prime}$ surface correlates to \ce{C(^3P_0) + NO(^2\Pi_{1/2})}, whereas the $^2A^{\prime\prime}$ and $^4A^{\prime\prime}$ surfaces correlate to \ce{C(^3P_1) + NO(^2\Pi_{1/2})}. Therefore, the numerators read
\begin{equation}
 q_{\ce{CNO}(^2A^{\prime})} = 2,
 \label{q_CNO_2A}
\end{equation}
\begin{equation}
 q_{\ce{CNO}(^2A^{\prime\prime})} = 2\,e^{-23.6/T},
 \label{q_CNO_2App}
\end{equation}
\begin{equation}
 q_{\ce{CNO}(^4A^{\prime\prime})} = 4\,e^{-23.6/T}.
 \label{q_CNO}
\end{equation}

The electronic degeneracies, i.e.~Eqs.~ \eqref{ge_NO} -- \eqref{ge_CO4A}, are applied irrespective of the type of reaction. Additionally, the doublet states, i.e.~ $^{2}A^{\prime}$ and $^{2}A^{\prime\prime}$, are assumed to exhibit similar dynamics, therefore, the contribution of the $^2A^{\prime}$ surface is herein counted twice.

All trajectory calculations presented in the present article are initialised as follows, unless mentioned otherwise. Initial conditions are sampled from a Maxwell distribution for temperatures between 2500 and 20,000~K. The particles are initialised with a centre-of-mass separation of 30 Bohr between reactants. The impact parameter is sampled using stratified sampling over 8 circular strata, with stratum boundaries ranging from 0.25~Bohr to a maximum impact parameter of 17~Bohr. A minimum of 30,000 trajectories are performed for each rovibrational state. Trajectories are integrated with an adaptive-step Bulirsch--Stoer scheme and are terminated if the departing fragments reach a separation of 40 Bohr or if the trajectory time exceeds $5\times10^5$ atomic units. 

\subsection{Master Equations}

The state-to-state kinetic database constructed in Sec.~\ref{Sec:StSKineticDatabase} has been used to conduct a rovibrational master equation analysis of the four reactive asymptotes, i.e.~ \ce{NO(X^2\Pi) + C(^3P)}, \ce{CN(X^2\Sigma^{+}) + O(^3P)}, \ce{CO(X^1\Sigma^+) + N(^4S)} and \ce{CO(X^1\Sigma^+) + N(^2D)}. For brevity, the system of equations is reported in the Supplementary Material (see Eqs.~ \eqref{S-eq:ME_O} -- \eqref{S-eq:k_glob_D_CO}). The set of master equations is numerically integrated using the \texttt{PLAsma in Thermodynamic nOn-equilibrium (PLATO)} library,~\cite{Munafo2025PLATO} developed at the University of California, Irvine. In the master equations, the exothermic rate coefficients are used as a reference, whereas the endothermic rate coefficients are reconstructed from micro-reversibility. This choice is motivated by the better statistical sampling of the exothermic processes. Similarly, the recombination rate coefficients are computed from the dissociation rate coefficients using micro-reversibility.

 \section{Results and Discussion}

\subsection{Topology of the Potential Energy Surfaces} \label{Sec:Topology}

The theoretical studies have demonstrated that the three heterogeneous exchange reactions of the \ce{CNO} system, i.e.~\ref{NOC_Ex1}--\ref{CNO_Ex1}, evolve on a shared manifold of doublet and quartet electronic states. To provide deeper insight into the dynamics of these reactive paths, the present section assesses the topology of the three PESs and relates the energetical structure of each isomer that governs the exchange reaction.

Figure~\ref{fig:CorrelationDiagram} presents the correlation diagram of the \ce{CNO} system constructed for the electronic states developed by Galvão and coworkers~\cite{goncalves2018cno,alves2021dmbe,alves2023high,Alves2019Quasiclassical,Lu2024Theoretical,Galvao2025Electronic}. It is important to note that the energetical structures, relative to the atomisation limit, of the PESs are systematically lower by approximately 0.6 eV from those reported by Koner~\textit{et al.}~\cite{koner2018cno}. This discrepancy is primarily attributed to the basis and treatment of electron correlation utilised for the calculation of the \textit{ab initio} energies. The correlation diagram indicates the presence of three linear isomers and multiple stationary points evolving on the PESs of the \ce{CNO} system, giving rise to many different reactive channels available for exchange and dissociation processes. Interestingly, it is shown that the doublet state $^2A^{\prime\prime}$ presents a non-linear cyclic intermediate connecting asymptotically \ce{NO(X^2\Pi) + C(^3P)} to \ce{CO(X^1\Sigma^+) + N(^2D)} through the lowest energy structure of the symmetry states, i.e. \ce{NCO ($^2A^{\prime}$,$^2A^{\prime\prime}$)}, lying 7.097 eV below the \ce{NO(X^2\Pi) + C(^3P)} channel.

\begin{figure*}[!ht]
\centering
\begin{subfigure}[t]{0.32\textwidth}
 \centering
 \includegraphics[width=\textwidth]{Figures/2A1OCN.pdf}
 \caption{\ce{NCO($^2A^{\prime}$)}.}
\end{subfigure}
\begin{subfigure}[t]{0.32\textwidth}
 \centering
 \includegraphics[width=\textwidth]{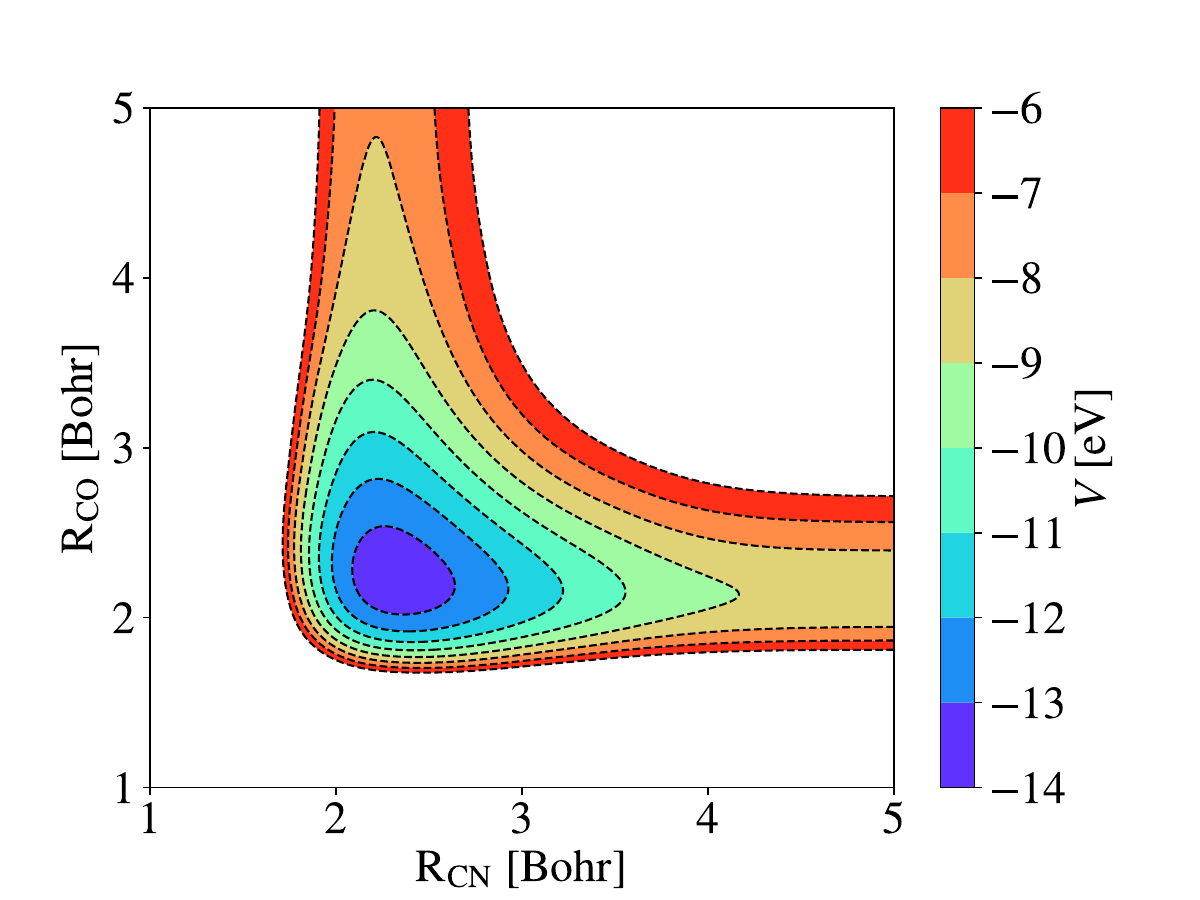}
 \caption{\ce{NCO($^2A^{\prime\prime}$)}.}
\end{subfigure}
\begin{subfigure}[t]{0.32\textwidth}
 \centering
 \includegraphics[width=\textwidth]{Figures/4A2OCN.pdf}
 \caption{\ce{NCO($^4A^{\prime\prime}$)}.}
\end{subfigure}

\begin{subfigure}[t]{0.32\textwidth}
 \centering
 \includegraphics[width=\textwidth]{Figures/2A1ONC.pdf}
 \caption{\ce{CNO($^2A^{\prime}$)}.}
\end{subfigure}
\begin{subfigure}[t]{0.32\textwidth}
 \centering
 \includegraphics[width=\textwidth]{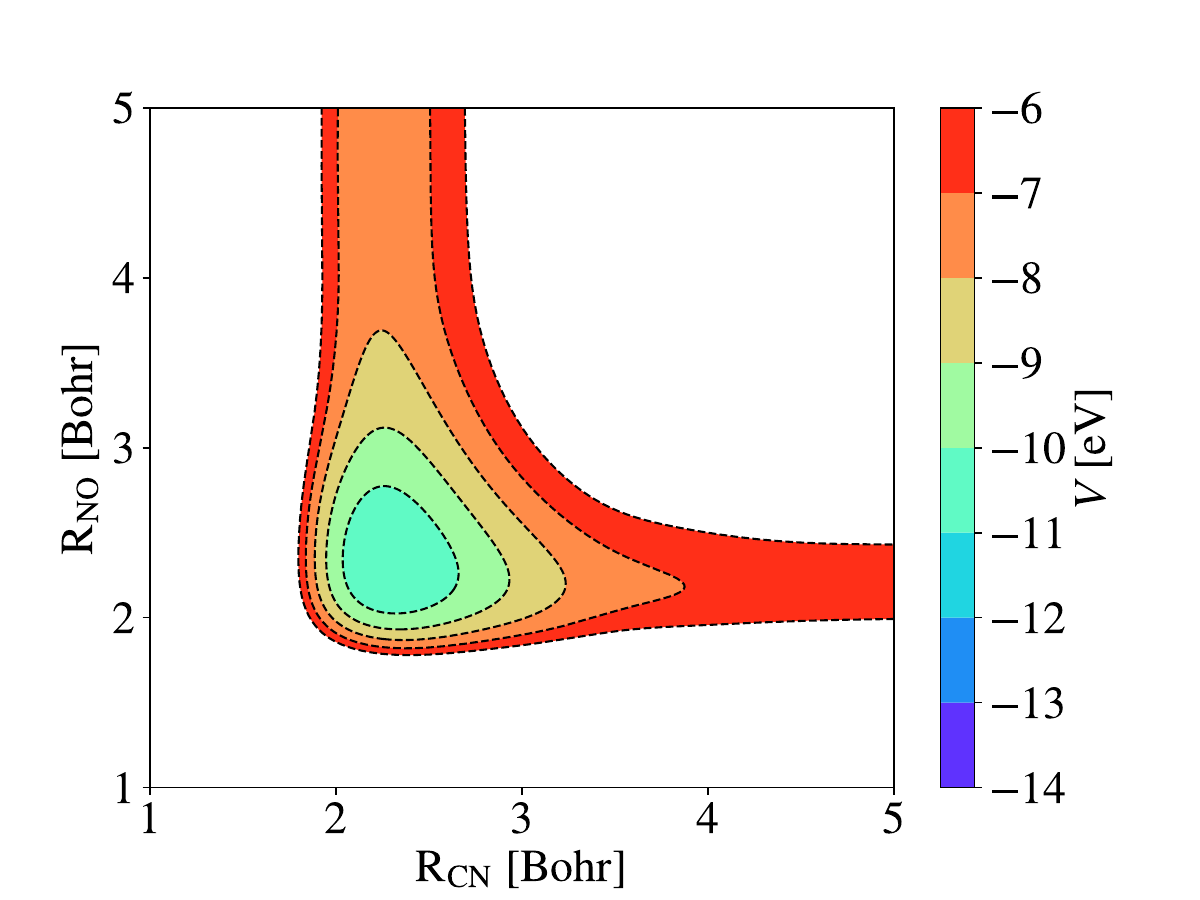}
 \caption{\ce{CNO($^2A^{\prime\prime}$)}.}
\end{subfigure}
\begin{subfigure}[t]{0.32\textwidth}
 \centering
 \includegraphics[width=\textwidth]{Figures/4A2ONC.pdf}
 \caption{\ce{CNO($^4A^{\prime\prime}$)}.}
\end{subfigure}

\begin{subfigure}[t]{0.32\textwidth}
 \centering
 \includegraphics[width=\textwidth]{Figures/2A1NOC.pdf}
 \caption{\ce{CON($^2A^{\prime}$)}.}
\end{subfigure}
\begin{subfigure}[t]{0.32\textwidth}
 \centering
 \includegraphics[width=\textwidth]{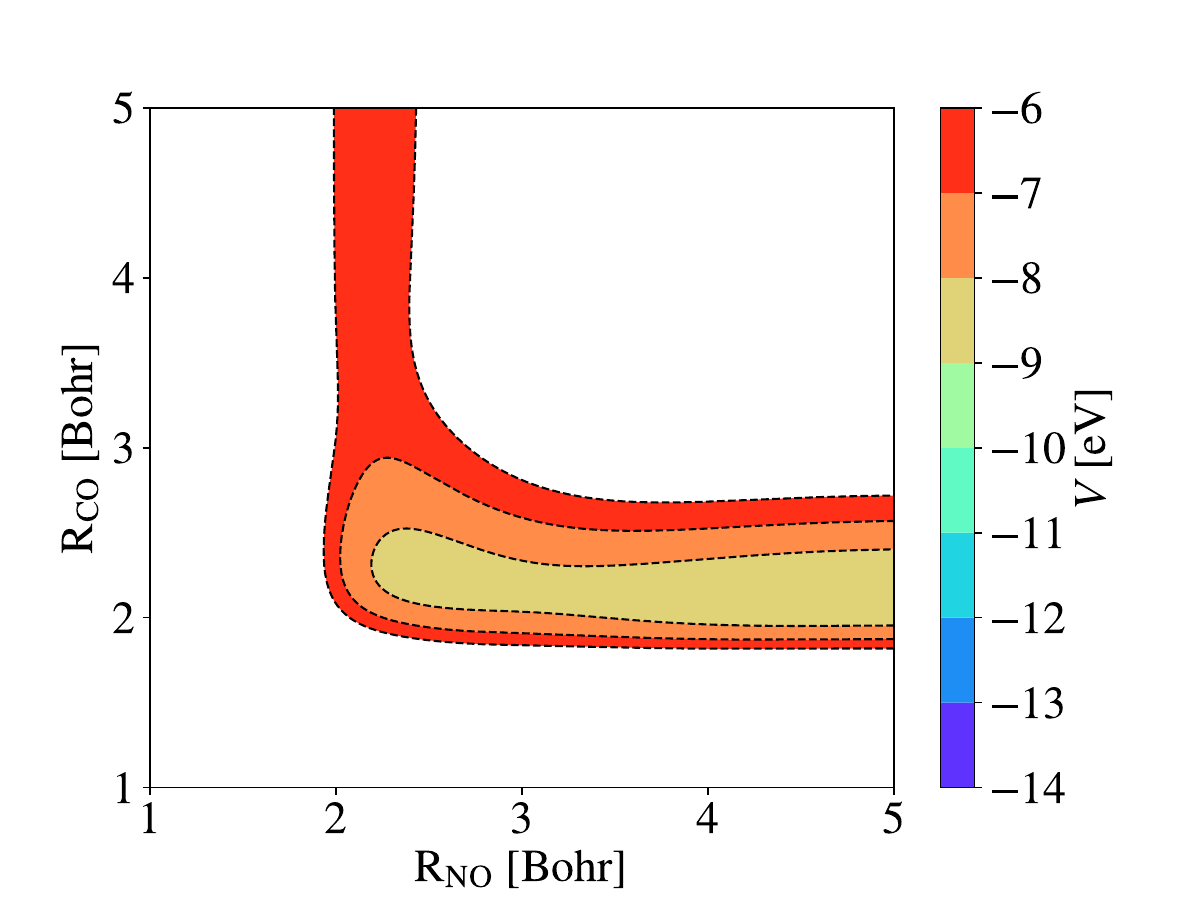}
 \caption{\ce{CON($^2A^{\prime\prime}$)}.}
\end{subfigure}
\begin{subfigure}[t]{0.32\textwidth}
 \centering
 \includegraphics[width=\textwidth]{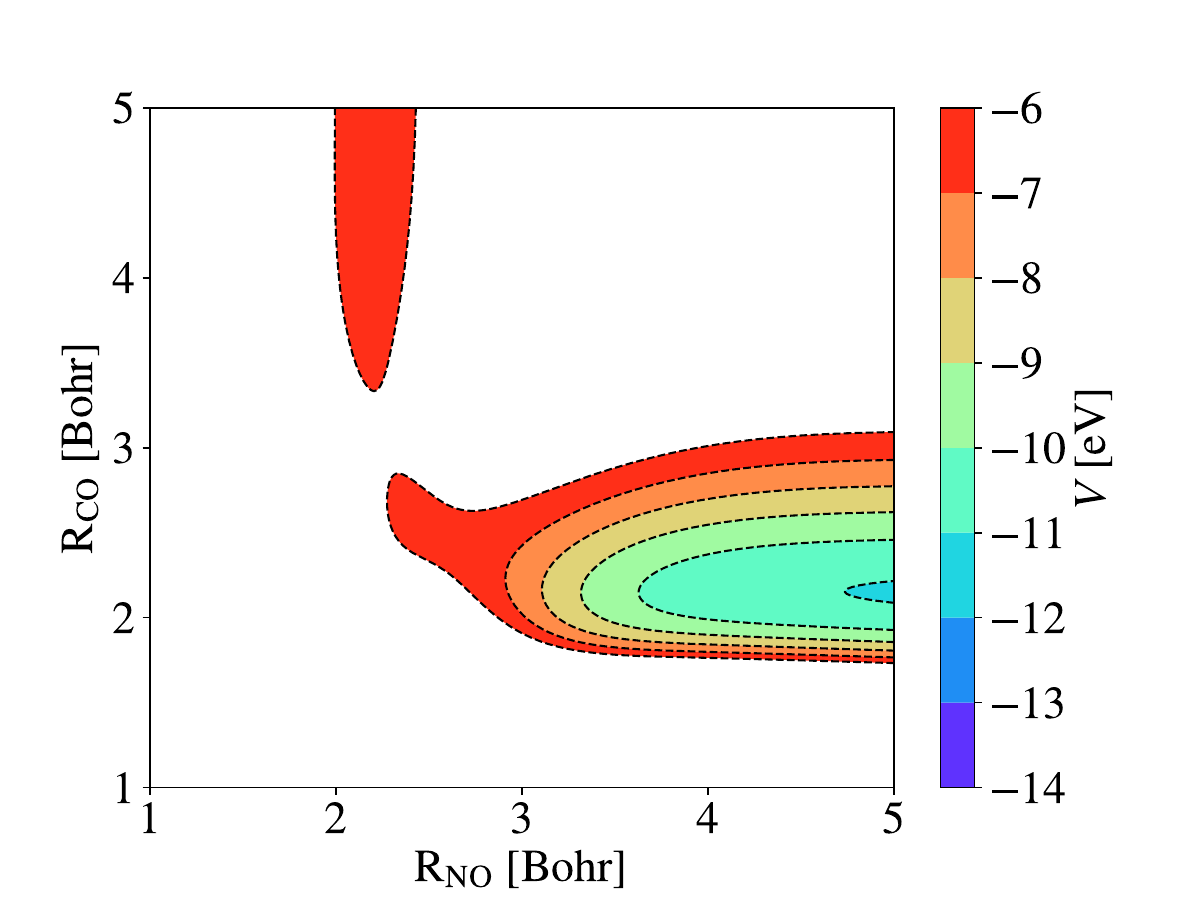}
 \caption{\ce{CON($^4A^{\prime\prime}$)}.}
\end{subfigure}

\caption{Contour plots of three PESs for the \ce{NCO}, \ce{CNO} and \ce{CON} isomers at an inter-bond angle fixed at $180^{\circ}$. The zero of energy is set to the atomisation energy, i.e.~ \ce{C(^3P) + N(^4S) + O(^3P)}.}
\label{TopologyPES}
\end{figure*}

Figure~\ref{TopologyPES} presents the contour plots of the PESs for the three isomers of the \ce{CNO} system on the doublet and quartet states at an inter-bond angle fixed to $180^{\circ}$. For all three isomers, the doublet states present a similar topology that strongly differs from that of the quartet state. The \ce{NCO} isomer exhibits the deepest wells on the doublet surfaces, i.e. the global minima of the \ce{CNO} system, providing a barrierless path for~\ref{CNO_Ex1} towards the exothermic \ce{CO(X^1\Sigma^+) + N(^4S,^2D)} asymptote. The \ce{CNO} isomer governs~\ref{NOC_Ex1} through two barrierless pathways, which exhibit moderately deep wells on the doublet surfaces and a quasi-monotonic profile on the quartet state. The \ce{CON} isomer drives the formation of \ce{N(^2D)} through~\ref{NOC_Ex2} on the doublet states, which exhibit similar topologies that strongly differ from that of the quartet state, the latter being responsible for the formation of \ce{N(^4S)} through~\ref{NOC_Ex2}. These observations are supported by the correlation diagram, which highlights the significant differences in the reactivity of the doublet and quartet states. Considering that the two doublet surfaces share similar topologies across all three isomers, the construction of the rovibrational state-specific kinetic database can be reduced to two PESs, i.e. $^2A^{\prime}$ and $^4A^{\prime\prime}$, thereby supporting the assumption regarding the number of PESs considered in the present study.

\subsection{Rovibrational State-specific Database}

\subsubsection{Inelastic Processes}

\begin{figure*}[!ht]
\centering

\begin{subfigure}[t]{0.49\textwidth}
 \centering
 \includegraphics[width=\textwidth, trim=1.5cm 2cm 2cm 2cm, clip]{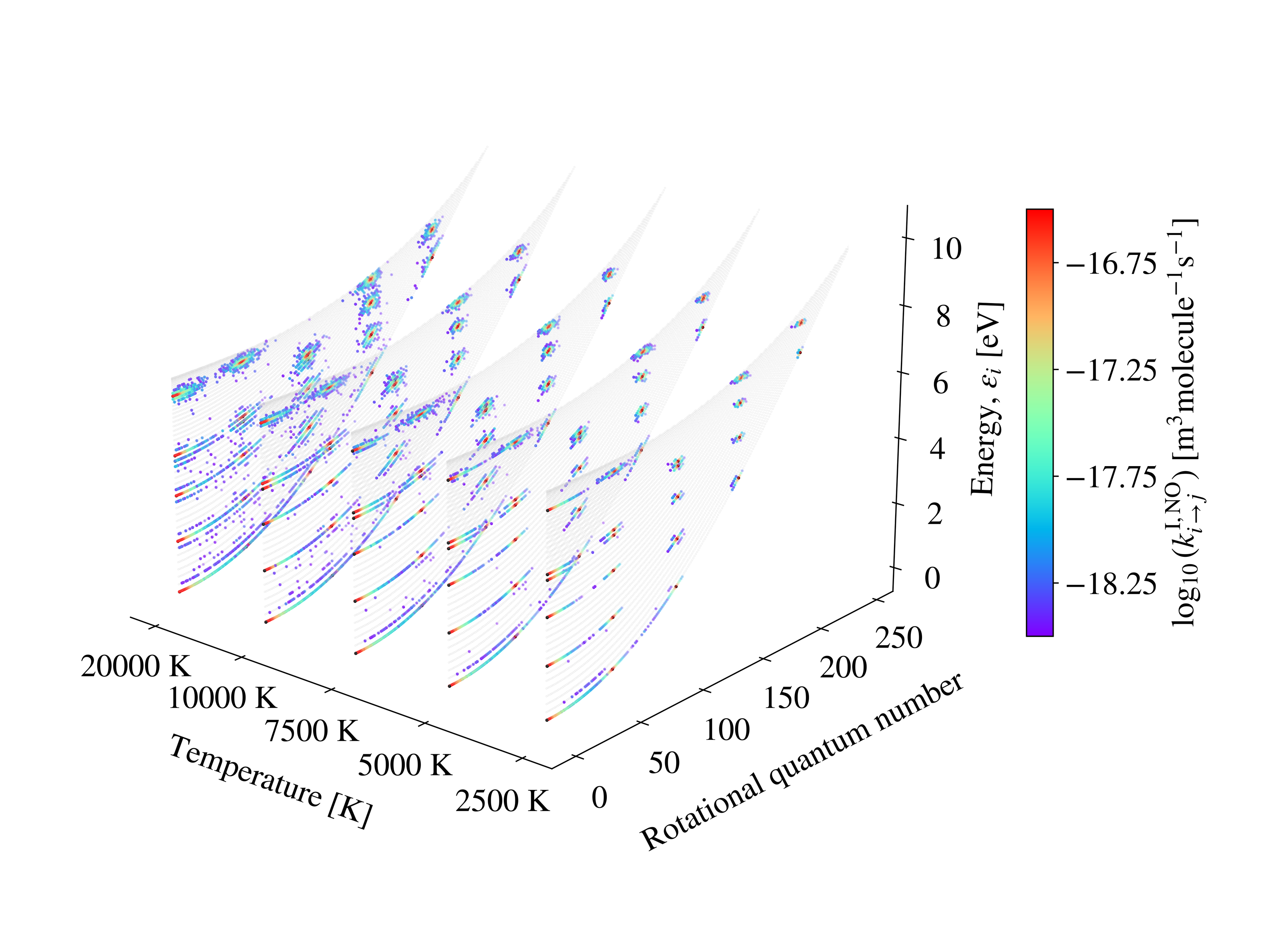}
 \caption{\ce{NO(X^2\Pi) + C(^3P)}.}
\end{subfigure}
\begin{subfigure}[t]{0.49\textwidth}
 \centering
 \includegraphics[width=\textwidth, trim=1.5cm 2cm 2cm 2cm, clip]{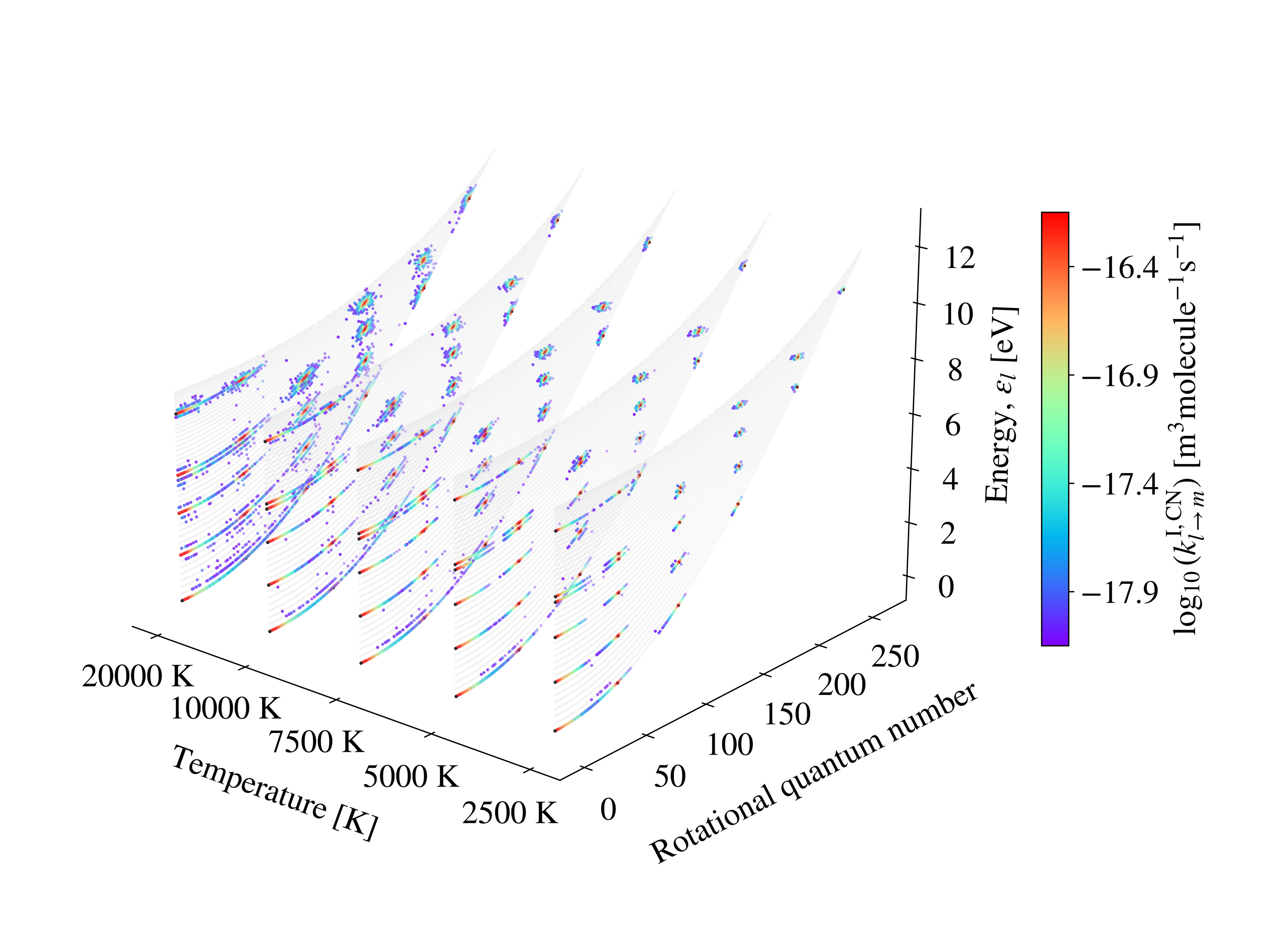}
 \caption{\ce{CN(X^2\Sigma^{+}) + O(^3P)}.}
\end{subfigure}

\begin{subfigure}[t]{0.49\textwidth}
 \centering
 \includegraphics[width=\textwidth, trim=1.5cm 2cm 2cm 2cm, clip]{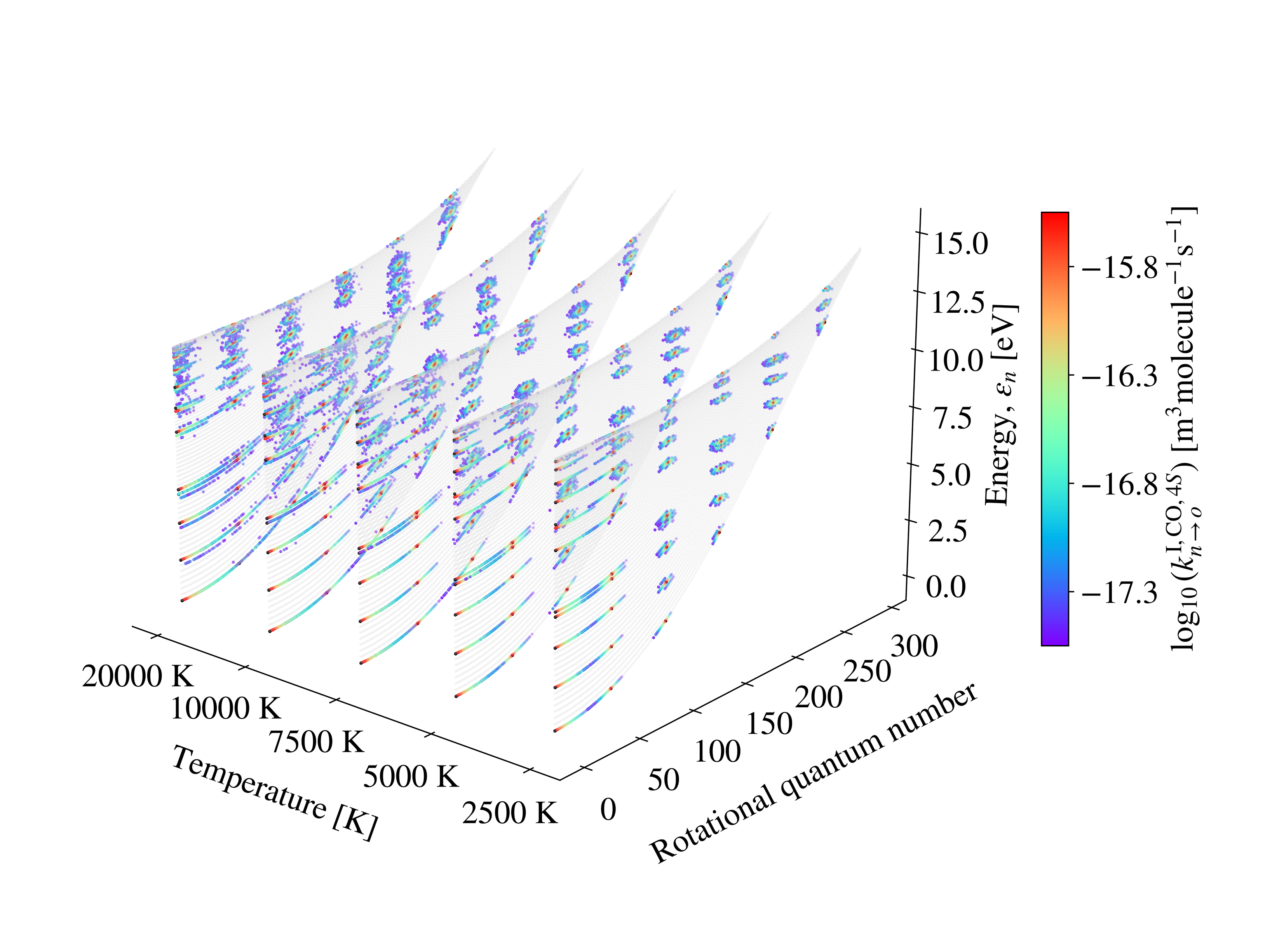}
 \caption{\ce{CO(X^1\Sigma^{+}) + N(^4S)}.}
\end{subfigure}
\begin{subfigure}[t]{0.49\textwidth}
 \centering
 \includegraphics[width=\textwidth, trim=1.5cm 2cm 2cm 2cm, clip]{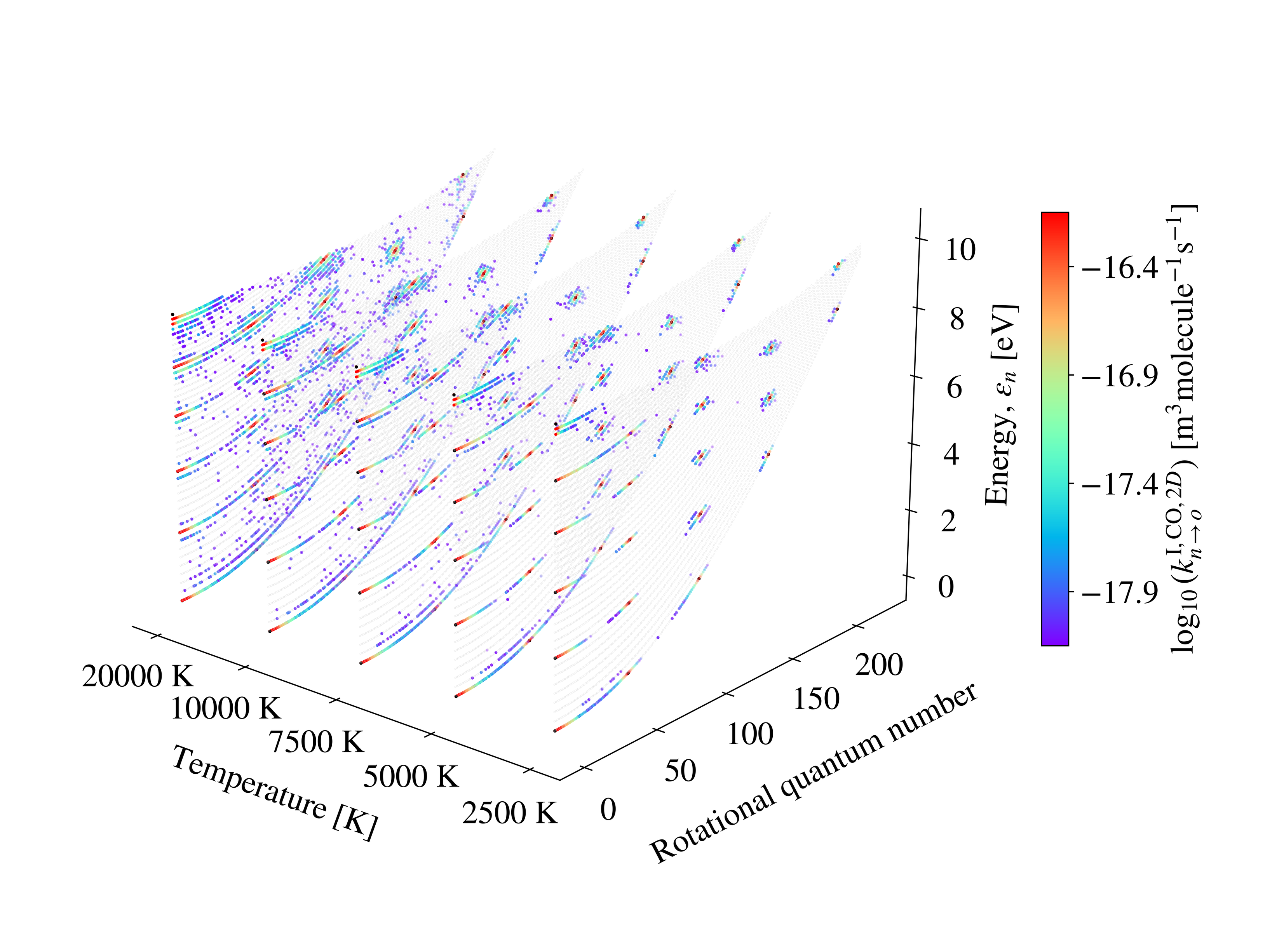}
 \caption{\ce{CO(X^1\Sigma^{+}) + N(^2D)}.}
\end{subfigure}

\caption{Distribution of the inelastic rate coefficients at five temperatures overlaid with the internal energy distribution. The grey dots indicate that the rate coefficients are below or above a threshold value. The black dots represent the initial states.}
\label{EnergyTransferMap}
\end{figure*}

Figure~\ref{EnergyTransferMap} shows the inelastic rate coefficients distribution across the rovibrational states at five temperatures, i.e.~ $T$ = 2500, 5000, 7500, 10,000 and 20,000~K. The energy-transfer rate coefficients of several rovibrational states, i.e.~black dots, are overlaid on the internal energies of \ce{NO(X^2\Pi)}, \ce{CN(X^2\Sigma^{+})} and \ce{CO(X^1\Sigma^{+})}. Considering that the \ce{CO(X^1\Sigma^{+}) + N(^2D)} collisional system originates from the first excited state of atomic nitrogen, i.e.~ $\varepsilon^{\ce{N(^2D)}} - \varepsilon^{\ce{N(^4S)}} \approx 2.4$ ~eV, the \ce{CO(X^1\Sigma^{+})} levels located within this energy of the centrifugal barrier lie above the atomisation limit of the $^2A^{\prime}$ surface when paired with \ce{N(^2D)}. These levels therefore carry no statistics in the \ce{CO(X^1\Sigma^{+}) + N(^2D)} database, and they are excluded from the rovibrational basis of the corresponding master equation. The effective diatomic potential of \ce{CO} is cropped accordingly in the graphical representations shown in the following sections.

For all thermal conditions, the collisional systems exhibit similar energy transfer behaviour. Specifically, low-lying states predominantly proceed within the same vibrational strands and hence result mainly in rotational excitation, whereas high-lying states transfer energy across different vibrational levels, thereby resulting in rovibrational excitation. Similar energy-transfer mechanisms are reported for other chemical systems,~\cite{jo2022rovibrational,Sharma2022HCN,jo2023rovibrational} thereby evidencing similarities between chemical systems that may be leveraged for model reduction.~\cite{Sharma2023Efficient} 

 \subsubsection{Dissociation Processes}

\begin{figure*}[!ht]
\centering
\begin{subfigure}[t]{0.49\textwidth}
 \centering
 \includegraphics[width=\textwidth, trim=1.5cm 1cm 1cm 2cm, clip]{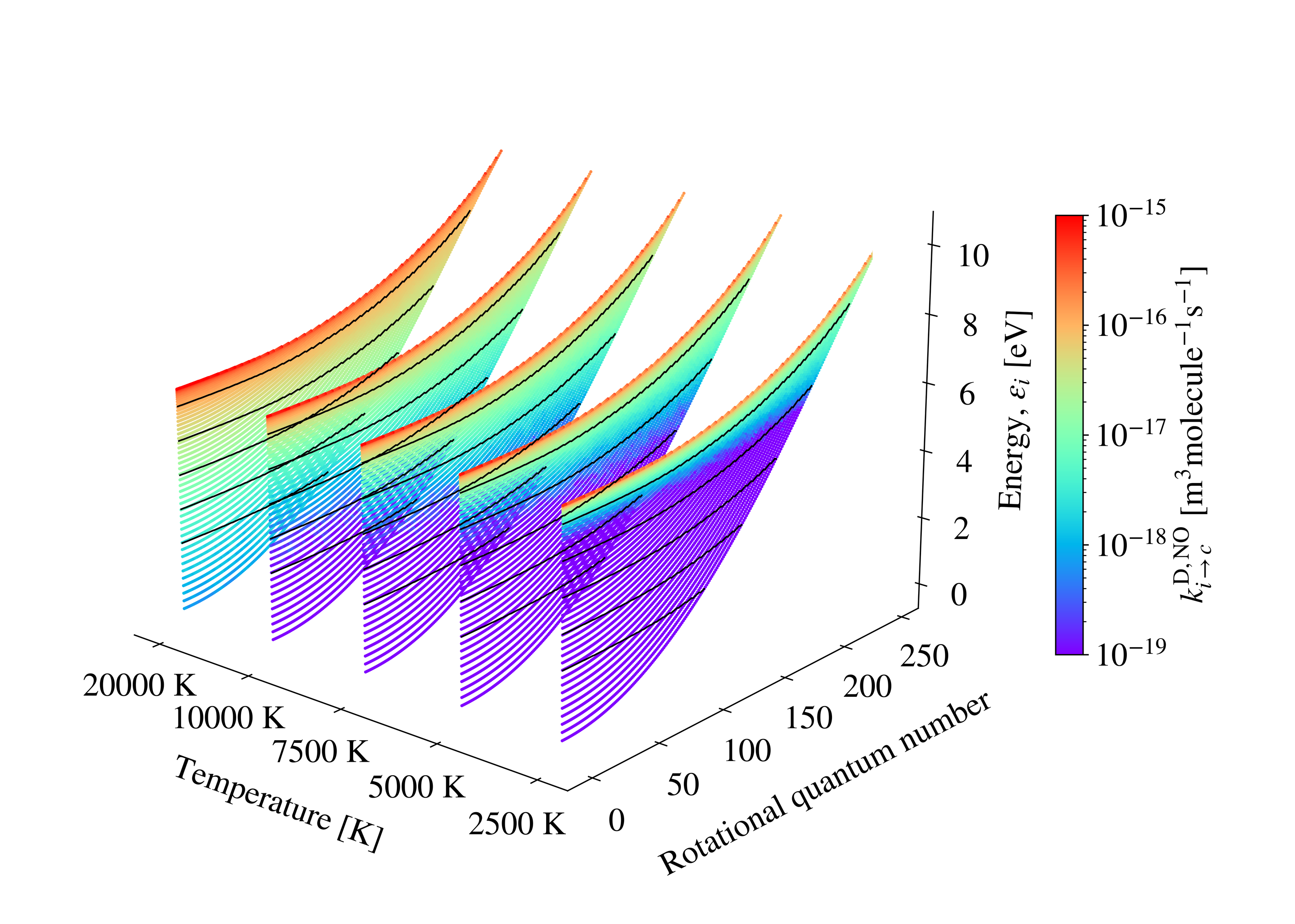}
 \caption{\ce{NO(X^2\Pi) + C(^3P)}.}
\end{subfigure}
\begin{subfigure}[t]{0.49\textwidth}
 \centering
 \includegraphics[width=\textwidth, trim=1.5cm 1cm 1cm 2cm, clip]{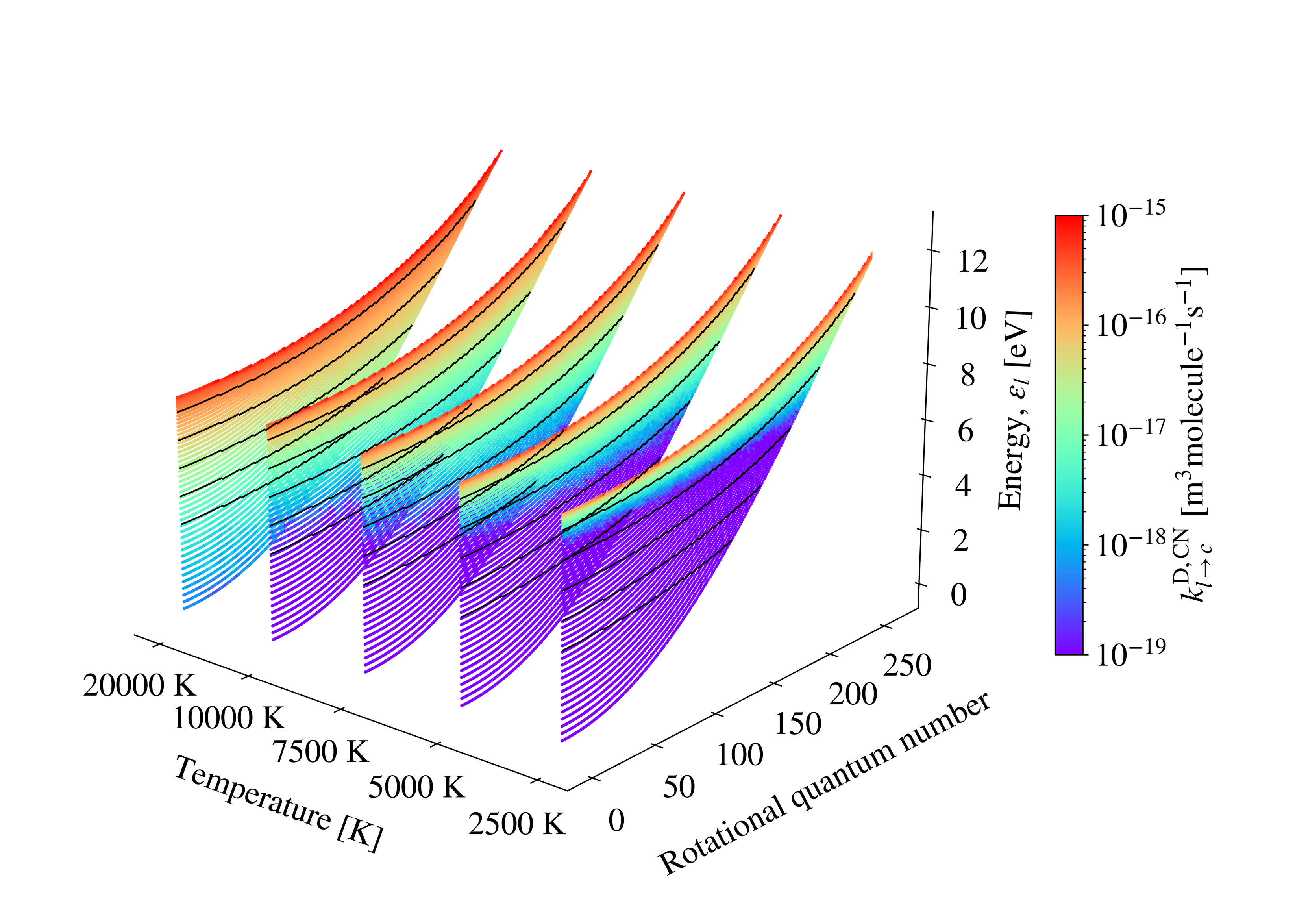}
 \caption{\ce{CN(X^2\Sigma^{+}) + O(^3P)}.}
\end{subfigure}

\begin{subfigure}[t]{0.49\textwidth}
 \centering
 \includegraphics[width=\textwidth, trim=1.5cm 1cm 1cm 2cm, clip]{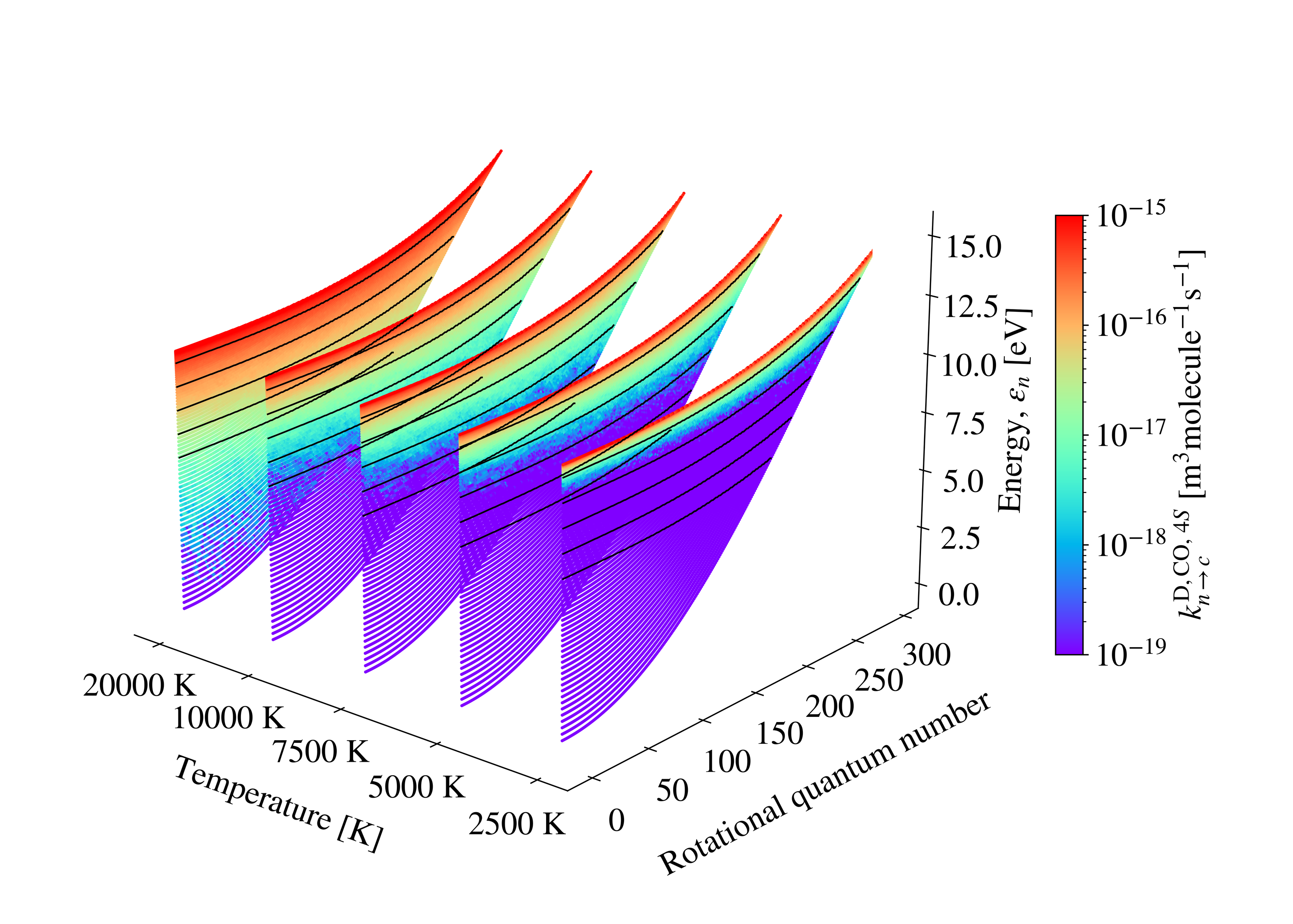}
 \caption{\ce{CO(X^1\Sigma^{+}) + N(^4S)}.}
\end{subfigure}
\begin{subfigure}[t]{0.49\textwidth}
 \centering
 \includegraphics[width=\textwidth, trim=1.5cm 1cm 1cm 2cm, clip]{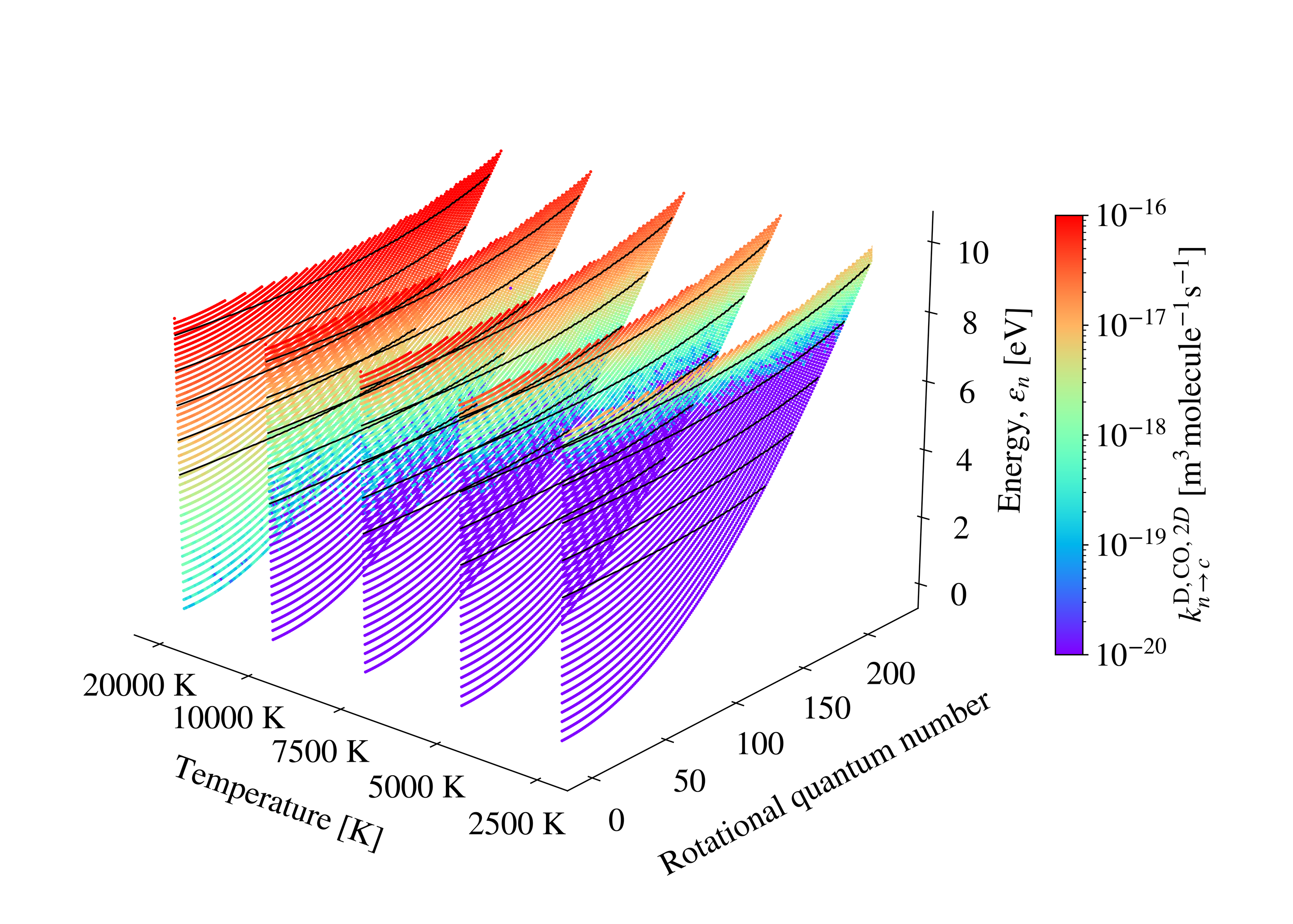}
 \caption{\ce{CO(X^1\Sigma^{+}) + N(^2D)}.}
\end{subfigure}

\caption{State-to-state total dissociation rate coefficients at five temperatures. The black lines indicate isolines from the centrifugal barrier.}
\label{fig:Dissociation}
\end{figure*}

Figure~\ref{fig:Dissociation} presents the total state-to-state dissociation rate coefficients. The energy deficit, defined as the distance of a rovibrational state from the centrifugal barrier in energy space, i.e.~ $\varepsilon^{\text{D}}_{i} = V_{\text{max}}(J) - \varepsilon_{i}$, is represented by five isolines across the internal energy distribution.

Figure~\ref{fig:Dissociation} indicates that the dissociation processes of all collisional systems are primarily governed by high-lying rovibrational states. Specifically, these states are characterised by large outer turning points and high internal energies, enabling the system to overcome the atomisation energy. These observations are consistent with previous studies for other systems relevant to Earth-atmosphere re-entry.~\cite{jo2022rovibrational,panesi2013rovibrational,Macdonald2018constructionI,jo2023rovibrational,rodriguez2026no2} 

The total state-specific dissociation rate coefficients exhibit a strong dependence on the energy deficit. These findings suggest that reduced-order representations of the kinetic database for the \ce{CNO} system may be constructed following methodologies previously developed for symmetric systems.~\cite{Sharma2023Efficient} 

%

\subsubsection{Heterogeneous Exchange Processes}

The analysis of exchange processes in heterogeneous collisional systems involves multiple reactive channels. To identify the preferred reaction pathway of a departing fragment, it is convenient to define a metric commonly referred to as the branching ratio. Considering a trajectory departing from the \ce{NO + C} asymptote, the branching ratios associated with the \ce{CN + O}, \ce{CO + N(^4S)}, and \ce{CO + N(^2D)} products are defined, respectively, as
\begin{equation}
 \Omega^{\ce{NO},\ce{CN}}_i = \frac{ \sum\limits_{l} k^{\text{E}, \ce{NO}}_{i\rightarrow l}}{\sum\limits_{l} k^{\text{E}, \ce{NO}}_{i\rightarrow l} + \sum\limits_{n} k^{\text{E}, \ce{NO}}_{i\rightarrow n,0} + \sum\limits_{n} k^{\text{E}, \ce{NO}}_{i\rightarrow n,1}},
\end{equation}

\begin{equation}
 \Omega^{\ce{NO},\ce{CO + N(^4S)}}_i = \frac{ \sum\limits_{n} k^{\text{E}, \ce{NO}}_{i\rightarrow n,0}}{\sum\limits_{l} k^{\text{E}, \ce{NO}}_{i\rightarrow l} + \sum\limits_{n} k^{\text{E}, \ce{NO}}_{i\rightarrow n,0} + \sum\limits_{n} k^{\text{E}, \ce{NO}}_{i\rightarrow n,1}},
\end{equation}

\begin{equation}
 \Omega^{\ce{NO},\ce{CO + N(^2D)}}_i = \frac{ \sum\limits_{n} k^{\text{E}, \ce{NO}}_{i\rightarrow n,1}}{\sum\limits_{l} k^{\text{E}, \ce{NO}}_{i\rightarrow l} + \sum\limits_{n} k^{\text{E}, \ce{NO}}_{i\rightarrow n,0} + \sum\limits_{n} k^{\text{E}, \ce{NO}}_{i\rightarrow n,1}},
\end{equation}
with analogous definitions for the other entrance channels.

\begin{figure*}[!ht]
\centering
\begin{subfigure}[t]{0.49\textwidth}
 \centering
 \includegraphics[width=\textwidth, trim=1.5cm 0.5cm 1cm 2cm, clip]{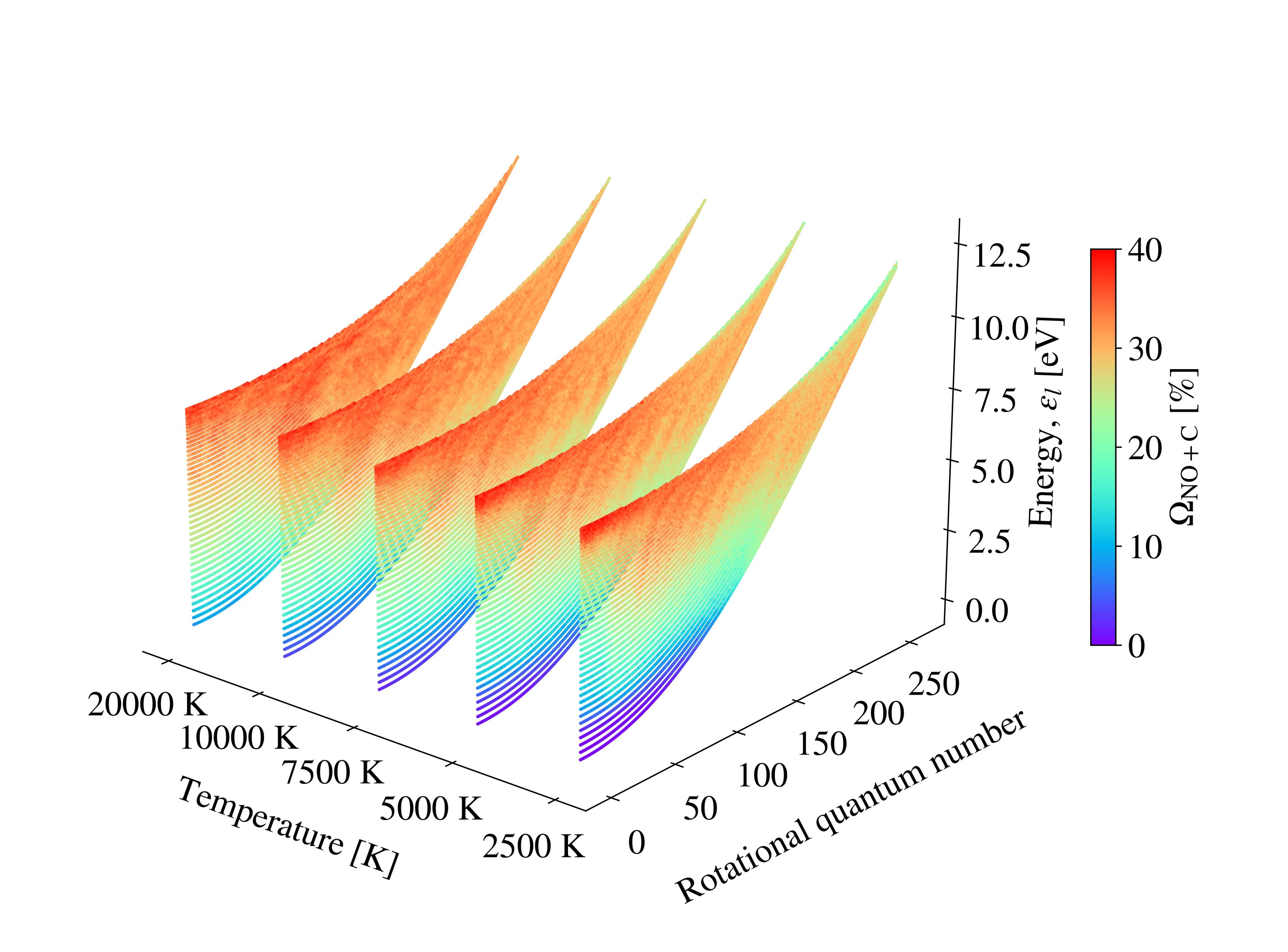}
 \caption{\ce{CN(X^2\Sigma^+) +O(^3P) <=> NO(X^2\Pi) + C(^3P)}.}
\end{subfigure}
\begin{subfigure}[t]{0.49\textwidth}
 \centering
 \includegraphics[width=\textwidth, trim=1.5cm 0.5cm 1cm 2cm, clip]{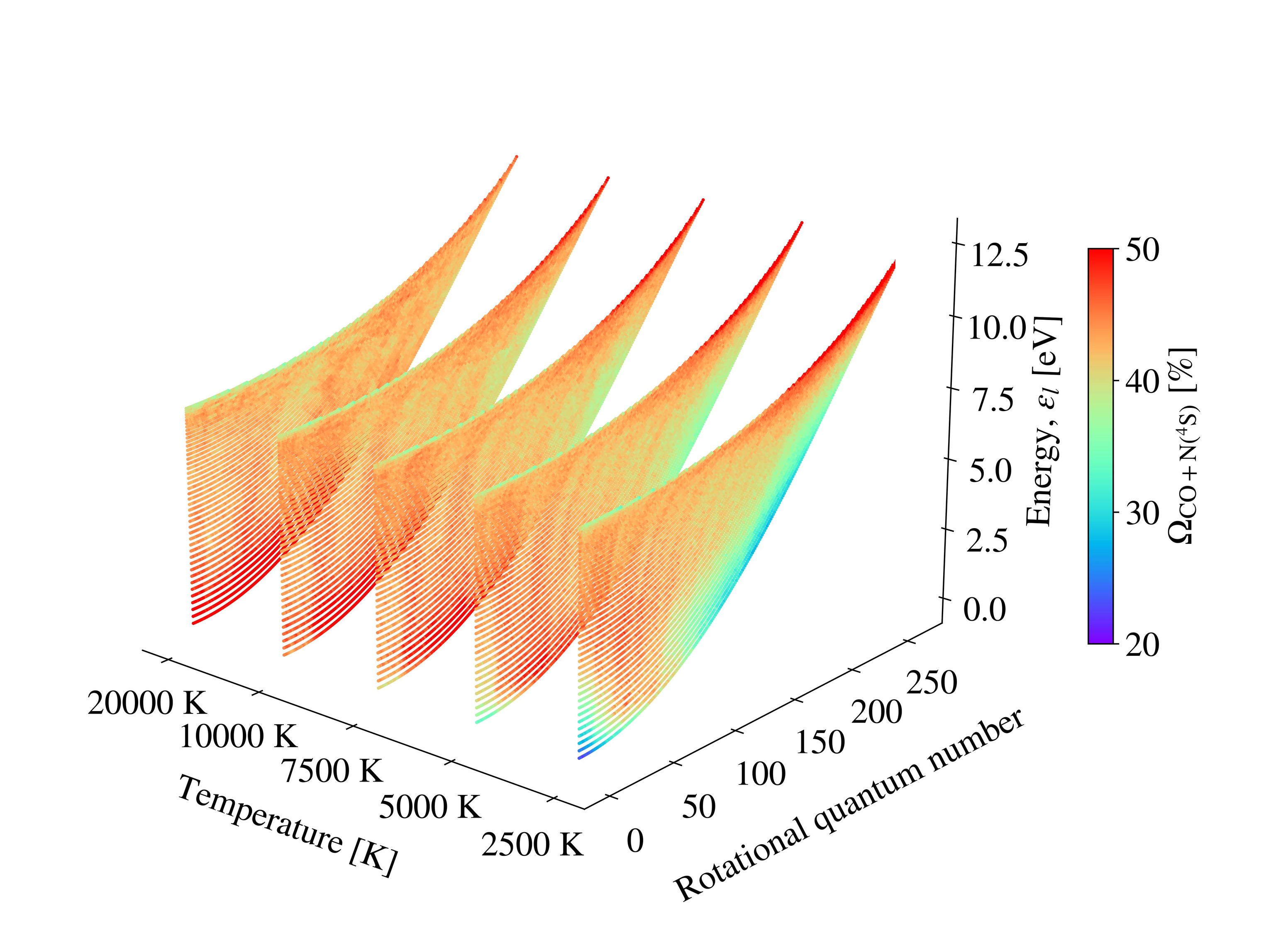}
 \caption{\ce{CN(X^2\Sigma^+) +O(^3P) <=> CO(X^1\Sigma^+) + N(^4S)}.}
\end{subfigure}
\begin{subfigure}[t]{0.49\textwidth}
 \centering
 \includegraphics[width=\textwidth, trim=1.5cm 0.5cm 1cm 2cm, clip]{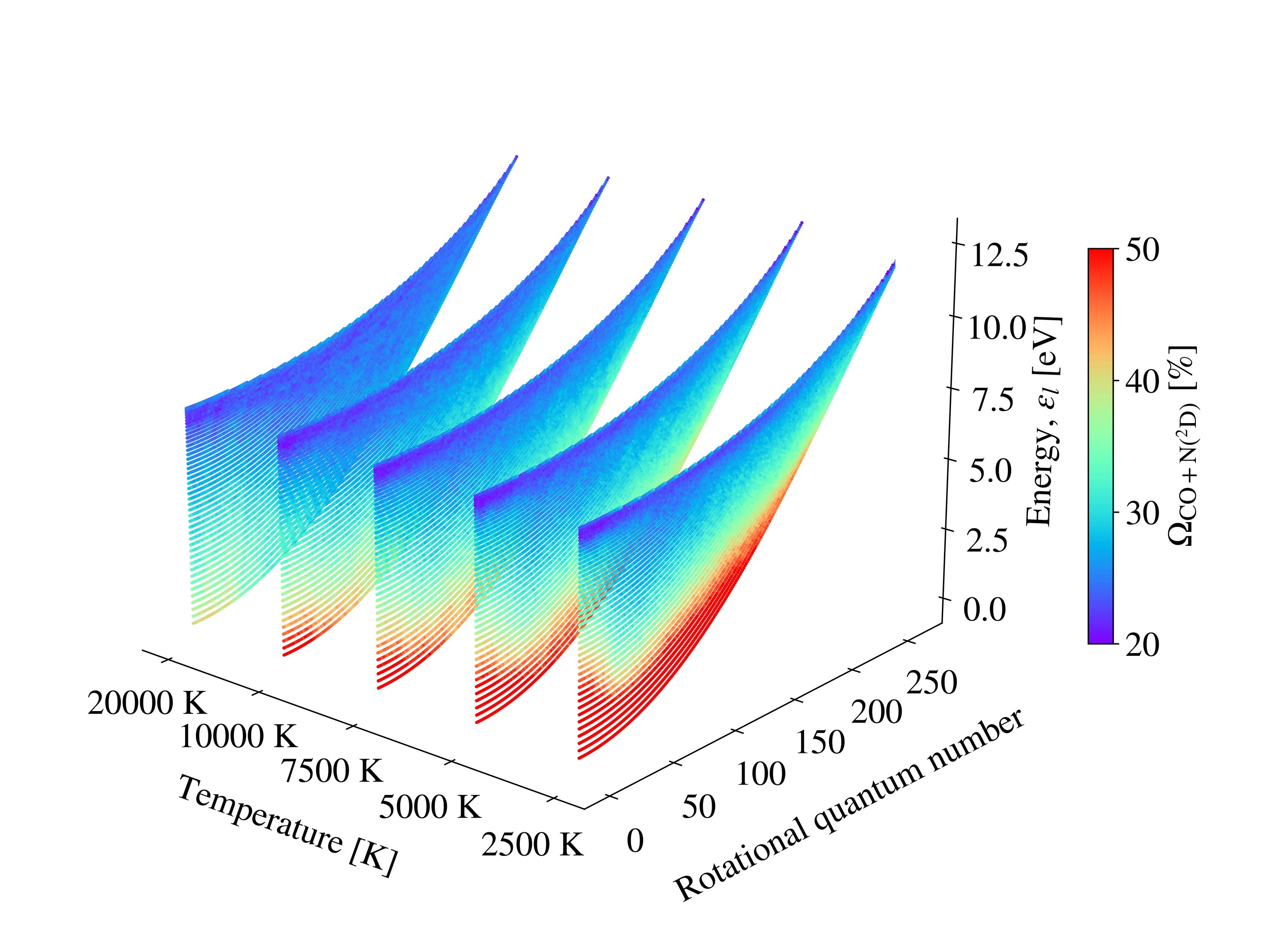}
 \caption{\ce{CN(X^2\Sigma^+) +O(^3P) <=> CO(X^1\Sigma^+) + N(^2D)}.}
\end{subfigure}

\caption{Branching ratio for the \ce{CN(X^2\Sigma^+) +O(^3P)} asymptote overlaid with the rovibrational state distribution of \ce{CN(X^2\Sigma^+)} at five temperatures.}
\label{DistributionExchange}
\end{figure*}

Figure~\ref{DistributionExchange} presents the branching ratio of the \ce{CN(X^2\Sigma^+) +O(^3P)} collisional system at five temperatures. For brevity, the branching ratios of the other collisional systems are reported in the Supplementary Material (see Figs.~\ref{S-BranchingRatioNOC} -- \ref{S-BranchingRatioCON2D}). The state-to-state branching ratio is herein included to highlight the governing exchange process for each collisional system.

Figure~\ref{DistributionExchange} indicates that the kinetics of the \ce{CN(X^2\Sigma^+) + O(^3P)} collisional system are primarily governed by the production of \ce{CO(X^1\Sigma^+) + N(^4S,^2D)}. Indeed, the correlation diagram (see Fig.~\ref{fig:CorrelationDiagram}) demonstrates that the formation of \ce{CO(X^1\Sigma^+) + N(^4S,^2D)} occurs through an exothermic barrierless process, hence favouring its formation over the \ce{NO(X^2\Pi) + C(^3P)} asymptote, which requires an endothermic pathway for its production. Note that, since \ce{CO(X^1\Sigma^+) + N(^4S)} is reached only on the quartet surface and \ce{CO(X^1\Sigma^+) + N(^2D)} only on the doublet surfaces, the branching between the two nitrogen states in Figs.~\ref{DistributionExchange} (b) and (c) directly reflects the symmetry state on which the exchange proceeds. Under low thermal conditions, the low-lying rovibrational states of \ce{CN(X^2\Sigma^+)} proceed primarily towards \ce{CO(X^1\Sigma^+) + N(^2D)}, i.e.~on the doublet state, whereas this preference vanishes with increasing temperature, the \ce{CO(X^1\Sigma^+) + N(^4S)} and \ce{CO(X^1\Sigma^+) + N(^2D)} channels contributing in comparable proportions at $T$ = 20,000~K. In contrast, the high-lying states favour the \ce{CO(X^1\Sigma^+) + N(^4S)} channel, i.e.~the quartet state, over \ce{CO(X^1\Sigma^+) + N(^2D)}. For increasing temperatures, the formation of the latter collisional system governs the heterogeneous exchange reactions.

For the \ce{NO(X^2\Pi) + C(^3P)} collisional system, it is shown that, at low temperatures, the low-lying states primarily proceed through the formation of \ce{CN(X^2\Sigma^+) +O(^3P)}, whereas high-lying states proceed through the formation of \ce{CO(X^1\Sigma^+) + N(^4S)} (see Fig.~\ref{S-BranchingRatioNOC}). For increasing temperatures, the formation of \ce{CO(X^1\Sigma^+) + N(^4S,^2D)} exceeds that of \ce{CN(X^2\Sigma^+) +O(^3P)} above approximately 5000~K (see Fig.~\ref{ExoRateCoefficients}), although the latter remains the largest single channel over the whole temperature range. Among the \ce{CO(X^1\Sigma^+) + N(^4S,^2D)} channels, the ground-state atomic nitrogen asymptote accounts for about 60 \% of the \ce{CO(X^1\Sigma^+)} production. The favoured formation of ground-state atomic nitrogen is primarily attributed to the direct reactive pathway on the $^4A^{\prime\prime}$ state, whereas the formation of the first excited-state nitrogen atom must proceed through a local minimum, i.e.~ \ce{CON($^2A^{\prime}, ^2A^{\prime\prime}$)} on the $^2A^{\prime}$ state (see Fig.~\ref{fig:CorrelationDiagram}). However, these branching fractions do not reproduce the shock-tube measurements of Dean~\textit{et al.},~\cite{Dean1991ShockTube} which indicate that about 60 \% of the \ce{NO(X^2\Pi) + C(^3P)} reaction proceeds towards \ce{CO(X^1\Sigma^+) + N} and about 40 \% towards \ce{CN(X^2\Sigma^+) + O(^3P)}, independently of the temperature between 2430 and 4040~K. Over this temperature range, the rate coefficients of the present study (see Tab.~\ref{S-ArrheniusCoefficientFit}) indicate that the \ce{CO(X^1\Sigma^+) + N(^4S,^2D)} channels account for 35 \% to 48 \% of the reaction, of which the \ce{CO(X^1\Sigma^+) + N(^4S)} channel accounts for 22 \% to 28 \%. A similar underestimation of the \ce{CO(X^1\Sigma^+)} branching fraction was reported by Koner~\textit{et al.}~\cite{koner2018cno} for trajectories that neglect nonadiabatic transitions.

For the \ce{CO(X^1\Sigma^+) + N(^4S)} collisional system, the heterogeneous exchange reactions predominantly proceed towards the formation of \ce{CN(X^2\Sigma^+) + O(^3P)}. This behaviour is primarily attributed to the significantly lower reaction energy associated with the \ce{CN(X^2\Sigma^+) + O(^3P)} asymptote, for which the transition state remains readily accessible over a broad range of collision energies. In contrast, the formation of \ce{NO(X^2\Pi) + C(^3P)} requires substantially larger energies to overcome the corresponding reactive pathway, hence restricting its contribution to the kinetics of the \ce{CO(X^1\Sigma^+) + N(^4S)} collisional system.

For the \ce{CO(X^1\Sigma^+) + N(^2D)} collisional system, the exchange reactions preferentially proceed towards the formation of \ce{CN(X^2\Sigma^+) + O(^3P)}. This behaviour is primarily attributed to the significantly lower endothermicity of this channel, i.e.~approximately 1.0~eV against 2.2~eV for the \ce{NO(X^2\Pi) + C(^3P)} asymptote, which makes the former channel accessible over a broader range of collision energies. The preference for the \ce{CN(X^2\Sigma^+) + O(^3P)} channel is most pronounced under low thermal conditions and for the low-lying states, and reduces to a factor of about four for the thermal rate coefficients at $T$ = 10,000~K (see Fig.~\ref{EndoRateCoefficients} and Tab.~\ref{S-ArrheniusCoefficientFit}), as the \ce{NO(X^2\Pi) + C(^3P)} channel becomes energetically accessible to the high-lying states.

\subsection{Master Equation Analysis}

In this section, master equation analyses are conducted for conditions representative of hypersonic flows. The investigation consists of extracting the moments of the population distributions to retrieve quantities of interest for the development of a reduced-order representation of the kinetic database. Additionally, high-speed and high-temperature computational fluid dynamics (CFD) simulation tools often rely on the assumption that chemical reactions occur within a QSS regime. Therefore, the validity of this assumption in the context of asymmetric collisional systems is herein assessed.

For this purpose, a series of isothermal reactor simulations is conducted. These numerical experiments involve an ideal reactor immersed in an isothermal heat bath maintained at a constant temperature $T$. The volume of the reactor is kept constant. The reactor is initialised with an initial pressure of $P$ = 1000~Pa, an initial internal temperature of $T_{\text{int},0} = 300$ ~K and a gas mixture composition of $50\%$ molecular species and $50\%$ atomic species, resulting in a total number density of the reactor at $2.41\times 10^{23}$~m$^{-3}$. The translational modes are initialised from a Maxwell distribution at the isothermal heat bath temperature. The rovibrational energies of the molecular species are sampled from a Boltzmann distribution at the initial internal temperature of the reactor. The numerical experiments are conducted at five heat bath temperatures ranging between 2500 and 20,000 K. The kinetic processes, i.e.~inelastic, dissociation and exchange, of all three collisional systems are investigated sequentially.

\subsubsection{Inelastic Processes}

 The internal energy relaxation of the three collisional systems under the sole action of inelastic processes, i.e.~\ref{NOC_Inelastic} -- \ref{CON_Inelastic}, is first investigated in detail. The dissociation and exchange mechanisms are herein neglected, i.e.~\ref{NOC_Ex1} -- \ref{CON_Dissociation}.

\begin{figure*}[!ht]
\centering
\begin{subfigure}[t]{0.24\textwidth}
 \centering
 \includegraphics[width=\textwidth]{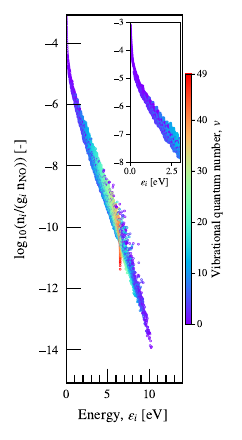}
 \caption{\ce{NO(X^2\Pi) + C(^3P)}.}
 \label{Pop_NO_C}
\end{subfigure}
\begin{subfigure}[t]{0.24\textwidth}
 \centering
 \includegraphics[width=\textwidth]{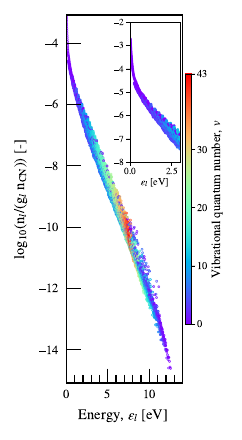}
 \caption{\ce{CN(X^2\Sigma^{+}) + O(^3P)}.}
 \label{Pop_CN_O}
\end{subfigure}
\begin{subfigure}[t]{0.24\textwidth}
 \centering
 \includegraphics[width=\textwidth]{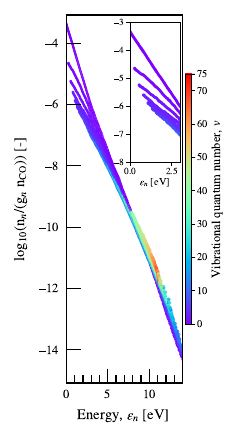}
 \caption{\ce{CO(X^1\Sigma^{+}) + N(^4S)}.}
\end{subfigure}
\begin{subfigure}[t]{0.24\textwidth}
 \centering
 \includegraphics[width=\textwidth]{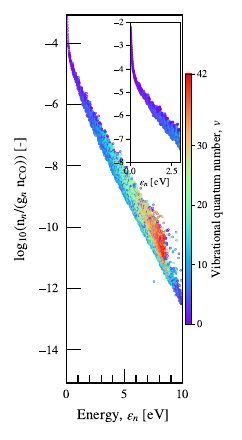}
 \caption{\ce{CO(X^1\Sigma^{+}) + N(^2D)}.}
\end{subfigure}
\caption{Rovibrational distribution of the \ce{NO(X^2\Pi) + C(^3P)}, \ce{CN(X^2\Sigma^{+}) + O(^3P)} and \ce{CO(X^1\Sigma^{+}) + N(^4S,^2D)} collisional systems at $T_t$~=~10,000~K and $T_v$~=~2000~K (dissociation and exchange processes excluded). The distributions are coloured by the vibrational quantum number.}
\label{Distribution_2000K}
\end{figure*}

Figure~\ref{Distribution_2000K} shows a comparison of the internal energy relaxation across all collisional systems. The comparison is performed at a translational temperature of $T_t$ ~=~10,000~K and a vibrational temperature of approximately $T_v$ ~=~2000~K, corresponding to a relaxation of about 10 \% of the equilibrium vibrational energy, the rotational and vibrational temperatures being defined by matching the average rotational and vibrational energies of the master equation solution with those of a two-temperature Boltzmann distribution. Note that Figs.~\ref{Pop_NO_C} and~\ref{Pop_CN_O} report the global rovibrational distributions of the \ce{NO(X^2\Pi) + C(^3P)} and \ce{CN(X^2\Sigma^{+}) + O(^3P)} systems, i.e.~summed over the contributions of the doublet and quartet states, whereas the \ce{CO(X^1\Sigma^{+}) + N(^4S,^2D)} systems each evolve on a single symmetry state. For completeness, the electronic state-specific distributions are reported in the Supplementary Material (see Figs.~\ref{S-PES_Specific_NO_C_2000K} and~\ref{S-PES_Specific_CN_O_2000K}).

For a similar electronic state, the collisional systems exhibit similar rovibrational distributions. Specifically, the energy transfer processes of the \ce{NO(X^2\Pi) + C(^3P)} and \ce{CN(X^2\Sigma^{+}) + O(^3P)} systems evolving on the quartet states appear to follow a similar distribution to the \ce{CO(X^1\Sigma^{+}) + N(^4S)} system. Note that only the \ce{CO(X^1\Sigma^{+}) + N(^4S)} system exhibits distinct vibrational-specific strands during the early stages of the relaxation (see the inset of Fig.~\ref{Distribution_2000K} (c)), whereas the quartet-state distributions of the \ce{NO(X^2\Pi) + C(^3P)} and \ce{CN(X^2\Sigma^{+}) + O(^3P)} systems only display a weak signature of such strands (see Figs.~\ref{S-PES_Specific_NO_C_2000K} (b) and~\ref{S-PES_Specific_CN_O_2000K} (b)). Similar strand-like structures were also reported for other collisional systems, e.g.~ \ce{N2O}.~ \cite{jo2022rovibrational} 

Similarly, the energy transfer processes of the \ce{NO(X^2\Pi) + C(^3P)} and \ce{CN(X^2\Sigma^{+}) + O(^3P)} systems evolving on the doublet states follow a similar distribution to the \ce{CO(X^1\Sigma^{+}) + N(^2D)} system. Indeed, the low-lying states appear to have equilibrated with each other and hence no longer exhibit vibrational-specific strands.

These observations suggest the possibility of developing a generalised reduced-order modelling approach capable of describing the kinetics of both symmetric and asymmetric systems, provided that the discrepancies between the different symmetry states are accurately represented.

\begin{figure*}[!ht]
\centering
\begin{subfigure}[t]{0.49\textwidth}
 \centering
 \includegraphics[width=\textwidth]{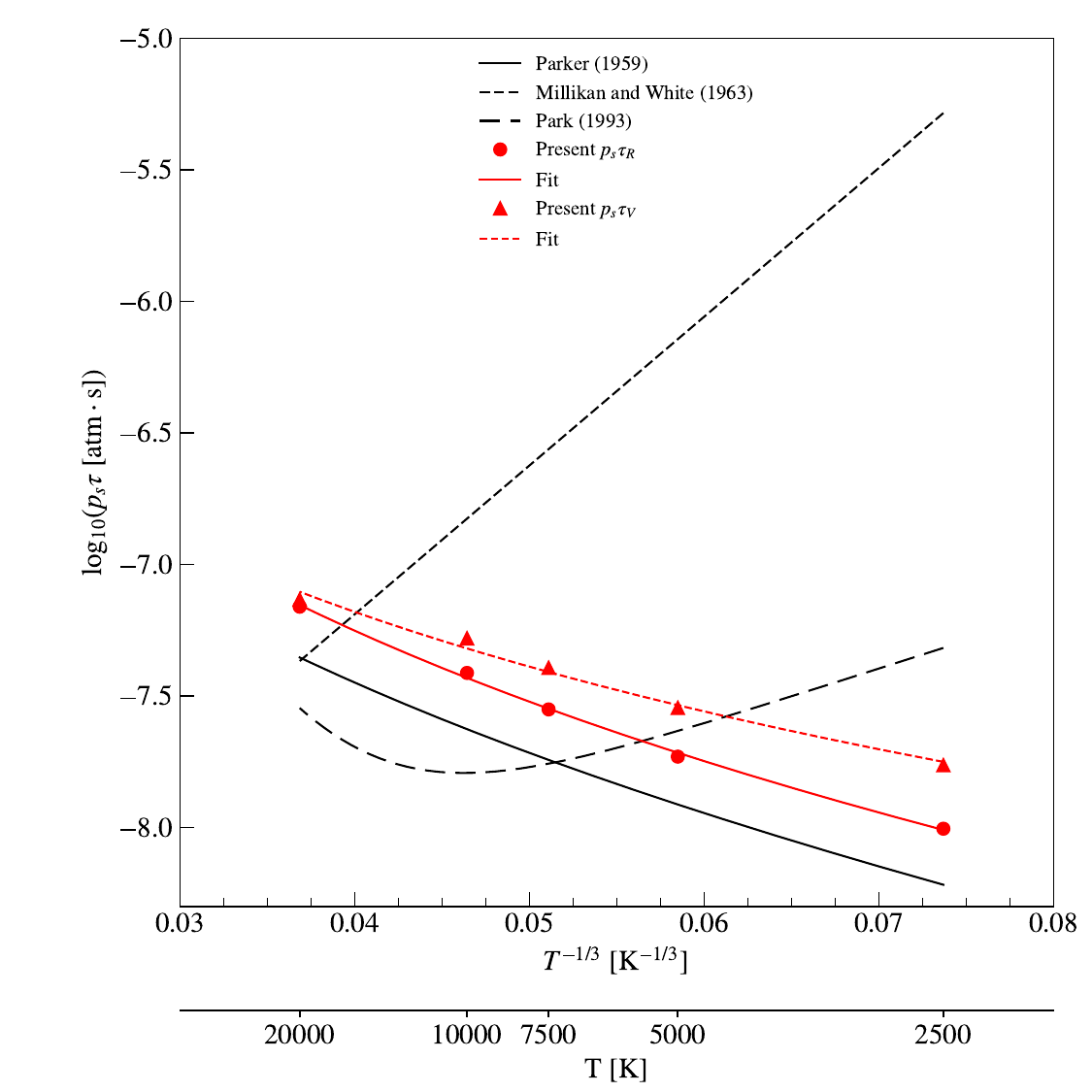}
 \caption{\ce{NO(X^2\Pi) + C(^3P)}.}
 \label{Tau_NOC}
\end{subfigure}
\begin{subfigure}[t]{0.49\textwidth}
 \centering
 \includegraphics[width=\textwidth]{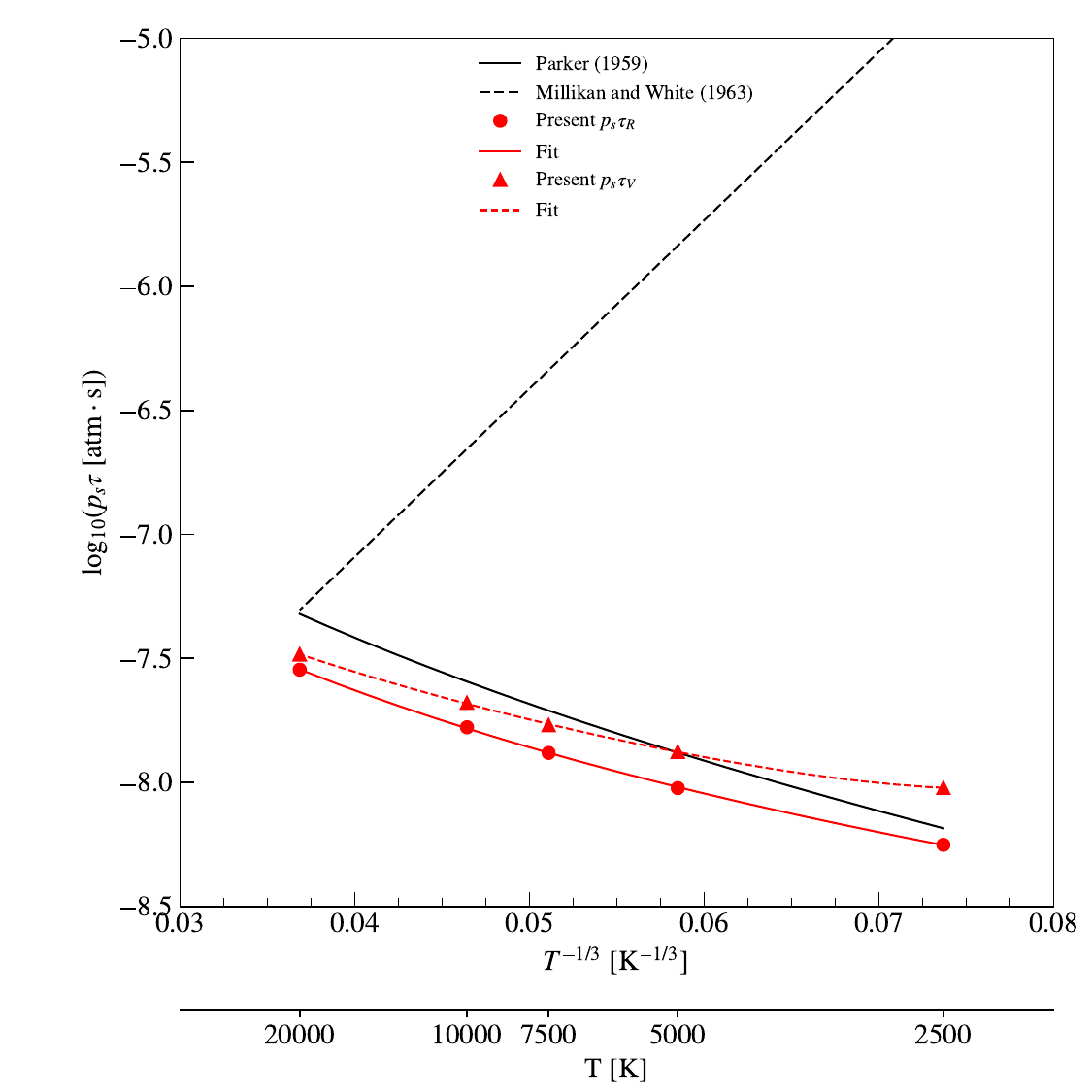}
 \caption{\ce{CN(X^2\Sigma^{+}) + O(^3P)}.}
 \label{Tau_CNO}
\end{subfigure}
\begin{subfigure}[t]{0.49\textwidth}
 \centering
 \includegraphics[width=\textwidth]{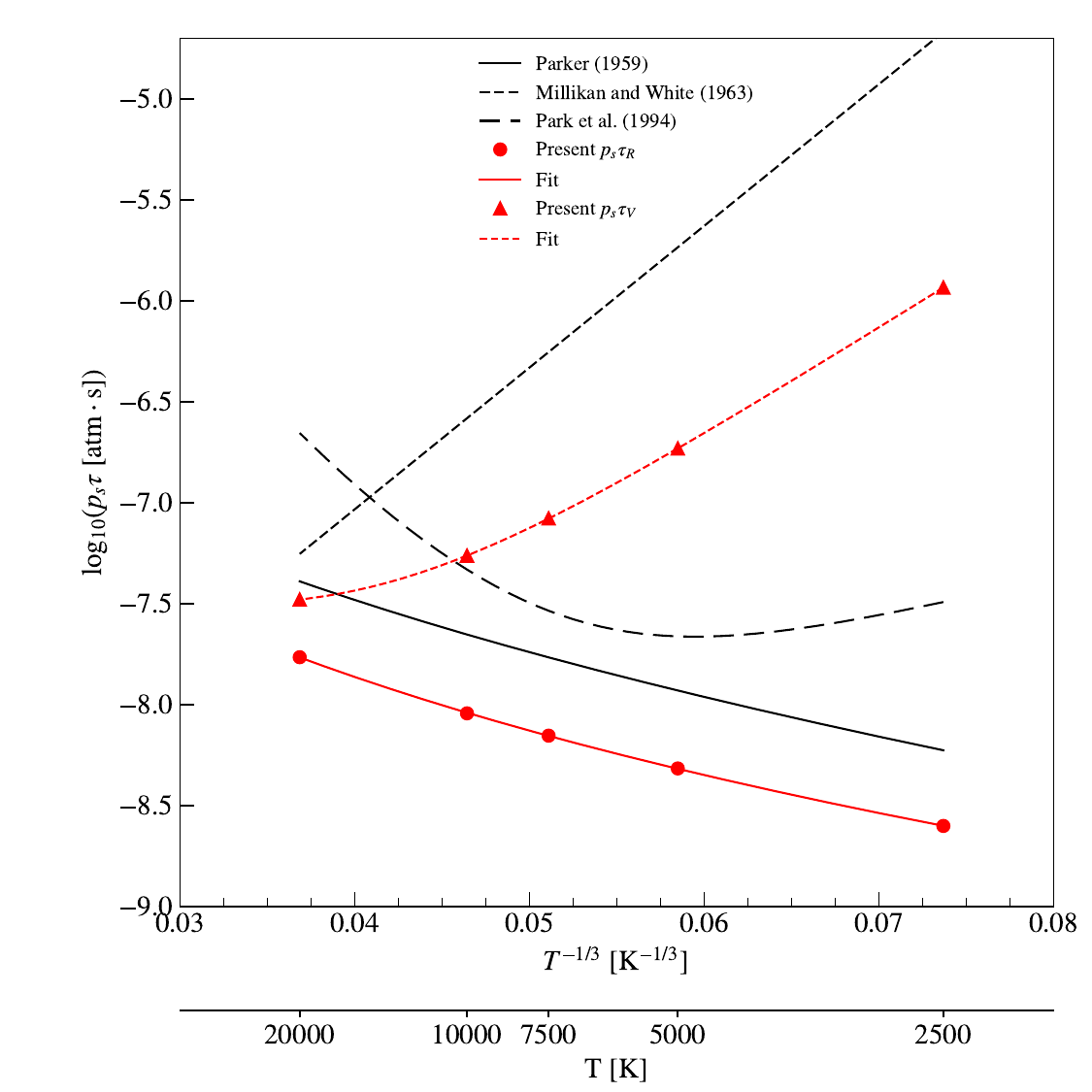}
 \caption{\ce{CO(X^1\Sigma^{+}) + N(^4S)}.}
\end{subfigure}
\begin{subfigure}[t]{0.49\textwidth}
 \centering
 \includegraphics[width=\textwidth]{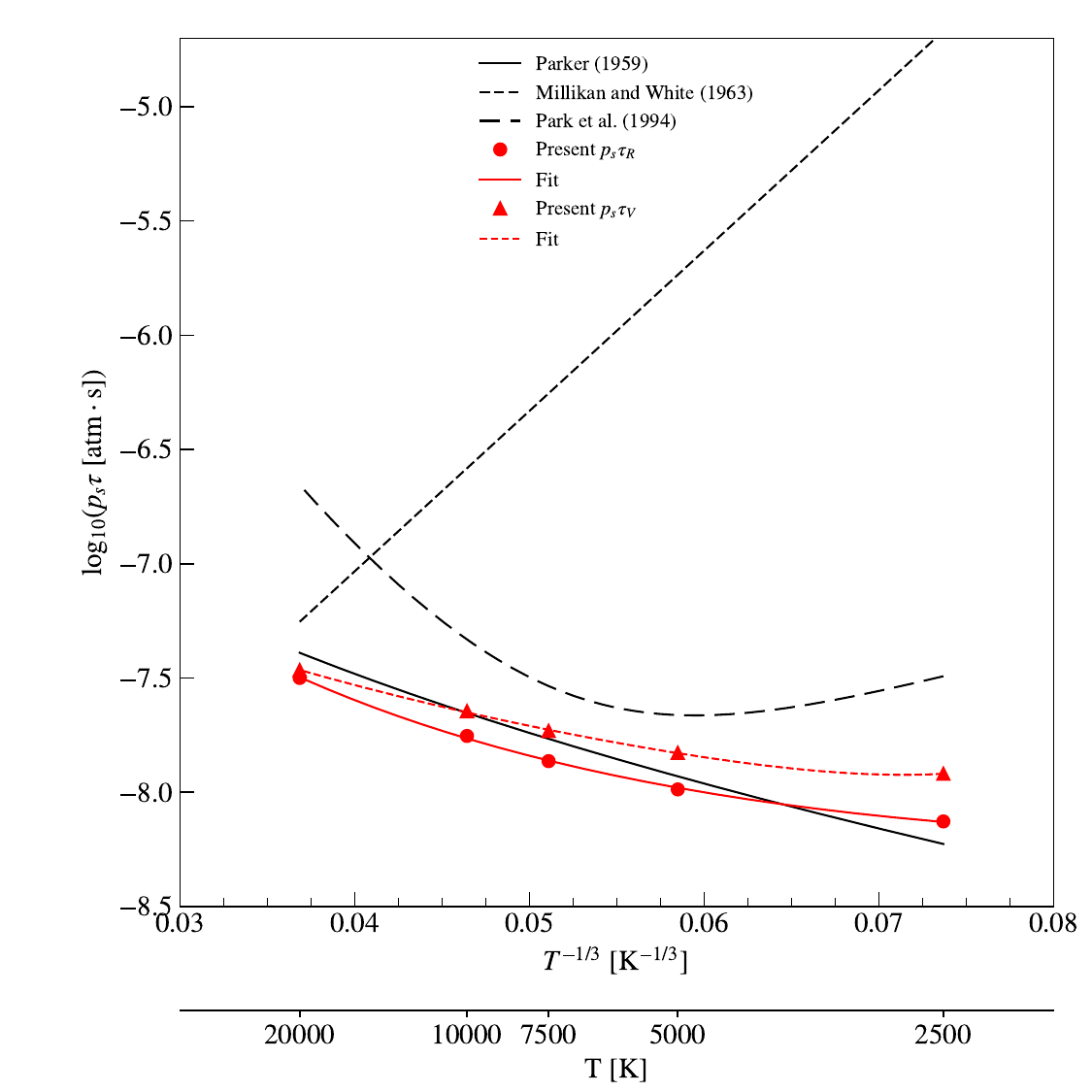}
 \caption{\ce{CO(X^1\Sigma^{+}) + N(^2D)}.}
\end{subfigure}

\caption{Comparison of the rotational and vibrational relaxation times with existing semi-empirical models\cite{Parker1959RotationalandVibrational,Millikan1963Relaxation,park1993review,park1994review} (dissociation and exchange processes excluded).}
\label{RelaxationTimes}
\end{figure*}

Figure~\ref{RelaxationTimes} compares the global rotational ($p_s\tau_{\text{r}}$) and vibrational ($p_s\tau_{\text{v}}$) relaxation times against a series of semi-empirical models,~ \cite{Parker1959RotationalandVibrational,Millikan1963Relaxation,park1993review,park1994review} where $p_s$ is the partial pressure of the atomic collision partner. The relaxation times are defined as the e-folding times of the average rotational and vibrational energies, i.e.~the time at which the average energy has covered a fraction $1 - 1/e$ of its total variation between the initial and the equilibrium values. For asymmetric systems, the energetic structures of the PESs, i.e.~linear and non-linear intermediates, and local minima, play a significant role in the inelastic processes. Therefore, it must be emphasised that the relaxation times reported in Figs.~\ref{Tau_NOC} and~\ref{Tau_CNO} are obtained from the master equations solved with the total rate coefficients, i.e.~the electronically weighted combination of the $^2A^{\prime}$ and $^4A^{\prime\prime}$ contributions, and therefore reflect the competition between the two symmetry states. For brevity, the symmetry-state specific relaxation times for the \ce{NO(X^2\Pi) + C(^3P)} and \ce{CN(X^2\Sigma^{+}) + O(^3P)} asymptotes, obtained by solving the master equations with the rate coefficients of a single surface, are reported in the Supplementary Material (see Fig.~\ref{S-PESSPecific_TauRTauV}).

The relaxation times obtained from the master equation analysis are curve-fitted using a modified Millikan--White-type exponential form~\cite{Millikan1963Relaxation} for the rotational relaxation time, and the Millikan and White expression, including the high-temperature correction term introduced by Park,~ \cite{park1993review} for the vibrational relaxation time. For brevity, the functional forms and recommended fitting coefficients are reported in the Supplementary Material (see Tab.~\ref{S-tab:FittingParametersTauRTauV}).

Figure~\ref{RelaxationTimes} shows that the Parker model~\cite{Parker1959RotationalandVibrational} closely reproduces the rotational relaxation times of all collisional systems. In contrast, the Millikan--White correlation~\cite{Millikan1963Relaxation} poorly describes the vibrational relaxation times, overestimating them by up to approximately three orders of magnitude at $T$ = 2500~K. Similarly, the vibrational relaxation model of Park,~\cite{park1993review,park1994review} which combines the Millikan--White form with tabulated parameters and a high-temperature correction, remains within a factor of about seven for the \ce{NO(X^2\Pi) + C(^3P)} and \ce{CO(X^1\Sigma^{+}) + N(^2D)} systems, whereas it underestimates the vibrational relaxation time of the \ce{CO(X^1\Sigma^{+}) + N(^4S)} system by more than one order of magnitude at $T$ = 2500~K. These observations highlight known shortcomings of such semi-empirical models when describing relaxation times obtained from first-principles calculations.~ \cite{Torres2025Characteristic} 

Additionally, Fig.~\ref{RelaxationTimes} indicates that, with the exception of the \ce{CO(X^1\Sigma^{+}) + N(^4S)} system, the rotational and vibrational relaxation times exhibit comparable time scales and unconventional temperature dependences, suggesting that the assumption of both energy modes being fully decoupled is invalid. For \ce{CO(X^1\Sigma^{+}) + N(^4S)}, the vibrational relaxation time exceeds its rotational counterpart by more than two orders of magnitude at $T$ = 2500~K and follows a Millikan--White-type dependence. Interestingly, similar conclusions were also reported for another asymmetric system~\cite{Sharma2022HCN}. Such behaviour is mainly attributed to the topology of the PESs, allowing trajectories to transition through these energetic structures before returning to their original reactive channel. In contrast, the slower vibrational relaxation of the \ce{CO(X^1\Sigma^{+}) + N(^4S)} system is attributed to the absence of a bound intermediate below the \ce{CO(X^1\Sigma^{+}) + N(^4S)} entrance channel, the \ce{NCO}($^4A^{\prime\prime}$) minimum, so that the energy transfer proceeds through direct, impulsive collisions rather than through a long-lived complex.

 \subsubsection{Dissociation Processes}

\begin{figure*}[!ht]
\centering
\begin{subfigure}[t]{0.49\textwidth}
 \centering
 \includegraphics[width=\textwidth]{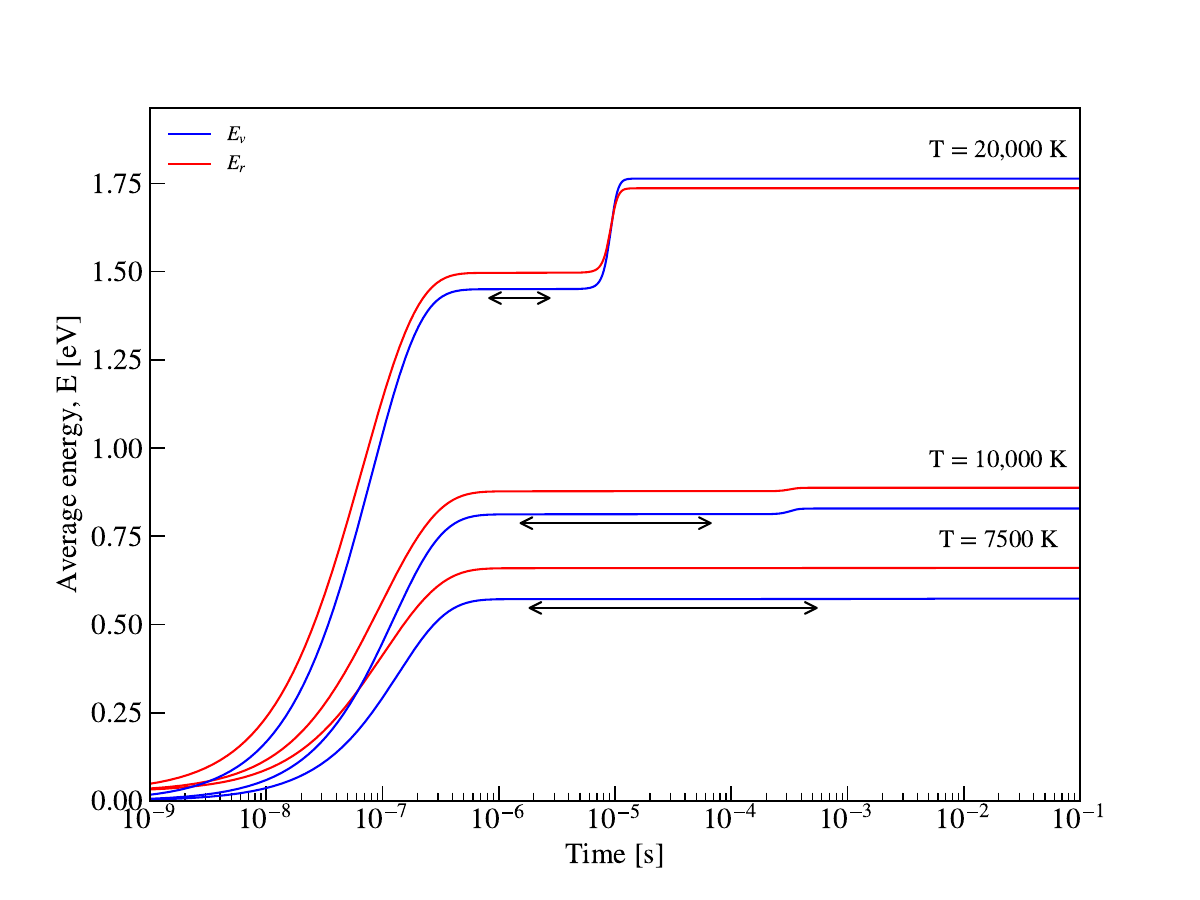}
 \caption{Average rotational and vibrational energies.}
\end{subfigure}
\begin{subfigure}[t]{0.49\textwidth}
 \centering
 \includegraphics[width=\textwidth]{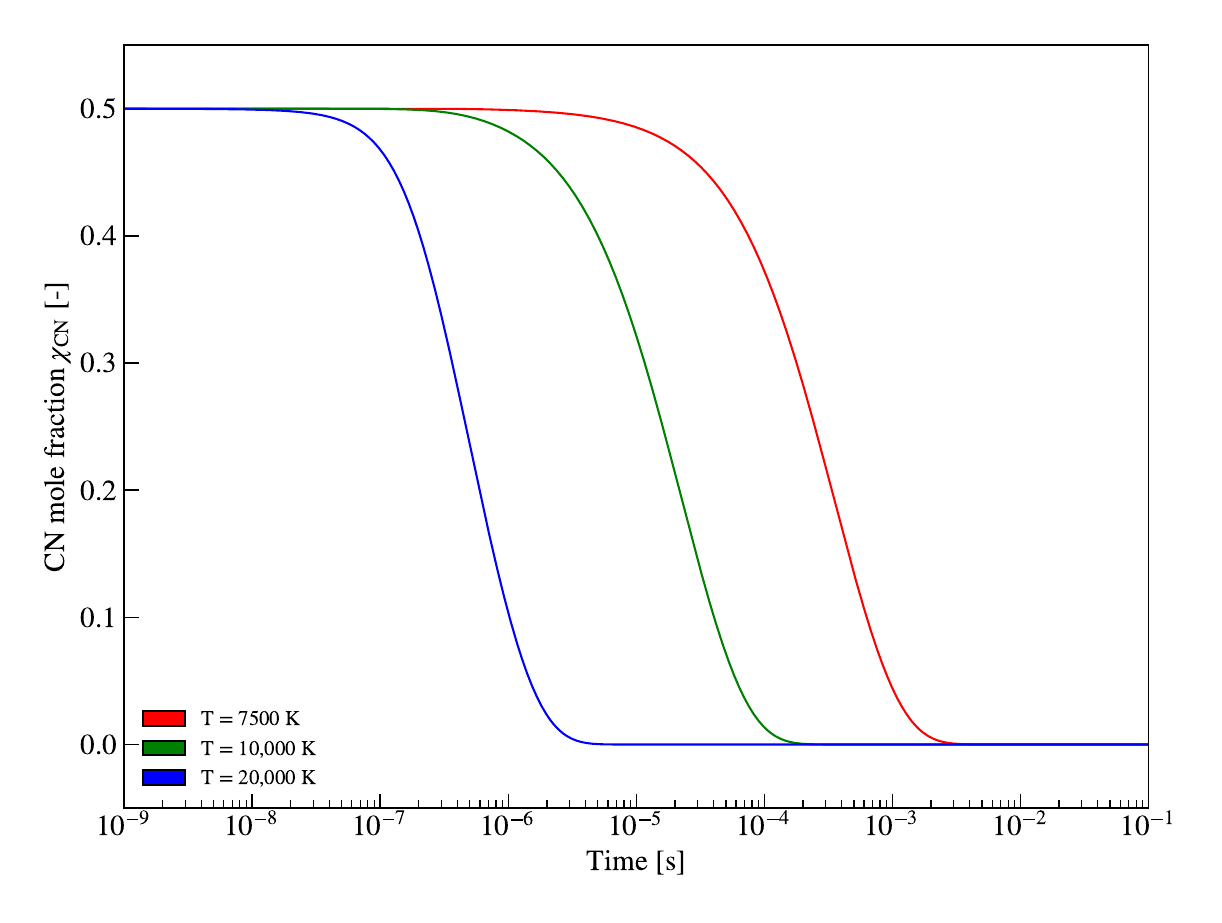}
 \caption{Mole fraction.}
\end{subfigure}
\caption{Temporal evolution of the average rotational and vibrational energies of \ce{CN(X^2\Sigma^{+})} and of its mole fraction at $T$ = 7500, 10,000 and 20,000~K (heterogeneous exchange reactions excluded).}
\label{AverageEnergy_MoleFraction}
\end{figure*}

The dissociation and the recombination processes are now included in addition to the inelastic processes. Figure~\ref{AverageEnergy_MoleFraction} shows the evolution of the average rotational and vibrational energies of \ce{CN(X^2\Sigma^{+})}, and of its corresponding mole fraction. For brevity, the present section only reports results for the \ce{CN + O} collisional system, while similar results for the other two collisional systems are reported in the Supplementary Material (see Figs.~\ref{S-DissEnergyMolar_NOC} -- \ref{S-DissEnergyMolar_CON2D}). 

%

\begin{figure*}
 \centering
 \includegraphics[width=\textwidth]{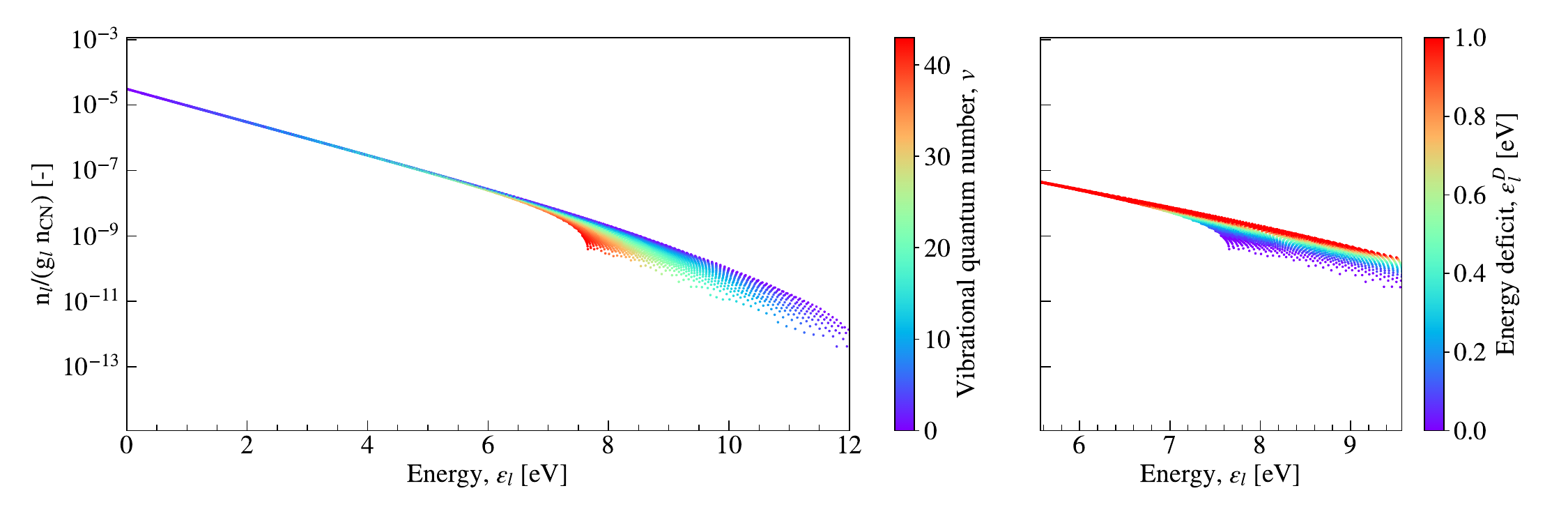}
 \caption{Rovibrational distribution of the \ce{CN(X^2\Sigma^{+}) + O(^3P)} collisional system at the midpoint of the QSS regime of the total dissociation at $T$ = 10,000~K (heterogeneous exchange reactions excluded). The distributions are coloured by the vibrational quantum number and energy deficit.}
 \label{DistributionAmidQSS}
\end{figure*}

To gain further physical insight into the dissociation processes, the rovibrational distribution function is examined at the midpoint of the QSS regime. Figure~\ref{DistributionAmidQSS} presents, in the left panel, the population distribution of \ce{CN(X^2\Sigma^{+})} coloured by its vibrational quantum number, and in the right panel, a zoom of the rovibrational states near the rotationless dissociation limit ($J = 0, v = v_{\text{max}}$), coloured by the state-specific energy deficit. For brevity, the present section only reports results for the aforementioned collisional system. Similar results for the other collisional systems are reported in the Supplementary Material (see Fig.~\ref{S-MidQSS_Others}).

For all collisional systems, it is shown that the high-lying states are not in equilibrium with each other and that dissociation is not governed by vibrational-specific dynamics. Instead, the right panels of Figs.~\ref{DistributionAmidQSS} and~\ref{S-MidQSS_Others} show that the population of the high-lying states is strongly correlated with the state-specific energy deficit, suggesting that the dissociation mechanism is primarily driven by the energetic distance of a state from the centrifugal barrier.

\begin{figure}
 \centering
 \includegraphics[width=\columnwidth]{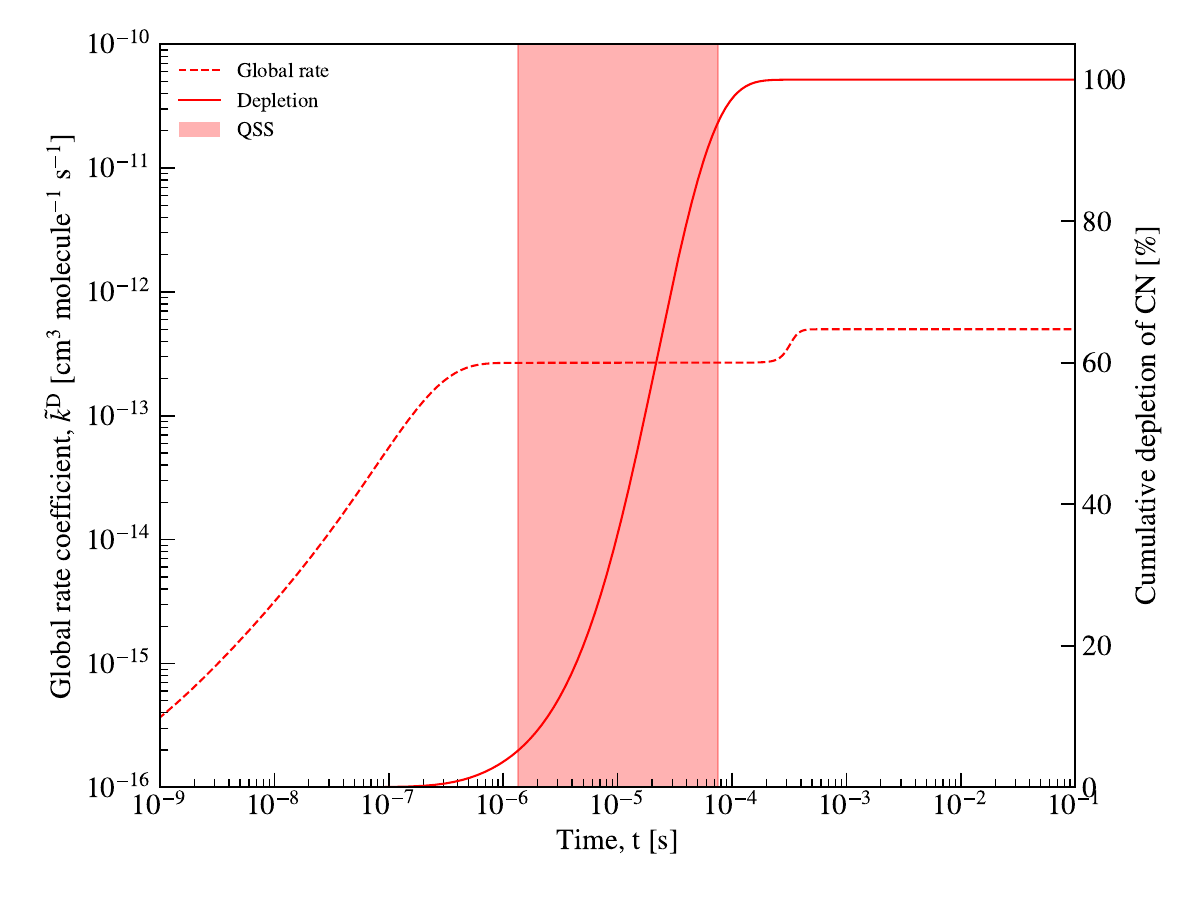}
 \caption{Time-cumulative depletion and global dissociation rate coefficient of \ce{CN(X^2\Sigma^{+}) + O(^3P)} at $T$ = 10,000~K.}
 \label{ValidityQSS}
\end{figure}

Finally, the validity of the QSS assumption for modelling dissociation reactions of asymmetric chemical systems is herein investigated. Figure~\ref{ValidityQSS} shows the time-cumulative depletion and the global dissociation rate coefficient of \ce{CN(X^2\Sigma^{+}) + O(^3P)} at $T$ = 10,000~K. The global dissociation rate coefficients are defined in the Supplementary Material, i.e.~Eqs.~ \eqref{S-eq:k_glob_D_NO} -- \eqref{S-eq:k_glob_D_CO}. Similar results for the other collisional systems are reported in the Supplementary Material (see Fig.~\ref{S-CumulativeProduction}).

Figure~\ref{ValidityQSS} indicates that, at $T$ = 10,000~K, the QSS regime accounts for the majority of the net depletion of the diatomic species. Specifically, approximately 78 \%, 89 \%, 72 \% and 93 \% of this depletion occur during the QSS regime for the \ce{NO(X^2\Pi) + C(^3P)}, \ce{CN(X^2\Sigma^{+}) + O(^3P)}, \ce{CO(X^1\Sigma^+) + N(^4S)} and \ce{CO(X^1\Sigma^+) + N(^2D)} systems, respectively. However, the QSS assumption strongly depends on the thermal conditions. Specifically, at $T$ = 5000~K, a large part of the depletion occurs outside the QSS regime, the fraction occurring within this regime being approximately 53 \%, 34 \%, 10 \% and 42 \% for the \ce{NO(X^2\Pi) + C(^3P)}, \ce{CN(X^2\Sigma^{+}) + O(^3P)}, \ce{CO(X^1\Sigma^+) + N(^4S)} and \ce{CO(X^1\Sigma^+) + N(^2D)} systems, respectively. These observations indicate that the QSS assumption holds only for moderate to high thermal conditions.

\begin{figure}[!ht]
\centering
\includegraphics[width=0.45\textwidth]{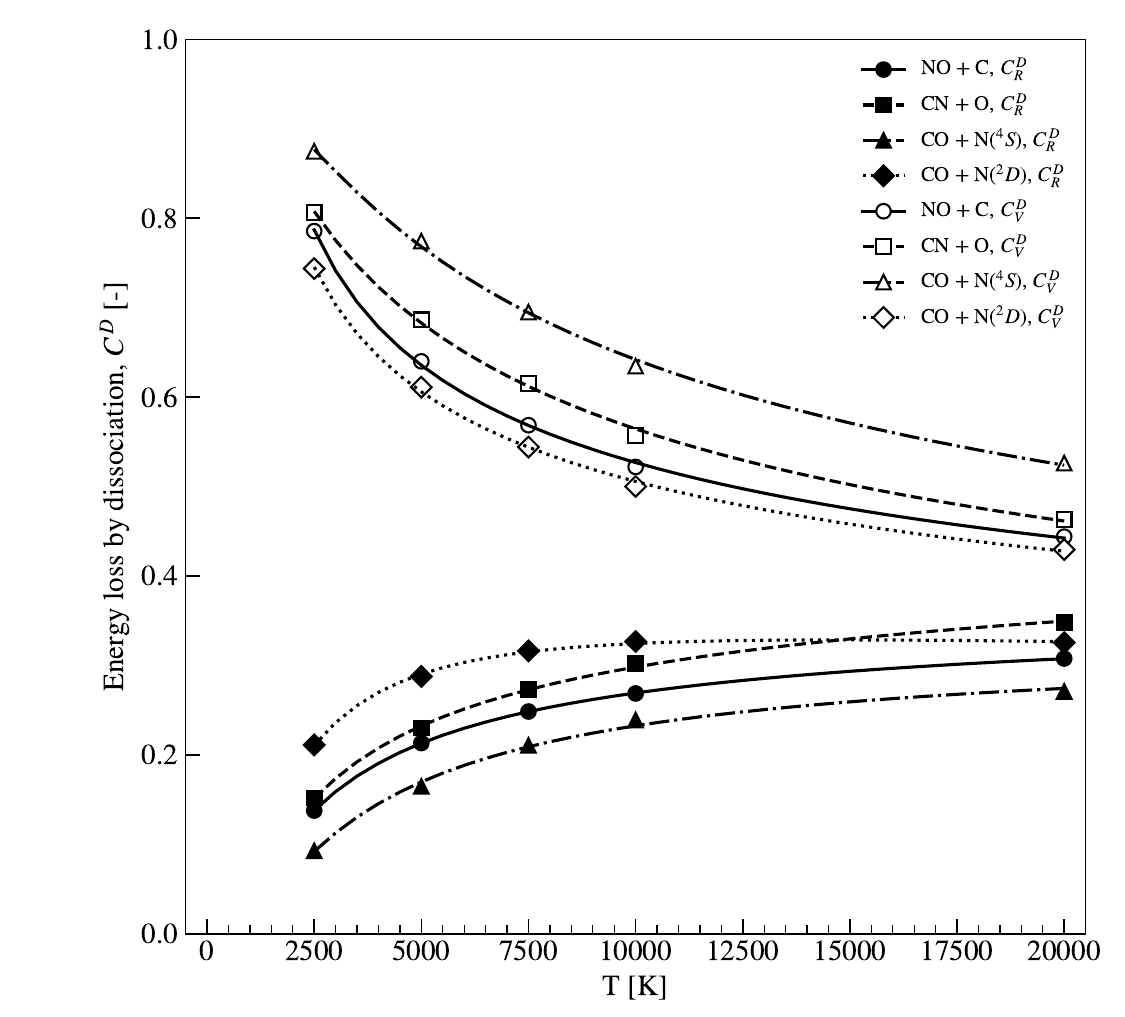}
\caption{Rotational and vibrational energy loss ratios due to dissociation as a function of the translational temperature.}
\label{EnergyLoss}
\end{figure}

Figure~\ref{EnergyLoss} shows the temperature dependence of the internal energy loss coefficients due to dissociation. These coefficients are extracted at the midpoint of the QSS regime of the total dissociation and follow the definition proposed by Panesi~\textit{et al.} \cite{panesi2013rovibrational} To facilitate the use of the present calculations in CFD simulations, the rotational and vibrational energy loss ratios due to dissociation have been curve-fitted onto a functional form with fitting parameters reported in the Supplementary Material (see Tabs.~\ref{S-RotEnergyLoss} and~\ref{S-VibEnergyLoss}). The rotational and vibrational energy loss coefficients of the \ce{NO(X^2\Pi) + C(^3P)} and \ce{CN(X^2\Sigma^{+}) + O(^3P)} systems appear to be comparable. This observation is consistent with the previous analysis, where both systems exhibit similar kinetics and possess comparable dissociation energies. In contrast, the \ce{CO(X^1\Sigma^+) + N(^4S, ^2D)} systems exhibit markedly different behaviour. In particular, \ce{CO(X^1\Sigma^+) + N(^2D)} displays a significantly smaller vibrational energy loss but a larger rotational energy loss than its \ce{N(^4S)} counterpart over the whole temperature range (see Tabs.~\ref{S-RotEnergyLoss} and~\ref{S-VibEnergyLoss}). This behaviour originates from the electronic energy carried by the first excited state of atomic nitrogen, which reduces the energy required for the collisional system to access the dissociation asymptote, i.e.~ \ce{C(^3P) + O(^3P) + N(^4S)}, thereby enabling dissociation from lower-lying vibrational levels and reducing the fraction of vibrational energy removed per dissociation.

\subsubsection{Heterogeneous Exchange Processes}

 The validity of the QSS assumption is assessed for the exchange processes. Considering that the \ce{CNO} system presents four reactive asymptotes, each heterogeneous exchange pathway is investigated separately. For the same reasons as in the previous sections, only the results for the \ce{NO(X^2\Pi) + C(^3P) <=> CN(X^2\Sigma^{+}) + O(^3P)} exchange reaction are herein presented, with the results of the other exchange reactions reported in the Supplementary Material.

\begin{figure*}[!ht]
\centering
\begin{subfigure}[t]{0.49\textwidth}
 \centering
 \includegraphics[width=\textwidth, trim=1.5cm 1cm 1cm 2cm, clip]{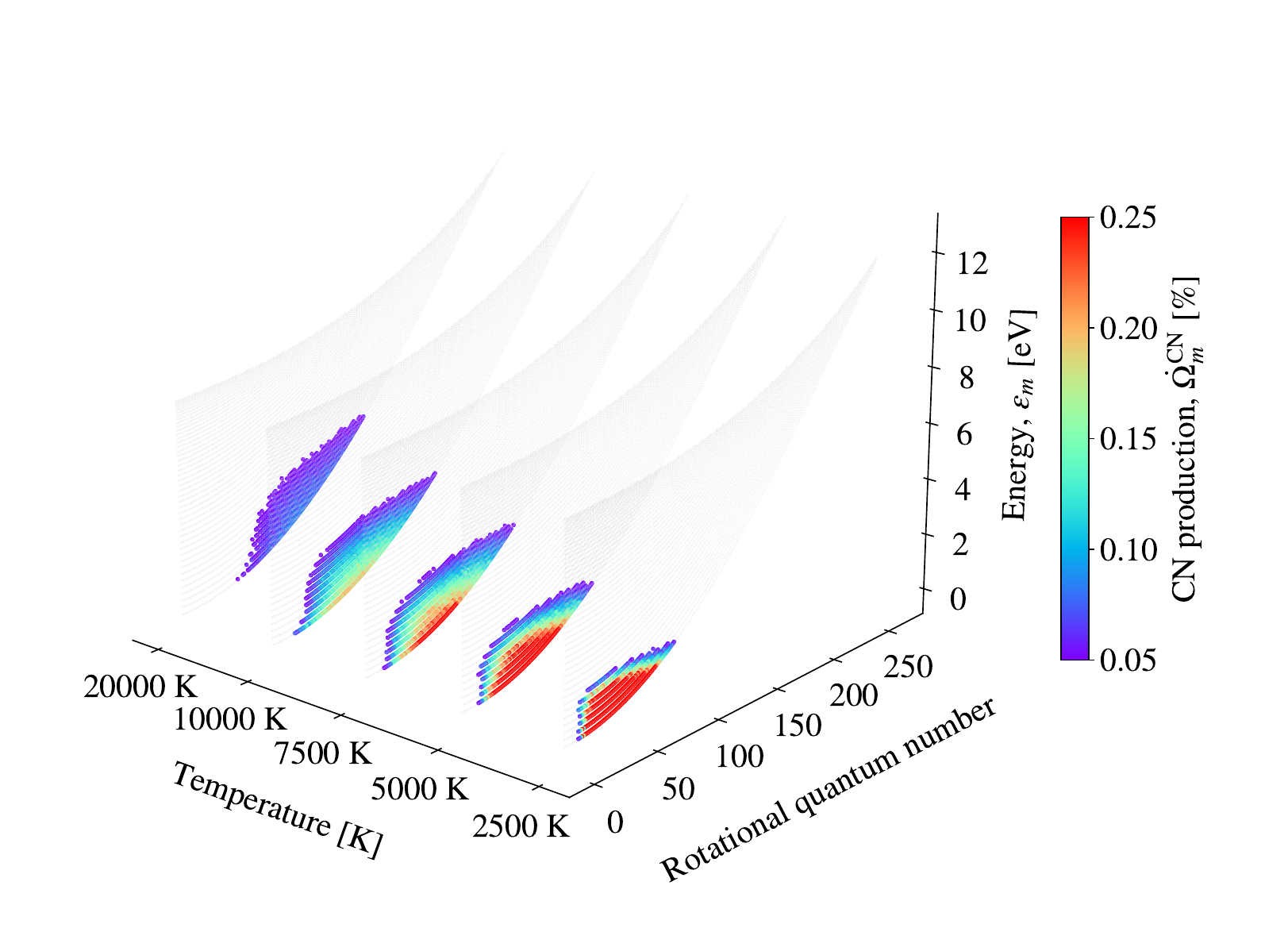}
 \caption{\ce{CN(X^2\Sigma^{+})} production.}
\end{subfigure}
\begin{subfigure}[t]{0.49\textwidth}
 \centering
 \includegraphics[width=\textwidth, trim=1.5cm 1cm 1cm 2cm, clip]{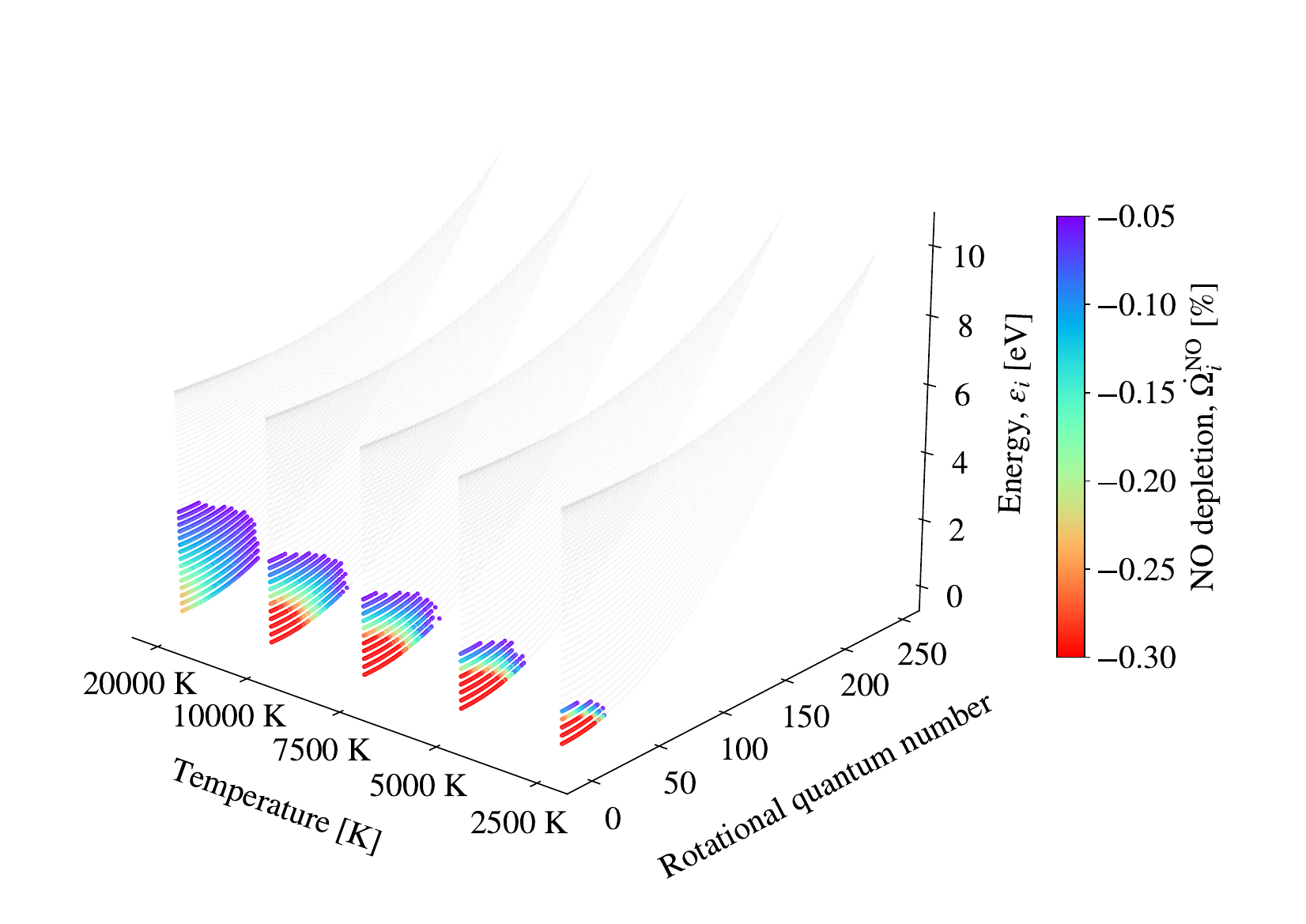}
 \caption{\ce{NO(X^2\Pi)} depletion.}
\end{subfigure}
\caption{Distribution of the exchange production and depletion rates for \ce{NO(X^2\Pi) + C(^3P) <=> CN(X^2\Sigma^{+}) + O(^3P)} at peak production time for five temperatures. The grey dots indicate that the production and depletion rates are below or above the threshold.}
\label{ProductionDepletion3D_NOC_CNO}
\end{figure*}

To gain further insight into the production and depletion mechanisms of \ce{CN(X^2\Sigma^{+})} and \ce{NO(X^2\Pi)}, respectively, through the aforementioned reaction, rovibrational state-resolved production and depletion rates are considered following the definition proposed by Jo~\textit{et al.} \ (see Eqs.~(27) and~(28) of Ref.~ \onlinecite{jo2022rovibrational}). These rates are evaluated at the time corresponding to the peak production for five thermal conditions and are presented in Fig.~\ref{ProductionDepletion3D_NOC_CNO}. It is shown that the low-lying states are the primary contributors to the conversion of \ce{NO(X^2\Pi)} into \ce{CN(X^2\Sigma^{+})}. Likewise, the production of \ce{CN(X^2\Sigma^{+})} is predominantly concentrated around its low-lying rovibrational states. Under low thermal conditions, the states contributing to both depletion and production remain strongly clustered within the low-energy region. However, as the temperature increases, the contributing region becomes more diffuse, involving a larger number of rovibrational states. The same conclusions hold for the other exchange processes reported in the Supplementary Material (see Figs.~\ref{S-ProdDep_CNO_CON4S} -- \ref{S-ProdDep_NOC_CON2D}).

\begin{figure*}[!ht]
\centering
\includegraphics[width=\textwidth]{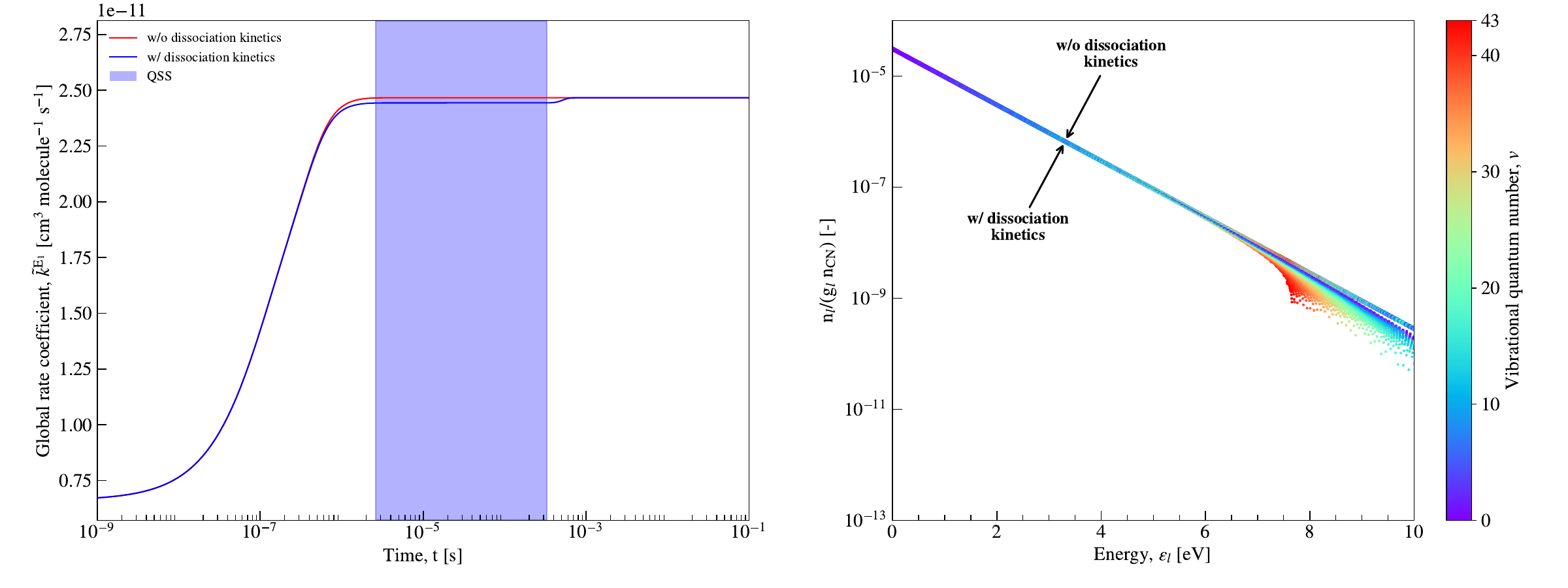}
\caption{Comparison of the global rate coefficients of the \ce{CN(X^2\Sigma^{+}) + O(^3P) -> NO(X^2\Pi) + C(^3P)} exchange reaction (left panel) and of the corresponding rovibrational distributions of \ce{CN(X^2\Sigma^{+})} at the midpoint of the QSS regime (right panel), with and without dissociation kinetics, at $T$ = 10,000~K. The initial mole fractions are $\chi_{\ce{CN}} = \chi_{\ce{O}} = 0.5$.}
\label{GlobalRates_Exchange}
\end{figure*}

Figure~\ref{GlobalRates_Exchange} presents the global exchange rate coefficient (left panel) and the corresponding rovibrational population distributions at the midpoint of the QSS regime (right panel), with the dissociation kinetics enabled or disabled. Note that the results of Fig.~\ref{GlobalRates_Exchange} are obtained for the reverse reactor, i.e.~ $\chi_{\ce{CN}} = \chi_{\ce{O}} = 0.5$, whereas Fig.~\ref{ProductionDepletion3D_NOC_CNO} refers to the \ce{NO(X^2\Pi) + C(^3P)} reactor, the corresponding results of which are reported in the Supplementary Material (see Fig.~\ref{S-GlobalRateExchange_NOC}). In the absence of dissociation kinetics, the global exchange rate coefficient increases monotonically towards its thermal value without an intermediate plateau distinct from thermal equilibrium, i.e.~no QSS regime is established by the exchange kinetics. When dissociation kinetics are included, a QSS plateau is formed, which suggests that the QSS regime is mainly governed by the dissociation kinetics; however, its value differs from the thermal one by about 1 \% at $T$ = 10,000~K. Similar observations can be made from the rovibrational population distributions extracted at the same time instant. Specifically, in the absence of dissociation kinetics, the rovibrational distribution indicates that the collisional system has already reached thermal equilibrium, whereas, when dissociation kinetics are included, the system exhibits a nonequilibrium rovibrational distribution. Comparable behaviour is also observed for the other exchange reactions reported in the Supplementary Material (see Figs.~\ref{S-GlobalRateExchange_NOC} -- \ref{S-GlobalRateExchange_CON2D}). 

\subsection{Rate Coefficients}

\begin{figure*}[t]
\centering
\begin{subfigure}[t]{0.32\textwidth}
 \centering
 \includegraphics[width=\textwidth]{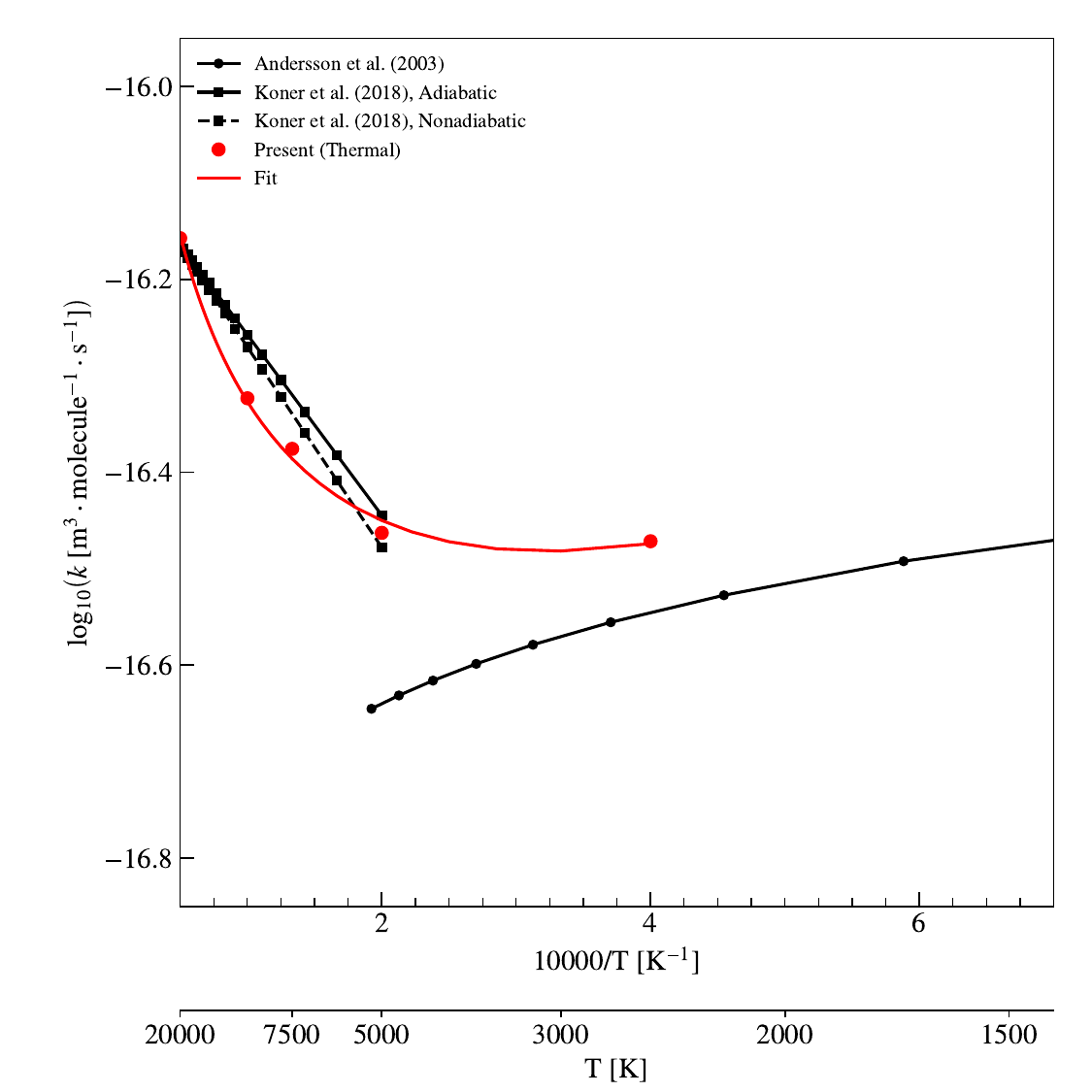}
 \caption{\small \mbox{\ce{NO(X^2\Pi) + C(^3P) ->}} \mbox{\ce{CN(X^2\Sigma^{+}) + O(^3P)}}.}
 \label{Exo1}
\end{subfigure}
\hfill
\begin{subfigure}[t]{0.32\textwidth}
 \centering
 \includegraphics[width=\textwidth]{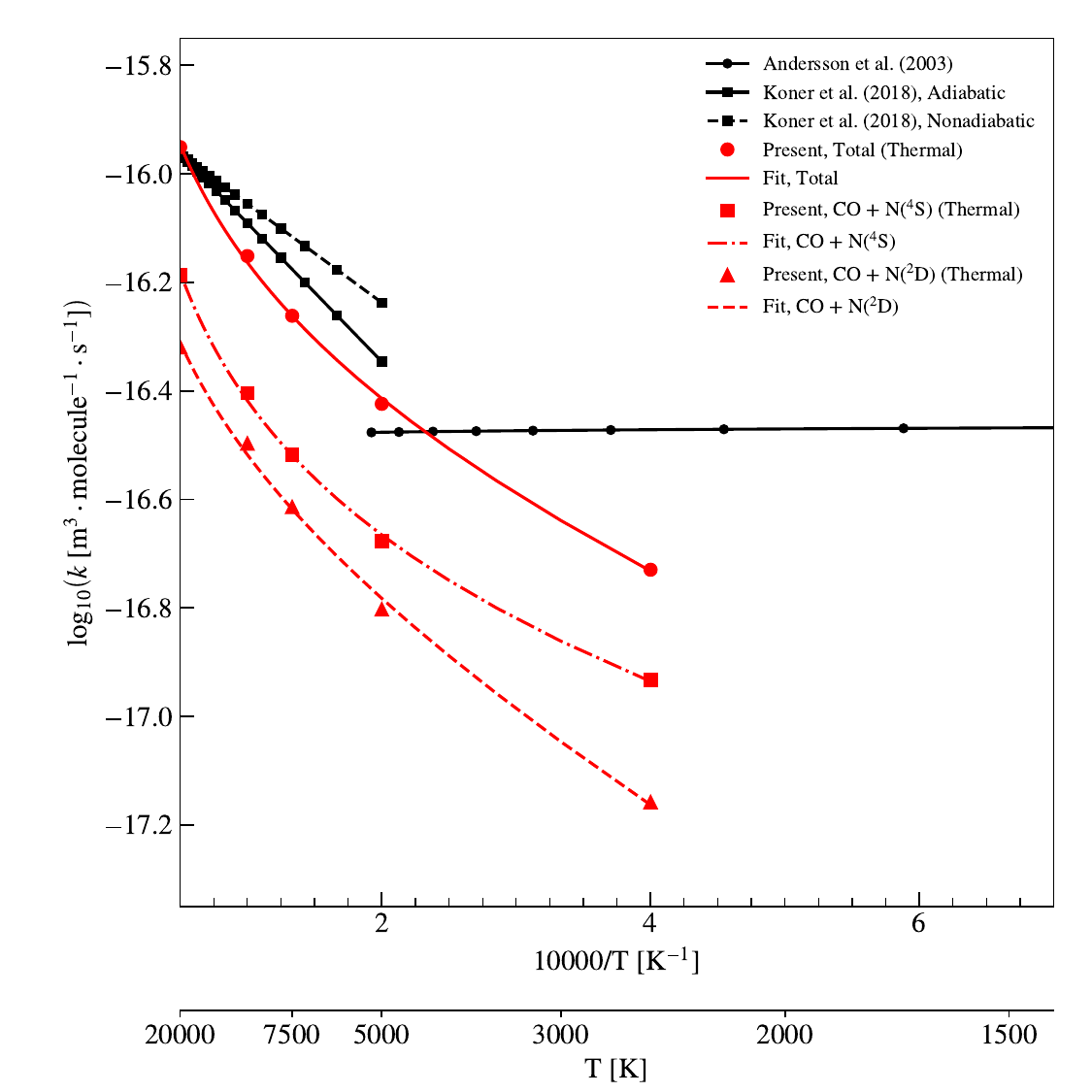}
 \caption{\small \mbox{\ce{NO(X^2\Pi) + C(^3P) ->}} \mbox{\ce{CO(X^1\Sigma^{+}) + N(^4S,^2D)}}.}
 \label{Exo2}
\end{subfigure}
\hfill
\begin{subfigure}[t]{0.32\textwidth}
 \centering
 \includegraphics[width=\textwidth]{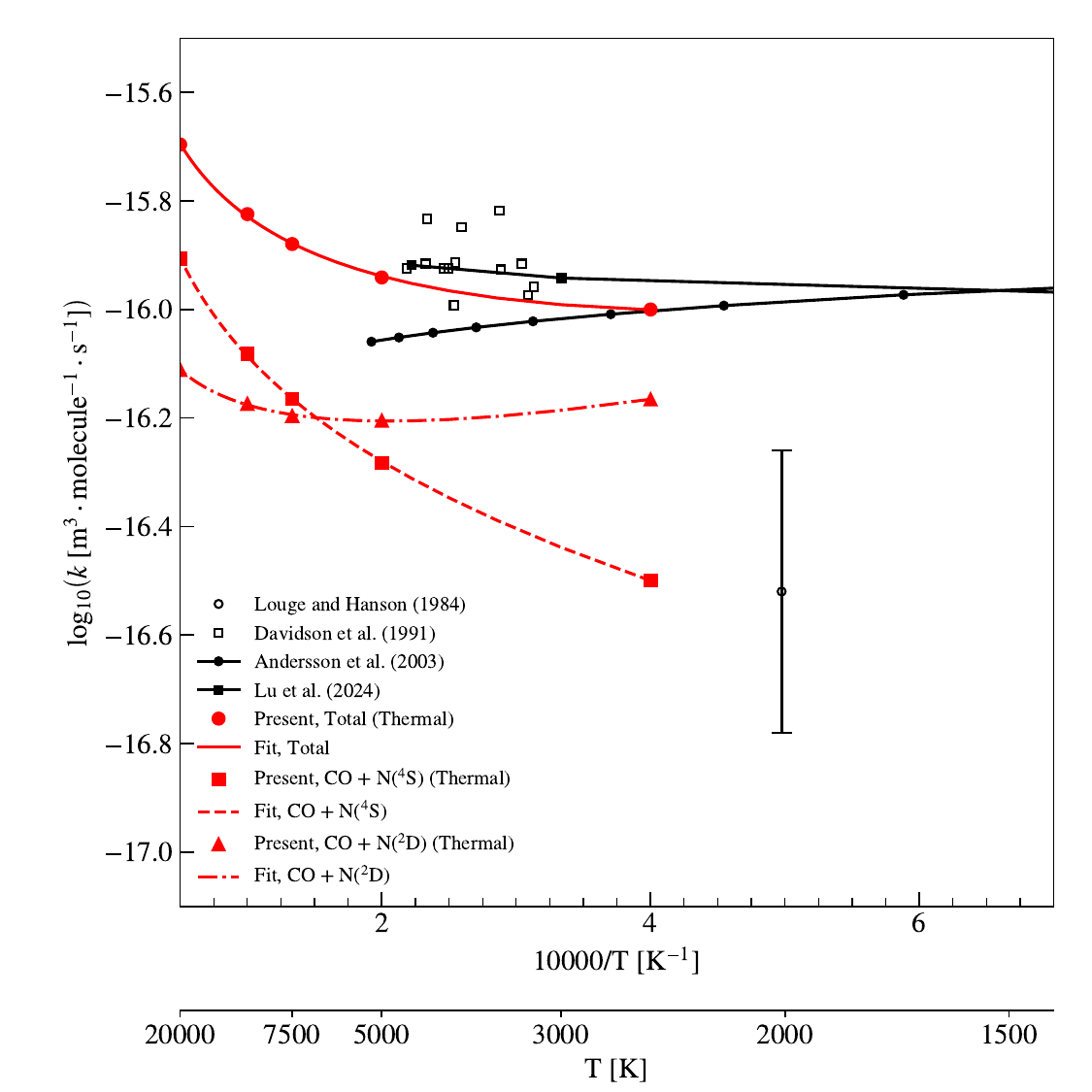}
 \caption{\small \mbox{\ce{CN(X^2\Sigma^{+}) + O(^3P) ->}} \mbox{\ce{CO(X^1\Sigma^{+}) + N(^4S,^2D)}}.}
 \label{Exo3}
\end{subfigure}
\caption{Comparison of the thermal rate coefficients for the exothermic exchange reactions against experiments\cite{Louge1984Cyanogen,Davidson1991ShockCN} and QCT calculations.~\cite{Andersson2003Computational,koner2018cno,Lu2024Theoretical}}
\label{ExoRateCoefficients}
\end{figure*}

\begin{figure*}[t]
\centering
\begin{subfigure}[t]{0.32\textwidth}
 \centering
 \includegraphics[width=\textwidth]{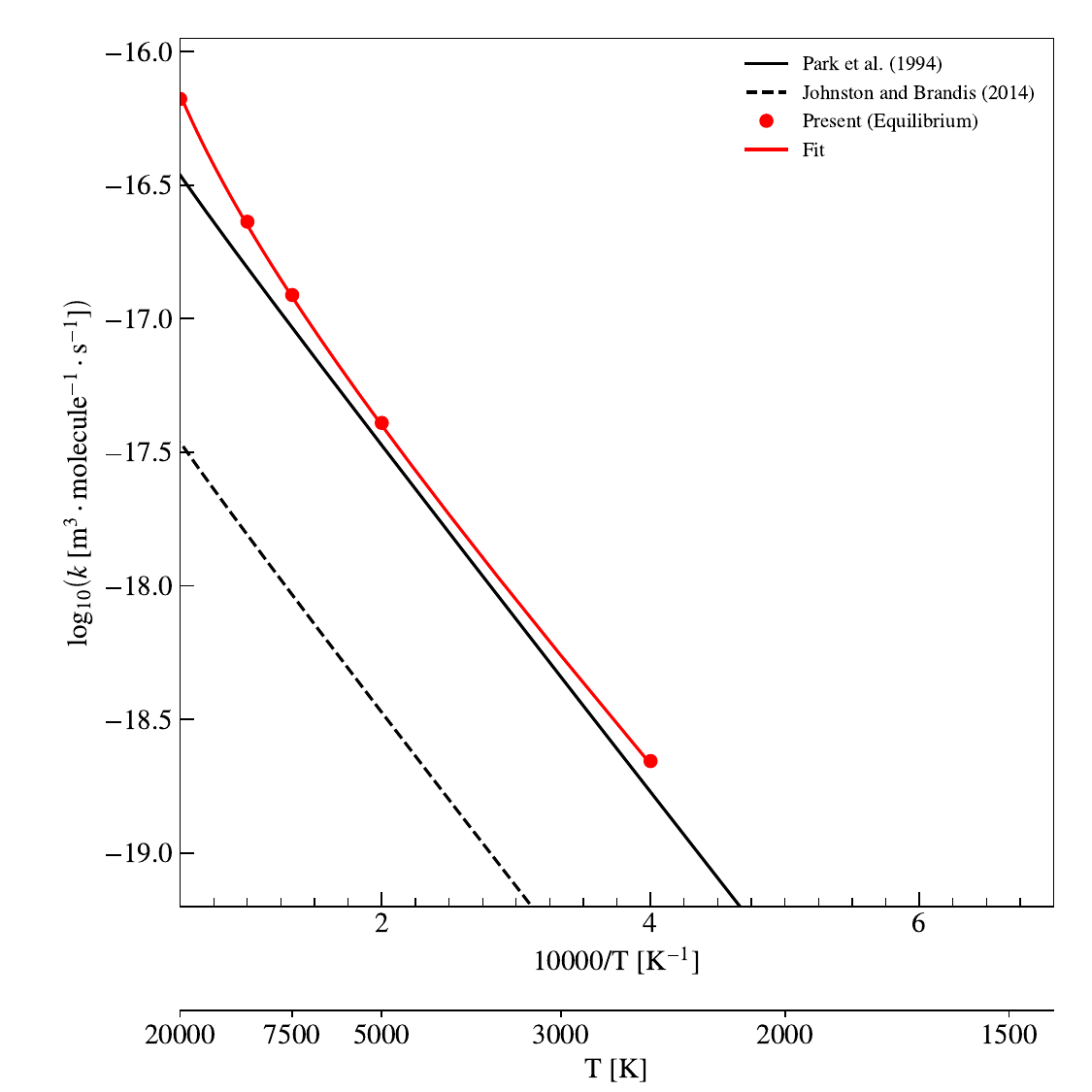}
 \caption{\small \mbox{\ce{CN(X^2\Sigma^{+}) + O(^3P) ->}} \mbox{\ce{NO(X^2\Pi) + C(^3P)}}.}
 \label{Endo1}
\end{subfigure}
\hfill
\begin{subfigure}[t]{0.32\textwidth}
 \centering
 \includegraphics[width=\textwidth]{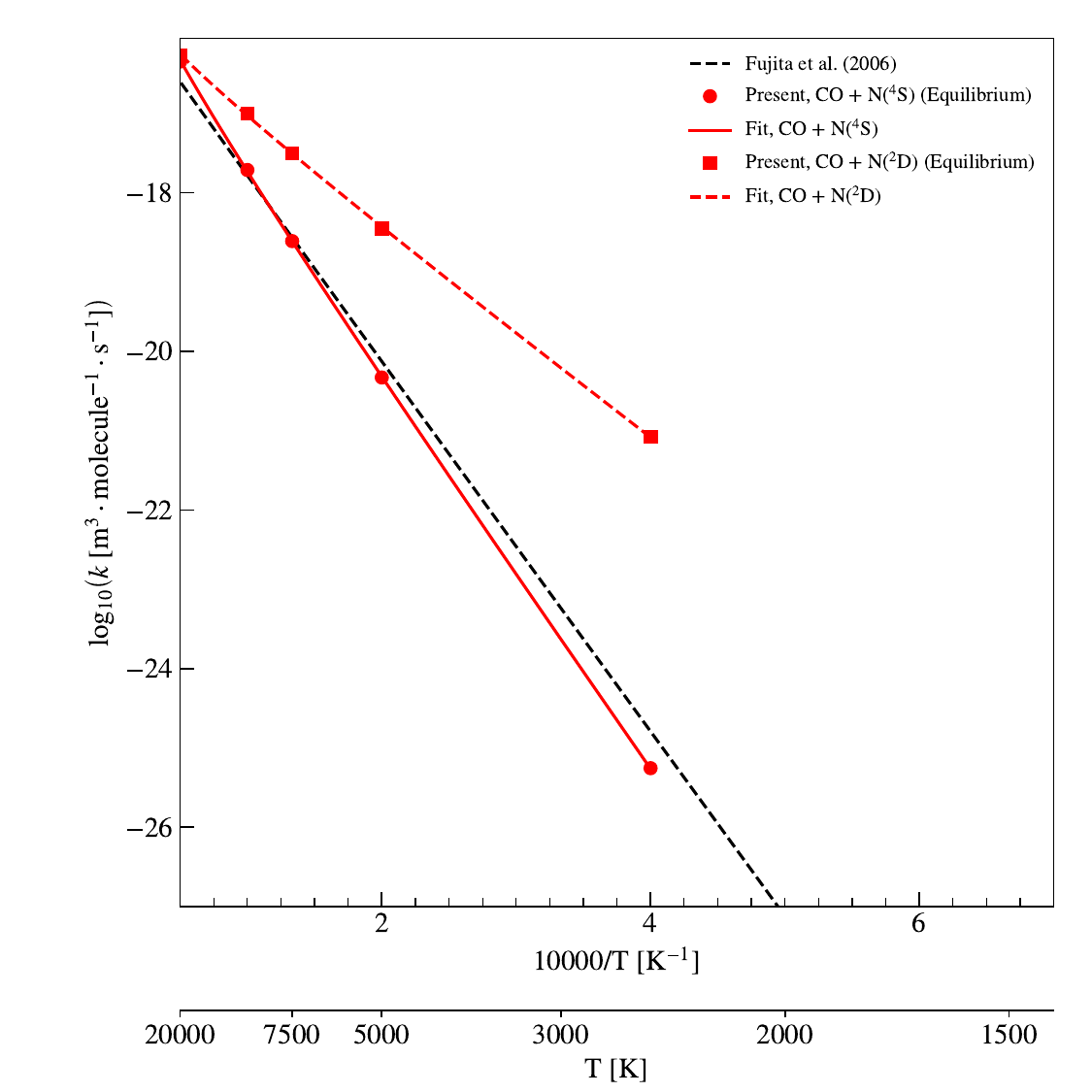}
 \caption{\small \mbox{\ce{CO(X^1\Sigma^{+}) + N(^4S,^2D) ->}} \mbox{\ce{NO(X^2\Pi) + C(^3P)}}.}
 \label{Endo2}
\end{subfigure}
\hfill
\begin{subfigure}[t]{0.32\textwidth}
 \centering
 \includegraphics[width=\textwidth]{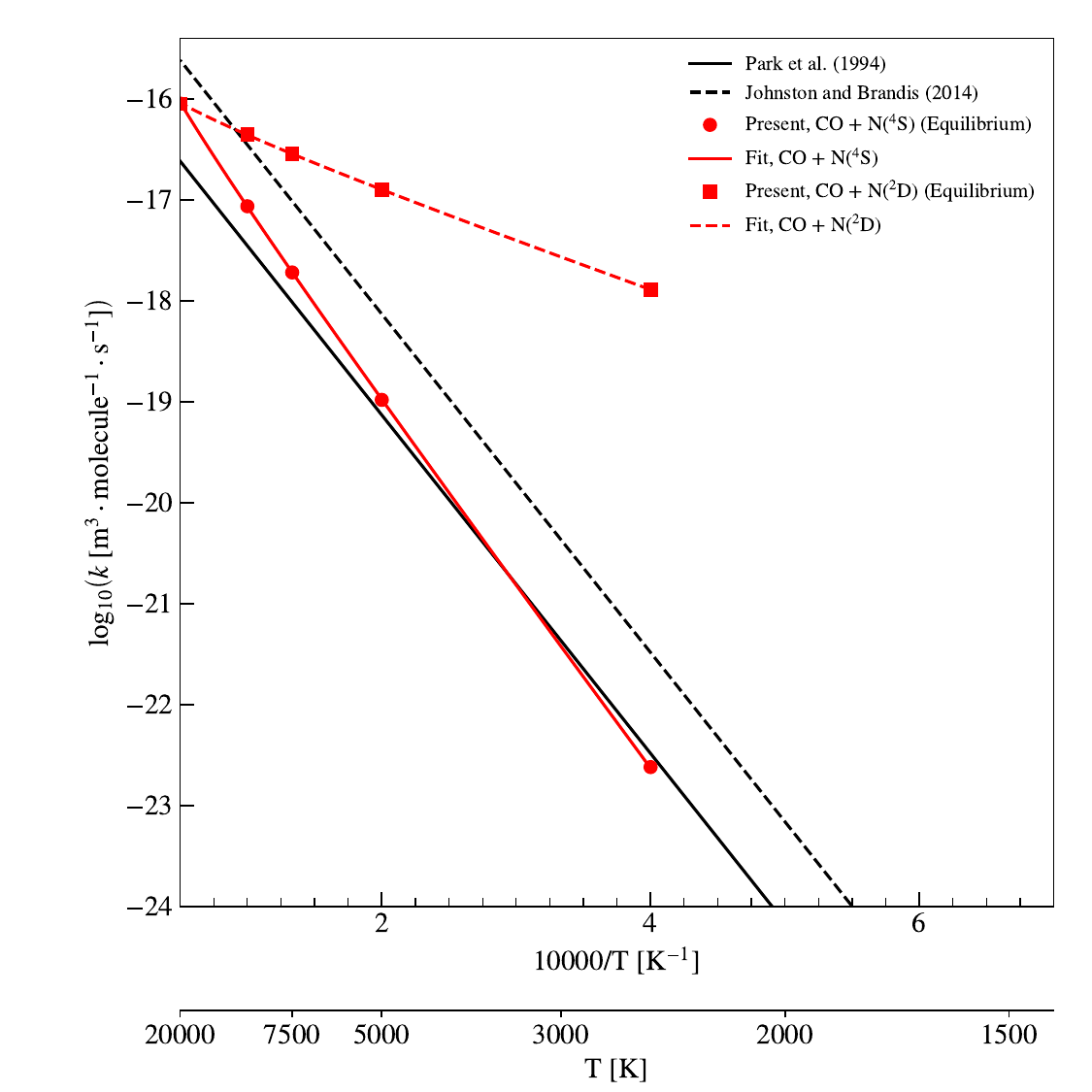}
 \caption{\small \mbox{\ce{CO(X^1\Sigma^{+}) + N(^4S,^2D) ->}} \mbox{\ce{CN(X^2\Sigma^{+}) + O(^3P)}}.}
 \label{Endo3}
\end{subfigure}
\caption{Comparison of the thermal rate coefficients for the endothermic exchange reactions against the chemistry models.~\cite{park1994review,Fujita2006Impact,Johnston2014Modeling}}
\label{EndoRateCoefficients}
\end{figure*}

\begin{figure*}[t]
\centering
\begin{subfigure}[t]{0.32\textwidth}
 \centering
 \includegraphics[width=\textwidth]{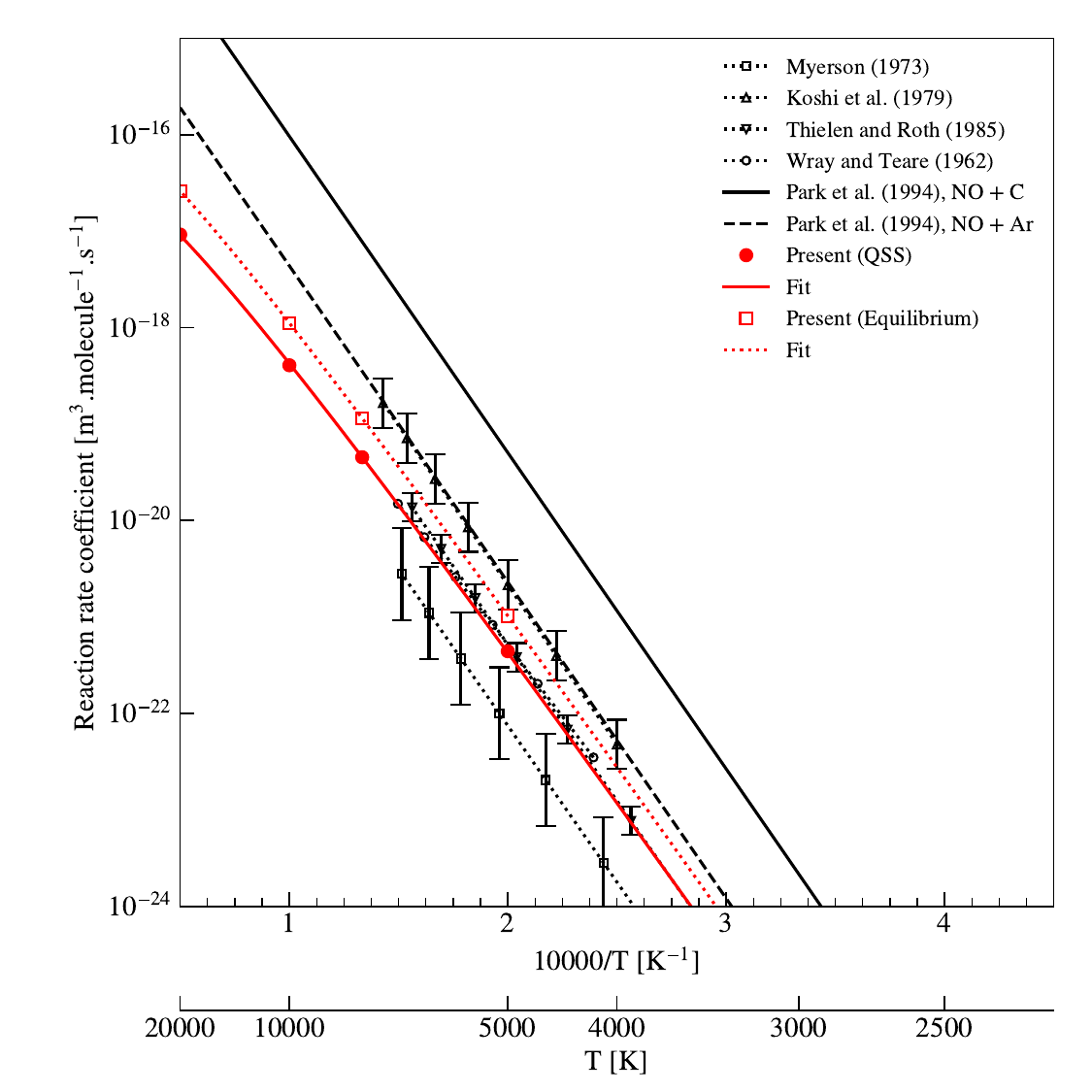}
 \caption{\small \mbox{\ce{NO(X^2\Pi) + C(^3P) ->}} \mbox{\ce{N(^4S) + O(^3P) + C(^3P)}}.}
 \label{Disso1}
\end{subfigure}
\hfill
\begin{subfigure}[t]{0.32\textwidth}
 \centering
 \includegraphics[width=\textwidth]{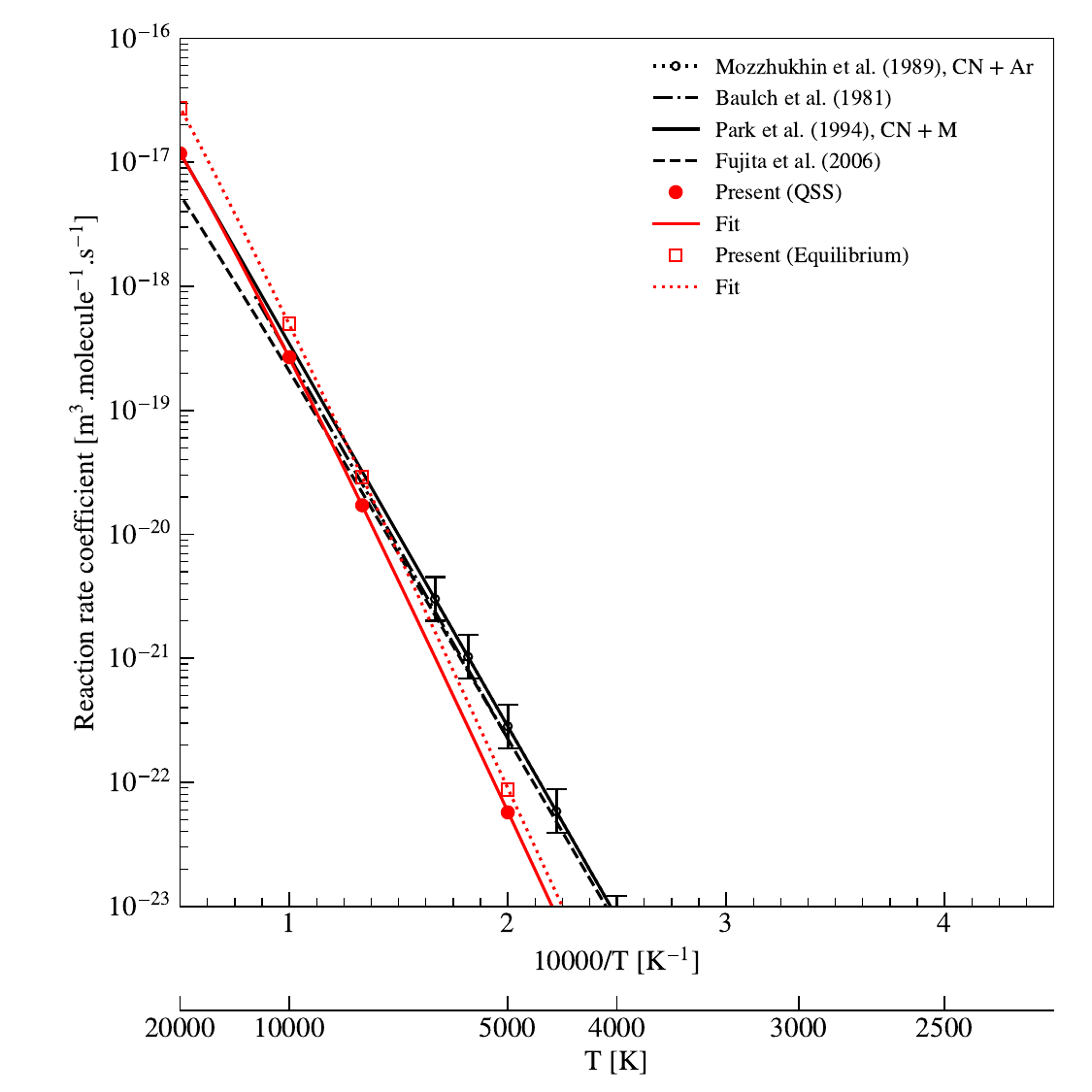}
 \caption{\small \mbox{\ce{CN(X^2\Sigma^{+}) + O(^3P) ->}} \mbox{\ce{C(^3P) + N(^4S) + O(^3P)}}.}
 \label{Disso2}
\end{subfigure}
\hfill
\begin{subfigure}[t]{0.32\textwidth}
 \centering
 \includegraphics[width=\textwidth]{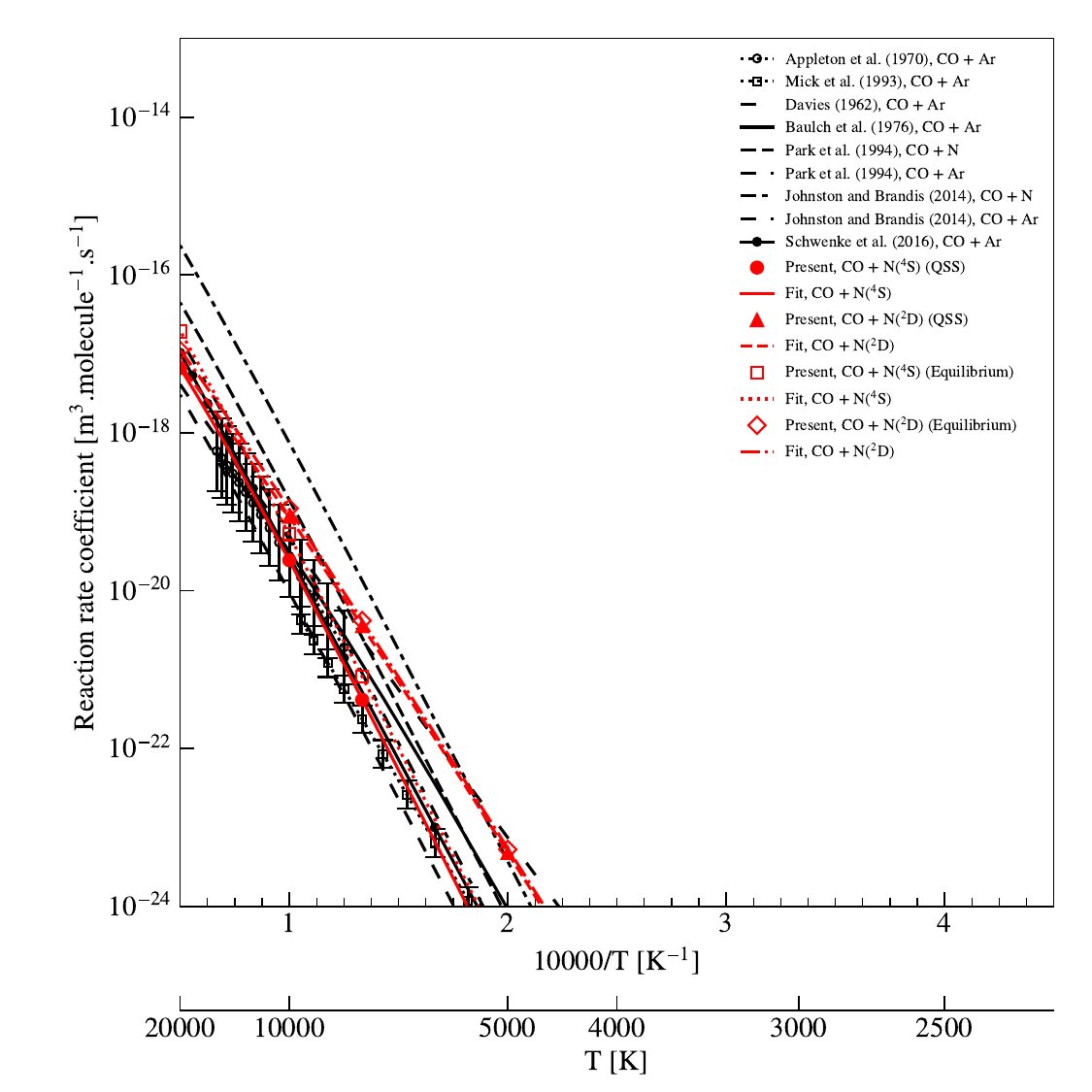}
 \caption{\small \mbox{\ce{CO(X^1\Sigma^{+}) + N(^4S,^2D) ->}} \mbox{\ce{C(^3P) + O(^3P) + N(^4S)}}.}
 \label{Disso3}
\end{subfigure}
\caption{Comparison of the QSS and equilibrium rate coefficients for the dissociation reactions against experiments,~\cite{Thielen1985Resonance,Wray1962Shock,Koshi1979Dissociation,Myerson1973Shock,Mozzhukhin1989CNDissociation,Appleton1970Shock,Mick1993Atomic} chemistry models\cite{park1994review,Baulch1976Evaluated,Baulch1981Evaluated,Davies1962Radiative,Fujita2006Impact,Johnston2014Modeling} and QCT calculations.~\cite{schwenke2016collisional}}
\label{DissoRateCoefficients}
\end{figure*}

This section presents the macroscopic rate coefficients for the exchange and dissociation reactions of the \ce{NO(X^2\Pi) + C(^3P)}, \ce{CN(X^2\Sigma^{+}) + O(^3P)} and \ce{CO(X^1\Sigma^{+}) + N(^4S,^2D)} collisional systems. Figures~\ref{ExoRateCoefficients}, \ref{EndoRateCoefficients} and~\ref{DissoRateCoefficients} summarise the thermal exchange rate coefficients and the QSS and equilibrium dissociation rate coefficients for these systems and compare them against the most representative studies available in the literature. Specifically, for the endothermic exchange reactions, comparisons are made against chemistry models.~\cite{park1994review,Fujita2006Impact,Johnston2014Modeling} For the exothermic exchange reactions, comparisons are performed against both experiments \cite{Louge1984Cyanogen,Davidson1991ShockCN} and QCT calculations.~\cite{Andersson2003Computational,koner2018cno,Lu2024Theoretical} For the dissociation reactions, comparisons are made against experiments,~\cite{Thielen1985Resonance,Wray1962Shock,Koshi1979Dissociation,Myerson1973Shock,Mozzhukhin1989CNDissociation,Appleton1970Shock,Mick1993Atomic} chemistry models \cite{park1994review,Baulch1976Evaluated,Baulch1981Evaluated,Davies1962Radiative,Fujita2006Impact,Johnston2014Modeling} and QCT calculations.~\cite{schwenke2016collisional}

Since the exchange reactions proceed at their thermal rate coefficients (see Fig.~\ref{GlobalRates_Exchange}), only thermal rate coefficients are reported hereafter, whereas QSS rate coefficients are reported for the dissociation reactions with their thermal counterparts being tabulated in the Supplementary Material for completeness. To facilitate the use of the present calculations by the high-enthalpy reactive gas flow community, these rate coefficients have been curve-fitted using a modified Arrhenius formulation and the corresponding fitting parameters are provided in the Supplementary Material (see Tab.~\ref{S-ArrheniusCoefficientFit}).

\begin{figure*}[t]
\centering
\begin{subfigure}[t]{0.32\textwidth}
 \centering
 \includegraphics[width=\textwidth]{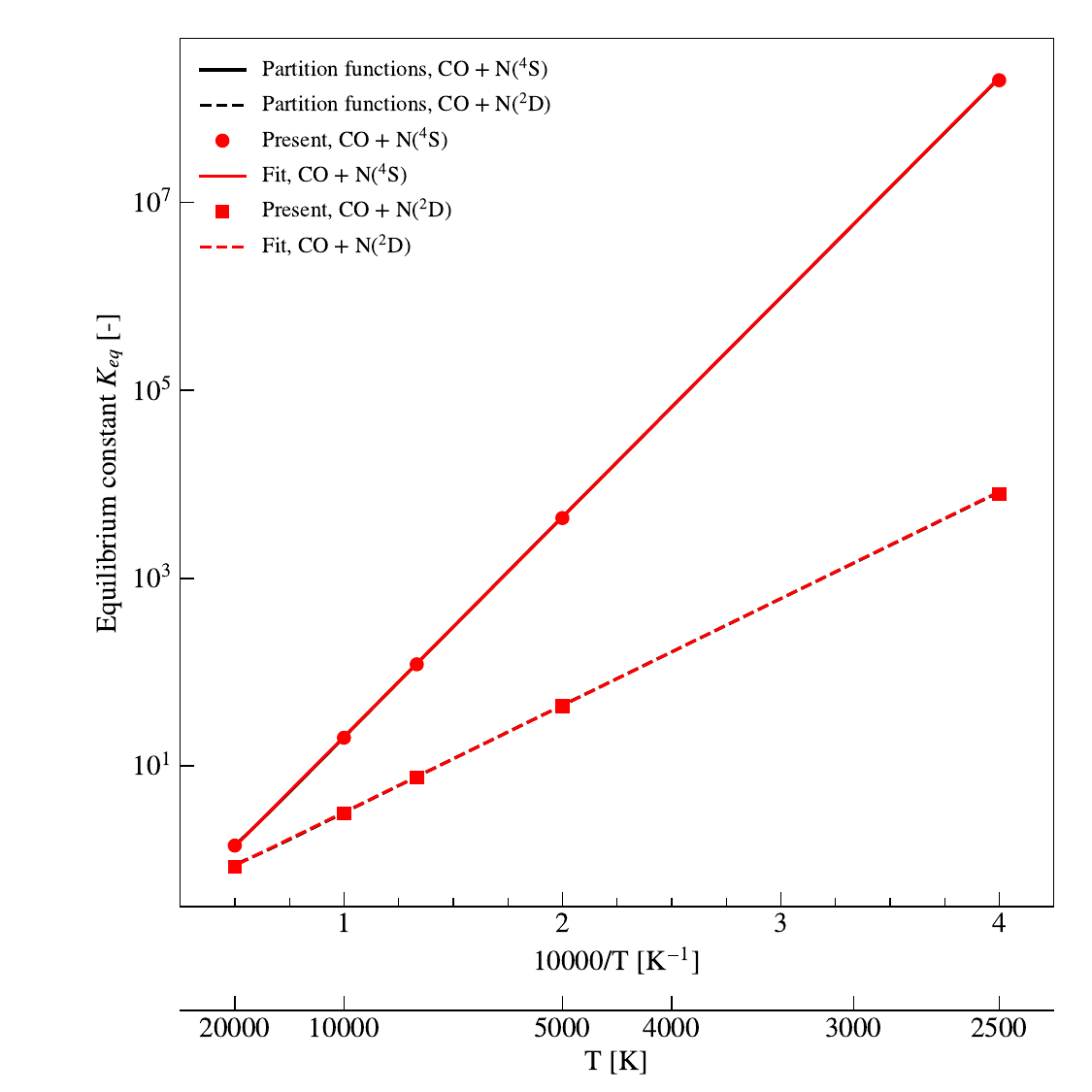}
 \caption{\small \mbox{\ce{NO(X^2\Pi) + C(^3P) <=>}} \mbox{\ce{CO(X^1\Sigma^{+}) + N(^4S,^2D)}}.}
 \label{Keq1}
\end{subfigure}
\hfill
\begin{subfigure}[t]{0.32\textwidth}
 \centering
 \includegraphics[width=\textwidth]{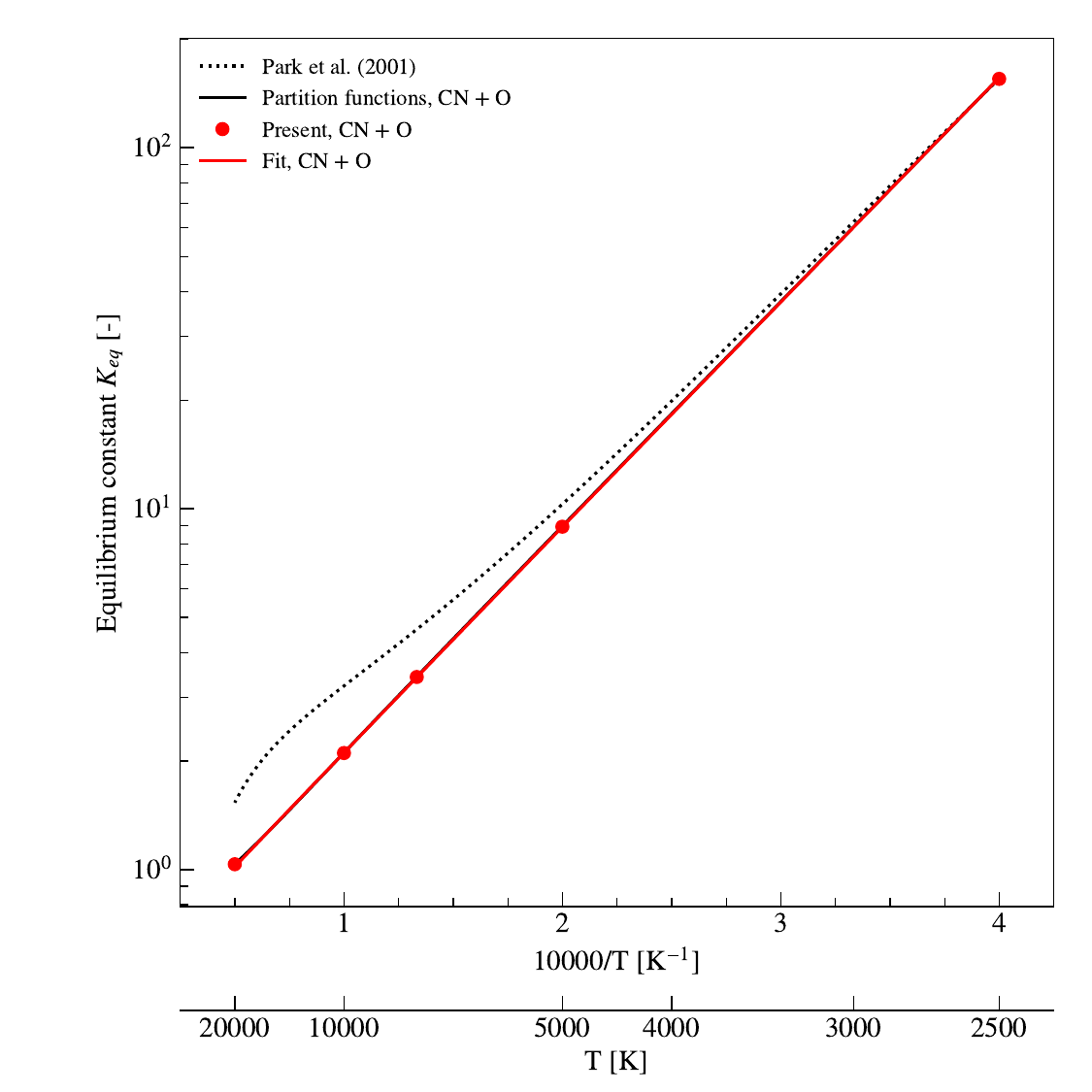}
 \caption{\small \mbox{\ce{NO(X^2\Pi) + C(^3P) <=>}} \mbox{\ce{CN(X^2\Sigma^{+}) + O(^3P)}}.}
 \label{Keq2}
\end{subfigure}
\hfill
\begin{subfigure}[t]{0.32\textwidth}
 \centering
 \includegraphics[width=\textwidth]{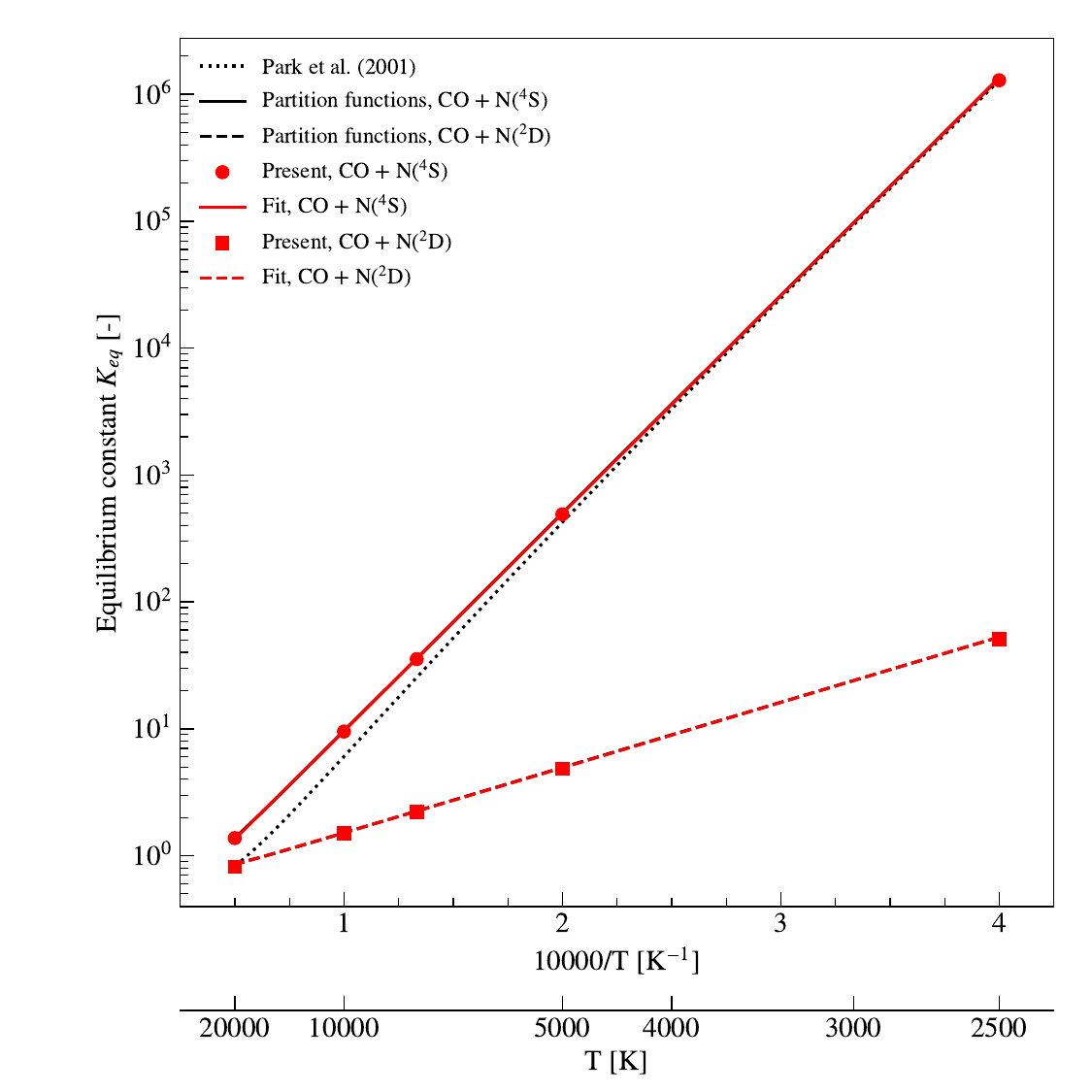}
 \caption{\small \mbox{\ce{CN(X^2\Sigma^{+}) + O(^3P) <=>}} \mbox{\ce{CO(X^1\Sigma^{+}) + N(^4S,^2D)}}.}
 \label{Keq3}
\end{subfigure}
\caption{Comparison of the macroscopic equilibrium constants for the exchange reactions against the chemistry model\cite{Park2001Chemical}.}
\label{EquilibriumConstant}
\end{figure*}

Figure~\ref{ExoRateCoefficients} indicates that the exothermic exchange rate coefficients are in reasonable agreement with both experimental measurements and previous QCT studies. Interestingly, the present calculations exhibit discrepancies with the QCT results of Koner~\textit{et al.}~\cite{koner2018cno} and Andersson~\textit{et al.}~\cite{Andersson2003Computational}, which are attributed to the different levels of theory employed in the calculation of the \textit{ab initio} energies. Additionally, the adiabatic and nonadiabatic rate coefficients of Koner~\textit{et al.},~\cite{koner2018cno} are nearly indistinguishable over the temperature range considered. This comparison supports the assumption that nonadiabatic transitions may be neglected for high-enthalpy applications (see Sec.~\ref{Sec:Nonadiabatic}). Furthermore, the \ce{NO(X^2\Pi) + C(^3P) -> CN(X^2\Sigma^{+}) + O(^3P)} and \ce{CN(X^2\Sigma^{+}) + O(^3P) -> CO(X^1\Sigma^{+}) + N(^4S,^2D)} reactions exhibit a weak dependence on temperature, whereas the \ce{NO(X^2\Pi) + C(^3P) -> CO(X^1\Sigma^{+}) + N(^4S,^2D)} reaction increases by a factor of about six over the same range. This weak dependence is expected, as these reactions correspond to barrierless exothermic processes, allowing them to proceed at temperatures as low as 15~K.~\cite{Chastaing2000Direct}

For the endothermic reactions, Fig.~\ref{EndoRateCoefficients} shows that the semi-empirical chemistry models~\cite{park1994review,Fujita2006Impact,Johnston2014Modeling} fail to accurately reproduce the present calculations. This discrepancy is primarily attributed to the fact that these models were calibrated using collisional systems different from those considered.

The thermodynamic consistency of the present database is further assessed by comparing the macroscopic equilibrium constants for the exchange reactions against the chemistry model of Park~\cite{Park2001Chemical} and theoretical calculations. For consistency, only the electronic states involved in the \ce{CNO} collisional system are considered. Figure~\ref{EquilibriumConstant} shows that the equilibrium constants, computed as the ratio of the forward and backward rate coefficients reported in the present study, are in good agreement with the baseline database. Since the endothermic rate coefficients are reconstructed from micro-reversibility, this comparison primarily verifies the consistency of the underlying thermodynamic data and of the fitting procedure.

For the dissociation processes, Fig.~\ref{DissoRateCoefficients} shows that the equilibrium rate coefficients exceed their QSS counterparts by up to a factor of about three, due to the depletion of the high-lying states, which govern the dissociation, in the QSS distribution. For the dissociation rate coefficients, Fig.~\ref{DissoRateCoefficients} indicates good agreement with previous studies for the \ce{CN(X^2\Sigma^{+}) + O(^3P)} and \ce{CO(X^1\Sigma^{+}) + N(^4S,^2D)} systems. In contrast, the \ce{NO(X^2\Pi) + C(^3P)} rate coefficient of Fig.~\ref{Disso1} follows the \ce{NO + Ar} measurements of Myerson\cite{Myerson1973Shock} and of Wray and Teare,\cite{Wray1962Shock} lying about one order of magnitude below the \ce{NO + Ar} model of Park\textit{ et al.}\cite{park1994review} and about two orders of magnitude below their \ce{NO + C} model, the latter being obtained by applying an empirical enhancement factor to the \ce{NO + Ar} rate coefficient. However, it must be emphasised that, to the authors' best knowledge, no independent experimental measurements and/or first-principles calculations currently exist for the specific collisional systems considered herein. Consequently, comparisons are made against measurements involving inert gases, e.g. \ce{NO + Ar},~\cite{Thielen1985Resonance,Wray1962Shock,Koshi1979Dissociation,Myerson1973Shock} \ce{CN + Ar} ~\cite{Mozzhukhin1989CNDissociation} and \ce{CO + Ar},~\cite{Mick1993Atomic,Appleton1970Shock} and against the semi-empirical chemistry models established for heavy-particle collisions. \cite{park1994review,Fujita2006Impact,Johnston2014Modeling} 

\section{Conclusions}

The present article constructs a rovibrational state-specific framework to gain physical insights into the kinetic processes of \ce{NO(X^2\Pi) + C(^3P)}, \ce{CN(X^2\Sigma^{+}) + O(^3P)}, and \ce{CO(X^1\Sigma^+) + N(^4S,^2D)} by means of master equation analysis. The kinetic database is constructed using quasi-classical trajectory calculations, with trajectories propagated on the $^2A^{\prime}$ and $^4A^{\prime\prime}$ PESs developed by Galvão and co-workers~\cite{goncalves2018cno,alves2021dmbe,alves2023high,Alves2019Quasiclassical,Lu2024Theoretical,Galvao2025Electronic}. The topology analysis of the PESs demonstrates that the doublet states, i.e. $^2A^{\prime}$ and $^2A^{\prime\prime}$, share a similar reactivity which strongly differs from that of the quartet state, i.e. $^4A^{\prime\prime}$, thereby supporting the assumption that the construction of the kinetic database can be restricted to the $^2A^{\prime}$ and $^4A^{\prime\prime}$ PESs.

The rovibrational state-specific database indicated several governing mechanisms across the inelastic, exchange, and dissociation processes. For inelastic processes, it is shown that low-lying states predominantly evolve within the same vibrational strands, resulting in rotational excitation, whereas high-lying states transfer energy across different vibrational levels, leading to rovibrational excitation. For dissociation processes, it is shown that these processes are primarily governed by high-lying rovibrational states, and that dissociation is strongly correlated with the state-specific energy deficit rather than being governed by vibrational-specific dynamics.
For heterogeneous exchange processes, it is shown that \ce{NO(X^2\Pi) + C(^3P)} predominantly proceeds through the formation of \ce{CN(X^2\Sigma^{+}) + O(^3P)} at low temperatures and \ce{CO(X^1\Sigma^+) + N(^4S,^2D)} at high temperatures, with a predominance for the formation of ground-state atomic nitrogen among the \ce{CO(X^1\Sigma^{+})}-forming channels, i.e.~about 60\%. This behaviour is primarily attributed to the direct reactive pathway on the $^4A^{\prime\prime}$ state, whereas formation of the first excited-state nitrogen atom must proceed via a local minimum on the $^2A^{\prime}$ state. However, the fraction of the reaction proceeding towards the \ce{CO(X^1\Sigma^+) + N(^4S,^2D)} channels remains lower than the shock-tube measurements of Dean~\textit{et al.},~\cite{Dean1991ShockTube} and \ce{CN(X^2\Sigma^{+}) + O(^3P)} remains the predominant channel over the temperature range of these measurements. This lower fraction is consistent with the \ce{CO(X^1\Sigma^{+})} branching fractions predicted by past QCT calculations that neglect nonadiabatic transitions.~\cite{koner2018cno}

The master equation analysis demonstrates that, with the exception of the \ce{CO(X^1\Sigma^+) + N(^4S)} system, the rotational and vibrational relaxation times exhibit comparable time scales and unconventional temperature dependences, suggesting that the assumption of both energy modes being fully decoupled is invalid. For \ce{CO(X^1\Sigma^+) + N(^4S)}, the vibrational relaxation time follows the conventional Millikan--White trend and exceeds its rotational counterpart by more than two orders of magnitude at the lowest temperature considered. It is also shown that the Parker model closely reproduces the rotational relaxation times, whereas the Millikan--White correlation strongly overestimates the vibrational relaxation times at low temperatures, highlighting the limitations of the semi-empirical models when describing relaxation times obtained from first-principles calculations. Additionally, the validity of the QSS assumption is assessed for both the dissociation and exchange processes. At $T$ = 10,000~K, between 72 \% and 93 \% of the depletion of the diatomic species occurs within the QSS regime for all collisional systems. For decreasing temperatures, a large fraction of the depletion occurs outside the QSS regime, which questions the validity of the QSS assumption under low thermal conditions. In contrast, no QSS regime is established for the exchange processes, with these reactions proceeding without the formation of a plateau unless dissociation kinetics are included. This demonstrates that the QSS assumption is not valid for modelling exchange processes, thereby questioning its use in high-speed and high-temperature computational fluid dynamics tools for asymmetric collisional systems. Since no quasi-steady state is established for these reactions, the thermal rate coefficients reported herein should be used instead.

Finally, thermal rate coefficients for the exchange reactions and QSS and equilibrium rate coefficients for the dissociation reactions are reported and compared against the most representative studies available in the literature. The exchange and dissociation rate coefficients are shown to be in reasonable agreement with experimental measurements and QCT calculations, whereas semi-empirical chemistry models are shown to poorly reproduce the present results for endothermic reactions. 

Throughout this article, consistent similarities are identified across inelastic, exchange, and dissociation processes in the three collisional systems. These similarities provide a promising physical basis for the development of a generalised reduced-order modelling approach capable of describing both symmetric and asymmetric systems, provided that discrepancies between different symmetry states are properly accounted for.

\section*{Acknowledgments}
This work has been supported by NASA under grant number 80NSSC21K1117. Any opinions, findings, and conclusions or recommendations expressed in this material are those of the author(s) and do not necessarily reflect the views of the National Aeronautics and Space Administration. B.R.L.G. acknowledges Conselho Nacional de Desenvolvimento Científico e Tecnológico (CNPq), grant 305211/2024-2.

\bibliography{reference}

\begin{thebibliography}{4}%
\makeatletter
\providecommand \@ifxundefined [1]{%
 \@ifx{#1\undefined}
}%
\providecommand \@ifnum [1]{%
 \ifnum #1\expandafter \@firstoftwo
 \else \expandafter \@secondoftwo
 \fi
}%
\providecommand \@ifx [1]{%
 \ifx #1\expandafter \@firstoftwo
 \else \expandafter \@secondoftwo
 \fi
}%
\providecommand \natexlab [1]{#1}%
\providecommand \enquote  [1]{``#1''}%
\providecommand \bibnamefont  [1]{#1}%
\providecommand \bibfnamefont [1]{#1}%
\providecommand \citenamefont [1]{#1}%
\providecommand \href@noop [0]{\@secondoftwo}%
\providecommand \href [0]{\begingroup \@sanitize@url \@href}%
\providecommand \@href[1]{\@@startlink{#1}\@@href}%
\providecommand \@@href[1]{\endgroup#1\@@endlink}%
\providecommand \@sanitize@url [0]{\catcode `\\12\catcode `\$12\catcode
  `\&12\catcode `\#12\catcode `\^12\catcode `\_12\catcode `\%12\relax}%
\providecommand \@@startlink[1]{}%
\providecommand \@@endlink[0]{}%
\providecommand \url  [0]{\begingroup\@sanitize@url \@url }%
\providecommand \@url [1]{\endgroup\@href {#1}{\urlprefix }}%
\providecommand \urlprefix  [0]{URL }%
\providecommand \Eprint [0]{\href }%
\providecommand \doibase [0]{http://dx.doi.org/}%
\providecommand \selectlanguage [0]{\@gobble}%
\providecommand \bibinfo  [0]{\@secondoftwo}%
\providecommand \bibfield  [0]{\@secondoftwo}%
\providecommand \translation [1]{[#1]}%
\providecommand \BibitemOpen [0]{}%
\providecommand \bibitemStop [0]{}%
\providecommand \bibitemNoStop [0]{.\EOS\space}%
\providecommand \EOS [0]{\spacefactor3000\relax}%
\providecommand \BibitemShut  [1]{\csname bibitem#1\endcsname}%
\let\auto@bib@innerbib\@empty
\bibitem [{\citenamefont {Panesi}\ \emph {et~al.}(2013)\citenamefont {Panesi},
  \citenamefont {Jaffe}, \citenamefont {Schwenke},\ and\ \citenamefont
  {Magin}}]{panesi2013rovibrational}%
  \BibitemOpen
  \bibfield  {author} {\bibinfo {author} {\bibfnamefont {M.}~\bibnamefont
  {Panesi}}, \bibinfo {author} {\bibfnamefont {R.~L.}\ \bibnamefont {Jaffe}},
  \bibinfo {author} {\bibfnamefont {D.~W.}\ \bibnamefont {Schwenke}}, \ and\
  \bibinfo {author} {\bibfnamefont {T.~E.}\ \bibnamefont {Magin}},\ }\bibfield
  {title} {\enquote {\bibinfo {title} {Rovibrational internal energy transfer
  and dissociation of
  {N}$_2$({$^{1}{{\Sigma}}^{+}_{\text{g}}$})-{N}($^{4}\text{S}_{\text{u}}$)
  system in hypersonic flows},}\ }\href {\doibase 10.1063/1.4774412} {\bibfield
   {journal} {\bibinfo  {journal} {The Journal of Chemical Physics}\ }\textbf
  {\bibinfo {volume} {138}},\ \bibinfo {pages} {044312} (\bibinfo {year}
  {2013})}\BibitemShut {NoStop}%
\bibitem [{\citenamefont {Millikan}\ and\ \citenamefont
  {White}(1963)}]{Millikan1963Relaxation}%
  \BibitemOpen
  \bibfield  {author} {\bibinfo {author} {\bibfnamefont {R.~C.}\ \bibnamefont
  {Millikan}}\ and\ \bibinfo {author} {\bibfnamefont {D.~R.}\ \bibnamefont
  {White}},\ }\bibfield  {title} {\enquote {\bibinfo {title} {Systematics of
  vibrational relaxation},}\ }\href {\doibase 10.1063/1.1734182} {\bibfield
  {journal} {\bibinfo  {journal} {The Journal of Chemical Physics}\ }\textbf
  {\bibinfo {volume} {39}},\ \bibinfo {pages} {3209--3213} (\bibinfo {year}
  {1963})}\BibitemShut {NoStop}%
\bibitem [{\citenamefont {Park}(1993)}]{park1993review}%
  \BibitemOpen
  \bibfield  {author} {\bibinfo {author} {\bibfnamefont {C.}~\bibnamefont
  {Park}},\ }\bibfield  {title} {\enquote {\bibinfo {title} {Review of
  chemical-kinetic problems of future {NASA} missions, {I: E}arth entries},}\
  }\href {\doibase 10.2514/3.431} {\bibfield  {journal} {\bibinfo  {journal}
  {Journal of Thermophysics and Heat Transfer}\ }\textbf {\bibinfo {volume}
  {7}},\ \bibinfo {pages} {385--398} (\bibinfo {year} {1993})}\BibitemShut
  {NoStop}%
\bibitem [{\citenamefont {Jo}\ \emph {et~al.}(2022)\citenamefont {Jo},
  \citenamefont {Venturi}, \citenamefont {Sharma}, \citenamefont {Munaf{\`o}},\
  and\ \citenamefont {Panesi}}]{jo2022rovibrational}%
  \BibitemOpen
  \bibfield  {author} {\bibinfo {author} {\bibfnamefont {S.~M.}\ \bibnamefont
  {Jo}}, \bibinfo {author} {\bibfnamefont {S.}~\bibnamefont {Venturi}},
  \bibinfo {author} {\bibfnamefont {M.~P.}\ \bibnamefont {Sharma}}, \bibinfo
  {author} {\bibfnamefont {A.}~\bibnamefont {Munaf{\`o}}}, \ and\ \bibinfo
  {author} {\bibfnamefont {M.}~\bibnamefont {Panesi}},\ }\bibfield  {title}
  {\enquote {\bibinfo {title} {Rovibrational-specific {QCT} and master equation
  study on {N}$_2$({X$^1{{\Sigma}}_g^{+}$}) + {O}($^3${P}) and {NO}({X$^2
  {{\Pi}}$}) + {N}($^4${S}) systems in high-energy collisions},}\ }\href
  {\doibase 10.1021/acs.jpca.1c10346} {\bibfield  {journal} {\bibinfo
  {journal} {The Journal of Physical Chemistry A}\ }\textbf {\bibinfo {volume}
  {126}},\ \bibinfo {pages} {3273--3290} (\bibinfo {year} {2022})}\BibitemShut
  {NoStop}%
\end{thebibliography}%


\begin{thebibliography}{107}%
\makeatletter
\providecommand \@ifxundefined [1]{%
 \@ifx{#1\undefined}
}%
\providecommand \@ifnum [1]{%
 \ifnum #1\expandafter \@firstoftwo
 \else \expandafter \@secondoftwo
 \fi
}%
\providecommand \@ifx [1]{%
 \ifx #1\expandafter \@firstoftwo
 \else \expandafter \@secondoftwo
 \fi
}%
\providecommand \natexlab [1]{#1}%
\providecommand \enquote  [1]{``#1''}%
\providecommand \bibnamefont  [1]{#1}%
\providecommand \bibfnamefont [1]{#1}%
\providecommand \citenamefont [1]{#1}%
\providecommand \href@noop [0]{\@secondoftwo}%
\providecommand \href [0]{\begingroup \@sanitize@url \@href}%
\providecommand \@href[1]{\@@startlink{#1}\@@href}%
\providecommand \@@href[1]{\endgroup#1\@@endlink}%
\providecommand \@sanitize@url [0]{\catcode `\\12\catcode `\$12\catcode
  `\&12\catcode `\#12\catcode `\^12\catcode `\_12\catcode `\%12\relax}%
\providecommand \@@startlink[1]{}%
\providecommand \@@endlink[0]{}%
\providecommand \url  [0]{\begingroup\@sanitize@url \@url }%
\providecommand \@url [1]{\endgroup\@href {#1}{\urlprefix }}%
\providecommand \urlprefix  [0]{URL }%
\providecommand \Eprint [0]{\href }%
\providecommand \doibase [0]{http://dx.doi.org/}%
\providecommand \selectlanguage [0]{\@gobble}%
\providecommand \bibinfo  [0]{\@secondoftwo}%
\providecommand \bibfield  [0]{\@secondoftwo}%
\providecommand \translation [1]{[#1]}%
\providecommand \BibitemOpen [0]{}%
\providecommand \bibitemStop [0]{}%
\providecommand \bibitemNoStop [0]{.\EOS\space}%
\providecommand \EOS [0]{\spacefactor3000\relax}%
\providecommand \BibitemShut  [1]{\csname bibitem#1\endcsname}%
\let\auto@bib@innerbib\@empty
\bibitem [{\citenamefont {Smoot}, \citenamefont {Hill},\ and\ \citenamefont
  {Xu}(1998)}]{Smoot1998Reburning}%
  \BibitemOpen
  \bibfield  {author} {\bibinfo {author} {\bibfnamefont {L.~D.}\ \bibnamefont
  {Smoot}}, \bibinfo {author} {\bibfnamefont {S.~C.}\ \bibnamefont {Hill}}, \
  and\ \bibinfo {author} {\bibfnamefont {H.}~\bibnamefont {Xu}},\ }\bibfield
  {title} {\enquote {\bibinfo {title} {{NO}$_x$ control through reburning},}\
  }\href {\doibase 10.1016/S0360-1285(97)00022-1} {\bibfield  {journal}
  {\bibinfo  {journal} {Progress in Energy and Combustion Science}\ }\textbf
  {\bibinfo {volume} {24}},\ \bibinfo {pages} {385--408} (\bibinfo {year}
  {1998})}\BibitemShut {NoStop}%
\bibitem [{\citenamefont {Lamoureux}\ \emph {et~al.}(2016)\citenamefont
  {Lamoureux}, \citenamefont {{El Merhubi}}, \citenamefont {Pillier},
  \citenamefont {{de Persis}},\ and\ \citenamefont
  {Desgroux}}]{Lamoureux2016Modeling}%
  \BibitemOpen
  \bibfield  {author} {\bibinfo {author} {\bibfnamefont {N.}~\bibnamefont
  {Lamoureux}}, \bibinfo {author} {\bibfnamefont {H.}~\bibnamefont {{El
  Merhubi}}}, \bibinfo {author} {\bibfnamefont {L.}~\bibnamefont {Pillier}},
  \bibinfo {author} {\bibfnamefont {S.}~\bibnamefont {{de Persis}}}, \ and\
  \bibinfo {author} {\bibfnamefont {P.}~\bibnamefont {Desgroux}},\ }\bibfield
  {title} {\enquote {\bibinfo {title} {Modeling of {NO} formation in low
  pressure premixed flames},}\ }\href {\doibase
  10.1016/j.combustflame.2015.11.007} {\bibfield  {journal} {\bibinfo
  {journal} {Combustion and Flame}\ }\textbf {\bibinfo {volume} {163}},\
  \bibinfo {pages} {557--575} (\bibinfo {year} {2016})}\BibitemShut {NoStop}%
\bibitem [{\citenamefont {Herbst}\ and\ \citenamefont
  {Klemperer}(1973)}]{Herbst1973formation}%
  \BibitemOpen
  \bibfield  {author} {\bibinfo {author} {\bibfnamefont {E.}~\bibnamefont
  {Herbst}}\ and\ \bibinfo {author} {\bibfnamefont {W.}~\bibnamefont
  {Klemperer}},\ }\bibfield  {title} {\enquote {\bibinfo {title} {The formation
  and depletion of molecules in dense interstellar clouds},}\ }\href {\doibase
  10.1086/152436} {\bibfield  {journal} {\bibinfo  {journal} {Astrophysical
  Journal}\ }\textbf {\bibinfo {volume} {185}},\ \bibinfo {pages} {505--534}
  (\bibinfo {year} {1973})}\BibitemShut {NoStop}%
\bibitem [{\citenamefont {Boger}\ and\ \citenamefont
  {Sternberg}(2005)}]{Boger2005dense}%
  \BibitemOpen
  \bibfield  {author} {\bibinfo {author} {\bibfnamefont {G.~I.}\ \bibnamefont
  {Boger}}\ and\ \bibinfo {author} {\bibfnamefont {A.}~\bibnamefont
  {Sternberg}},\ }\bibfield  {title} {\enquote {\bibinfo {title} {{CN} and
  {HCN} in dense interstellar clouds},}\ }\href {\doibase 10.1086/432864}
  {\bibfield  {journal} {\bibinfo  {journal} {The Astrophysical Journal}\
  }\textbf {\bibinfo {volume} {632}},\ \bibinfo {pages} {302} (\bibinfo {year}
  {2005})}\BibitemShut {NoStop}%
\bibitem [{\citenamefont {Park}(1990)}]{park1989nonequilibrium}%
  \BibitemOpen
  \bibfield  {author} {\bibinfo {author} {\bibfnamefont {C.}~\bibnamefont
  {Park}},\ }\href@noop {} {\emph {\bibinfo {title} {Nonequilibrium Hypersonic
  Aerothermodynamics}}},\ \bibinfo {edition} {1st}\ ed.\ (\bibinfo  {publisher}
  {John Wiley \& Sons},\ \bibinfo {year} {1990})\BibitemShut {NoStop}%
\bibitem [{\citenamefont {Park}\ \emph {et~al.}(1994)\citenamefont {Park},
  \citenamefont {Howe}, \citenamefont {Jaffe},\ and\ \citenamefont
  {Candler}}]{park1994review}%
  \BibitemOpen
  \bibfield  {author} {\bibinfo {author} {\bibfnamefont {C.}~\bibnamefont
  {Park}}, \bibinfo {author} {\bibfnamefont {J.~T.}\ \bibnamefont {Howe}},
  \bibinfo {author} {\bibfnamefont {R.~L.}\ \bibnamefont {Jaffe}}, \ and\
  \bibinfo {author} {\bibfnamefont {G.~V.}\ \bibnamefont {Candler}},\
  }\bibfield  {title} {\enquote {\bibinfo {title} {Review of chemical-kinetic
  problems of future {NASA} missions, {II: M}ars entries},}\ }\href {\doibase
  10.2514/3.496} {\bibfield  {journal} {\bibinfo  {journal} {Journal of
  Thermophysics and Heat Transfer}\ }\textbf {\bibinfo {volume} {8}},\ \bibinfo
  {pages} {9--23} (\bibinfo {year} {1994})}\BibitemShut {NoStop}%
\bibitem [{\citenamefont {Park}\ and\ \citenamefont
  {Ahn}(1999)}]{Park1999Stagnation}%
  \BibitemOpen
  \bibfield  {author} {\bibinfo {author} {\bibfnamefont {C.}~\bibnamefont
  {Park}}\ and\ \bibinfo {author} {\bibfnamefont {H.}~\bibnamefont {Ahn}},\
  }\bibfield  {title} {\enquote {\bibinfo {title} {Stagnation-point heat
  transfer rates for {Pioneer-Venus} probes},}\ }\href {\doibase
  10.2514/2.6426} {\bibfield  {journal} {\bibinfo  {journal} {Journal of
  Thermophysics and Heat Transfer}\ }\textbf {\bibinfo {volume} {13}},\
  \bibinfo {pages} {33--41} (\bibinfo {year} {1999})}\BibitemShut {NoStop}%
\bibitem [{\citenamefont {Park}, \citenamefont {Jaffe},\ and\ \citenamefont
  {Partridge}(2001)}]{Park2001Chemical}%
  \BibitemOpen
  \bibfield  {author} {\bibinfo {author} {\bibfnamefont {C.}~\bibnamefont
  {Park}}, \bibinfo {author} {\bibfnamefont {R.~L.}\ \bibnamefont {Jaffe}}, \
  and\ \bibinfo {author} {\bibfnamefont {H.}~\bibnamefont {Partridge}},\
  }\bibfield  {title} {\enquote {\bibinfo {title} {Chemical-kinetic parameters
  of hyperbolic {Earth} entry},}\ }\href {\doibase 10.2514/2.6582} {\bibfield
  {journal} {\bibinfo  {journal} {Journal of Thermophysics and Heat Transfer}\
  }\textbf {\bibinfo {volume} {15}},\ \bibinfo {pages} {76--90} (\bibinfo
  {year} {2001})}\BibitemShut {NoStop}%
\bibitem [{\citenamefont {Tibere-Inglesse}\ \emph {et~al.}(2024)\citenamefont
  {Tibere-Inglesse}, \citenamefont {West}, \citenamefont {Jelloian},
  \citenamefont {Minesi}, \citenamefont {Spearrin}, \citenamefont {Clarke},
  \citenamefont {di~Mare}, \citenamefont {McGilvray},\ and\ \citenamefont
  {Cruden}}]{TibereInglesse2024Reconstruction}%
  \BibitemOpen
  \bibfield  {author} {\bibinfo {author} {\bibfnamefont {A.~C.}\ \bibnamefont
  {Tibere-Inglesse}}, \bibinfo {author} {\bibfnamefont {T.~K.}\ \bibnamefont
  {West}}, \bibinfo {author} {\bibfnamefont {C.~C.}\ \bibnamefont {Jelloian}},
  \bibinfo {author} {\bibfnamefont {N.~Q.}\ \bibnamefont {Minesi}}, \bibinfo
  {author} {\bibfnamefont {M.~R.}\ \bibnamefont {Spearrin}}, \bibinfo {author}
  {\bibfnamefont {J.}~\bibnamefont {Clarke}}, \bibinfo {author} {\bibfnamefont
  {L.}~\bibnamefont {di~Mare}}, \bibinfo {author} {\bibfnamefont
  {M.}~\bibnamefont {McGilvray}}, \ and\ \bibinfo {author} {\bibfnamefont
  {B.~A.}\ \bibnamefont {Cruden}},\ }\bibfield  {title} {\enquote {\bibinfo
  {title} {Reconstruction of {M}ars 2020 backshell radiative heating via
  infrared emission spectroscopy of {CO}$_2$/{N}$_2$/{A}r shockwaves},}\ }\href
  {\doibase 10.2514/1.T6916} {\bibfield  {journal} {\bibinfo  {journal}
  {Journal of Thermophysics and Heat Transfer}\ }\textbf {\bibinfo {volume}
  {38}},\ \bibinfo {pages} {568--585} (\bibinfo {year} {2024})}\BibitemShut
  {NoStop}%
\bibitem [{\citenamefont {Fawley}\ \emph {et~al.}(2025)\citenamefont {Fawley},
  \citenamefont {Putnam}, \citenamefont {D’Souza}, \citenamefont {McDaniel},\
  and\ \citenamefont {Borner}}]{Fawley2025Magnetohydrodynamic}%
  \BibitemOpen
  \bibfield  {author} {\bibinfo {author} {\bibfnamefont {D.~M.}\ \bibnamefont
  {Fawley}}, \bibinfo {author} {\bibfnamefont {Z.~R.}\ \bibnamefont {Putnam}},
  \bibinfo {author} {\bibfnamefont {S.}~\bibnamefont {D’Souza}}, \bibinfo
  {author} {\bibfnamefont {R.}~\bibnamefont {McDaniel}}, \ and\ \bibinfo
  {author} {\bibfnamefont {A.}~\bibnamefont {Borner}},\ }\bibfield  {title}
  {\enquote {\bibinfo {title} {Magnetohydrodynamic and aerodynamic assessment
  of a {70°} spherecone at {Mars} and {Venus}},}\ }\href {\doibase
  10.2514/1.A35945} {\bibfield  {journal} {\bibinfo  {journal} {Journal of
  Spacecraft and Rockets}\ }\textbf {\bibinfo {volume} {62}},\ \bibinfo {pages}
  {494--509} (\bibinfo {year} {2025})}\BibitemShut {NoStop}%
\bibitem [{\citenamefont {Johnston}\ and\ \citenamefont
  {Brandis}(2014)}]{Johnston2014Modeling}%
  \BibitemOpen
  \bibfield  {author} {\bibinfo {author} {\bibfnamefont {C.~O.}\ \bibnamefont
  {Johnston}}\ and\ \bibinfo {author} {\bibfnamefont {A.~M.}\ \bibnamefont
  {Brandis}},\ }\bibfield  {title} {\enquote {\bibinfo {title} {Modeling of
  nonequilibrium {CO} fourth-positive and {CN} violet emission in
  {CO}$_2$–{N}$_2$ gases},}\ }\href {\doibase 10.1016/j.jqsrt.2014.08.025}
  {\bibfield  {journal} {\bibinfo  {journal} {Journal of Quantitative
  Spectroscopy and Radiative Transfer}\ }\textbf {\bibinfo {volume} {149}},\
  \bibinfo {pages} {303--317} (\bibinfo {year} {2014})}\BibitemShut {NoStop}%
\bibitem [{\citenamefont {Jaffe}\ \emph {et~al.}(2017)\citenamefont {Jaffe},
  \citenamefont {Schwenke}, \citenamefont {Chaban}, \citenamefont {Prabhu},
  \citenamefont {Johnston},\ and\ \citenamefont {Panesi}}]{Jaffe2017Mars}%
  \BibitemOpen
  \bibfield  {author} {\bibinfo {author} {\bibfnamefont {R.~L.}\ \bibnamefont
  {Jaffe}}, \bibinfo {author} {\bibfnamefont {D.~W.}\ \bibnamefont {Schwenke}},
  \bibinfo {author} {\bibfnamefont {G.~M.}\ \bibnamefont {Chaban}}, \bibinfo
  {author} {\bibfnamefont {D.~K.}\ \bibnamefont {Prabhu}}, \bibinfo {author}
  {\bibfnamefont {C.~O.}\ \bibnamefont {Johnston}}, \ and\ \bibinfo {author}
  {\bibfnamefont {M.}~\bibnamefont {Panesi}},\ }\bibfield  {title} {\enquote
  {\bibinfo {title} {On the development of a new nonequilibrium chemistry model
  for {M}ars entry},}\ }in\ \href {\doibase 10.2514/6.2017-1372} {\emph
  {\bibinfo {booktitle} {55th AIAA Aerospace Sciences Meeting}}}\ (\bibinfo
  {address} {Grapevine, TX, AIAA Paper 2017-1372},\ \bibinfo {year}
  {2017})\BibitemShut {NoStop}%
\bibitem [{\citenamefont {Borner}\ \emph {et~al.}(2026)\citenamefont {Borner},
  \citenamefont {Gallis}, \citenamefont {Parai}, \citenamefont {Avice},
  \citenamefont {Petkov}, \citenamefont {Swaminathan-Gopalan}, \citenamefont
  {Sotin},\ and\ \citenamefont {Rabinovitch}}]{Borner2026Noble}%
  \BibitemOpen
  \bibfield  {author} {\bibinfo {author} {\bibfnamefont {A.}~\bibnamefont
  {Borner}}, \bibinfo {author} {\bibfnamefont {M.~A.}\ \bibnamefont {Gallis}},
  \bibinfo {author} {\bibfnamefont {R.}~\bibnamefont {Parai}}, \bibinfo
  {author} {\bibfnamefont {G.}~\bibnamefont {Avice}}, \bibinfo {author}
  {\bibfnamefont {M.~P.}\ \bibnamefont {Petkov}}, \bibinfo {author}
  {\bibfnamefont {K.}~\bibnamefont {Swaminathan-Gopalan}}, \bibinfo {author}
  {\bibfnamefont {C.}~\bibnamefont {Sotin}}, \ and\ \bibinfo {author}
  {\bibfnamefont {J.}~\bibnamefont {Rabinovitch}},\ }\bibfield  {title}
  {\enquote {\bibinfo {title} {Noble gas fractionation predictions for high
  speed sampling in the upper atmosphere of {Venus}},}\ }\href {\doibase
  10.1016/j.icarus.2025.116800} {\bibfield  {journal} {\bibinfo  {journal}
  {Icarus}\ }\textbf {\bibinfo {volume} {444}},\ \bibinfo {pages} {116800}
  (\bibinfo {year} {2026})}\BibitemShut {NoStop}%
\bibitem [{\citenamefont {Cabrera}\ and\ \citenamefont
  {West}(2026)}]{Cabrera2026Pioneer}%
  \BibitemOpen
  \bibfield  {author} {\bibinfo {author} {\bibfnamefont {J.~V.~V.}\
  \bibnamefont {Cabrera}}\ and\ \bibinfo {author} {\bibfnamefont {T.~K.}\
  \bibnamefont {West}},\ }\bibfield  {title} {\enquote {\bibinfo {title}
  {{P}ioneer {V}enus {L}arge {P}robe stagnation point entry heating with
  coupled ablation},}\ }\href {\doibase 10.2514/1.A36431} {\bibfield  {journal}
  {\bibinfo  {journal} {Journal of Spacecraft and Rockets}\ }\textbf {\bibinfo
  {volume} {63}},\ \bibinfo {pages} {360--368} (\bibinfo {year}
  {2026})}\BibitemShut {NoStop}%
\bibitem [{\citenamefont {Meurisse}\ \emph {et~al.}(2027)\citenamefont
  {Meurisse}, \citenamefont {Borner}, \citenamefont {Naughton},\ and\
  \citenamefont {Gentry}}]{meurisse2026venusaerosol}%
  \BibitemOpen
  \bibfield  {author} {\bibinfo {author} {\bibfnamefont {J.~B.~E.}\
  \bibnamefont {Meurisse}}, \bibinfo {author} {\bibfnamefont {A.}~\bibnamefont
  {Borner}}, \bibinfo {author} {\bibfnamefont {C.}~\bibnamefont {Naughton}}, \
  and\ \bibinfo {author} {\bibfnamefont {D.~M.}\ \bibnamefont {Gentry}},\
  }\bibfield  {title} {\enquote {\bibinfo {title} {Multi-physics modeling of
  aerosol capture for {Venus} entry probes},}\ }\href {\doibase
  10.1016/j.ast.2026.114003} {\bibfield  {journal} {\bibinfo  {journal}
  {Aerospace Science and Technology}\ }\textbf {\bibinfo {volume} {180}},\
  \bibinfo {pages} {114003} (\bibinfo {year} {2027})}\BibitemShut {NoStop}%
\bibitem [{\citenamefont {Martin}\ \emph {et~al.}(2015)\citenamefont {Martin},
  \citenamefont {Cozmuta}, \citenamefont {Wright},\ and\ \citenamefont
  {Boyd}}]{martin2015kinetic}%
  \BibitemOpen
  \bibfield  {author} {\bibinfo {author} {\bibfnamefont {A.}~\bibnamefont
  {Martin}}, \bibinfo {author} {\bibfnamefont {I.}~\bibnamefont {Cozmuta}},
  \bibinfo {author} {\bibfnamefont {M.~J.}\ \bibnamefont {Wright}}, \ and\
  \bibinfo {author} {\bibfnamefont {I.~D.}\ \bibnamefont {Boyd}},\ }\bibfield
  {title} {\enquote {\bibinfo {title} {Kinetic rates for gas-phase chemistry of
  phenolic-based carbon ablator in atmospheric air},}\ }\href {\doibase
  10.2514/1.T4184} {\bibfield  {journal} {\bibinfo  {journal} {Journal of
  Thermophysics and Heat Transfer}\ }\textbf {\bibinfo {volume} {29}},\
  \bibinfo {pages} {222--240} (\bibinfo {year} {2015})}\BibitemShut {NoStop}%
\bibitem [{\citenamefont {Prata}, \citenamefont {Schwartzentruber},\ and\
  \citenamefont {Minton}(2022)}]{prata2022aircarbon}%
  \BibitemOpen
  \bibfield  {author} {\bibinfo {author} {\bibfnamefont {K.~S.}\ \bibnamefont
  {Prata}}, \bibinfo {author} {\bibfnamefont {T.~E.}\ \bibnamefont
  {Schwartzentruber}}, \ and\ \bibinfo {author} {\bibfnamefont {T.~K.}\
  \bibnamefont {Minton}},\ }\bibfield  {title} {\enquote {\bibinfo {title}
  {Air--carbon ablation model for hypersonic flight from molecular-beam
  data},}\ }\href {\doibase 10.2514/1.J060516} {\bibfield  {journal} {\bibinfo
  {journal} {AIAA Journal}\ }\textbf {\bibinfo {volume} {60}},\ \bibinfo
  {pages} {627--640} (\bibinfo {year} {2022})}\BibitemShut {NoStop}%
\bibitem [{\citenamefont {Rao}\ \emph {et~al.}(2025)\citenamefont {Rao},
  \citenamefont {Crespo}, \citenamefont {Valentini}, \citenamefont {Taylor},
  \citenamefont {Murray}, \citenamefont {Vaughan}, \citenamefont {Alviani},\
  and\ \citenamefont {Kulakhmetov}}]{Rao2024OSPREE}%
  \BibitemOpen
  \bibfield  {author} {\bibinfo {author} {\bibfnamefont {A.~P.}\ \bibnamefont
  {Rao}}, \bibinfo {author} {\bibfnamefont {J.~D.}\ \bibnamefont {Crespo}},
  \bibinfo {author} {\bibfnamefont {P.}~\bibnamefont {Valentini}}, \bibinfo
  {author} {\bibfnamefont {J.~B.}\ \bibnamefont {Taylor}}, \bibinfo {author}
  {\bibfnamefont {V.~J.}\ \bibnamefont {Murray}}, \bibinfo {author}
  {\bibfnamefont {E.~I.}\ \bibnamefont {Vaughan}}, \bibinfo {author}
  {\bibfnamefont {R.}~\bibnamefont {Alviani}}, \ and\ \bibinfo {author}
  {\bibfnamefont {M.}~\bibnamefont {Kulakhmetov}},\ }\bibfield  {title}
  {\enquote {\bibinfo {title} {Developing the {OSPREE} payload for
  spectroscopic measurements of a {M}ach 25+ plasma sheath},}\ }\href {\doibase
  10.2514/1.A36325} {\bibfield  {journal} {\bibinfo  {journal} {Journal of
  Spacecraft and Rockets}\ }\textbf {\bibinfo {volume} {62}},\ \bibinfo {pages}
  {1486--1495} (\bibinfo {year} {2025})}\BibitemShut {NoStop}%
\bibitem [{\citenamefont {Rao}\ \emph {et~al.}(2026)\citenamefont {Rao},
  \citenamefont {Crespo}, \citenamefont {Valentini}, \citenamefont {Murray},
  \citenamefont {Vaughan}, \citenamefont {Davis}, \citenamefont {Alviani},\
  and\ \citenamefont {Kulakhmetov}}]{Rao2026OSPREE}%
  \BibitemOpen
  \bibfield  {author} {\bibinfo {author} {\bibfnamefont {A.~P.}\ \bibnamefont
  {Rao}}, \bibinfo {author} {\bibfnamefont {J.~D.}\ \bibnamefont {Crespo}},
  \bibinfo {author} {\bibfnamefont {P.}~\bibnamefont {Valentini}}, \bibinfo
  {author} {\bibfnamefont {V.~J.}\ \bibnamefont {Murray}}, \bibinfo {author}
  {\bibfnamefont {E.~I.}\ \bibnamefont {Vaughan}}, \bibinfo {author}
  {\bibfnamefont {Z.~S.}\ \bibnamefont {Davis}}, \bibinfo {author}
  {\bibfnamefont {R.}~\bibnamefont {Alviani}}, \ and\ \bibinfo {author}
  {\bibfnamefont {M.}~\bibnamefont {Kulakhmetov}},\ }\bibfield  {title}
  {\enquote {\bibinfo {title} {Return of {OSPREE}: {I}n situ spectroscopic
  measurements during atmospheric reentry},}\ }\href {\doibase
  10.2514/1.A36508} {\bibfield  {journal} {\bibinfo  {journal} {Journal of
  Spacecraft and Rockets}\ }\textbf {\bibinfo {volume} {63}},\ \bibinfo {pages}
  {951--964} (\bibinfo {year} {2026})}\BibitemShut {NoStop}%
\bibitem [{\citenamefont {Meyers}\ \emph {et~al.}(2026)\citenamefont {Meyers},
  \citenamefont {Singh}, \citenamefont {Munaf{\`o}},\ and\ \citenamefont
  {Panesi}}]{Meyers2026PlasmatronX}%
  \BibitemOpen
  \bibfield  {author} {\bibinfo {author} {\bibfnamefont {J.~M.}\ \bibnamefont
  {Meyers}}, \bibinfo {author} {\bibfnamefont {A.}~\bibnamefont {Singh}},
  \bibinfo {author} {\bibfnamefont {A.}~\bibnamefont {Munaf{\`o}}}, \ and\
  \bibinfo {author} {\bibfnamefont {M.}~\bibnamefont {Panesi}},\ }\bibfield
  {title} {\enquote {\bibinfo {title} {Gas surface interaction reactant and
  product characterization in high enthalpy air plasma flows via
  ns{TALIF}/{LIF} imaging in the {UIUC} 350 {kW} {P}lasmatron {X} facility},}\
  }in\ \href {\doibase 10.2514/6.2026-2463} {\emph {\bibinfo {booktitle} {AIAA
  SCITECH 2026 Forum}}}\ (\bibinfo {address} {Orlando, FL, AIAA Paper
  2026-2463},\ \bibinfo {year} {2026})\BibitemShut {NoStop}%
\bibitem [{\citenamefont {Kumar}\ \emph {et~al.}(2027)\citenamefont {Kumar},
  \citenamefont {Munaf{\`o}}, \citenamefont {Vollmer}, \citenamefont {Franco},
  \citenamefont {Stephani}, \citenamefont {Bodony},\ and\ \citenamefont
  {Panesi}}]{Kumar2027Coils}%
  \BibitemOpen
  \bibfield  {author} {\bibinfo {author} {\bibfnamefont {S.}~\bibnamefont
  {Kumar}}, \bibinfo {author} {\bibfnamefont {A.}~\bibnamefont {Munaf{\`o}}},
  \bibinfo {author} {\bibfnamefont {B.}~\bibnamefont {Vollmer}}, \bibinfo
  {author} {\bibfnamefont {M.}~\bibnamefont {Franco}}, \bibinfo {author}
  {\bibfnamefont {K.~A.}\ \bibnamefont {Stephani}}, \bibinfo {author}
  {\bibfnamefont {D.~J.}\ \bibnamefont {Bodony}}, \ and\ \bibinfo {author}
  {\bibfnamefont {M.}~\bibnamefont {Panesi}},\ }\bibfield  {title} {\enquote
  {\bibinfo {title} {From coils to surface recession: {M}ultiphysics simulation
  of ablation in {ICP} wind tunnels},}\ }\href {\doibase
  10.1016/j.ijheatmasstransfer.2026.129637} {\bibfield  {journal} {\bibinfo
  {journal} {International Journal of Heat and Mass Transfer}\ }\textbf
  {\bibinfo {volume} {273}},\ \bibinfo {pages} {129637} (\bibinfo {year}
  {2027})}\BibitemShut {NoStop}%
\bibitem [{\citenamefont {Dubrin}\ \emph {et~al.}(1964)\citenamefont {Dubrin},
  \citenamefont {MacKay}, \citenamefont {Pandow},\ and\ \citenamefont
  {Wolfgang}}]{Dubrin1964Reactions}%
  \BibitemOpen
  \bibfield  {author} {\bibinfo {author} {\bibfnamefont {J.}~\bibnamefont
  {Dubrin}}, \bibinfo {author} {\bibfnamefont {C.}~\bibnamefont {MacKay}},
  \bibinfo {author} {\bibfnamefont {M.~L.}\ \bibnamefont {Pandow}}, \ and\
  \bibinfo {author} {\bibfnamefont {R.}~\bibnamefont {Wolfgang}},\ }\bibfield
  {title} {\enquote {\bibinfo {title} {Reactions of atomic carbon with
  $\pi$-bonded inorganic molecules},}\ }\href {\doibase
  10.1016/0022-1902(64)80156-1} {\bibfield  {journal} {\bibinfo  {journal}
  {Journal of Inorganic and Nuclear Chemistry}\ }\textbf {\bibinfo {volume}
  {26}},\ \bibinfo {pages} {2113--2122} (\bibinfo {year} {1964})}\BibitemShut
  {NoStop}%
\bibitem [{\citenamefont {Braun}\ \emph {et~al.}(1969)\citenamefont {Braun},
  \citenamefont {Bass}, \citenamefont {Davis},\ and\ \citenamefont
  {Simmons}}]{Braun1969Flash}%
  \BibitemOpen
  \bibfield  {author} {\bibinfo {author} {\bibfnamefont {W.}~\bibnamefont
  {Braun}}, \bibinfo {author} {\bibfnamefont {A.~M.}\ \bibnamefont {Bass}},
  \bibinfo {author} {\bibfnamefont {D.~D.}\ \bibnamefont {Davis}}, \ and\
  \bibinfo {author} {\bibfnamefont {J.~D.}\ \bibnamefont {Simmons}},\
  }\bibfield  {title} {\enquote {\bibinfo {title} {Flash photolysis of carbon
  suboxide: absolute rate constants for reactions of {C}($^3${P}) and
  {C}($^1${D}) with {H}$_2$, {N}$_2$, {CO}, {NO}, {O}$_2$ and {CH}$_4$},}\
  }\href {\doibase 10.1098/rspa.1969.0168} {\bibfield  {journal} {\bibinfo
  {journal} {Proceedings of the Royal Society of London. A. Mathematical and
  Physical Sciences}\ }\textbf {\bibinfo {volume} {312}},\ \bibinfo {pages}
  {417--434} (\bibinfo {year} {1969})}\BibitemShut {NoStop}%
\bibitem [{\citenamefont {Donovan}\ and\ \citenamefont
  {Husain}(1970)}]{donovan1970recent}%
  \BibitemOpen
  \bibfield  {author} {\bibinfo {author} {\bibfnamefont {R.~J.}\ \bibnamefont
  {Donovan}}\ and\ \bibinfo {author} {\bibfnamefont {D.}~\bibnamefont
  {Husain}},\ }\bibfield  {title} {\enquote {\bibinfo {title} {Recent advances
  in the chemistry of electronically excited atoms},}\ }\href {\doibase
  10.1021/cr60266a003} {\bibfield  {journal} {\bibinfo  {journal} {Chemical
  Reviews}\ }\textbf {\bibinfo {volume} {70}},\ \bibinfo {pages} {489--516}
  (\bibinfo {year} {1970})}\BibitemShut {NoStop}%
\bibitem [{\citenamefont {Husain}\ and\ \citenamefont
  {Kirsch}(1971{\natexlab{a}})}]{Husain1971Study}%
  \BibitemOpen
  \bibfield  {author} {\bibinfo {author} {\bibfnamefont {D.}~\bibnamefont
  {Husain}}\ and\ \bibinfo {author} {\bibfnamefont {L.~J.}\ \bibnamefont
  {Kirsch}},\ }\bibfield  {title} {\enquote {\bibinfo {title} {Study of
  electronically excited carbon atoms{,} {C}(2$^1$d$_2$){,} by the attenuation
  of atomic emission{,} (3$^1$p$^0_1$→2$^1$d$_2$). part 1.—collisional
  deactivation of {C}(2$^1$d$_2$) by the noble gases},}\ }\href {\doibase
  10.1039/TF9716702886} {\bibfield  {journal} {\bibinfo  {journal} {Trans.
  Faraday Soc.}\ }\textbf {\bibinfo {volume} {67}},\ \bibinfo {pages}
  {2886--2895} (\bibinfo {year} {1971}{\natexlab{a}})}\BibitemShut {NoStop}%
\bibitem [{\citenamefont {Husain}\ and\ \citenamefont
  {Kirsch}(1971{\natexlab{b}})}]{Husain1971Kinetic}%
  \BibitemOpen
  \bibfield  {author} {\bibinfo {author} {\bibfnamefont {D.}~\bibnamefont
  {Husain}}\ and\ \bibinfo {author} {\bibfnamefont {L.~J.}\ \bibnamefont
  {Kirsch}},\ }\bibfield  {title} {\enquote {\bibinfo {title} {Kinetic
  investigation of {C}($^3${P}$_j$) by photoelectric measurement of the
  attenuation of atomic emission in the vacuum ultra-violet},}\ }\href
  {\doibase 10.1016/0009-2614(71)80087-8} {\bibfield  {journal} {\bibinfo
  {journal} {Chemical Physics Letters}\ }\textbf {\bibinfo {volume} {8}},\
  \bibinfo {pages} {543--546} (\bibinfo {year}
  {1971}{\natexlab{b}})}\BibitemShut {NoStop}%
\bibitem [{\citenamefont {Husain}\ and\ \citenamefont
  {Young}(1975)}]{Husain1975Kinetic}%
  \BibitemOpen
  \bibfield  {author} {\bibinfo {author} {\bibfnamefont {D.}~\bibnamefont
  {Husain}}\ and\ \bibinfo {author} {\bibfnamefont {A.~N.}\ \bibnamefont
  {Young}},\ }\bibfield  {title} {\enquote {\bibinfo {title} {Kinetic
  investigation of ground state carbon atoms, {C}($^3${P}$_j$)},}\ }\href
  {\doibase 10.1039/F29757100525} {\bibfield  {journal} {\bibinfo  {journal}
  {J. Chem. Soc., Faraday Trans. 2}\ }\textbf {\bibinfo {volume} {71}},\
  \bibinfo {pages} {525--531} (\bibinfo {year} {1975})}\BibitemShut {NoStop}%
\bibitem [{\citenamefont {Husain}\ and\ \citenamefont
  {Newton}(1982)}]{Husain1982Collisional}%
  \BibitemOpen
  \bibfield  {author} {\bibinfo {author} {\bibfnamefont {D.}~\bibnamefont
  {Husain}}\ and\ \bibinfo {author} {\bibfnamefont {D.~P.}\ \bibnamefont
  {Newton}},\ }\bibfield  {title} {\enquote {\bibinfo {title} {Collisional
  quenching of electronically excited carbon atoms{,} c[2p$^2$($^1$s$_0$)]},}\
  }\href {\doibase 10.1039/F29827800051} {\bibfield  {journal} {\bibinfo
  {journal} {Journal of the Chemical Society, Faraday Transactions 2: Molecular
  and Chemical Physics}\ }\textbf {\bibinfo {volume} {78}},\ \bibinfo {pages}
  {51--71} (\bibinfo {year} {1982})}\BibitemShut {NoStop}%
\bibitem [{\citenamefont {Krause}(1981)}]{Krause1981Velocity}%
  \BibitemOpen
  \bibfield  {author} {\bibinfo {author} {\bibfnamefont {H.~F.}\ \bibnamefont
  {Krause}},\ }\bibfield  {title} {\enquote {\bibinfo {title}
  {Velocity-selected molecular beam study of rotational excitation in the
  reaction {C + NO → CN + O}},}\ }\href {\doibase
  10.1016/0009-2614(81)85556-X} {\bibfield  {journal} {\bibinfo  {journal}
  {Chemical Physics Letters}\ }\textbf {\bibinfo {volume} {78}},\ \bibinfo
  {pages} {78--80} (\bibinfo {year} {1981})}\BibitemShut {NoStop}%
\bibitem [{\citenamefont {Dorthe}\ \emph {et~al.}(1985)\citenamefont {Dorthe},
  \citenamefont {Costes}, \citenamefont {Naulin}, \citenamefont
  {Joussot‐Dubien}, \citenamefont {Vaucamps},\ and\ \citenamefont
  {Nouchi}}]{Dorthe1985Reactive}%
  \BibitemOpen
  \bibfield  {author} {\bibinfo {author} {\bibfnamefont {G.}~\bibnamefont
  {Dorthe}}, \bibinfo {author} {\bibfnamefont {M.}~\bibnamefont {Costes}},
  \bibinfo {author} {\bibfnamefont {C.}~\bibnamefont {Naulin}}, \bibinfo
  {author} {\bibfnamefont {J.}~\bibnamefont {Joussot‐Dubien}}, \bibinfo
  {author} {\bibfnamefont {C.}~\bibnamefont {Vaucamps}}, \ and\ \bibinfo
  {author} {\bibfnamefont {G.}~\bibnamefont {Nouchi}},\ }\bibfield  {title}
  {\enquote {\bibinfo {title} {Reactive scattering using pulsed crossed
  supersonic molecular beams. example of the {C+NO→CN+O} and
  {C+N$_2$O→CN+NO} reactions},}\ }\href {\doibase 10.1063/1.449171}
  {\bibfield  {journal} {\bibinfo  {journal} {The Journal of Chemical Physics}\
  }\textbf {\bibinfo {volume} {83}},\ \bibinfo {pages} {3171--3172} (\bibinfo
  {year} {1985})}\BibitemShut {NoStop}%
\bibitem [{\citenamefont {Sekiya}, \citenamefont {Tsuji},\ and\ \citenamefont
  {Nishimura}(1986)}]{Sekiya1986Vibrational}%
  \BibitemOpen
  \bibfield  {author} {\bibinfo {author} {\bibfnamefont {H.}~\bibnamefont
  {Sekiya}}, \bibinfo {author} {\bibfnamefont {M.}~\bibnamefont {Tsuji}}, \
  and\ \bibinfo {author} {\bibfnamefont {Y.}~\bibnamefont {Nishimura}},\
  }\bibfield  {title} {\enquote {\bibinfo {title} {Vibrational distributions of
  {CN}({X$^2{{\Sigma}}^{+}$}) resulting from the {C}($^3${P})+{NO} and
  {C}($^3${P})+{N}$_2${O} reactions at 300 {K}},}\ }\href {\doibase
  10.1063/1.450214} {\bibfield  {journal} {\bibinfo  {journal} {The Journal of
  Chemical Physics}\ }\textbf {\bibinfo {volume} {84}},\ \bibinfo {pages}
  {3739--3744} (\bibinfo {year} {1986})}\BibitemShut {NoStop}%
\bibitem [{\citenamefont {Naulin}\ \emph {et~al.}(1988)\citenamefont {Naulin},
  \citenamefont {Costes}, \citenamefont {Benseddik},\ and\ \citenamefont
  {Dorthe}}]{Naulin1988Kinematic}%
  \BibitemOpen
  \bibfield  {author} {\bibinfo {author} {\bibfnamefont {C.}~\bibnamefont
  {Naulin}}, \bibinfo {author} {\bibfnamefont {M.}~\bibnamefont {Costes}},
  \bibinfo {author} {\bibfnamefont {A.}~\bibnamefont {Benseddik}}, \ and\
  \bibinfo {author} {\bibfnamefont {G.}~\bibnamefont {Dorthe}},\ }\bibfield
  {title} {\enquote {\bibinfo {title} {Kinematic effects on laser-induced
  fluorescence measurements performed in reactive crossed beam experiments},}\
  }\href {\doibase 10.1155/LC.8.283} {\bibfield  {journal} {\bibinfo  {journal}
  {Laser Chemistry}\ }\textbf {\bibinfo {volume} {8}},\ \bibinfo {pages}
  {353067} (\bibinfo {year} {1988})}\BibitemShut {NoStop}%
\bibitem [{\citenamefont {Costes}\ \emph {et~al.}(1989)\citenamefont {Costes},
  \citenamefont {Naulin}, \citenamefont {Dorthe}, \citenamefont {Daleau},
  \citenamefont {Joussot-Dubien}, \citenamefont {Lalaude}, \citenamefont
  {Vinckert}, \citenamefont {Destor}, \citenamefont {Vaucamps},\ and\
  \citenamefont {Nouchi}}]{Costes1989pulsed}%
  \BibitemOpen
  \bibfield  {author} {\bibinfo {author} {\bibfnamefont {M.}~\bibnamefont
  {Costes}}, \bibinfo {author} {\bibfnamefont {C.}~\bibnamefont {Naulin}},
  \bibinfo {author} {\bibfnamefont {G.}~\bibnamefont {Dorthe}}, \bibinfo
  {author} {\bibfnamefont {G.}~\bibnamefont {Daleau}}, \bibinfo {author}
  {\bibfnamefont {J.}~\bibnamefont {Joussot-Dubien}}, \bibinfo {author}
  {\bibfnamefont {C.}~\bibnamefont {Lalaude}}, \bibinfo {author} {\bibfnamefont
  {M.}~\bibnamefont {Vinckert}}, \bibinfo {author} {\bibfnamefont
  {A.}~\bibnamefont {Destor}}, \bibinfo {author} {\bibfnamefont
  {C.}~\bibnamefont {Vaucamps}}, \ and\ \bibinfo {author} {\bibfnamefont
  {G.}~\bibnamefont {Nouchi}},\ }\bibfield  {title} {\enquote {\bibinfo {title}
  {A pulsed crossed supersonic molecular beam apparatus to study the dynamics
  of refractory atom reactions},}\ }\href {\doibase
  10.1088/0022-3735/22/12/013} {\bibfield  {journal} {\bibinfo  {journal}
  {Journal of Physics E: Scientific Instruments}\ }\textbf {\bibinfo {volume}
  {22}},\ \bibinfo {pages} {1017} (\bibinfo {year} {1989})}\BibitemShut
  {NoStop}%
\bibitem [{\citenamefont {Costes}\ \emph {et~al.}(1990)\citenamefont {Costes},
  \citenamefont {Naulin}, \citenamefont {Dorthe},\ and\ \citenamefont
  {Moudden}}]{Costes1990Reactive}%
  \BibitemOpen
  \bibfield  {author} {\bibinfo {author} {\bibfnamefont {M.}~\bibnamefont
  {Costes}}, \bibinfo {author} {\bibfnamefont {C.}~\bibnamefont {Naulin}},
  \bibinfo {author} {\bibfnamefont {G.}~\bibnamefont {Dorthe}}, \ and\ \bibinfo
  {author} {\bibfnamefont {Z.}~\bibnamefont {Moudden}},\ }\bibfield  {title}
  {\enquote {\bibinfo {title} {Reactive bimolecular collisions studied with
  combined pulsed lasers and pulsed, crossed, supersonic molecular beams},}\
  }\href {\doibase 10.1155/1990/32023} {\bibfield  {journal} {\bibinfo
  {journal} {Laser Chemistry}\ }\textbf {\bibinfo {volume} {10}},\ \bibinfo
  {pages} {032023} (\bibinfo {year} {1990})}\BibitemShut {NoStop}%
\bibitem [{\citenamefont {Costes}, \citenamefont {Naulin},\ and\ \citenamefont
  {Dorthe}(1990)}]{Costes1990dissociation}%
  \BibitemOpen
  \bibfield  {author} {\bibinfo {author} {\bibfnamefont {M.}~\bibnamefont
  {Costes}}, \bibinfo {author} {\bibfnamefont {C.}~\bibnamefont {Naulin}}, \
  and\ \bibinfo {author} {\bibfnamefont {G.}~\bibnamefont {Dorthe}},\
  }\bibfield  {title} {\enquote {\bibinfo {title} {{The dissociation energy of
  the CN radical determined from the CN internal energy release of the C + NO
  yielding CN + O reaction}},}\ }\href@noop {} {\bibfield  {journal} {\bibinfo
  {journal} {Astronomy and Astrophysics}\ }\textbf {\bibinfo {volume} {232}},\
  \bibinfo {pages} {270--276} (\bibinfo {year} {1990})}\BibitemShut {NoStop}%
\bibitem [{\citenamefont {Dorthe}\ \emph {et~al.}(1991)\citenamefont {Dorthe},
  \citenamefont {Caubet}, \citenamefont {Vias}, \citenamefont {Barrere},\ and\
  \citenamefont {Marchais}}]{Dorthe1991Fast}%
  \BibitemOpen
  \bibfield  {author} {\bibinfo {author} {\bibfnamefont {G.}~\bibnamefont
  {Dorthe}}, \bibinfo {author} {\bibfnamefont {P.}~\bibnamefont {Caubet}},
  \bibinfo {author} {\bibfnamefont {T.}~\bibnamefont {Vias}}, \bibinfo {author}
  {\bibfnamefont {B.}~\bibnamefont {Barrere}}, \ and\ \bibinfo {author}
  {\bibfnamefont {J.}~\bibnamefont {Marchais}},\ }\bibfield  {title} {\enquote
  {\bibinfo {title} {Fast flow studies of atomic carbon kinetics at room
  temperature},}\ }\href {\doibase 10.1021/j100166a038} {\bibfield  {journal}
  {\bibinfo  {journal} {The Journal of Physical Chemistry}\ }\textbf {\bibinfo
  {volume} {95}},\ \bibinfo {pages} {5109--5116} (\bibinfo {year}
  {1991})}\BibitemShut {NoStop}%
\bibitem [{\citenamefont {Naulin}, \citenamefont {Costes},\ and\ \citenamefont
  {Dorthe}(1991)}]{Naulin1991ReactionDynamics}%
  \BibitemOpen
  \bibfield  {author} {\bibinfo {author} {\bibfnamefont {C.}~\bibnamefont
  {Naulin}}, \bibinfo {author} {\bibfnamefont {M.}~\bibnamefont {Costes}}, \
  and\ \bibinfo {author} {\bibfnamefont {G.}~\bibnamefont {Dorthe}},\
  }\bibfield  {title} {\enquote {\bibinfo {title} {The {C}({$^3${P}$_J$}) +
  {NO}({X}{$ ^2{{\Pi}}_r$}) $\rightarrow$ {CN}({X}{$^2{{\Sigma}}^+$}) +
  {O}({{$^3${P}$_J$}}) reaction dynamics studied at 0.06 and 0.23 {eV} relative
  translational energy in pulsed crossed supersonic molecular beams},}\ }\href
  {\doibase 10.1016/0301-0104(91)80065-P} {\bibfield  {journal} {\bibinfo
  {journal} {Chemical Physics}\ }\textbf {\bibinfo {volume} {153}},\ \bibinfo
  {pages} {519--530} (\bibinfo {year} {1991})}\BibitemShut {NoStop}%
\bibitem [{\citenamefont {Costes}\ \emph {et~al.}(1993)\citenamefont {Costes},
  \citenamefont {Naulin}, \citenamefont {Ghanem},\ and\ \citenamefont
  {Dorthe}}]{Costes1993Translational}%
  \BibitemOpen
  \bibfield  {author} {\bibinfo {author} {\bibfnamefont {M.}~\bibnamefont
  {Costes}}, \bibinfo {author} {\bibfnamefont {C.}~\bibnamefont {Naulin}},
  \bibinfo {author} {\bibfnamefont {N.}~\bibnamefont {Ghanem}}, \ and\ \bibinfo
  {author} {\bibfnamefont {G.}~\bibnamefont {Dorthe}},\ }\bibfield  {title}
  {\enquote {\bibinfo {title} {Translational energy dependence of the reactive
  cross-section of the {C} + {NO} → {CN} + {O} reaction studied in pulsed{,}
  crossed{,} supersonic molecular beams},}\ }\href {\doibase
  10.1039/FT9938901501} {\bibfield  {journal} {\bibinfo  {journal} {Journal of
  the Chemical Society, Faraday Transactions}\ }\textbf {\bibinfo {volume}
  {89}},\ \bibinfo {pages} {1501--1504} (\bibinfo {year} {1993})}\BibitemShut
  {NoStop}%
\bibitem [{\citenamefont {Becker}, \citenamefont {Brockmann},\ and\
  \citenamefont {Wiesen}(1988)}]{Becker1988Spectroscopic}%
  \BibitemOpen
  \bibfield  {author} {\bibinfo {author} {\bibfnamefont {K.~H.}\ \bibnamefont
  {Becker}}, \bibinfo {author} {\bibfnamefont {K.~J.}\ \bibnamefont
  {Brockmann}}, \ and\ \bibinfo {author} {\bibfnamefont {P.}~\bibnamefont
  {Wiesen}},\ }\bibfield  {title} {\enquote {\bibinfo {title} {Spectroscopic
  identification of {C}($^3${P}) atoms in halogenomethane + {H} flame systems
  and measurements of {C}($^3${P}) reaction rate constants by two-photon
  laser-induced fluorescence},}\ }\href {\doibase 10.1039/F29888400455}
  {\bibfield  {journal} {\bibinfo  {journal} {J. Chem. Soc., Faraday Trans. 2}\
  }\textbf {\bibinfo {volume} {84}},\ \bibinfo {pages} {455--461} (\bibinfo
  {year} {1988})}\BibitemShut {NoStop}%
\bibitem [{\citenamefont {Bergeat}\ \emph {et~al.}(1999)\citenamefont
  {Bergeat}, \citenamefont {Calvo}, \citenamefont {Dorthe},\ and\ \citenamefont
  {Loison}}]{Bergeat1999FastFlow}%
  \BibitemOpen
  \bibfield  {author} {\bibinfo {author} {\bibfnamefont {A.}~\bibnamefont
  {Bergeat}}, \bibinfo {author} {\bibfnamefont {T.}~\bibnamefont {Calvo}},
  \bibinfo {author} {\bibfnamefont {G.}~\bibnamefont {Dorthe}}, \ and\ \bibinfo
  {author} {\bibfnamefont {J.~C.}\ \bibnamefont {Loison}},\ }\bibfield  {title}
  {\enquote {\bibinfo {title} {Fast-flow study of the {C} + {NO} and {C} +
  {O}$_2$ reactions},}\ }\href {\doibase 10.1016/S0009-2614(99)00586-2}
  {\bibfield  {journal} {\bibinfo  {journal} {Chemical Physics Letters}\
  }\textbf {\bibinfo {volume} {308}},\ \bibinfo {pages} {7--12} (\bibinfo
  {year} {1999})}\BibitemShut {NoStop}%
\bibitem [{\citenamefont {Chastaing}, \citenamefont {Le~Picard},\ and\
  \citenamefont {Sims}(2000)}]{Chastaing2000Direct}%
  \BibitemOpen
  \bibfield  {author} {\bibinfo {author} {\bibfnamefont {D.}~\bibnamefont
  {Chastaing}}, \bibinfo {author} {\bibfnamefont {S.~D.}\ \bibnamefont
  {Le~Picard}}, \ and\ \bibinfo {author} {\bibfnamefont {I.~R.}\ \bibnamefont
  {Sims}},\ }\bibfield  {title} {\enquote {\bibinfo {title} {Direct kinetic
  measurements on reactions of atomic carbon, {C}({$^3$P}), with {O}$_2$ and
  {NO} at temperatures down to 15 {K}},}\ }\href {\doibase 10.1063/1.481448}
  {\bibfield  {journal} {\bibinfo  {journal} {The Journal of Chemical Physics}\
  }\textbf {\bibinfo {volume} {112}},\ \bibinfo {pages} {8466--8469} (\bibinfo
  {year} {2000})}\BibitemShut {NoStop}%
\bibitem [{\citenamefont {Dean}, \citenamefont {Hanson},\ and\ \citenamefont
  {Bowman}(1991)}]{Dean1991ShockTube}%
  \BibitemOpen
  \bibfield  {author} {\bibinfo {author} {\bibfnamefont {A.~J.}\ \bibnamefont
  {Dean}}, \bibinfo {author} {\bibfnamefont {R.~K.}\ \bibnamefont {Hanson}}, \
  and\ \bibinfo {author} {\bibfnamefont {C.~T.}\ \bibnamefont {Bowman}},\
  }\bibfield  {title} {\enquote {\bibinfo {title} {A shock tube study of
  reactions of carbon atoms and methylidyne with nitric oxide including product
  channel measurements},}\ }\href {\doibase 10.1021/j100161a042} {\bibfield
  {journal} {\bibinfo  {journal} {The Journal of Physical Chemistry}\ }\textbf
  {\bibinfo {volume} {95}},\ \bibinfo {pages} {3180--3189} (\bibinfo {year}
  {1991})}\BibitemShut {NoStop}%
\bibitem [{\citenamefont {Koner}, \citenamefont {Bemish},\ and\ \citenamefont
  {Meuwly}(2018)}]{koner2018cno}%
  \BibitemOpen
  \bibfield  {author} {\bibinfo {author} {\bibfnamefont {D.}~\bibnamefont
  {Koner}}, \bibinfo {author} {\bibfnamefont {R.~J.}\ \bibnamefont {Bemish}}, \
  and\ \bibinfo {author} {\bibfnamefont {M.}~\bibnamefont {Meuwly}},\
  }\bibfield  {title} {\enquote {\bibinfo {title} {The {C}($^3${P}) + {NO}({X}
  $^2{{\Pi}}$) $\rightarrow$ {O}($^3${P}) + {CN}({X} $^2{{\Sigma}}^+$),
  {N}($^2${D})/{N}($^4${S}) + {CO}({X} $^1{{\Sigma}}^+$) reaction: {R}ates,
  branching ratios, and final states from 15 {K} to 20 000 {K}},}\ }\href
  {\doibase 10.1063/1.5046906} {\bibfield  {journal} {\bibinfo  {journal} {The
  Journal of Chemical Physics}\ }\textbf {\bibinfo {volume} {149}},\ \bibinfo
  {pages} {094305} (\bibinfo {year} {2018})}\BibitemShut {NoStop}%
\bibitem [{\citenamefont {Boden}\ and\ \citenamefont
  {Thrush}(1968)}]{Boden1968Kinetics}%
  \BibitemOpen
  \bibfield  {author} {\bibinfo {author} {\bibfnamefont {J.~C.}\ \bibnamefont
  {Boden}}\ and\ \bibinfo {author} {\bibfnamefont {B.~A.}\ \bibnamefont
  {Thrush}},\ }\bibfield  {title} {\enquote {\bibinfo {title} {Kinetics of
  reactions involving {CN} emission {IV}. {S}tudy of the reactions of {CN} by
  electronic absorption spectroscopy},}\ }\href {\doibase
  10.1098/rspa.1968.0108} {\bibfield  {journal} {\bibinfo  {journal}
  {Proceedings of the Royal Society of London. Series A. Mathematical and
  Physical Sciences}\ }\textbf {\bibinfo {volume} {305}},\ \bibinfo {pages}
  {107--123} (\bibinfo {year} {1968})}\BibitemShut {NoStop}%
\bibitem [{\citenamefont {Schmatjko}\ and\ \citenamefont
  {Wolfrum}(1978)}]{Schmatjko1978Reaktionen}%
  \BibitemOpen
  \bibfield  {author} {\bibinfo {author} {\bibfnamefont {K.~J.}\ \bibnamefont
  {Schmatjko}}\ and\ \bibinfo {author} {\bibfnamefont {J.}~\bibnamefont
  {Wolfrum}},\ }\bibfield  {title} {\enquote {\bibinfo {title} {{Reaktionen}
  von {Molekülen} in definierten {Schwingungszuständen}: {VI}.
  {Energieverteilung} in den {Reaktionen} {CN}(v) + {O}($^3${P}),
  {O}$^{*}_2$},}\ }\href {\doibase 10.1002/bbpc.197800076} {\bibfield
  {journal} {\bibinfo  {journal} {Berichte der Bunsengesellschaft für
  physikalische Chemie}\ }\textbf {\bibinfo {volume} {82}},\ \bibinfo {pages}
  {419--428} (\bibinfo {year} {1978})}\BibitemShut {NoStop}%
\bibitem [{\citenamefont {Louge}\ and\ \citenamefont
  {Hanson}(1984)}]{Louge1984Cyanogen}%
  \BibitemOpen
  \bibfield  {author} {\bibinfo {author} {\bibfnamefont {M.~Y.}\ \bibnamefont
  {Louge}}\ and\ \bibinfo {author} {\bibfnamefont {R.~K.}\ \bibnamefont
  {Hanson}},\ }\bibfield  {title} {\enquote {\bibinfo {title} {Shock tube study
  of cyanogen oxidation kinetics},}\ }\href {\doibase 10.1002/kin.550160306}
  {\bibfield  {journal} {\bibinfo  {journal} {International Journal of Chemical
  Kinetics}\ }\textbf {\bibinfo {volume} {16}},\ \bibinfo {pages} {231--250}
  (\bibinfo {year} {1984})}\BibitemShut {NoStop}%
\bibitem [{\citenamefont {Davidson}\ \emph {et~al.}(1991)\citenamefont
  {Davidson}, \citenamefont {Dean}, \citenamefont {Dirosa},\ and\ \citenamefont
  {Hanson}}]{Davidson1991ShockCN}%
  \BibitemOpen
  \bibfield  {author} {\bibinfo {author} {\bibfnamefont {D.~F.}\ \bibnamefont
  {Davidson}}, \bibinfo {author} {\bibfnamefont {A.~J.}\ \bibnamefont {Dean}},
  \bibinfo {author} {\bibfnamefont {M.~D.}\ \bibnamefont {Dirosa}}, \ and\
  \bibinfo {author} {\bibfnamefont {R.~K.}\ \bibnamefont {Hanson}},\ }\bibfield
   {title} {\enquote {\bibinfo {title} {Shock tube measurements of the
  reactions of {CN} with {O} and {O}$_2$},}\ }\href {\doibase
  10.1002/kin.550231107} {\bibfield  {journal} {\bibinfo  {journal}
  {International Journal of Chemical Kinetics}\ }\textbf {\bibinfo {volume}
  {23}},\ \bibinfo {pages} {1035--1050} (\bibinfo {year} {1991})}\BibitemShut
  {NoStop}%
\bibitem [{\citenamefont {Titarchuk}\ and\ \citenamefont
  {Halpern}(1995)}]{Titarchuk1995Kinetics}%
  \BibitemOpen
  \bibfield  {author} {\bibinfo {author} {\bibfnamefont {T.~A.}\ \bibnamefont
  {Titarchuk}}\ and\ \bibinfo {author} {\bibfnamefont {J.~B.}\ \bibnamefont
  {Halpern}},\ }\bibfield  {title} {\enquote {\bibinfo {title} {Kinetics of
  {CN} reactions with {O}($^3${P}) and {SO}$_2$},}\ }\href {\doibase
  10.1016/0009-2614(94)01338-V} {\bibfield  {journal} {\bibinfo  {journal}
  {Chemical Physics Letters}\ }\textbf {\bibinfo {volume} {232}},\ \bibinfo
  {pages} {192--196} (\bibinfo {year} {1995})}\BibitemShut {NoStop}%
\bibitem [{\citenamefont {Lindackers}, \citenamefont {Burmeister},\ and\
  \citenamefont {Roth}(1990)}]{Lindackers1990High}%
  \BibitemOpen
  \bibfield  {author} {\bibinfo {author} {\bibfnamefont {D.}~\bibnamefont
  {Lindackers}}, \bibinfo {author} {\bibfnamefont {M.}~\bibnamefont
  {Burmeister}}, \ and\ \bibinfo {author} {\bibfnamefont {P.}~\bibnamefont
  {Roth}},\ }\bibfield  {title} {\enquote {\bibinfo {title} {High-temperature
  kinetics of the reaction {CN} + {CO}$_2$},}\ }\href {\doibase
  10.1016/0010-2180(90)90023-K} {\bibfield  {journal} {\bibinfo  {journal}
  {Combustion and Flame}\ }\textbf {\bibinfo {volume} {81}},\ \bibinfo {pages}
  {251--259} (\bibinfo {year} {1990})}\BibitemShut {NoStop}%
\bibitem [{\citenamefont {Halvick}, \citenamefont {Rayez},\ and\ \citenamefont
  {Evleth}(1984)}]{Halvick1984Theoretical}%
  \BibitemOpen
  \bibfield  {author} {\bibinfo {author} {\bibfnamefont {P.}~\bibnamefont
  {Halvick}}, \bibinfo {author} {\bibfnamefont {J.~C.}\ \bibnamefont {Rayez}},
  \ and\ \bibinfo {author} {\bibfnamefont {E.~M.}\ \bibnamefont {Evleth}},\
  }\bibfield  {title} {\enquote {\bibinfo {title} {Theoretical approach to the
  reaction {C}($^3${P})+{NO}({X}{$^2{{\Pi}}$})},}\ }\href {\doibase
  10.1063/1.447704} {\bibfield  {journal} {\bibinfo  {journal} {The Journal of
  Chemical Physics}\ }\textbf {\bibinfo {volume} {81}},\ \bibinfo {pages}
  {728--737} (\bibinfo {year} {1984})}\BibitemShut {NoStop}%
\bibitem [{\citenamefont {Abrahamsson}\ \emph {et~al.}(2008)\citenamefont
  {Abrahamsson}, \citenamefont {Andersson}, \citenamefont {Marković},\ and\
  \citenamefont {Nyman}}]{Abrahamsson2008New}%
  \BibitemOpen
  \bibfield  {author} {\bibinfo {author} {\bibfnamefont {E.}~\bibnamefont
  {Abrahamsson}}, \bibinfo {author} {\bibfnamefont {S.}~\bibnamefont
  {Andersson}}, \bibinfo {author} {\bibfnamefont {N.}~\bibnamefont
  {Marković}}, \ and\ \bibinfo {author} {\bibfnamefont {G.}~\bibnamefont
  {Nyman}},\ }\bibfield  {title} {\enquote {\bibinfo {title} {A new reaction
  path for the {C} + {NO} reaction: dynamics on the {$^4A^{\prime\prime}$}
  potential-energy surface},}\ }\href {\doibase 10.1039/B803787F} {\bibfield
  {journal} {\bibinfo  {journal} {Phys. Chem. Chem. Phys.}\ }\textbf {\bibinfo
  {volume} {10}},\ \bibinfo {pages} {4400--4409} (\bibinfo {year}
  {2008})}\BibitemShut {NoStop}%
\bibitem [{\citenamefont {Rayez}\ \emph {et~al.}(1986)\citenamefont {Rayez},
  \citenamefont {Halvick}, \citenamefont {Rayez},\ and\ \citenamefont
  {Duguay}}]{Rayez1986Three}%
  \BibitemOpen
  \bibfield  {author} {\bibinfo {author} {\bibfnamefont {J.~C.}\ \bibnamefont
  {Rayez}}, \bibinfo {author} {\bibfnamefont {P.}~\bibnamefont {Halvick}},
  \bibinfo {author} {\bibfnamefont {M.~T.}\ \bibnamefont {Rayez}}, \ and\
  \bibinfo {author} {\bibfnamefont {B.}~\bibnamefont {Duguay}},\ }\bibfield
  {title} {\enquote {\bibinfo {title} {Three-atom indirect exchange reactions.
  {I.} {1D} {QCT} study of the topological factors influencing the energetic
  distribution on the products},}\ }\href {\doibase
  10.1016/0301-0104(86)85075-3} {\bibfield  {journal} {\bibinfo  {journal}
  {Chemical Physics}\ }\textbf {\bibinfo {volume} {101}},\ \bibinfo {pages}
  {401--412} (\bibinfo {year} {1986})}\BibitemShut {NoStop}%
\bibitem [{\citenamefont {Halvick}\ \emph {et~al.}(1987)\citenamefont
  {Halvick}, \citenamefont {Rayez}, \citenamefont {Rayez},\ and\ \citenamefont
  {Duguay}}]{Halvick1987Three}%
  \BibitemOpen
  \bibfield  {author} {\bibinfo {author} {\bibfnamefont {P.}~\bibnamefont
  {Halvick}}, \bibinfo {author} {\bibfnamefont {J.~C.}\ \bibnamefont {Rayez}},
  \bibinfo {author} {\bibfnamefont {M.~T.}\ \bibnamefont {Rayez}}, \ and\
  \bibinfo {author} {\bibfnamefont {B.}~\bibnamefont {Duguay}},\ }\bibfield
  {title} {\enquote {\bibinfo {title} {Three-atom indirect exchange reactions.
  {II}. dynamical behaviours explained by a simple model},}\ }\href {\doibase
  10.1016/0301-0104(87)85051-6} {\bibfield  {journal} {\bibinfo  {journal}
  {Chemical Physics}\ }\textbf {\bibinfo {volume} {114}},\ \bibinfo {pages}
  {375--387} (\bibinfo {year} {1987})}\BibitemShut {NoStop}%
\bibitem [{\citenamefont {Persson}, \citenamefont {Roos},\ and\ \citenamefont
  {Simonson}(1995)}]{Persson1995theoretical}%
  \BibitemOpen
  \bibfield  {author} {\bibinfo {author} {\bibfnamefont {J.~B.}\ \bibnamefont
  {Persson}}, \bibinfo {author} {\bibfnamefont {B.~O.}\ \bibnamefont {Roos}}, \
  and\ \bibinfo {author} {\bibfnamefont {M.}~\bibnamefont {Simonson}},\
  }\bibfield  {title} {\enquote {\bibinfo {title} {A theoretical investigation
  of the two lowest potential energy surfaces for the reaction
  {C}($^3${P})+{NO}({$^2{{\Pi}}$})},}\ }\href {\doibase
  10.1016/0009-2614(94)01501-L} {\bibfield  {journal} {\bibinfo  {journal}
  {Chemical Physics Letters}\ }\textbf {\bibinfo {volume} {234}},\ \bibinfo
  {pages} {382--389} (\bibinfo {year} {1995})}\BibitemShut {NoStop}%
\bibitem [{\citenamefont {Simonson}\ \emph {et~al.}(1995)\citenamefont
  {Simonson}, \citenamefont {Marković}, \citenamefont {Nordholm},\ and\
  \citenamefont {Persson}}]{Simonson1995Quasiclassical}%
  \BibitemOpen
  \bibfield  {author} {\bibinfo {author} {\bibfnamefont {M.}~\bibnamefont
  {Simonson}}, \bibinfo {author} {\bibfnamefont {N.}~\bibnamefont {Marković}},
  \bibinfo {author} {\bibfnamefont {S.}~\bibnamefont {Nordholm}}, \ and\
  \bibinfo {author} {\bibfnamefont {B.~J.}\ \bibnamefont {Persson}},\
  }\bibfield  {title} {\enquote {\bibinfo {title} {Quasiclassical trajectory
  study of the {C} + {NO} reaction on a new potential energy surface},}\ }\href
  {\doibase 10.1016/0301-0104(95)00236-H} {\bibfield  {journal} {\bibinfo
  {journal} {Chemical Physics}\ }\textbf {\bibinfo {volume} {200}},\ \bibinfo
  {pages} {141--160} (\bibinfo {year} {1995})}\BibitemShut {NoStop}%
\bibitem [{\citenamefont {Andersson}, \citenamefont {Marković},\ and\
  \citenamefont {Nyman}(2000{\natexlab{a}})}]{Andersson2000QCT}%
  \BibitemOpen
  \bibfield  {author} {\bibinfo {author} {\bibfnamefont {S.}~\bibnamefont
  {Andersson}}, \bibinfo {author} {\bibfnamefont {N.}~\bibnamefont
  {Marković}}, \ and\ \bibinfo {author} {\bibfnamefont {G.}~\bibnamefont
  {Nyman}},\ }\bibfield  {title} {\enquote {\bibinfo {title} {Quasi-classical
  trajectory simulations of {C} + {NO} crossed molecular beam experiments},}\
  }\href {\doibase 10.1016/S0301-0104(00)00214-7} {\bibfield  {journal}
  {\bibinfo  {journal} {Chemical Physics}\ }\textbf {\bibinfo {volume} {259}},\
  \bibinfo {pages} {99--108} (\bibinfo {year}
  {2000}{\natexlab{a}})}\BibitemShut {NoStop}%
\bibitem [{\citenamefont {Andersson}, \citenamefont {Marković},\ and\
  \citenamefont {Nyman}(2000{\natexlab{b}})}]{Andersson2000Improved}%
  \BibitemOpen
  \bibfield  {author} {\bibinfo {author} {\bibfnamefont {S.}~\bibnamefont
  {Andersson}}, \bibinfo {author} {\bibfnamefont {N.}~\bibnamefont
  {Marković}}, \ and\ \bibinfo {author} {\bibfnamefont {G.}~\bibnamefont
  {Nyman}},\ }\bibfield  {title} {\enquote {\bibinfo {title} {An improved
  potential energy surface for the {C} + {NO} reaction},}\ }\href {\doibase
  10.1039/A908183F} {\bibfield  {journal} {\bibinfo  {journal} {Phys. Chem.
  Chem. Phys.}\ }\textbf {\bibinfo {volume} {2}},\ \bibinfo {pages} {613--620}
  (\bibinfo {year} {2000}{\natexlab{b}})}\BibitemShut {NoStop}%
\bibitem [{\citenamefont {Andersson}, \citenamefont {Markovi{\' c}},\ and\
  \citenamefont {Nyman}(2003)}]{Andersson2003Computational}%
  \BibitemOpen
  \bibfield  {author} {\bibinfo {author} {\bibfnamefont {S.}~\bibnamefont
  {Andersson}}, \bibinfo {author} {\bibfnamefont {N.}~\bibnamefont {Markovi{\'
  c}}}, \ and\ \bibinfo {author} {\bibfnamefont {G.}~\bibnamefont {Nyman}},\
  }\bibfield  {title} {\enquote {\bibinfo {title} {Computational studies of the
  kinetics of the {C} + {NO} and {O} + {CN} reactions},}\ }\href {\doibase
  10.1021/jp0222604} {\bibfield  {journal} {\bibinfo  {journal} {The Journal of
  Physical Chemistry A}\ }\textbf {\bibinfo {volume} {107}},\ \bibinfo {pages}
  {5439--5447} (\bibinfo {year} {2003})}\BibitemShut {NoStop}%
\bibitem [{\citenamefont {Abrahamsson}\ \emph {et~al.}(2009)\citenamefont
  {Abrahamsson}, \citenamefont {Andersson}, \citenamefont {Marković},\ and\
  \citenamefont {Nyman}}]{Abrahamsson2009Dynamics}%
  \BibitemOpen
  \bibfield  {author} {\bibinfo {author} {\bibfnamefont {E.}~\bibnamefont
  {Abrahamsson}}, \bibinfo {author} {\bibfnamefont {S.}~\bibnamefont
  {Andersson}}, \bibinfo {author} {\bibfnamefont {N.}~\bibnamefont
  {Marković}}, \ and\ \bibinfo {author} {\bibfnamefont {G.}~\bibnamefont
  {Nyman}},\ }\bibfield  {title} {\enquote {\bibinfo {title} {Dynamics of the
  {O} + {CN} reaction and {N} + {CO} scattering on two coupled surfaces},}\
  }\href {\doibase 10.1021/jp904954k} {\bibfield  {journal} {\bibinfo
  {journal} {The Journal of Physical Chemistry A}\ }\textbf {\bibinfo {volume}
  {113}},\ \bibinfo {pages} {14824--14830} (\bibinfo {year}
  {2009})}\BibitemShut {NoStop}%
\bibitem [{\citenamefont {Abrahamsson}\ \emph {et~al.}(2006)\citenamefont
  {Abrahamsson}, \citenamefont {Andersson}, \citenamefont {Nyman},\ and\
  \citenamefont {Marković}}]{Abrahamsson2006Classical}%
  \BibitemOpen
  \bibfield  {author} {\bibinfo {author} {\bibfnamefont {E.}~\bibnamefont
  {Abrahamsson}}, \bibinfo {author} {\bibfnamefont {S.}~\bibnamefont
  {Andersson}}, \bibinfo {author} {\bibfnamefont {G.}~\bibnamefont {Nyman}}, \
  and\ \bibinfo {author} {\bibfnamefont {N.}~\bibnamefont {Marković}},\
  }\bibfield  {title} {\enquote {\bibinfo {title} {Classical and quantum
  dynamics of the {O}+{CN} reaction},}\ }\href {\doibase
  10.1016/j.chemphys.2005.11.012} {\bibfield  {journal} {\bibinfo  {journal}
  {Chemical Physics}\ }\textbf {\bibinfo {volume} {324}},\ \bibinfo {pages}
  {507--514} (\bibinfo {year} {2006})}\BibitemShut {NoStop}%
\bibitem [{\citenamefont {Frankcombe}\ and\ \citenamefont
  {Andersson}(2012)}]{Frankcombe2012Adiabatic}%
  \BibitemOpen
  \bibfield  {author} {\bibinfo {author} {\bibfnamefont {T.~J.}\ \bibnamefont
  {Frankcombe}}\ and\ \bibinfo {author} {\bibfnamefont {S.}~\bibnamefont
  {Andersson}},\ }\bibfield  {title} {\enquote {\bibinfo {title} {An adiabatic
  capture theory and quasiclassical trajectory study of {C} + {NO} and {O} +
  {CN} on the {$^2A^{\prime}$}, {$^2A^{\prime\prime}$}, and
  {$^4A^{\prime\prime}$} potential energy surfaces},}\ }\href {\doibase
  10.1021/jp3018869} {\bibfield  {journal} {\bibinfo  {journal} {The Journal of
  Physical Chemistry A}\ }\textbf {\bibinfo {volume} {116}},\ \bibinfo {pages}
  {4705--4711} (\bibinfo {year} {2012})}\BibitemShut {NoStop}%
\bibitem [{\citenamefont {Gonçalves}\ \emph {et~al.}(2018)\citenamefont
  {Gonçalves}, \citenamefont {Galvão}, \citenamefont {Mota}, \citenamefont
  {Braga},\ and\ \citenamefont {Varandas}}]{goncalves2018cno}%
  \BibitemOpen
  \bibfield  {author} {\bibinfo {author} {\bibfnamefont {C.~E.~M.}\
  \bibnamefont {Gonçalves}}, \bibinfo {author} {\bibfnamefont {B.~R.~L.}\
  \bibnamefont {Galvão}}, \bibinfo {author} {\bibfnamefont {V.~C.}\
  \bibnamefont {Mota}}, \bibinfo {author} {\bibfnamefont {J.~P.}\ \bibnamefont
  {Braga}}, \ and\ \bibinfo {author} {\bibfnamefont {A.~J.~C.}\ \bibnamefont
  {Varandas}},\ }\bibfield  {title} {\enquote {\bibinfo {title} {Accurate
  {Explicit-Correlation-MRCI-Based} {DMBE} potential-energy surface for
  ground-state {CNO}},}\ }\href {\doibase 10.1021/acs.jpca.8b01881} {\bibfield
  {journal} {\bibinfo  {journal} {The Journal of Physical Chemistry A}\
  }\textbf {\bibinfo {volume} {122}},\ \bibinfo {pages} {4198--4207} (\bibinfo
  {year} {2018})}\BibitemShut {NoStop}%
\bibitem [{\citenamefont {Alves}\ \emph {et~al.}(2021)\citenamefont {Alves},
  \citenamefont {Gonçalves}, \citenamefont {Braga}, \citenamefont {Mota},
  \citenamefont {Varandas},\ and\ \citenamefont {Galvão}}]{alves2021dmbe}%
  \BibitemOpen
  \bibfield  {author} {\bibinfo {author} {\bibfnamefont {M.~O.}\ \bibnamefont
  {Alves}}, \bibinfo {author} {\bibfnamefont {C.~E.~M.}\ \bibnamefont
  {Gonçalves}}, \bibinfo {author} {\bibfnamefont {J.~P.}\ \bibnamefont
  {Braga}}, \bibinfo {author} {\bibfnamefont {V.~C.}\ \bibnamefont {Mota}},
  \bibinfo {author} {\bibfnamefont {A.~J.~C.}\ \bibnamefont {Varandas}}, \ and\
  \bibinfo {author} {\bibfnamefont {B.~R.~L.}\ \bibnamefont {Galvão}},\
  }\bibfield  {title} {\enquote {\bibinfo {title} {Accurate {DMBE}
  potential-energy surface for {CNO}($^2${A}$''$) and rate coefficients in
  {C}($^3${P}) + {NO} collisions},}\ }\href {\doibase 10.1063/5.0031199}
  {\bibfield  {journal} {\bibinfo  {journal} {The Journal of Chemical Physics}\
  }\textbf {\bibinfo {volume} {154}},\ \bibinfo {pages} {034303} (\bibinfo
  {year} {2021})}\BibitemShut {NoStop}%
\bibitem [{\citenamefont {Alves}\ \emph {et~al.}(2023)\citenamefont {Alves},
  \citenamefont {Mota}, \citenamefont {Braga}, \citenamefont {Varandas},
  \citenamefont {Guo},\ and\ \citenamefont {Galvão}}]{alves2023high}%
  \BibitemOpen
  \bibfield  {author} {\bibinfo {author} {\bibfnamefont {M.~O.}\ \bibnamefont
  {Alves}}, \bibinfo {author} {\bibfnamefont {V.~C.}\ \bibnamefont {Mota}},
  \bibinfo {author} {\bibfnamefont {J.~P.}\ \bibnamefont {Braga}}, \bibinfo
  {author} {\bibfnamefont {A.~J.~C.}\ \bibnamefont {Varandas}}, \bibinfo
  {author} {\bibfnamefont {H.}~\bibnamefont {Guo}}, \ and\ \bibinfo {author}
  {\bibfnamefont {B.~R.~L.}\ \bibnamefont {Galvão}},\ }\bibfield  {title}
  {\enquote {\bibinfo {title} {High-accuracy {DMBE} potential energy surface
  for {CNO}($^4${A}$''$) and the rate coefficients for the {C} + {NO} reaction
  in the $^2${A}$'$, $^2${A}$''$, and $^4${A}$''$ states},}\ }\href {\doibase
  10.1063/5.0180918} {\bibfield  {journal} {\bibinfo  {journal} {The Journal of
  Chemical Physics}\ }\textbf {\bibinfo {volume} {159}},\ \bibinfo {pages}
  {214307} (\bibinfo {year} {2023})}\BibitemShut {NoStop}%
\bibitem [{\citenamefont {Alves}\ \emph {et~al.}(2019)\citenamefont {Alves},
  \citenamefont {Gon{\c{c}}alves}, \citenamefont {Braga}, \citenamefont {Mota},
  \citenamefont {Varandas},\ and\ \citenamefont
  {Galvão}}]{Alves2019Quasiclassical}%
  \BibitemOpen
  \bibfield  {author} {\bibinfo {author} {\bibfnamefont {M.~V.}\ \bibnamefont
  {Alves}}, \bibinfo {author} {\bibfnamefont {C.~E.~M.}\ \bibnamefont
  {Gon{\c{c}}alves}}, \bibinfo {author} {\bibfnamefont {J.~P.}\ \bibnamefont
  {Braga}}, \bibinfo {author} {\bibfnamefont {V.~C.}\ \bibnamefont {Mota}},
  \bibinfo {author} {\bibfnamefont {A.~J.~C.}\ \bibnamefont {Varandas}}, \ and\
  \bibinfo {author} {\bibfnamefont {B.~R.~L.}\ \bibnamefont {Galvão}},\
  }\bibfield  {title} {\enquote {\bibinfo {title} {Quasiclassical study of the
  {C}($^3${P}) + {NO}({X}{$^2{{\Pi}}$}) and {O}($^3${P}) +
  {CN}({X}{$^2{{\Sigma}}^{+}$}) collisional processes on an accurate {DMBE}
  potential energy surface},}\ }\href {\doibase 10.1021/acs.jpca.9b03860}
  {\bibfield  {journal} {\bibinfo  {journal} {The Journal of Physical Chemistry
  A}\ }\textbf {\bibinfo {volume} {123}},\ \bibinfo {pages} {7195--7200}
  (\bibinfo {year} {2019})}\BibitemShut {NoStop}%
\bibitem [{\citenamefont {Lu}\ \emph {et~al.}(2024)\citenamefont {Lu},
  \citenamefont {Alves}, \citenamefont {Galvão},\ and\ \citenamefont
  {Guo}}]{Lu2024Theoretical}%
  \BibitemOpen
  \bibfield  {author} {\bibinfo {author} {\bibfnamefont {D.}~\bibnamefont
  {Lu}}, \bibinfo {author} {\bibfnamefont {M.~O.}\ \bibnamefont {Alves}},
  \bibinfo {author} {\bibfnamefont {B.~R.~L.}\ \bibnamefont {Galvão}}, \ and\
  \bibinfo {author} {\bibfnamefont {H.}~\bibnamefont {Guo}},\ }\bibfield
  {title} {\enquote {\bibinfo {title} {Theoretical study of the {O}($^3${P}) +
  {CN}({X}{$^2{{\Sigma}}^{+}$}) $\rightarrow$ {CO}({X}{$^1{{\Sigma}}^{+}$}) +
  {N}($^2${D})/{N}($^4${S}) reactions},}\ }\href {\doibase 10.1063/5.0188867}
  {\bibfield  {journal} {\bibinfo  {journal} {The Journal of Chemical Physics}\
  }\textbf {\bibinfo {volume} {160}},\ \bibinfo {pages} {064302} (\bibinfo
  {year} {2024})}\BibitemShut {NoStop}%
\bibitem [{\citenamefont {Galv{\~a}o}\ \emph {et~al.}(2025)\citenamefont
  {Galv{\~a}o}, \citenamefont {Alves}, \citenamefont {Lu},\ and\ \citenamefont
  {Guo}}]{Galvao2025Electronic}%
  \BibitemOpen
  \bibfield  {author} {\bibinfo {author} {\bibfnamefont {B.~R.~L.}\
  \bibnamefont {Galv{\~a}o}}, \bibinfo {author} {\bibfnamefont {M.~O.}\
  \bibnamefont {Alves}}, \bibinfo {author} {\bibfnamefont {D.}~\bibnamefont
  {Lu}}, \ and\ \bibinfo {author} {\bibfnamefont {H.}~\bibnamefont {Guo}},\
  }\bibfield  {title} {\enquote {\bibinfo {title} {Electronic quenching of
  {N}($^2${D}) in collision with {CO}({$^1{{\Sigma}}^{+}$}) via spin-forbidden
  transitions},}\ }\href {\doibase 10.1063/5.0260316} {\bibfield  {journal}
  {\bibinfo  {journal} {The Journal of Chemical Physics}\ }\textbf {\bibinfo
  {volume} {162}},\ \bibinfo {pages} {154301} (\bibinfo {year}
  {2025})}\BibitemShut {NoStop}%
\bibitem [{\citenamefont {Venturi}\ \emph {et~al.}(2020)\citenamefont
  {Venturi}, \citenamefont {Sharma}, \citenamefont {Lopez},\ and\ \citenamefont
  {Panesi}}]{Venturi2020Data}%
  \BibitemOpen
  \bibfield  {author} {\bibinfo {author} {\bibfnamefont {S.}~\bibnamefont
  {Venturi}}, \bibinfo {author} {\bibfnamefont {M.~P.}\ \bibnamefont {Sharma}},
  \bibinfo {author} {\bibfnamefont {B.}~\bibnamefont {Lopez}}, \ and\ \bibinfo
  {author} {\bibfnamefont {M.}~\bibnamefont {Panesi}},\ }\bibfield  {title}
  {\enquote {\bibinfo {title} {Data-inspired and physics-driven model reduction
  for dissociation: {A}pplication to the {O}$_2$ + {O} system},}\ }\href
  {\doibase 10.1021/acs.jpca.0c04516} {\bibfield  {journal} {\bibinfo
  {journal} {The Journal of Physical Chemistry A}\ }\textbf {\bibinfo {volume}
  {124}},\ \bibinfo {pages} {8359--8372} (\bibinfo {year} {2020})}\BibitemShut
  {NoStop}%
\bibitem [{\citenamefont {Zanardi}, \citenamefont {Venturi},\ and\
  \citenamefont {Panesi}(2023)}]{Zanardi2023Adaptive}%
  \BibitemOpen
  \bibfield  {author} {\bibinfo {author} {\bibfnamefont {I.}~\bibnamefont
  {Zanardi}}, \bibinfo {author} {\bibfnamefont {S.}~\bibnamefont {Venturi}}, \
  and\ \bibinfo {author} {\bibfnamefont {M.}~\bibnamefont {Panesi}},\
  }\bibfield  {title} {\enquote {\bibinfo {title} {Adaptive physics-informed
  neural operator for coarse-grained non-equilibrium flows},}\ }\href {\doibase
  10.1038/s41598-023-41039-y} {\bibfield  {journal} {\bibinfo  {journal}
  {Scientific Reports}\ }\textbf {\bibinfo {volume} {13}},\ \bibinfo {pages}
  {15497} (\bibinfo {year} {2023})}\BibitemShut {NoStop}%
\bibitem [{\citenamefont {Jacobsen}\ \emph {et~al.}(2024)\citenamefont
  {Jacobsen}, \citenamefont {Zanardi}, \citenamefont {Bhola}, \citenamefont
  {Duraisamy},\ and\ \citenamefont {Panesi}}]{Jacobsen2024Information}%
  \BibitemOpen
  \bibfield  {author} {\bibinfo {author} {\bibfnamefont {C.}~\bibnamefont
  {Jacobsen}}, \bibinfo {author} {\bibfnamefont {I.}~\bibnamefont {Zanardi}},
  \bibinfo {author} {\bibfnamefont {S.}~\bibnamefont {Bhola}}, \bibinfo
  {author} {\bibfnamefont {K.}~\bibnamefont {Duraisamy}}, \ and\ \bibinfo
  {author} {\bibfnamefont {M.}~\bibnamefont {Panesi}},\ }\bibfield  {title}
  {\enquote {\bibinfo {title} {Information theoretic clustering for
  coarse-grained modeling of non-equilibrium gas dynamics},}\ }\href {\doibase
  10.1016/j.jcp.2024.112977} {\bibfield  {journal} {\bibinfo  {journal}
  {Journal of Computational Physics}\ }\textbf {\bibinfo {volume} {507}},\
  \bibinfo {pages} {112977} (\bibinfo {year} {2024})}\BibitemShut {NoStop}%
\bibitem [{\citenamefont {Zanardi}\ \emph {et~al.}(2025)\citenamefont
  {Zanardi}, \citenamefont {Padovan}, \citenamefont {Bodony},\ and\
  \citenamefont {Panesi}}]{Zanardi2025PetrovGalerkin}%
  \BibitemOpen
  \bibfield  {author} {\bibinfo {author} {\bibfnamefont {I.}~\bibnamefont
  {Zanardi}}, \bibinfo {author} {\bibfnamefont {A.}~\bibnamefont {Padovan}},
  \bibinfo {author} {\bibfnamefont {D.~J.}\ \bibnamefont {Bodony}}, \ and\
  \bibinfo {author} {\bibfnamefont {M.}~\bibnamefont {Panesi}},\ }\bibfield
  {title} {\enquote {\bibinfo {title} {{P}etrov-{G}alerkin model reduction for
  thermochemical nonequilibrium gas mixtures},}\ }\href {\doibase
  10.1016/j.jcp.2025.113999} {\bibfield  {journal} {\bibinfo  {journal}
  {Journal of Computational Physics}\ }\textbf {\bibinfo {volume} {533}},\
  \bibinfo {pages} {113999} (\bibinfo {year} {2025})}\BibitemShut {NoStop}%
\bibitem [{\citenamefont {Kim}\ and\ \citenamefont
  {Jo}(2021)}]{Kim2021Modification}%
  \BibitemOpen
  \bibfield  {author} {\bibinfo {author} {\bibfnamefont {J.~G.}\ \bibnamefont
  {Kim}}\ and\ \bibinfo {author} {\bibfnamefont {S.~M.}\ \bibnamefont {Jo}},\
  }\bibfield  {title} {\enquote {\bibinfo {title} {Modification of
  chemical-kinetic parameters for 11-air species in re-entry flows},}\ }\href
  {\doibase 10.1016/j.ijheatmasstransfer.2021.120950} {\bibfield  {journal}
  {\bibinfo  {journal} {International Journal of Heat and Mass Transfer}\
  }\textbf {\bibinfo {volume} {169}},\ \bibinfo {pages} {120950} (\bibinfo
  {year} {2021})}\BibitemShut {NoStop}%
\bibitem [{\citenamefont {Torres}, \citenamefont {Geistfeld},\ and\
  \citenamefont {Schwartzentruber}(2024)}]{Torres2024High}%
  \BibitemOpen
  \bibfield  {author} {\bibinfo {author} {\bibfnamefont {E.}~\bibnamefont
  {Torres}}, \bibinfo {author} {\bibfnamefont {E.~C.}\ \bibnamefont
  {Geistfeld}}, \ and\ \bibinfo {author} {\bibfnamefont {T.~E.}\ \bibnamefont
  {Schwartzentruber}},\ }\bibfield  {title} {\enquote {\bibinfo {title}
  {High-temperature nonequilibrium air chemistry from first principles},}\
  }\href {\doibase 10.2514/1.T6863} {\bibfield  {journal} {\bibinfo  {journal}
  {Journal of Thermophysics and Heat Transfer}\ }\textbf {\bibinfo {volume}
  {38}},\ \bibinfo {pages} {260--291} (\bibinfo {year} {2024})}\BibitemShut
  {NoStop}%
\bibitem [{\citenamefont {Gimelshein}\ and\ \citenamefont
  {Wysong}(2019)}]{gimelshein2019nonequilibrium}%
  \BibitemOpen
  \bibfield  {author} {\bibinfo {author} {\bibfnamefont {S.~F.}\ \bibnamefont
  {Gimelshein}}\ and\ \bibinfo {author} {\bibfnamefont {I.~J.}\ \bibnamefont
  {Wysong}},\ }\bibfield  {title} {\enquote {\bibinfo {title} {Nonequilibrium
  air flow predictions with a high-fidelity direct simulation {{M}onte {C}arlo}
  approach},}\ }\href {\doibase 10.1103/PhysRevFluids.4.033405} {\bibfield
  {journal} {\bibinfo  {journal} {Physical Review Fluids}\ }\textbf {\bibinfo
  {volume} {4}},\ \bibinfo {pages} {033405} (\bibinfo {year}
  {2019})}\BibitemShut {NoStop}%
\bibitem [{\citenamefont {Civrais}, \citenamefont {White},\ and\ \citenamefont
  {Steijl}(2024{\natexlab{a}})}]{civrais2024qkpartI}%
  \BibitemOpen
  \bibfield  {author} {\bibinfo {author} {\bibfnamefont {C.~H.~B.}\
  \bibnamefont {Civrais}}, \bibinfo {author} {\bibfnamefont {C.}~\bibnamefont
  {White}}, \ and\ \bibinfo {author} {\bibfnamefont {R.}~\bibnamefont
  {Steijl}},\ }\bibfield  {title} {\enquote {\bibinfo {title} {Quantum kinetics
  chemistry models with an anharmonic oscillator model. {M}odel derivation and
  limitations},}\ }\href {\doibase 10.1063/5.0215706} {\bibfield  {journal}
  {\bibinfo  {journal} {Physics of Fluids}\ }\textbf {\bibinfo {volume} {36}},\
  \bibinfo {pages} {086120} (\bibinfo {year} {2024}{\natexlab{a}})}\BibitemShut
  {NoStop}%
\bibitem [{\citenamefont {Civrais}, \citenamefont {White},\ and\ \citenamefont
  {Steijl}(2024{\natexlab{b}})}]{civrais2024qkpartII}%
  \BibitemOpen
  \bibfield  {author} {\bibinfo {author} {\bibfnamefont {C.~H.~B.}\
  \bibnamefont {Civrais}}, \bibinfo {author} {\bibfnamefont {C.}~\bibnamefont
  {White}}, \ and\ \bibinfo {author} {\bibfnamefont {R.}~\bibnamefont
  {Steijl}},\ }\bibfield  {title} {\enquote {\bibinfo {title} {Quantum kinetics
  chemistry models with an anharmonic oscillator model. {M}odel extension and
  validation},}\ }\href {\doibase 10.1063/5.0238068} {\bibfield  {journal}
  {\bibinfo  {journal} {Physics of Fluids}\ }\textbf {\bibinfo {volume} {36}},\
  \bibinfo {pages} {116128} (\bibinfo {year} {2024}{\natexlab{b}})}\BibitemShut
  {NoStop}%
\bibitem [{\citenamefont {Civrais}\ \emph {et~al.}(2025)\citenamefont
  {Civrais}, \citenamefont {Pfeiffer}, \citenamefont {White},\ and\
  \citenamefont {Steijl}}]{civrais2025VibronicQK}%
  \BibitemOpen
  \bibfield  {author} {\bibinfo {author} {\bibfnamefont {C.~H.~B.}\
  \bibnamefont {Civrais}}, \bibinfo {author} {\bibfnamefont {M.}~\bibnamefont
  {Pfeiffer}}, \bibinfo {author} {\bibfnamefont {C.}~\bibnamefont {White}}, \
  and\ \bibinfo {author} {\bibfnamefont {R.}~\bibnamefont {Steijl}},\
  }\bibfield  {title} {\enquote {\bibinfo {title} {Development of a chemistry
  model for vibronically excited species in direct simulation {M}onte
  {C}arlo},}\ }\href {\doibase 10.1063/5.0295298} {\bibfield  {journal}
  {\bibinfo  {journal} {Physics of Fluids}\ }\textbf {\bibinfo {volume} {37}},\
  \bibinfo {pages} {106126} (\bibinfo {year} {2025})}\BibitemShut {NoStop}%
\bibitem [{\citenamefont {Minaev}, \citenamefont {Ivanova},\ and\ \citenamefont
  {Muldahmetov}(1989)}]{Minaev1989MINDO}%
  \BibitemOpen
  \bibfield  {author} {\bibinfo {author} {\bibfnamefont {B.~F.}\ \bibnamefont
  {Minaev}}, \bibinfo {author} {\bibfnamefont {N.~M.}\ \bibnamefont {Ivanova}},
  \ and\ \bibinfo {author} {\bibfnamefont {Z.~M.}\ \bibnamefont
  {Muldahmetov}},\ }\bibfield  {title} {\enquote {\bibinfo {title} {{MINDO/3}
  {CI} study of {NCO} spectrum and the chemiluminescent reaction {N} + {CO}
  $\rightarrow$ {NCO} + $h\nu$},}\ }\href {\doibase 10.1080/00387018908053945}
  {\bibfield  {journal} {\bibinfo  {journal} {Spectroscopy Letters}\ }\textbf
  {\bibinfo {volume} {22}},\ \bibinfo {pages} {901--923} (\bibinfo {year}
  {1989})}\BibitemShut {NoStop}%
\bibitem [{\citenamefont {Bernstein}(1979)}]{bernstein1979AtomMolecule}%
  \BibitemOpen
  \bibinfo {editor} {\bibfnamefont {R.~B.}\ \bibnamefont {Bernstein}},\ ed.,\
  \href {\doibase 10.1007/978-1-4613-2913-8} {\emph {\bibinfo {title}
  {Atom-Molecule Collision Theory: A Guide for the Experimentalist}}}\
  (\bibinfo  {publisher} {Plenum Press},\ \bibinfo {year} {1979})\BibitemShut
  {NoStop}%
\bibitem [{\citenamefont {Venturi}, \citenamefont {Jaffe},\ and\ \citenamefont
  {Panesi}(2020)}]{Venturi2020Bayesian}%
  \BibitemOpen
  \bibfield  {author} {\bibinfo {author} {\bibfnamefont {S.}~\bibnamefont
  {Venturi}}, \bibinfo {author} {\bibfnamefont {R.~L.}\ \bibnamefont {Jaffe}},
  \ and\ \bibinfo {author} {\bibfnamefont {M.}~\bibnamefont {Panesi}},\
  }\bibfield  {title} {\enquote {\bibinfo {title} {{Bayesian} machine learning
  approach to the quantification of uncertainties on \textit{ab initio}
  potential energy surfaces},}\ }\href {\doibase 10.1021/acs.jpca.0c02395}
  {\bibfield  {journal} {\bibinfo  {journal} {The Journal of Physical Chemistry
  A}\ }\textbf {\bibinfo {volume} {124}},\ \bibinfo {pages} {5129--5146}
  (\bibinfo {year} {2020})}\BibitemShut {NoStop}%
\bibitem [{\citenamefont {Schwenke}(1988)}]{Schwenke1988Calculations}%
  \BibitemOpen
  \bibfield  {author} {\bibinfo {author} {\bibfnamefont {D.~W.}\ \bibnamefont
  {Schwenke}},\ }\bibfield  {title} {\enquote {\bibinfo {title} {Calculations
  of rate constants for the three‐body recombination of {H}$_2$ in the
  presence of {H}$_2$},}\ }\href {\doibase 10.1063/1.455104} {\bibfield
  {journal} {\bibinfo  {journal} {The Journal of Chemical Physics}\ }\textbf
  {\bibinfo {volume} {89}},\ \bibinfo {pages} {2076--2091} (\bibinfo {year}
  {1988})}\BibitemShut {NoStop}%
\bibitem [{\citenamefont {Jo}\ \emph {et~al.}(2023)\citenamefont {Jo},
  \citenamefont {Venturi}, \citenamefont {Kim},\ and\ \citenamefont
  {Panesi}}]{jo2023rovibrational}%
  \BibitemOpen
  \bibfield  {author} {\bibinfo {author} {\bibfnamefont {S.~M.}\ \bibnamefont
  {Jo}}, \bibinfo {author} {\bibfnamefont {S.}~\bibnamefont {Venturi}},
  \bibinfo {author} {\bibfnamefont {J.~G.}\ \bibnamefont {Kim}}, \ and\
  \bibinfo {author} {\bibfnamefont {M.}~\bibnamefont {Panesi}},\ }\bibfield
  {title} {\enquote {\bibinfo {title} {Rovibrational internal energy transfer
  and dissociation of high-temperature oxygen mixture},}\ }\href {\doibase
  10.1063/5.0133463} {\bibfield  {journal} {\bibinfo  {journal} {The Journal of
  Chemical Physics}\ }\textbf {\bibinfo {volume} {158}},\ \bibinfo {pages}
  {064305} (\bibinfo {year} {2023})}\BibitemShut {NoStop}%
\bibitem [{\citenamefont {Macdonald}\ \emph
  {et~al.}(2018{\natexlab{a}})\citenamefont {Macdonald}, \citenamefont {Jaffe},
  \citenamefont {Schwenke},\ and\ \citenamefont
  {Panesi}}]{Macdonald2018constructionI}%
  \BibitemOpen
  \bibfield  {author} {\bibinfo {author} {\bibfnamefont {R.~L.}\ \bibnamefont
  {Macdonald}}, \bibinfo {author} {\bibfnamefont {R.~L.}\ \bibnamefont
  {Jaffe}}, \bibinfo {author} {\bibfnamefont {D.~W.}\ \bibnamefont {Schwenke}},
  \ and\ \bibinfo {author} {\bibfnamefont {M.}~\bibnamefont {Panesi}},\
  }\bibfield  {title} {\enquote {\bibinfo {title} {Construction of a
  coarse-grain quasi-classical trajectory method. {I.} {T}heory and application
  to {N}$_2$–{N}$_2$ system},}\ }\href {\doibase 10.1063/1.5011331}
  {\bibfield  {journal} {\bibinfo  {journal} {The Journal of Chemical Physics}\
  }\textbf {\bibinfo {volume} {148}},\ \bibinfo {pages} {054309} (\bibinfo
  {year} {2018}{\natexlab{a}})}\BibitemShut {NoStop}%
\bibitem [{\citenamefont {Macdonald}\ \emph
  {et~al.}(2018{\natexlab{b}})\citenamefont {Macdonald}, \citenamefont
  {Grover}, \citenamefont {Schwartzentruber},\ and\ \citenamefont
  {Panesi}}]{Macdonald2018constructionII}%
  \BibitemOpen
  \bibfield  {author} {\bibinfo {author} {\bibfnamefont {R.~L.}\ \bibnamefont
  {Macdonald}}, \bibinfo {author} {\bibfnamefont {M.~S.}\ \bibnamefont
  {Grover}}, \bibinfo {author} {\bibfnamefont {T.~E.}\ \bibnamefont
  {Schwartzentruber}}, \ and\ \bibinfo {author} {\bibfnamefont
  {M.}~\bibnamefont {Panesi}},\ }\bibfield  {title} {\enquote {\bibinfo {title}
  {Construction of a coarse-grain quasi-classical trajectory method. {II.}
  {C}omparison against the direct molecular simulation method},}\ }\href
  {\doibase 10.1063/1.5011332} {\bibfield  {journal} {\bibinfo  {journal} {The
  Journal of Chemical Physics}\ }\textbf {\bibinfo {volume} {148}},\ \bibinfo
  {pages} {054310} (\bibinfo {year} {2018}{\natexlab{b}})}\BibitemShut
  {NoStop}%
\bibitem [{\citenamefont {Jo}\ \emph {et~al.}(2022)\citenamefont {Jo},
  \citenamefont {Venturi}, \citenamefont {Sharma}, \citenamefont {Munaf{\`o}},\
  and\ \citenamefont {Panesi}}]{jo2022rovibrational}%
  \BibitemOpen
  \bibfield  {author} {\bibinfo {author} {\bibfnamefont {S.~M.}\ \bibnamefont
  {Jo}}, \bibinfo {author} {\bibfnamefont {S.}~\bibnamefont {Venturi}},
  \bibinfo {author} {\bibfnamefont {M.~P.}\ \bibnamefont {Sharma}}, \bibinfo
  {author} {\bibfnamefont {A.}~\bibnamefont {Munaf{\`o}}}, \ and\ \bibinfo
  {author} {\bibfnamefont {M.}~\bibnamefont {Panesi}},\ }\bibfield  {title}
  {\enquote {\bibinfo {title} {Rovibrational-specific {QCT} and master equation
  study on {N}$_2$({X$^1{{\Sigma}}_g^{+}$}) + {O}($^3${P}) and {NO}({X$^2
  {{\Pi}}$}) + {N}($^4${S}) systems in high-energy collisions},}\ }\href
  {\doibase 10.1021/acs.jpca.1c10346} {\bibfield  {journal} {\bibinfo
  {journal} {The Journal of Physical Chemistry A}\ }\textbf {\bibinfo {volume}
  {126}},\ \bibinfo {pages} {3273--3290} (\bibinfo {year} {2022})}\BibitemShut
  {NoStop}%
\bibitem [{\citenamefont {Civrais}\ \emph {et~al.}(2026)\citenamefont
  {Civrais}, \citenamefont {Galvão}, \citenamefont {Jo},\ and\ \citenamefont
  {Panesi}}]{civrais2026NOC}%
  \BibitemOpen
  \bibfield  {author} {\bibinfo {author} {\bibfnamefont {C.~H.~B.}\
  \bibnamefont {Civrais}}, \bibinfo {author} {\bibfnamefont {B.~R.~L.}\
  \bibnamefont {Galvão}}, \bibinfo {author} {\bibfnamefont {S.~M.}\
  \bibnamefont {Jo}}, \ and\ \bibinfo {author} {\bibfnamefont {M.}~\bibnamefont
  {Panesi}},\ }\bibfield  {title} {\enquote {\bibinfo {title}
  {Rovibrational-specific energy transfer and dissociation of
  {NO}({X}$^2{{\Pi}}_r$) + {C}($^3${P}) collisions at high temperatures},}\
  }in\ \href {\doibase 10.2514/6.2026-2643} {\emph {\bibinfo {booktitle} {AIAA
  SCITECH 2026 Forum}}}\ (\bibinfo {address} {Orlando, FL, AIAA Paper
  2026-2643},\ \bibinfo {year} {2026})\BibitemShut {NoStop}%
\bibitem [{\citenamefont {Priyadarshini}\ \emph {et~al.}(2022)\citenamefont
  {Priyadarshini}, \citenamefont {Jo}, \citenamefont {Venturi}, \citenamefont
  {Schwenke}, \citenamefont {Jaffe},\ and\ \citenamefont
  {Panesi}}]{Sharma2022HCN}%
  \BibitemOpen
  \bibfield  {author} {\bibinfo {author} {\bibfnamefont {M.~S.}\ \bibnamefont
  {Priyadarshini}}, \bibinfo {author} {\bibfnamefont {S.~M.}\ \bibnamefont
  {Jo}}, \bibinfo {author} {\bibfnamefont {S.}~\bibnamefont {Venturi}},
  \bibinfo {author} {\bibfnamefont {D.~W.}\ \bibnamefont {Schwenke}}, \bibinfo
  {author} {\bibfnamefont {R.~L.}\ \bibnamefont {Jaffe}}, \ and\ \bibinfo
  {author} {\bibfnamefont {M.}~\bibnamefont {Panesi}},\ }\bibfield  {title}
  {\enquote {\bibinfo {title} {Comprehensive study of {HCN}: {P}otential energy
  surfaces, state-to-state kinetics, and master equation analysis},}\ }\href
  {\doibase 10.1021/acs.jpca.2c03959} {\bibfield  {journal} {\bibinfo
  {journal} {The Journal of Physical Chemistry A}\ }\textbf {\bibinfo {volume}
  {126}},\ \bibinfo {pages} {8249--8265} (\bibinfo {year} {2022})}\BibitemShut
  {NoStop}%
\bibitem [{\citenamefont {Munaf\`{o}}\ and\ \citenamefont
  {Panesi}(2025)}]{Munafo2025PLATO}%
  \BibitemOpen
  \bibfield  {author} {\bibinfo {author} {\bibfnamefont {A.}~\bibnamefont
  {Munaf\`{o}}}\ and\ \bibinfo {author} {\bibfnamefont {M.}~\bibnamefont
  {Panesi}},\ }\bibfield  {title} {\enquote {\bibinfo {title}
  {{\texttt{PLATO}}: {A} high-fidelity library for multicomponent gases and
  plasmas},}\ }\href {\doibase 10.2514/1.T7097} {\bibfield  {journal} {\bibinfo
   {journal} {Journal of Thermophysics and Heat Transfer}\ }\textbf {\bibinfo
  {volume} {39}},\ \bibinfo {pages} {850--870} (\bibinfo {year}
  {2025})}\BibitemShut {NoStop}%
\bibitem [{\citenamefont {Priyadarshini}\ \emph {et~al.}(2023)\citenamefont
  {Priyadarshini}, \citenamefont {Venturi}, \citenamefont {Zanardi},\ and\
  \citenamefont {Panesi}}]{Sharma2023Efficient}%
  \BibitemOpen
  \bibfield  {author} {\bibinfo {author} {\bibfnamefont {M.~S.}\ \bibnamefont
  {Priyadarshini}}, \bibinfo {author} {\bibfnamefont {S.}~\bibnamefont
  {Venturi}}, \bibinfo {author} {\bibfnamefont {I.}~\bibnamefont {Zanardi}}, \
  and\ \bibinfo {author} {\bibfnamefont {M.}~\bibnamefont {Panesi}},\
  }\bibfield  {title} {\enquote {\bibinfo {title} {Efficient quasi-classical
  trajectory calculations by means of neural operator architectures},}\ }\href
  {\doibase 10.1039/D2CP05506F} {\bibfield  {journal} {\bibinfo  {journal}
  {Phys. Chem. Chem. Phys.}\ }\textbf {\bibinfo {volume} {25}},\ \bibinfo
  {pages} {13902--13912} (\bibinfo {year} {2023})}\BibitemShut {NoStop}%
\bibitem [{\citenamefont {Panesi}\ \emph {et~al.}(2013)\citenamefont {Panesi},
  \citenamefont {Jaffe}, \citenamefont {Schwenke},\ and\ \citenamefont
  {Magin}}]{panesi2013rovibrational}%
  \BibitemOpen
  \bibfield  {author} {\bibinfo {author} {\bibfnamefont {M.}~\bibnamefont
  {Panesi}}, \bibinfo {author} {\bibfnamefont {R.~L.}\ \bibnamefont {Jaffe}},
  \bibinfo {author} {\bibfnamefont {D.~W.}\ \bibnamefont {Schwenke}}, \ and\
  \bibinfo {author} {\bibfnamefont {T.~E.}\ \bibnamefont {Magin}},\ }\bibfield
  {title} {\enquote {\bibinfo {title} {Rovibrational internal energy transfer
  and dissociation of
  {N}$_2$({$^{1}{{\Sigma}}^{+}_{\text{g}}$})-{N}($^{4}\text{S}_{\text{u}}$)
  system in hypersonic flows},}\ }\href {\doibase 10.1063/1.4774412} {\bibfield
   {journal} {\bibinfo  {journal} {The Journal of Chemical Physics}\ }\textbf
  {\bibinfo {volume} {138}},\ \bibinfo {pages} {044312} (\bibinfo {year}
  {2013})}\BibitemShut {NoStop}%
\bibitem [{\citenamefont {Rodriguez}\ and\ \citenamefont
  {Macdonald}(2026)}]{rodriguez2026no2}%
  \BibitemOpen
  \bibfield  {author} {\bibinfo {author} {\bibfnamefont {L.~F.}\ \bibnamefont
  {Rodriguez}}\ and\ \bibinfo {author} {\bibfnamefont {R.~L.}\ \bibnamefont
  {Macdonald}},\ }\bibfield  {title} {\enquote {\bibinfo {title}
  {Rovibrational-specific kinetics database and master equation study for the
  {O}$_2$ + {N} and {NO} + {O} systems using ground electronic state {NO}$_2$
  potential energy surfaces},}\ }\href {\doibase 10.1021/acs.jpca.6c02327}
  {\bibfield  {journal} {\bibinfo  {journal} {The Journal of Physical Chemistry
  A}\ } (\bibinfo {year} {2026}),\ 10.1021/acs.jpca.6c02327},\ \bibinfo {note}
  {published online 16 September 2026}\BibitemShut {NoStop}%
\bibitem [{\citenamefont {Parker}(1959)}]{Parker1959RotationalandVibrational}%
  \BibitemOpen
  \bibfield  {author} {\bibinfo {author} {\bibfnamefont {J.~G.}\ \bibnamefont
  {Parker}},\ }\bibfield  {title} {\enquote {\bibinfo {title} {Rotational and
  vibrational relaxation in diatomic gases},}\ }\href {\doibase
  10.1063/1.1724417} {\bibfield  {journal} {\bibinfo  {journal} {The Physics of
  Fluids}\ }\textbf {\bibinfo {volume} {2}},\ \bibinfo {pages} {449--462}
  (\bibinfo {year} {1959})}\BibitemShut {NoStop}%
\bibitem [{\citenamefont {Millikan}\ and\ \citenamefont
  {White}(1963)}]{Millikan1963Relaxation}%
  \BibitemOpen
  \bibfield  {author} {\bibinfo {author} {\bibfnamefont {R.~C.}\ \bibnamefont
  {Millikan}}\ and\ \bibinfo {author} {\bibfnamefont {D.~R.}\ \bibnamefont
  {White}},\ }\bibfield  {title} {\enquote {\bibinfo {title} {Systematics of
  vibrational relaxation},}\ }\href {\doibase 10.1063/1.1734182} {\bibfield
  {journal} {\bibinfo  {journal} {The Journal of Chemical Physics}\ }\textbf
  {\bibinfo {volume} {39}},\ \bibinfo {pages} {3209--3213} (\bibinfo {year}
  {1963})}\BibitemShut {NoStop}%
\bibitem [{\citenamefont {Park}(1993)}]{park1993review}%
  \BibitemOpen
  \bibfield  {author} {\bibinfo {author} {\bibfnamefont {C.}~\bibnamefont
  {Park}},\ }\bibfield  {title} {\enquote {\bibinfo {title} {Review of
  chemical-kinetic problems of future {NASA} missions, {I: E}arth entries},}\
  }\href {\doibase 10.2514/3.431} {\bibfield  {journal} {\bibinfo  {journal}
  {Journal of Thermophysics and Heat Transfer}\ }\textbf {\bibinfo {volume}
  {7}},\ \bibinfo {pages} {385--398} (\bibinfo {year} {1993})}\BibitemShut
  {NoStop}%
\bibitem [{\citenamefont {Torres}\ and\ \citenamefont
  {Schwartzentruber}(2025)}]{Torres2025Characteristic}%
  \BibitemOpen
  \bibfield  {author} {\bibinfo {author} {\bibfnamefont {E.}~\bibnamefont
  {Torres}}\ and\ \bibinfo {author} {\bibfnamefont {T.~E.}\ \bibnamefont
  {Schwartzentruber}},\ }\bibfield  {title} {\enquote {\bibinfo {title}
  {Characteristic vibrational and rotational relaxation times for air species
  from first-principles calculations},}\ }\href {\doibase 10.2514/1.T7042}
  {\bibfield  {journal} {\bibinfo  {journal} {Journal of Thermophysics and Heat
  Transfer}\ }\textbf {\bibinfo {volume} {39}},\ \bibinfo {pages} {223--249}
  (\bibinfo {year} {2025})}\BibitemShut {NoStop}%
\bibitem [{\citenamefont {Fujita}, \citenamefont {Yamada},\ and\ \citenamefont
  {Ishii}(2006)}]{Fujita2006Impact}%
  \BibitemOpen
  \bibfield  {author} {\bibinfo {author} {\bibfnamefont {K.}~\bibnamefont
  {Fujita}}, \bibinfo {author} {\bibfnamefont {T.}~\bibnamefont {Yamada}}, \
  and\ \bibinfo {author} {\bibfnamefont {N.}~\bibnamefont {Ishii}},\ }\bibfield
   {title} {\enquote {\bibinfo {title} {Impact of ablation gas kinetics on
  hyperbolic entry radiative heating},}\ }in\ \href {\doibase
  10.2514/6.2006-1185} {\emph {\bibinfo {booktitle} {44th AIAA Aerospace
  Sciences Meeting and Exhibit}}}\ (\bibinfo {address} {Reno, NV, AIAA Paper
  2006-1185},\ \bibinfo {year} {2006})\BibitemShut {NoStop}%
\bibitem [{\citenamefont {Thielen}\ and\ \citenamefont
  {Roth}(1985)}]{Thielen1985Resonance}%
  \BibitemOpen
  \bibfield  {author} {\bibinfo {author} {\bibfnamefont {K.}~\bibnamefont
  {Thielen}}\ and\ \bibinfo {author} {\bibfnamefont {P.}~\bibnamefont {Roth}},\
  }\bibfield  {title} {\enquote {\bibinfo {title} {Resonance absorption
  measurements of {N} and {O} atoms in high temperature {NO} dissociation and
  formation kinetics},}\ }\href {\doibase 10.1016/S0082-0784(85)80558-0}
  {\bibfield  {journal} {\bibinfo  {journal} {Symposium (International) on
  Combustion}\ }\textbf {\bibinfo {volume} {20}},\ \bibinfo {pages} {685--693}
  (\bibinfo {year} {1985})}\BibitemShut {NoStop}%
\bibitem [{\citenamefont {Wray}\ and\ \citenamefont
  {Teare}(1962)}]{Wray1962Shock}%
  \BibitemOpen
  \bibfield  {author} {\bibinfo {author} {\bibfnamefont {K.~L.}\ \bibnamefont
  {Wray}}\ and\ \bibinfo {author} {\bibfnamefont {J.~D.}\ \bibnamefont
  {Teare}},\ }\bibfield  {title} {\enquote {\bibinfo {title} {Shock‐tube
  study of the kinetics of nitric oxide at high temperatures},}\ }\href
  {\doibase 10.1063/1.1732338} {\bibfield  {journal} {\bibinfo  {journal} {The
  Journal of Chemical Physics}\ }\textbf {\bibinfo {volume} {36}},\ \bibinfo
  {pages} {2582--2596} (\bibinfo {year} {1962})}\BibitemShut {NoStop}%
\bibitem [{\citenamefont {Koshi}\ \emph {et~al.}(1979)\citenamefont {Koshi},
  \citenamefont {Bando}, \citenamefont {Saito},\ and\ \citenamefont
  {Asaba}}]{Koshi1979Dissociation}%
  \BibitemOpen
  \bibfield  {author} {\bibinfo {author} {\bibfnamefont {M.}~\bibnamefont
  {Koshi}}, \bibinfo {author} {\bibfnamefont {S.}~\bibnamefont {Bando}},
  \bibinfo {author} {\bibfnamefont {M.}~\bibnamefont {Saito}}, \ and\ \bibinfo
  {author} {\bibfnamefont {T.}~\bibnamefont {Asaba}},\ }\bibfield  {title}
  {\enquote {\bibinfo {title} {Dissociation of nitric oxide in shock waves},}\
  }\href {\doibase 10.1016/S0082-0784(79)80056-9} {\bibfield  {journal}
  {\bibinfo  {journal} {Symposium (International) on Combustion}\ }\textbf
  {\bibinfo {volume} {17}},\ \bibinfo {pages} {553--562} (\bibinfo {year}
  {1979})}\BibitemShut {NoStop}%
\bibitem [{\citenamefont {Myerson}(1973)}]{Myerson1973Shock}%
  \BibitemOpen
  \bibfield  {author} {\bibinfo {author} {\bibfnamefont {A.~L.}\ \bibnamefont
  {Myerson}},\ }\bibfield  {title} {\enquote {\bibinfo {title} {Shock-tube atom
  kinetics of nitric oxide decomposition},}\ }\href {\doibase
  10.1016/S0082-0784(73)80023-2} {\bibfield  {journal} {\bibinfo  {journal}
  {Symposium (International) on Combustion}\ }\textbf {\bibinfo {volume}
  {14}},\ \bibinfo {pages} {219--228} (\bibinfo {year} {1973})},\ \bibinfo
  {note} {fourteenth Symposium (International) on Combustion}\BibitemShut
  {NoStop}%
\bibitem [{\citenamefont {Mozzhukhin}, \citenamefont {Burmeister},\ and\
  \citenamefont {Roth}(1989)}]{Mozzhukhin1989CNDissociation}%
  \BibitemOpen
  \bibfield  {author} {\bibinfo {author} {\bibfnamefont {E.}~\bibnamefont
  {Mozzhukhin}}, \bibinfo {author} {\bibfnamefont {M.}~\bibnamefont
  {Burmeister}}, \ and\ \bibinfo {author} {\bibfnamefont {P.}~\bibnamefont
  {Roth}},\ }\bibfield  {title} {\enquote {\bibinfo {title} {High temperature
  dissociation of {CN}},}\ }\href {\doibase 10.1002/bbpc.19890930114}
  {\bibfield  {journal} {\bibinfo  {journal} {Berichte der Bunsengesellschaft
  für physikalische Chemie}\ }\textbf {\bibinfo {volume} {93}},\ \bibinfo
  {pages} {70--75} (\bibinfo {year} {1989})}\BibitemShut {NoStop}%
\bibitem [{\citenamefont {Appleton}, \citenamefont {Steinberg},\ and\
  \citenamefont {Liquornik}(1970)}]{Appleton1970Shock}%
  \BibitemOpen
  \bibfield  {author} {\bibinfo {author} {\bibfnamefont {J.~P.}\ \bibnamefont
  {Appleton}}, \bibinfo {author} {\bibfnamefont {M.}~\bibnamefont {Steinberg}},
  \ and\ \bibinfo {author} {\bibfnamefont {D.~J.}\ \bibnamefont {Liquornik}},\
  }\bibfield  {title} {\enquote {\bibinfo {title} {Shock‐tube study of carbon
  monoxide dissociation using vacuum‐ultraviolet absorption},}\ }\href
  {\doibase 10.1063/1.1673286} {\bibfield  {journal} {\bibinfo  {journal} {The
  Journal of Chemical Physics}\ }\textbf {\bibinfo {volume} {52}},\ \bibinfo
  {pages} {2205--2221} (\bibinfo {year} {1970})}\BibitemShut {NoStop}%
\bibitem [{\citenamefont {Mick}, \citenamefont {Burmeister},\ and\
  \citenamefont {Roth}(1993)}]{Mick1993Atomic}%
  \BibitemOpen
  \bibfield  {author} {\bibinfo {author} {\bibfnamefont {H.}~\bibnamefont
  {Mick}}, \bibinfo {author} {\bibfnamefont {M.}~\bibnamefont {Burmeister}}, \
  and\ \bibinfo {author} {\bibfnamefont {P.}~\bibnamefont {Roth}},\ }\bibfield
  {title} {\enquote {\bibinfo {title} {Atomic resonance absorption spectroscopy
  measurements on high-temperature {CO} dissociation kinetics},}\ }\href
  {\doibase 10.2514/3.11602} {\bibfield  {journal} {\bibinfo  {journal} {AIAA
  Journal}\ }\textbf {\bibinfo {volume} {31}},\ \bibinfo {pages} {671--676}
  (\bibinfo {year} {1993})}\BibitemShut {NoStop}%
\bibitem [{\citenamefont {Baulch}, \citenamefont {Drysdale},\ and\
  \citenamefont {Horne}(1976)}]{Baulch1976Evaluated}%
  \BibitemOpen
  \bibfield  {author} {\bibinfo {author} {\bibfnamefont {D.~L.}\ \bibnamefont
  {Baulch}}, \bibinfo {author} {\bibfnamefont {D.~D.}\ \bibnamefont
  {Drysdale}}, \ and\ \bibinfo {author} {\bibfnamefont {D.}~\bibnamefont
  {Horne}},\ }\href@noop {} {\emph {\bibinfo {title} {Evaluated Kinetic Data
  for High Temperature Reactions. Vol. 2: Homogeneous Gas Phase Reactions of
  the {H}$_{2}$ - {N}$_{2}$ - {O}$_{2}$ System}}},\ Vol.~\bibinfo {volume} {3}\
  (\bibinfo  {publisher} {University of Leeds},\ \bibinfo {address}
  {Butterworths, London},\ \bibinfo {year} {1976})\BibitemShut {NoStop}%
\bibitem [{\citenamefont {Baulch}\ \emph {et~al.}(1981)\citenamefont {Baulch},
  \citenamefont {Duxbury}, \citenamefont {Grant},\ and\ \citenamefont
  {Montague}}]{Baulch1981Evaluated}%
  \BibitemOpen
  \bibfield  {author} {\bibinfo {author} {\bibfnamefont {D.~L.}\ \bibnamefont
  {Baulch}}, \bibinfo {author} {\bibfnamefont {J.}~\bibnamefont {Duxbury}},
  \bibinfo {author} {\bibfnamefont {S.~J.}\ \bibnamefont {Grant}}, \ and\
  \bibinfo {author} {\bibfnamefont {D.~C.}\ \bibnamefont {Montague}},\
  }\bibfield  {title} {\enquote {\bibinfo {title} {Evaluated kinetic data for
  high temperature reactions. {V}olume 4: Homogeneous gas phase reactions of
  halogen- and cyanide-containing species},}\ }\href@noop {} {\bibfield
  {journal} {\bibinfo  {journal} {Journal of Physical and Chemical Reference
  Data}\ }\textbf {\bibinfo {volume} {10}} (\bibinfo {year} {1981})},\ \bibinfo
  {note} {{Suppl.} 1}\BibitemShut {NoStop}%
\bibitem [{\citenamefont {Davies}(1962)}]{Davies1962Radiative}%
  \BibitemOpen
  \bibfield  {author} {\bibinfo {author} {\bibfnamefont {W.~O.}\ \bibnamefont
  {Davies}},\ }\href {https://ntrs.nasa.gov/citations/19630000268} {\enquote
  {\bibinfo {title} {Radiative transfer on {Entry} into {Mars} and {Venus}},}\
  }\bibinfo {type} {Tech. Rep.}\ \bibinfo {number} {CR-58574}\ (\bibinfo
  {institution} {National Aeronautics and Space Administration},\ \bibinfo
  {year} {1962})\BibitemShut {NoStop}%
\bibitem [{\citenamefont {Schwenke}, \citenamefont {Jaffe},\ and\ \citenamefont
  {Chaban}(2016)}]{schwenke2016collisional}%
  \BibitemOpen
  \bibfield  {author} {\bibinfo {author} {\bibfnamefont {D.~W.}\ \bibnamefont
  {Schwenke}}, \bibinfo {author} {\bibfnamefont {R.~L.}\ \bibnamefont {Jaffe}},
  \ and\ \bibinfo {author} {\bibfnamefont {G.~M.}\ \bibnamefont {Chaban}},\
  }\href {https://ntrs.nasa.gov/citations/20160012461} {\enquote {\bibinfo
  {title} {Collisional dissociation of {CO}: ab initio potential energy
  surfaces and quasiclassical trajectory rate coefficients},}\ }\bibinfo {type}
  {Tech. Rep.}\ \bibinfo {number} {ARC-E-DAA-TN30024}\ (\bibinfo  {institution}
  {NASA Ames Research Center},\ \bibinfo {year} {2016})\BibitemShut {NoStop}%
\end{thebibliography}%

\end{document}


\title{Rovibrational state-specific quasi-classical trajectory calculation and master equation analysis of the CNO system in high-energy collisions}

\author{C. H. B. Civrais}
\email{civraisc@mst.edu (Corresponding Author)}
\affiliation{Department of Mechanical and Aerospace Engineering, University of California, Irvine, CA 92697, USA}
\affiliation{Present institution: Mechanical and Aerospace Engineering Department, Missouri University of Science and Technology, Rolla, MO, 65401}

\author{Breno R. L. Galvão}
\affiliation{Centro Federal de Educação Tecnológica de Minas Gerais, CEFET-MG, Av. Amazonas 5253, (30421-169), Belo Horizonte, Minas Gerais, Brazil}

\author{Sung Min Jo}
\affiliation{UCF Center of Excellence in Hypersonics and Space Propulsion, University of Central Florida, Orlando, FL 32816, USA}

\author{Marco Panesi}
\email{mpanesi@uci.edu (Corresponding Author)}
\affiliation{Department of Mechanical and Aerospace Engineering, University of California, Irvine, CA 92697, USA}

\date{\today}

\maketitle

\section{Physical Modelling}
\subsection{Master Equation Analysis}

The system of rovibrational master equations reads:

\begin{widetext}
\begin{equation}
    \frac{d n_{\ce{O}}}{dt} =
    \sum\limits_{i \in \mathcal{I}_{\ce{NO}}} \dot{\omega}^{\text{D},\ce{NO}}_{i \rightarrow c}
    + \sum\limits_{n \in \mathcal{I}_{\ce{CO}}} \sum\limits_{p \in \mathcal{I}_{\text{N}}} \dot{\omega}^{\text{D},\ce{CO}}_{n,p \rightarrow c}
    + \sum\limits_{i \in \mathcal{I}_{\ce{NO}}} \sum\limits_{l \in \mathcal{I}_{\ce{CN}}} \dot{\omega}^{\text{E},\ce{NO}}_{i\rightarrow l}
    - \sum\limits_{l \in \mathcal{I}_{\ce{CN}}} \sum\limits_{n \in \mathcal{I}_{\ce{CO}}} \sum\limits_{p \in \mathcal{I}_{\text{N}}} \dot{\omega}^{\text{E},\ce{CN}}_{l\rightarrow n,p},
\label{eq:ME_O}
\end{equation}

\begin{equation}
    \frac{d n_{\ce{N(^4S)}}}{dt} =
    \sum\limits_{i \in \mathcal{I}_{\ce{NO}}} \dot{\omega}^{\text{D},\ce{NO}}_{i \rightarrow c}
    + \sum\limits_{l \in \mathcal{I}_{\ce{CN}}} \dot{\omega}^{\text{D},\ce{CN}}_{l \rightarrow c}
    + \sum\limits_{n \in \mathcal{I}_{\ce{CO}}} \dot{\omega}^{\text{D},\ce{CO}}_{n,1 \rightarrow c}
    + \sum\limits_{i \in \mathcal{I}_{\ce{NO}}} \sum\limits_{n \in \mathcal{I}_{\ce{CO}}} \dot{\omega}^{\text{E},\ce{NO}}_{i\rightarrow n,0}
    + \sum\limits_{l \in \mathcal{I}_{\ce{CN}}} \sum\limits_{n \in \mathcal{I}_{\ce{CO}}} \dot{\omega}^{\text{E},\ce{CN}}_{l\rightarrow n,0},
\label{eq:ME_N4S}
\end{equation}

\begin{equation}
    \frac{d n_{\ce{N(^2D)}}}{dt} =
    - \sum\limits_{n \in \mathcal{I}_{\ce{CO}}} \dot{\omega}^{\text{D},\ce{CO}}_{n,1 \rightarrow c}
    + \sum\limits_{i \in \mathcal{I}_{\ce{NO}}} \sum\limits_{n \in \mathcal{I}_{\ce{CO}}} \dot{\omega}^{\text{E},\ce{NO}}_{i\rightarrow n,1}
    + \sum\limits_{l \in \mathcal{I}_{\ce{CN}}} \sum\limits_{n \in \mathcal{I}_{\ce{CO}}} \dot{\omega}^{\text{E},\ce{CN}}_{l\rightarrow n,1},
\label{eq:ME_N2D}
\end{equation}

\begin{equation}
    \frac{d n_{\ce{C}}}{dt} =
    \sum\limits_{l \in \mathcal{I}_{\ce{CN}}} \dot{\omega}^{\text{D},\ce{CN}}_{l \rightarrow c}
    + \sum\limits_{n \in \mathcal{I}_{\ce{CO}}} \sum\limits_{p \in \mathcal{I}_{\text{N}}} \dot{\omega}^{\text{D},\ce{CO}}_{n,p \rightarrow c}
    - \sum\limits_{i \in \mathcal{I}_{\ce{NO}}} \sum\limits_{l \in \mathcal{I}_{\ce{CN}}} \dot{\omega}^{\text{E},\ce{NO}}_{i\rightarrow l}
    - \sum\limits_{i \in \mathcal{I}_{\ce{NO}}} \sum\limits_{n \in \mathcal{I}_{\ce{CO}}} \sum\limits_{p \in \mathcal{I}_{\text{N}}} \dot{\omega}^{\text{E},\ce{NO}}_{i\rightarrow n,p},
\label{eq:ME_C}
\end{equation}

\begin{equation}
    \frac{d n_i}{dt} = - \sum\limits_{j \in \mathcal{I}_{\ce{NO}}} \dot{\omega}^{\text{I},\ce{NO}}_{i \rightarrow j}
    - \sum\limits_{l \in \mathcal{I}_{\ce{CN}}} \dot{\omega}^{\text{E},\ce{NO}}_{i\rightarrow l}
    - \sum\limits_{n \in \mathcal{I}_{\ce{CO}}} \sum\limits_{p \in \mathcal{I}_{\text{N}}} \dot{\omega}^{\text{E},\ce{NO}}_{i\rightarrow n,p}
    - \dot{\omega}^{\text{D},\ce{NO}}_{i \rightarrow c},
\label{eq:ME_NOi}
\end{equation}

\begin{equation}
    \frac{d n_l}{dt} = - \sum\limits_{m \in \mathcal{I}_{\ce{CN}}} \dot{\omega}^{\text{I},\ce{CN}}_{l \rightarrow m}
    + \sum\limits_{i \in \mathcal{I}_{\ce{NO}}} \dot{\omega}^{\text{E},\ce{NO}}_{i\rightarrow l}
    - \sum\limits_{n \in \mathcal{I}_{\ce{CO}}} \sum\limits_{p \in \mathcal{I}_{\text{N}}} \dot{\omega}^{\text{E},\ce{CN}}_{l\rightarrow n,p}
    - \dot{\omega}^{\text{D},\ce{CN}}_{l \rightarrow c},
\label{eq:ME_CNl}
\end{equation}

\begin{equation}
    \frac{d n_n}{dt} = - \sum\limits_{p \in \mathcal{I}_{\text{N}}} \sum\limits_{o \in \mathcal{I}_{\ce{CO}}} \dot{\omega}^{\text{I},\ce{CO}}_{n,p \rightarrow o\!,p}
    + \sum\limits_{p \in \mathcal{I}_{\text{N}}} \sum\limits_{i \in \mathcal{I}_{\ce{NO}}} \dot{\omega}^{\text{E},\ce{NO}}_{i\rightarrow n,p}
    + \sum\limits_{p \in \mathcal{I}_{\text{N}}} \sum\limits_{l \in \mathcal{I}_{\ce{CN}}} \dot{\omega}^{\text{E},\ce{CN}}_{l\rightarrow n,p}
    - \sum\limits_{p \in \mathcal{I}_{\text{N}}} \dot{\omega}^{\text{D},\ce{CO}}_{n,p \rightarrow c},
\label{eq:ME_COm}
\end{equation}
\end{widetext}

where $t$ is the time and $\dot{\omega}^{\text{I}}$, $\dot{\omega}^{\text{E}}$, and $\dot{\omega}^{\text{D}}$ are the net mass production rates for the inelastic, exchange, and dissociation processes, respectively. It is noted that, consistent with the adiabatic correlation of the PESs to the ground-state atomisation limit, i.e.~\ce{C(^3P) + N(^4S) + O(^3P)}, all dissociation processes produce ground-state atoms; the dissociation of \ce{CO} in collision with \ce{N(^2D)} therefore also de-excites the nitrogen atom, i.e.~\ce{CO(n) + N(^2D) -> C(^3P) + O(^3P) + N(^4S)}. The net mass production rates are defined as,

\begin{widetext}


\begin{equation}
    \dot{\omega}^{\text{I},\ce{NO}}_{i \rightarrow j} = k^{\text{I}, \ce{NO}}_{i\rightarrow j} \left[ n_{i} - \frac{n_{j}}{K^{\text{I}}_{i,j} } \right] n_{\ce{C}} ,
\label{eq:omega_I_NO}
\end{equation}

\begin{equation}
    \dot{\omega}^{\text{I},\ce{CN}}_{l \rightarrow m} = k^{\text{I}, \ce{CN}}_{l\rightarrow m} \left[ n_{l} - \frac{n_{m}}{K^{\text{I}}_{l,m} } \right] n_{\ce{O}} ,
\label{eq:omega_I_CN}
\end{equation}

\begin{equation}
    \dot{\omega}^{\text{I},\ce{CO}}_{n,p \rightarrow o\!,p} = k^{\text{I}, \ce{CO}}_{n,p \rightarrow o\!,p} \left[ n_{n}\, n_{\text{N}(p)} - \frac{n_{o}\, n_{\text{N}(p)}}{K^{\text{I},\ce{CO}}_{n,p;\, o\!,p} } \right] ,
\label{eq:omega_I_CO}
\end{equation}


\begin{equation}
    \dot{\omega}^{\text{E},\ce{NO}}_{i\rightarrow l} = k^{\text{E}, \ce{NO}}_{i\rightarrow l} \left[ n_{i}\, n_{\ce{C}} - \frac{n_{l}\, n_{\ce{O}}}{K^{\text{E},\ce{NO}}_{i,l} }  \right]  ,
\label{eq:omega_E_NO_CN}
\end{equation}

\begin{equation}
    \dot{\omega}^{\text{E},\ce{NO}}_{i\rightarrow n,p} = k^{\text{E}, \ce{NO}}_{i\rightarrow n,p} \left[ n_{i}\, n_{\ce{C}} - \frac{n_{n}\, n_{\text{N}(p)}}{K^{\text{E},\ce{NO}}_{i,n,p} }  \right]  ,
\label{eq:omega_E_NO_CO}
\end{equation}

\begin{equation}
    \dot{\omega}^{\text{E},\ce{CN}}_{l\rightarrow n,p} = k^{\text{E}, \ce{CN}}_{l\rightarrow n,p} \left[ n_{l}\, n_{\ce{O}} - \frac{n_{n}\, n_{\text{N}(p)}}{K^{\text{E},\ce{CN}}_{l,n,p} }  \right]  ,
\label{eq:omega_E_CN_CO}
\end{equation}


\begin{equation}
    \dot{\omega}^{\text{D},\ce{NO}}_{i \rightarrow c} = k^{\text{D}, \ce{NO}}_{i\rightarrow c} \left[ n_{i} - \frac{n_{\ce{N(^4S)}}\, n_{\ce{O}}}{K^{\text{D},\ce{NO}}_{i} } \right] n_{\ce{C}} ,
\label{eq:omega_D_NO}
\end{equation}

\begin{equation}
    \dot{\omega}^{\text{D},\ce{CN}}_{l \rightarrow c} = k^{\text{D}, \ce{CN}}_{l\rightarrow c} \left[ n_{l} - \frac{n_{\ce{C}}\, n_{\ce{N(^4S)}}}{K^{\text{D},\ce{CN}}_{l} } \right] n_{\ce{O}} ,
\label{eq:omega_D_CN}
\end{equation}

\begin{equation}
    \dot{\omega}^{\text{D},\ce{CO}}_{n,p \rightarrow c} = k^{\text{D}, \ce{CO}}_{n,p\rightarrow c} \left[ n_{n}\, n_{\text{N}(p)} - \frac{n_{\ce{C}}\, n_{\ce{O}}\, n_{\ce{N(^4S)}}}{K^{\text{D},\ce{CO}}_{n,p} } \right] .
\label{eq:omega_D_CO}
\end{equation}

\end{widetext}


The equilibrium constants are obtained from detailed balance. For the inelastic processes,
\begin{widetext}

\begin{equation}
    K^{\text{I}}_{i,j} = \frac{g^{\ce{NO}}_j\, \exp\left(-\dfrac{\varepsilon_j}{kT} \right) }{g^{\ce{NO}}_i\, \exp\left(-\dfrac{\varepsilon_i}{kT} \right) } ,
\label{eq:K_I}
\end{equation}

\begin{equation}
    K^{\text{I}}_{l,m} = \frac{g^{\ce{CN}}_m\, \exp\left(-\dfrac{\varepsilon_m}{kT} \right) }{g^{\ce{CN}}_l\, \exp\left(-\dfrac{\varepsilon_l}{kT} \right) } ,
\label{eq:K_I_CN}
\end{equation}
and, 
\begin{equation}
    K^{\text{I},\ce{CO}}_{n,p;\, o\!,p} = \frac{g^{\ce{CO}}_{o}\, \exp\left(-\dfrac{\varepsilon_{o}}{kT} \right) }{g^{\ce{CO}}_{n}\, \exp\left(-\dfrac{\varepsilon_{n}}{kT} \right) } .
\label{eq:K_I_CO}
\end{equation}

For the exchange reactions,
\begin{equation}
    K^{\text{E},\ce{NO}}_{i,l} = \frac{g^{\ce{CN}}_l\, Q^{\ce{CN}}_t\, g^{\ce{O}}\, Q^{\ce{O}}_t }{g^{\ce{NO}}_i\, Q^{\ce{NO}}_t\, g^{\ce{C}}\, Q^{\ce{C}}_t } \exp\left(-\frac{\varepsilon^{\text{D},\ce{NO}}-\varepsilon^{\text{D},\ce{CN}} + \varepsilon_l - \varepsilon_i }{kT} \right) ,
\label{eq:K_E_NO_CN}
\end{equation}

\begin{equation}
    K^{\text{E},\ce{NO}}_{i,n,p} = \frac{g^{\ce{CO}}_n\, Q^{\ce{CO}}_t\, g^{\ce{N}(p)}\, Q^{\ce{N}(p)}_t }{g^{\ce{NO}}_i\, Q^{\ce{NO}}_t\, g^{\ce{C}}\, Q^{\ce{C}}_t } \exp\left(-\frac{\varepsilon^{\text{D},\ce{NO}} -\varepsilon^{\text{D},\ce{CO}} + \varepsilon_n - \varepsilon_i + \varepsilon_{\text{N}(p)} }{kT} \right) ,
\label{eq:K_E_NO_CO}
\end{equation}

\begin{equation}
    K^{\text{E},\ce{CN}}_{l,n,p} = \frac{g^{\ce{CO}}_n\, Q^{\ce{CO}}_t\, g^{\ce{N}(p)}\, Q^{\ce{N}(p)}_t }{g^{\ce{CN}}_l\, Q^{\ce{CN}}_t\, g^{\ce{O}}\, Q^{\ce{O}}_t } \exp\left(-\frac{\varepsilon^{\text{D},\ce{CN}} -\varepsilon^{\text{D},\ce{CO}} + \varepsilon_n - \varepsilon_l + \varepsilon_{\text{N}(p)} }{kT} \right) .
\label{eq:K_E_CN_CO}
\end{equation}

For the dissociation processes,
\begin{equation}
    K^{\text{D},\ce{NO}}_{i} = \frac{g^{\ce{N(^4S)}}\, Q^{\ce{N(^4S)}}_t\, g^{\ce{O}}\, Q^{\ce{O}}_t }{ g^{\ce{NO}}_i\, Q^{\ce{NO}}_t } \exp\left(-\frac{\varepsilon^{\text{D},\ce{NO}} - \varepsilon_i }{kT} \right) ,
\label{eq:K_D_NO}
\end{equation}

\begin{equation}
    K^{\text{D},\ce{CN}}_{l} = \frac{ g^{\ce{C}}\, Q^{\ce{C}}_t\, g^{\ce{N(^4S)}}\, Q^{\ce{N(^4S)}}_t }{ g^{\ce{CN}}_l\, Q^{\ce{CN}}_t } \exp\left(-\frac{\varepsilon^{\text{D},\ce{CN}}  - \varepsilon_l }{kT} \right) ,
\label{eq:K_D_CN}
\end{equation}

\begin{equation}
    K^{\text{D},\ce{CO}}_{n,p} = \frac{g^{\ce{N(^4S)}}\, Q^{\ce{N(^4S)}}_t\, g^{\ce{C}}\, Q^{\ce{C}}_t\, g^{\ce{O}}\, Q^{\ce{O}}_t }{g^{\ce{N}(p)}\, Q^{\ce{N}(p)}_t\,  g^{\ce{CO}}_n\, Q^{\ce{CO}}_t } \exp\left(-\frac{\varepsilon^{\text{D},\ce{CO}} - \varepsilon_{n} - \varepsilon_{\text{N}(p)} }{kT} \right) .
\label{eq:K_D_CO}
\end{equation}

\end{widetext}


The rovibrational degeneracies are given by,
\begin{equation}
    g^{\ce{NO}}_i = g^{\ce{NO}}_e\, g^{\ce{NO}}_{\text{nuc}}(2J_i + 1),
\label{eq:g_NO}
\end{equation}
\begin{equation}
    g^{\ce{CN}}_l = g^{\ce{CN}}_e\, g^{\ce{CN}}_{\text{nuc}}(2J_l + 1),
\label{eq:g_CN}
\end{equation}
\begin{equation}
    g^{\ce{CO}}_n = g^{\ce{CO}}_e\, g^{\ce{CO}}_{\text{nuc}}(2J_n + 1),
\label{eq:g_CO}
\end{equation}
where $g_e$ and $g_{\text{nuc}}$ denote the electronic and nuclear-spin degeneracies of the corresponding species, with $g^{\ce{NO}}_e = 4$, $g^{\ce{CN}}_e = 2$, and $g^{\ce{CO}}_e = 1$. The atomic factors $g^{\ce{C}}$, $g^{\ce{O}}$, $g^{\ce{N(^4S)}}$ and $g^{\ce{N(^2D)}}$ are the fine-structure-resolved electronic partition functions of the atoms adopted in the main text, i.e.~$g^{\ce{C}} = 1 + 3e^{-23.6/T} + 5e^{-62.4/T}$, $g^{\ce{O}} = 5 + 3e^{-228/T} + e^{-326/T}$, $g^{\ce{N(^4S)}} = 4$ and $g^{\ce{N(^2D)}} = 6 + 4e^{-12.53/T}$. Additionally, $Q_t$ denotes the translational partition function per unit volume, $\varepsilon^{\text{D}}$ the dissociation energy of the corresponding diatom measured from its lowest rovibrational state, and $\varepsilon_{\text{N}(p)}$ the electronic energy of atomic nitrogen, i.e.~$\varepsilon_{\text{N}(0)} = 0$ and $\varepsilon_{\text{N}(1)} = 2.384$~eV. Note that the spin-orbit splitting of \ce{NO(X^2\Pi)} is neglected in the rovibrational degeneracies, i.e.~$g^{\ce{NO}}_e = 4$, whereas the trajectories are weighted with $q_{\ce{NO(X^2\Pi)}} = 2 + 2e^{-172.4/T}$; the resulting discrepancy amounts to a few per cent at the lowest temperatures and falls below 1\% above approximately 10,000~K.


For the exchange reactions, the global exchange rate coefficients are computed from the microscopic state-specific rate coefficients and the rovibrational population distribution at a given time, and read,
\begin{equation}
    \tilde{k}^{\text{E}, \ce{NO}\rightarrow\ce{CN}} = \frac{1}{n_{\ce{NO}}} \sum_{i \in \mathcal{I}_{\ce{NO}}} \sum_{l \in \mathcal{I}_{\ce{CN}}} n_i\, k^{\text{E}, \ce{NO}}_{i\rightarrow l},
\label{eq:k_glob_E_NO_CN}
\end{equation}
\begin{equation}
    \tilde{k}^{\text{E}, \ce{NO}\rightarrow\ce{CO}} = \frac{1}{n_{\ce{NO}}} \sum_{i \in \mathcal{I}_{\ce{NO}}} \sum_{n \in \mathcal{I}_{\ce{CO}}} \sum_{p \in \mathcal{I}_{\text{N}}} n_i\, k^{\text{E}, \ce{NO}}_{i\rightarrow n,p},
\label{eq:k_glob_E_NO_CO}
\end{equation}
and
\begin{equation}
    \tilde{k}^{\text{E}, \ce{CN}\rightarrow\ce{CO}} = \frac{1}{n_{\ce{CN}}} \sum_{l \in \mathcal{I}_{\ce{CN}}} \sum_{n \in \mathcal{I}_{\ce{CO}}} \sum_{p \in \mathcal{I}_{\text{N}}} n_l\, k^{\text{E}, \ce{CN}}_{l\rightarrow n,p}.
\label{eq:k_glob_E_CN_CO}
\end{equation}
Since the reported quantities resolve the electronic state of atomic nitrogen, the global exchange rate coefficients towards \ce{CO(X^1\Sigma^+) + N(p)} are also defined per electronic state, i.e.
\begin{equation}
    \tilde{k}^{\text{E}, \ce{NO}\rightarrow\ce{CO},p} = \frac{1}{n_{\ce{NO}}} \sum_{i \in \mathcal{I}_{\ce{NO}}} \sum_{n \in \mathcal{I}_{\ce{CO}}} n_i\, k^{\text{E}, \ce{NO}}_{i\rightarrow n,p},
\label{eq:k_glob_E_NO_CO_p}
\end{equation}
and
\begin{equation}
    \tilde{k}^{\text{E}, \ce{CN}\rightarrow\ce{CO},p} = \frac{1}{n_{\ce{CN}}} \sum_{l \in \mathcal{I}_{\ce{CN}}} \sum_{n \in \mathcal{I}_{\ce{CO}}} n_l\, k^{\text{E}, \ce{CN}}_{l\rightarrow n,p},
\label{eq:k_glob_E_CN_CO_p}
\end{equation}
such that the summed rate coefficients are recovered as $\tilde{k}^{\text{E}, \ce{NO}\rightarrow\ce{CO}} = \sum_{p} \tilde{k}^{\text{E}, \ce{NO}\rightarrow\ce{CO},p}$ and $\tilde{k}^{\text{E}, \ce{CN}\rightarrow\ce{CO}} = \sum_{p} \tilde{k}^{\text{E}, \ce{CN}\rightarrow\ce{CO},p}$.

For the dissociation reactions, the global dissociation rate coefficients follow the definition of Panesi~\textit{et al.}~\cite{panesi2013rovibrational} which yields,
\begin{equation}
    \tilde{k}^{\text{D}, \ce{NO}} = \frac{1}{n_{\ce{NO}}} \sum_{i \in \mathcal{I}_{\ce{NO}}} n_i\, k^{\text{D}, \ce{NO}}_{i\rightarrow c},
\label{eq:k_glob_D_NO}
\end{equation}
\begin{equation}
    \tilde{k}^{\text{D}, \ce{CN}} = \frac{1}{n_{\ce{CN}}} \sum_{l \in \mathcal{I}_{\ce{CN}}} n_l\, k^{\text{D}, \ce{CN}}_{l\rightarrow c},
\label{eq:k_glob_D_CN}
\end{equation}
\begin{equation}
    \tilde{k}^{\text{D}, \ce{CO}} = \frac{1}{n_{\ce{CO}}} \sum_{n \in \mathcal{I}_{\ce{CO}}}\sum_{p \in \mathcal{I}_{\ce{N}}} n_n\, k^{\text{D}, \ce{CO}}_{n,p\rightarrow c}.
\label{eq:k_glob_D_CO}
\end{equation}

\section{Results and Discussion}

\subsection{Rovibrational State-specific Database}

\subsubsection{Heterogeneous Exchange Processes}

\begin{figure*}[!ht]
\centering
\begin{subfigure}[t]{0.49\textwidth}
     \centering
     \includegraphics[width=\textwidth, trim=1.5cm 0.5cm 1cm 2cm, clip]{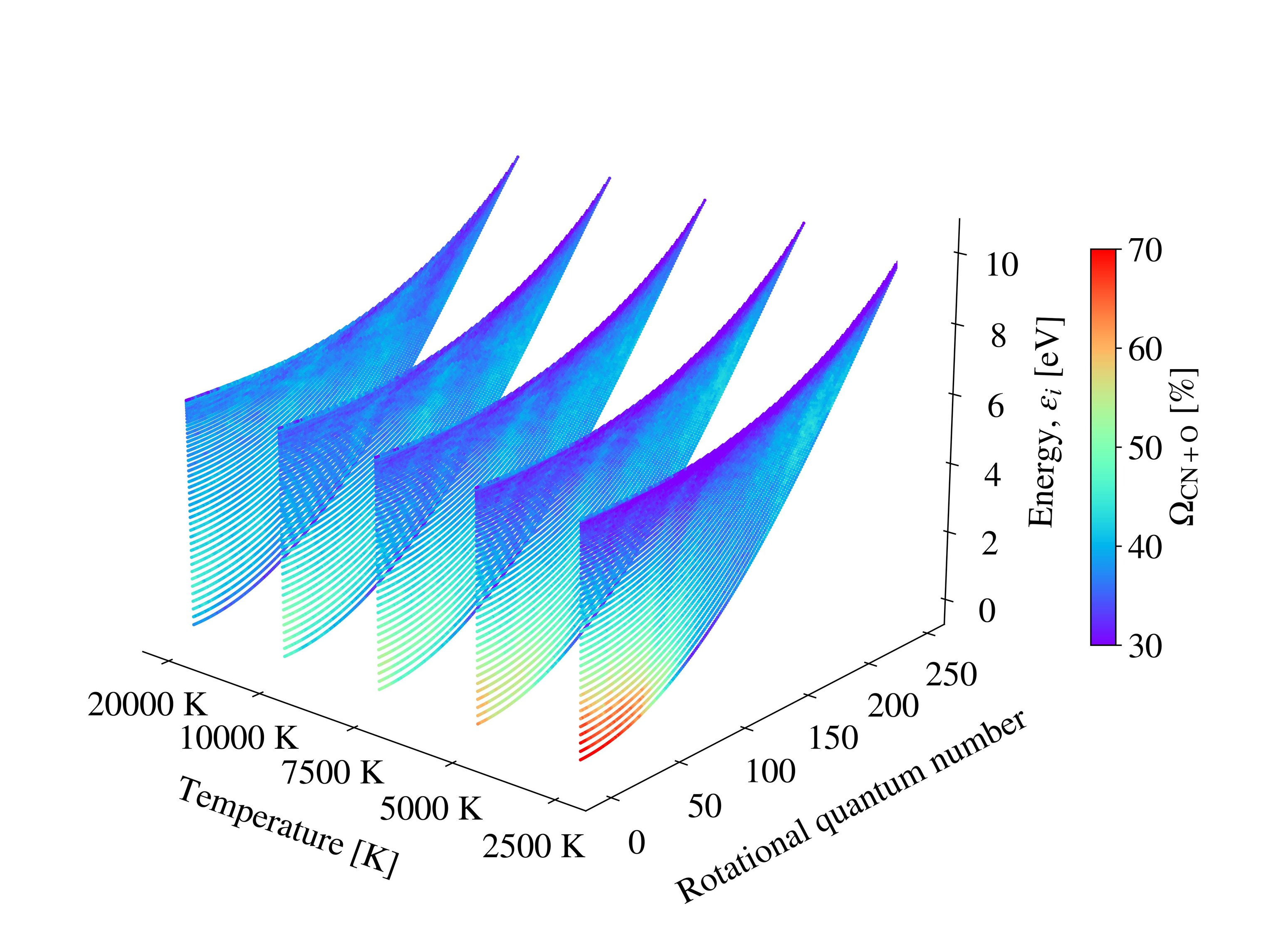}
     \caption{\ce{NO(X^2\Pi) + C(^3P) <=> CN(X^2\Sigma^+) +O(^3P)}.}
\end{subfigure}
\begin{subfigure}[t]{0.49\textwidth}
     \centering
     \includegraphics[width=\textwidth, trim=1.5cm 0.5cm 1cm 2cm, clip]{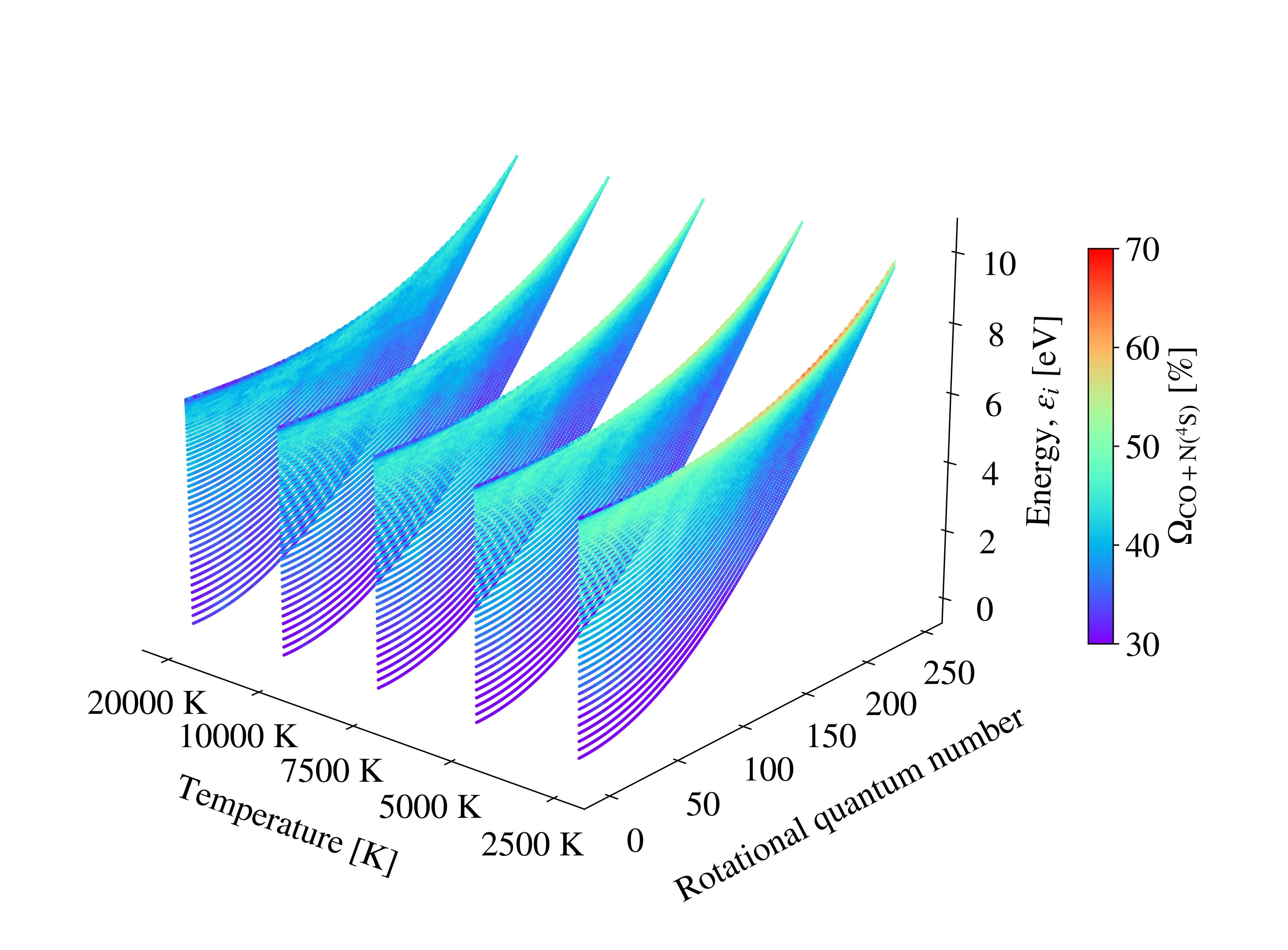}
     \caption{\ce{NO(X^2\Pi) + C(^3P) <=> CO(X^1\Sigma^+) + N(^4S)}.}
\end{subfigure}
\begin{subfigure}[t]{0.49\textwidth}
     \centering
     \includegraphics[width=\textwidth, trim=1.5cm 0.5cm 1cm 2cm, clip]{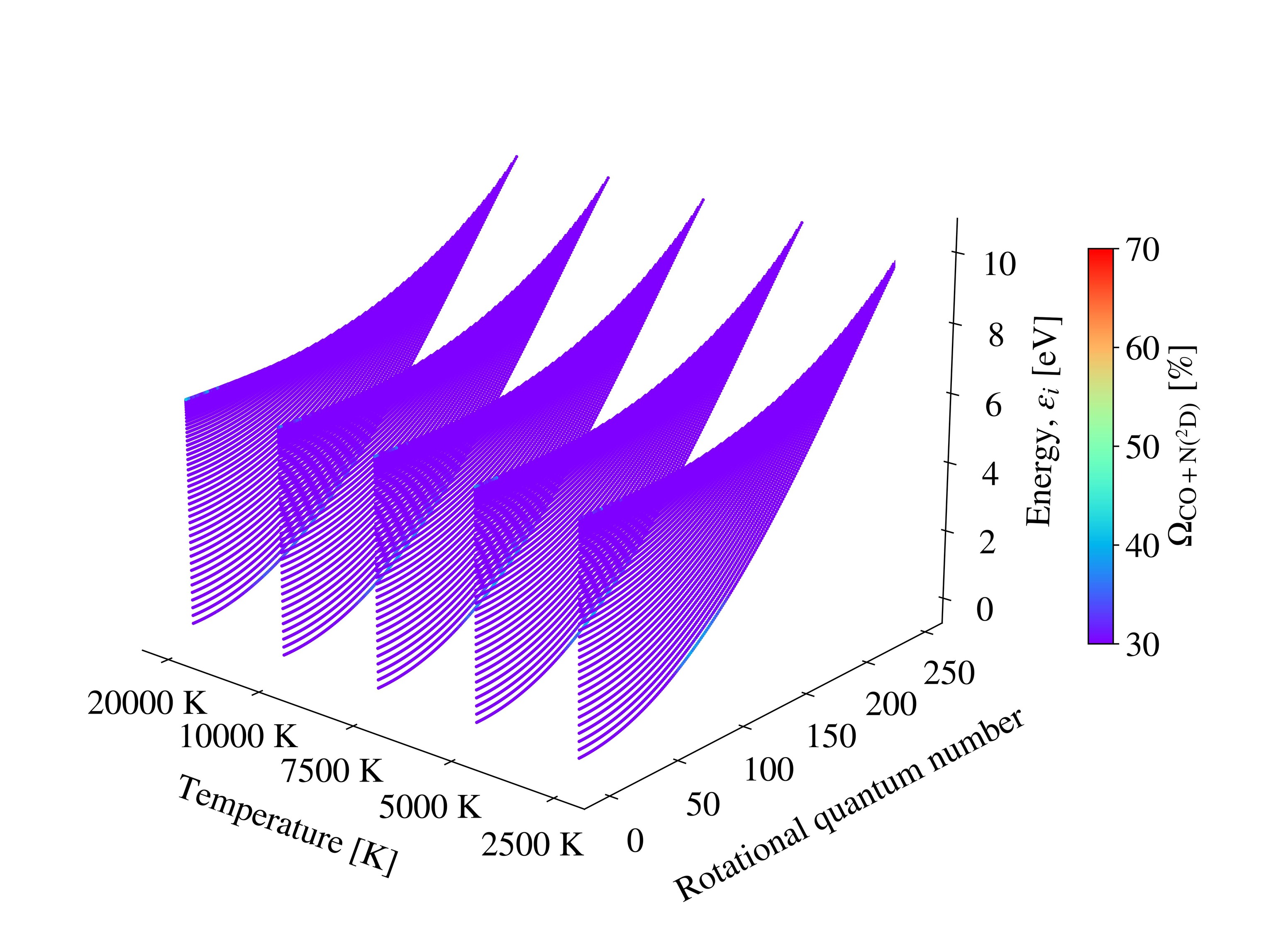}
     \caption{\ce{NO(X^2\Pi) + C(^3P) <=> CO(X^1\Sigma^+) + N(^2D)}.}
\end{subfigure}

\caption{Branching ratio for the \ce{NO(X^2\Pi) + C(^3P)} asymptote overlaid with the rovibrational state distribution of \ce{NO(X^2\Pi)} at five temperatures.}
\label{BranchingRatioNOC}
\end{figure*}

\begin{figure*}[!ht]
\centering
\begin{subfigure}[t]{0.49\textwidth}
     \centering
     \includegraphics[width=\textwidth, trim=1.5cm 0.5cm 1cm 2cm, clip]{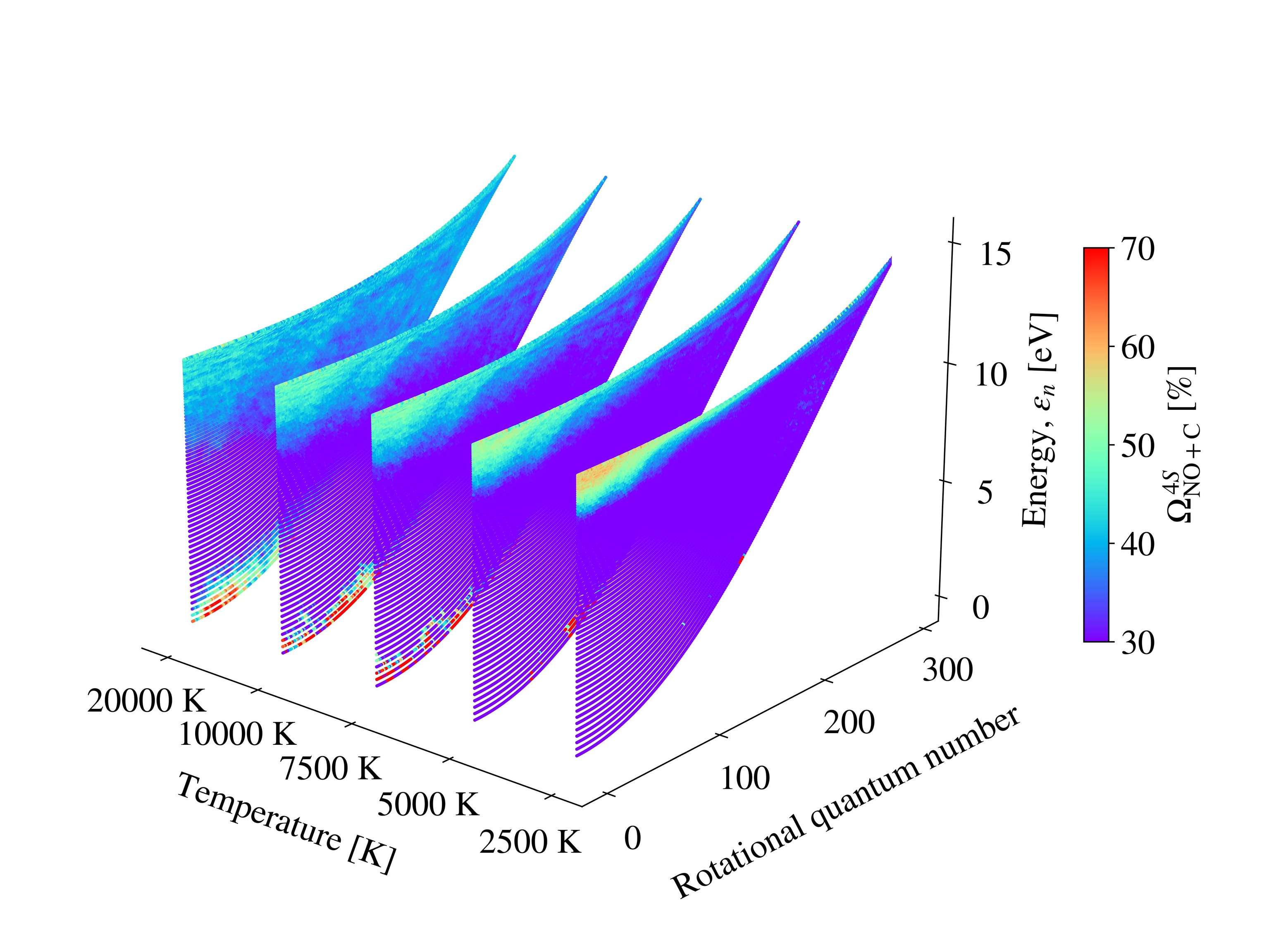}
     \caption{\ce{CO(X^1\Sigma^+) + N(^4S) <=> NO(X^2\Pi) + C(^3P)}.}
\end{subfigure}
\begin{subfigure}[t]{0.49\textwidth}
     \centering
     \includegraphics[width=\textwidth, trim=1.5cm 0.5cm 1cm 2cm, clip]{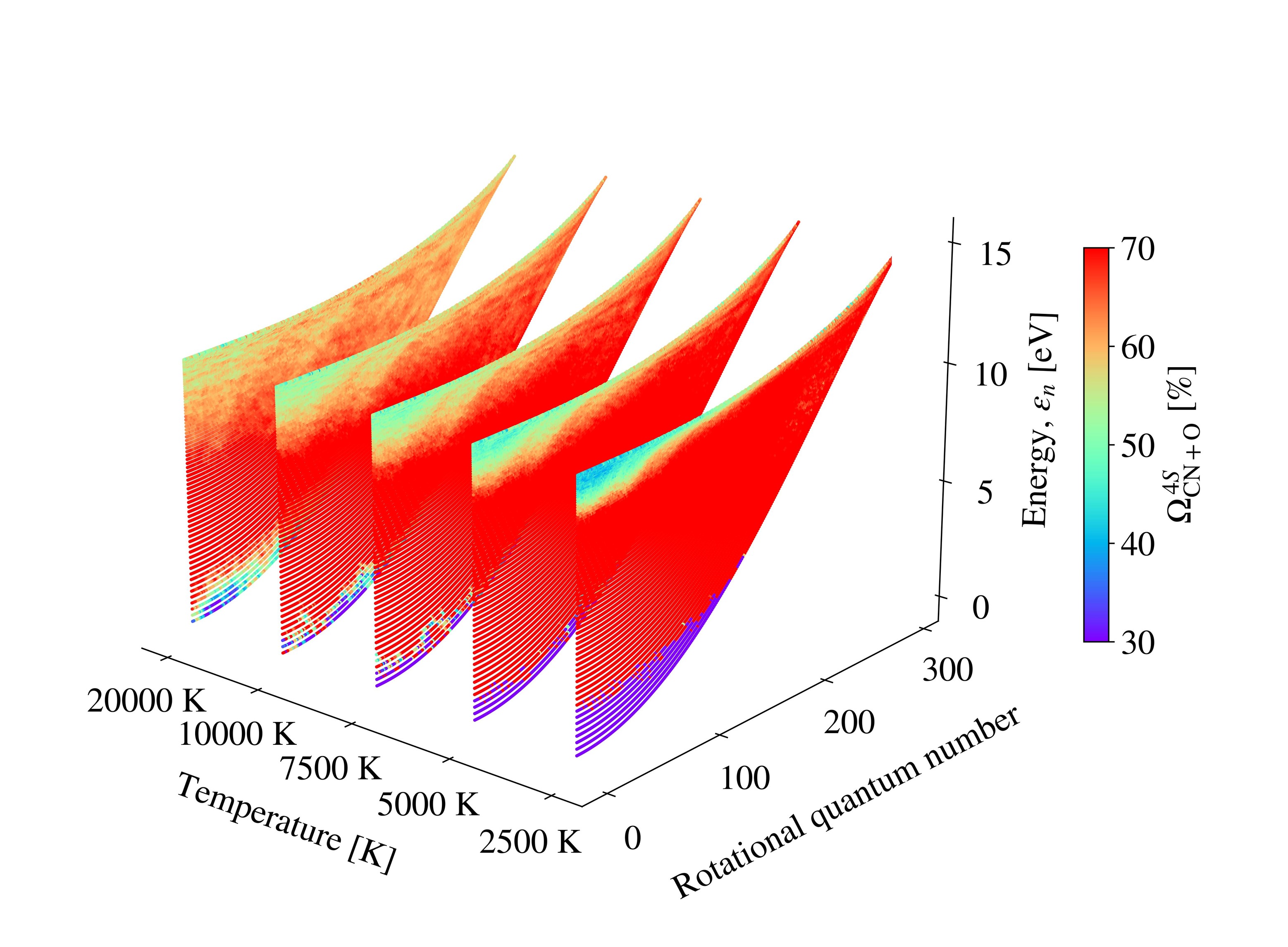}
     \caption{\ce{CO(X^1\Sigma^+) + N(^4S) <=> CN(X^2\Sigma^+) +O(^3P)}.}
\end{subfigure}

\caption{Branching ratio for the \ce{CO(X^1\Sigma^+) + N(^4S)} asymptote overlaid with the rovibrational state distribution of \ce{CO(X^1\Sigma^+)} at five temperatures.}
\label{BranchingRatioCON4S}
\end{figure*}

\begin{figure*}[!ht]
\centering
\begin{subfigure}[t]{0.49\textwidth}
     \centering
     \includegraphics[width=\textwidth, trim=1.5cm 0.5cm 1cm 2cm, clip]{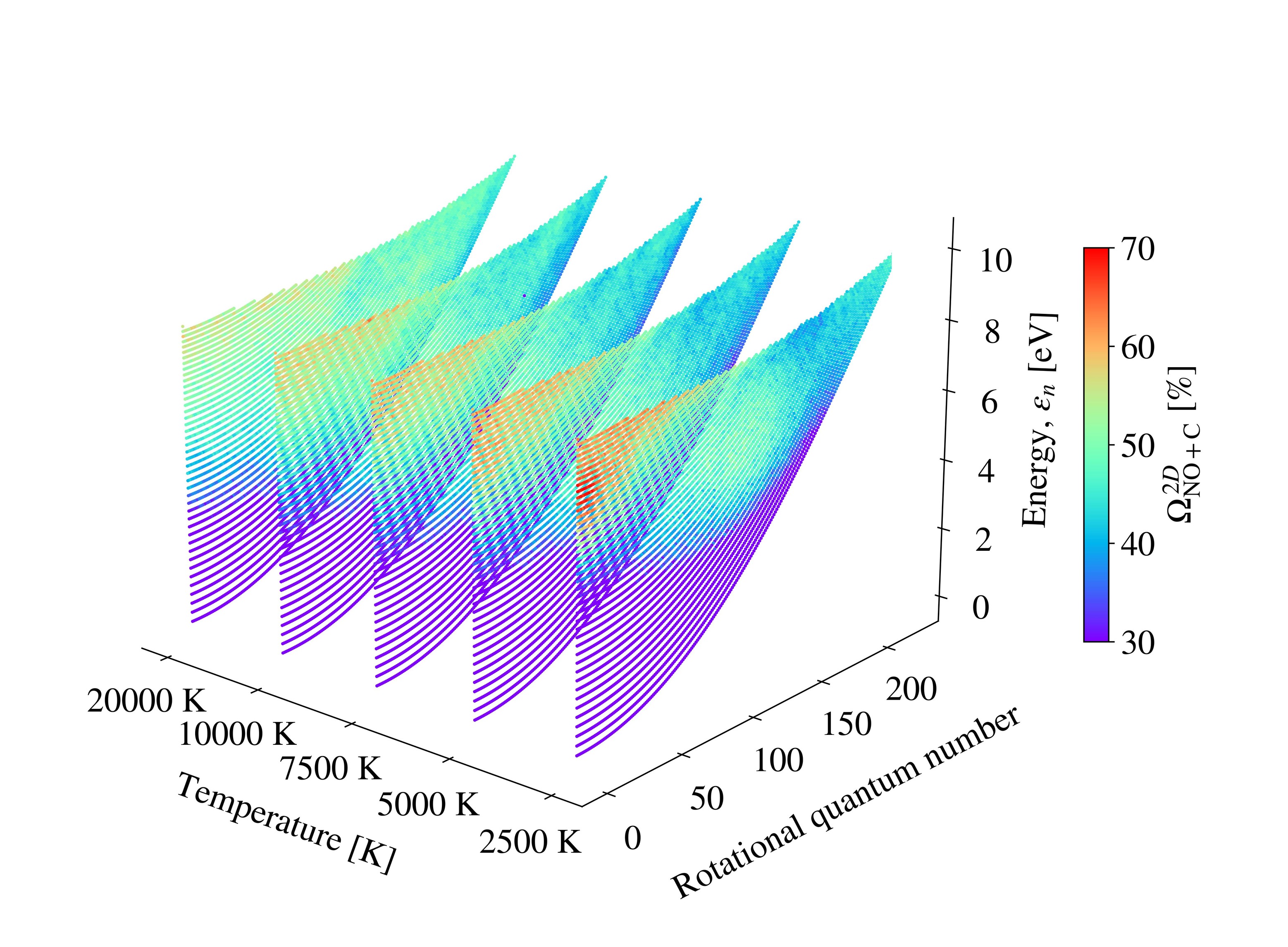}
     \caption{\ce{CO(X^1\Sigma^+) + N(^2D) <=> NO(X^2\Pi) + C(^3P)}.}
\end{subfigure}
\begin{subfigure}[t]{0.49\textwidth}
     \centering
     \includegraphics[width=\textwidth, trim=1.5cm 0.5cm 1cm 2cm, clip]{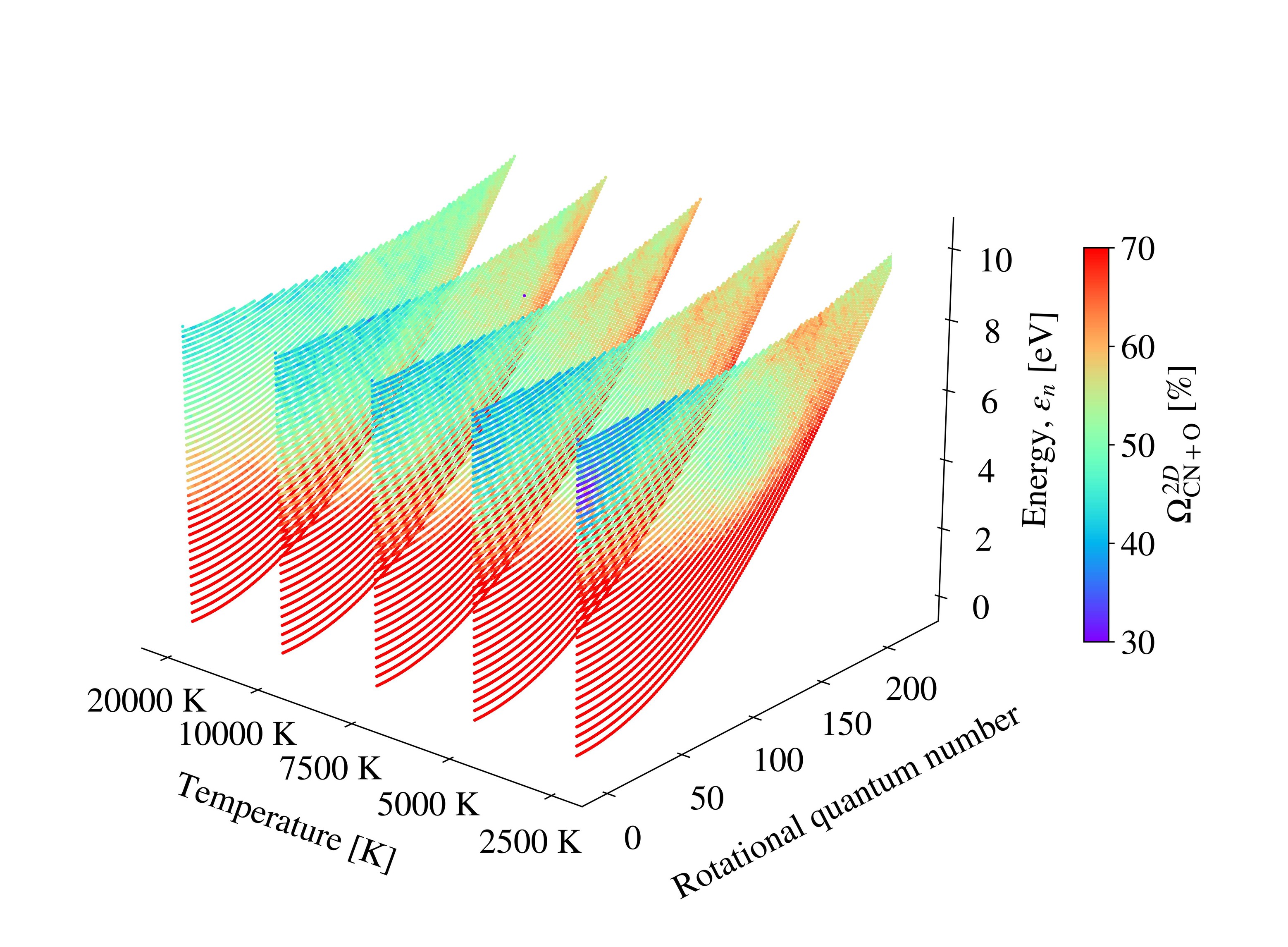}
     \caption{\ce{CO(X^1\Sigma^+) + N(^2D) <=> CN(X^2\Sigma^+) +O(^3P)}.}
\end{subfigure}
\caption{Branching ratio for the \ce{CO(X^1\Sigma^+) + N(^2D)} asymptote overlaid with the rovibrational state distribution at five temperatures.}
\label{BranchingRatioCON2D}
\end{figure*}

\subsection{Master Equation Analysis}

\subsubsection{Inelastic Processes}

\begin{figure*}[!ht]
\centering
\begin{subfigure}[t]{0.49\textwidth}
     \centering
     \includegraphics[width=\textwidth]{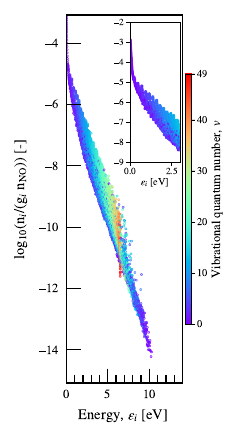}
     \caption{$^2A^{\prime}$.}
\end{subfigure}
\begin{subfigure}[t]{0.49\textwidth}
     \centering
     \includegraphics[width=\textwidth]{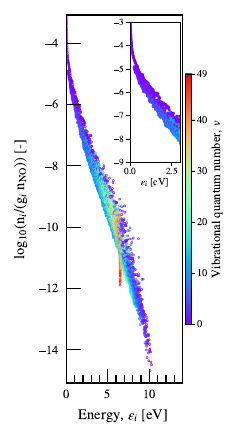}
     \caption{$^4A^{\prime\prime}$.}
\end{subfigure}
\caption{Potential energy surface-specific rovibrational distribution of the \ce{NO(X^2\Pi) + C(^3P)} collisional system at $T_t$~=~10,000~K and $T_v$~=~2000~K (dissociation and exchange processes excluded). The distributions are coloured by the vibrational quantum number.}
\label{PES_Specific_NO_C_2000K}
\end{figure*}

\begin{figure*}[!ht]
\centering
\begin{subfigure}[t]{0.49\textwidth}
     \centering
     \includegraphics[width=\textwidth]{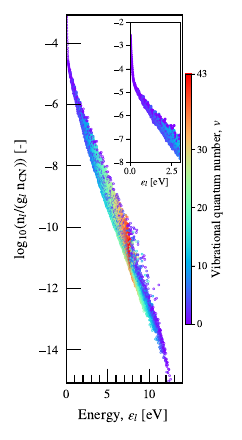}
     \caption{$^2A^{\prime}$.}
\end{subfigure}
\begin{subfigure}[t]{0.49\textwidth}
     \centering
     \includegraphics[width=\textwidth]{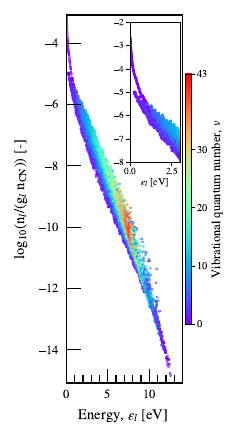}
     \caption{$^4A^{\prime\prime}$.}
\end{subfigure}
\caption{Potential energy surface-specific rovibrational distribution of the \ce{CN(X^2\Sigma^{+}) + O(^3P)} collisional system at $T_t$~=~10,000~K and $T_v$~=~2000~K (dissociation and exchange processes excluded). The distributions are coloured by the vibrational quantum number.}
\label{PES_Specific_CN_O_2000K}
\end{figure*}

\begin{figure*}
    \centering
    \begin{subfigure}[t]{0.49\textwidth}
         \centering
         \includegraphics[width=\linewidth]{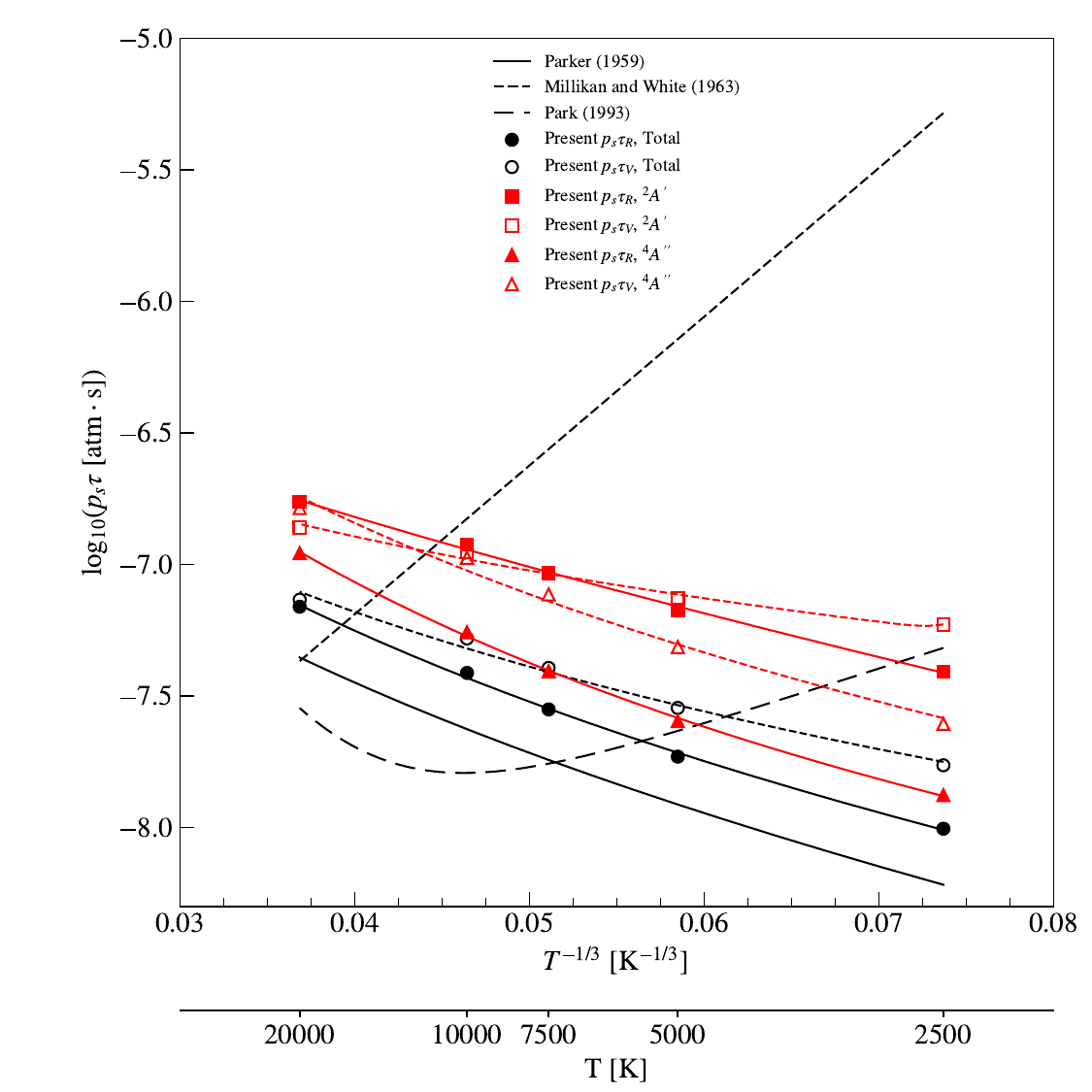}
         \caption{\ce{NO(X^2\Pi) + C(^3P)}}
    \end{subfigure}
    \begin{subfigure}[t]{0.49\textwidth}
         \centering
    \includegraphics[width=\linewidth]{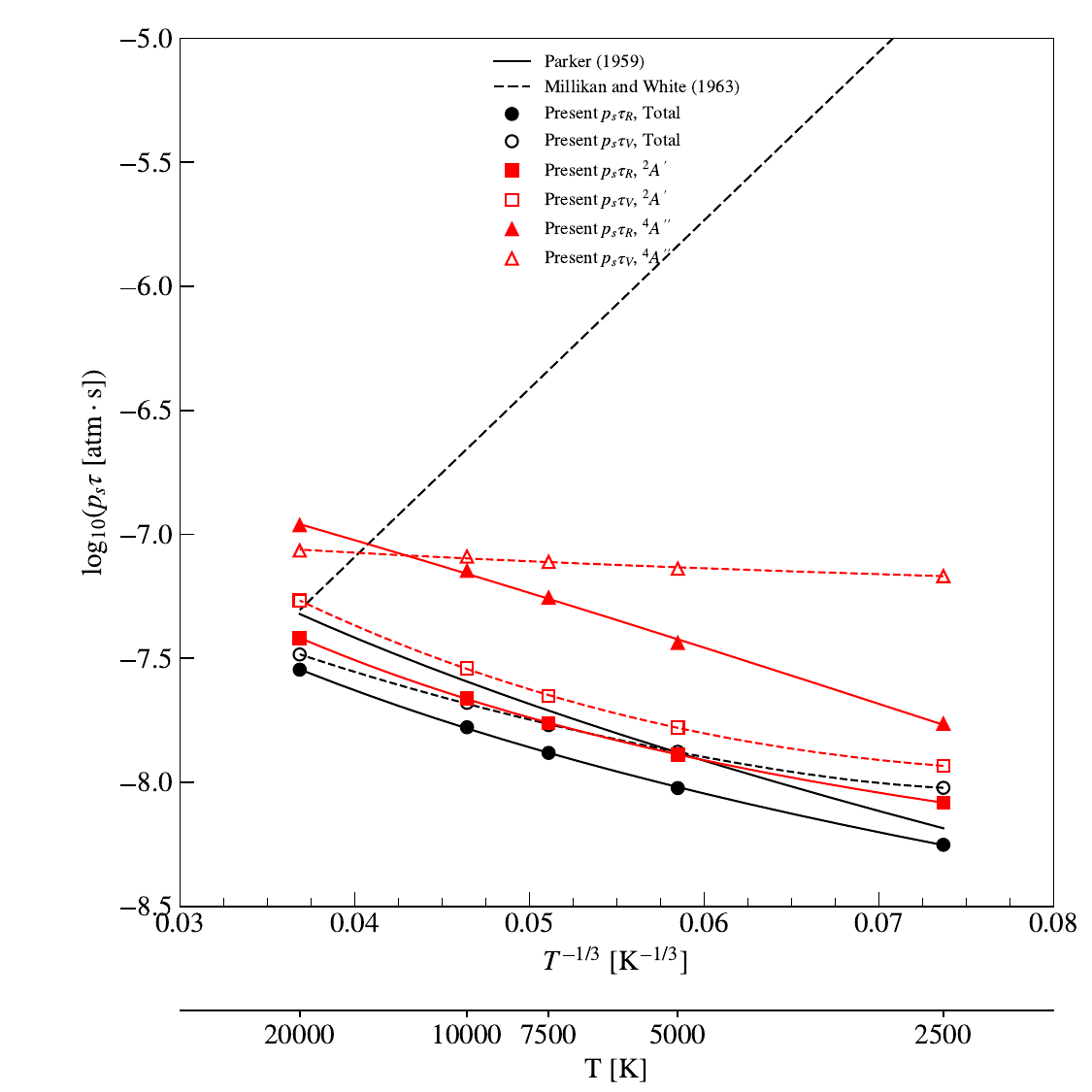}
         \caption{\ce{CN(X^2\Sigma^{+}) + O(^3P)}}
    \end{subfigure}
    \caption{Potential energy surfaces specific relaxation times for \ce{NO(X^2\Pi) + C(^3P)} and \ce{CN(X^2\Sigma^{+}) + O(^3P)}.}
    \label{PESSPecific_TauRTauV}
\end{figure*}

The relaxation times obtained from the master equation analysis are curve-fitted onto a modified Millikan--White-type exponential form for the rotational relaxation time and onto the Millikan--White form,\cite{Millikan1963Relaxation} with the high-temperature correction term proposed by Park,~\cite{park1993review} for the vibrational relaxation time. Therefore, the rotational relaxation time, expressed in atm~s, is defined as,
\begin{equation}
    p_s \tau_{r} = T^{n_r}\, e^{A_r T^{-\frac{1}{3}} + B_r},
\end{equation}
where $A_r$, $n_r$ and $B_r$ are tunable parameters and $p_s$ is the partial pressure of the atomic collision partner, and the vibrational relaxation time follows the Millikan--White ($\tau_{\text{MW}}$) formalism with the high-temperature correction term ($\tau_{\text{P}}$),
\begin{equation}
    \tau_{v} = \tau_{\text{MW}} + \tau_{\text{P}},
\end{equation}
where
\begin{equation}
    p_s\, \tau_{\text{MW}} = e^{A_v \left(T^{-\frac{1}{3}} - B_v  \right) - 18.42},
\end{equation}
with $p_s$ the partial pressure of the atomic collision partner, expressed in atm, and
\begin{equation}
   \tau_{\text{P}} = \frac{1}{n_s \sigma_v \sqrt{\frac{8kT}{\pi \mu}} },
\end{equation}
where $A_v$ and $B_v$ are tunable parameters, $n_s = p_s/k_{\text{B}}T$ is the number density of the atomic collision partner, $\mu$ the reduced mass of the colliding pair, and $\sigma_v = \alpha_v T^{\beta_v}$ a limiting cross-section with tunable parameters $\alpha_v$ and $\beta_v$. Note that $\alpha_v$ is expressed in m$^2$~K$^{-\beta_v}$, so that $\sigma_v$ is expressed in m$^2$, and that $\tau_{\text{P}}$ is evaluated in s at the number density of interest before being added to $\tau_{\text{MW}}$. The tunable parameters are summarised in Tab.~\ref{tab:FittingParametersTauRTauV} and are valid between 2500 and 20,000~K. For the \ce{NO(X^2\Pi) + C(^3P)} system, the Millikan--White term represents less than 1 \% of the vibrational relaxation time at all temperatures, and the vibrational relaxation time is therefore given by $\tau_{\text{P}}$ alone. Both relaxation times are normalised by the atomic collision partner, i.e.~by its partial pressure in $\tau_{\text{MW}}$ and by its number density in $\tau_{\text{P}}$, following the convention adopted for the rovibrational state-specific databases of Panesi~\textit{et al.}\cite{panesi2013rovibrational} and Jo~\textit{et al.}\cite{jo2022rovibrational} The coefficients are therefore pairwise quantities and can be introduced directly into the usual mixture rule $\langle \tau_s \rangle^{-1} = \sum_q x_q / \tau_{s,q}$.

\begin{table*}[ht!]
    \centering
    \caption{Curve fit parameters for the rotational and vibrational relaxation times for the \ce{NO(X^2 \Pi) + C(^3P)}, \ce{CN(X^2 \Sigma^{+}) + O(^3P)} and \ce{CO(X^1 \Sigma^{+}) + N(^4S,^2D)} collisional systems.}
    \label{tab:FittingParametersTauRTauV}
    \begin{tabular*}{\textwidth}{@{\extracolsep{\fill}}ccccccc@{}}
        Colliding pair & $A_r$ [-] & $n_r$ [-] & $B_r$ [-] & $A_v$ [-] & $B_v$ [-] & $\sigma_v = \alpha_v T^{\beta_v}$ [m$^2$] \\
        \hline
        \ce{NO(X^2 \Pi) + C(^3P)} & $-6.704$ & $0.8279$ & $-24.427$ & --- & --- & $4.185\times 10^{-20} T^{-0.216}$ \\
        \ce{CN(X^2 \Sigma^{+}) + O(^3P)} & $1.815$ & $0.8171$ & $-25.530$ & $117.305$ & $0.0911$ & $6.576\times 10^{-20} T^{-0.166}$ \\
        \ce{CO(X^1 \Sigma^{+}) + N(^4S)} & $-3.462$ & $0.8624$ & $-26.297$ & $124.577$ & $0.0356$ & $8.263\times 10^{-21} T^{0.084}$ \\
        \ce{CO(X^1 \Sigma^{+}) + N(^2D)} & $48.485$ & $1.5609$ & $-34.507$ & $136.561$ & $0.0835$ & $3.911\times 10^{-20} T^{-0.121}$ \\
    \end{tabular*}
\end{table*}

\subsubsection{Dissociation Processes}

\begin{figure*}[!ht]
\centering
\begin{subfigure}[t]{0.49\textwidth}
     \centering
     \includegraphics[width=\textwidth]{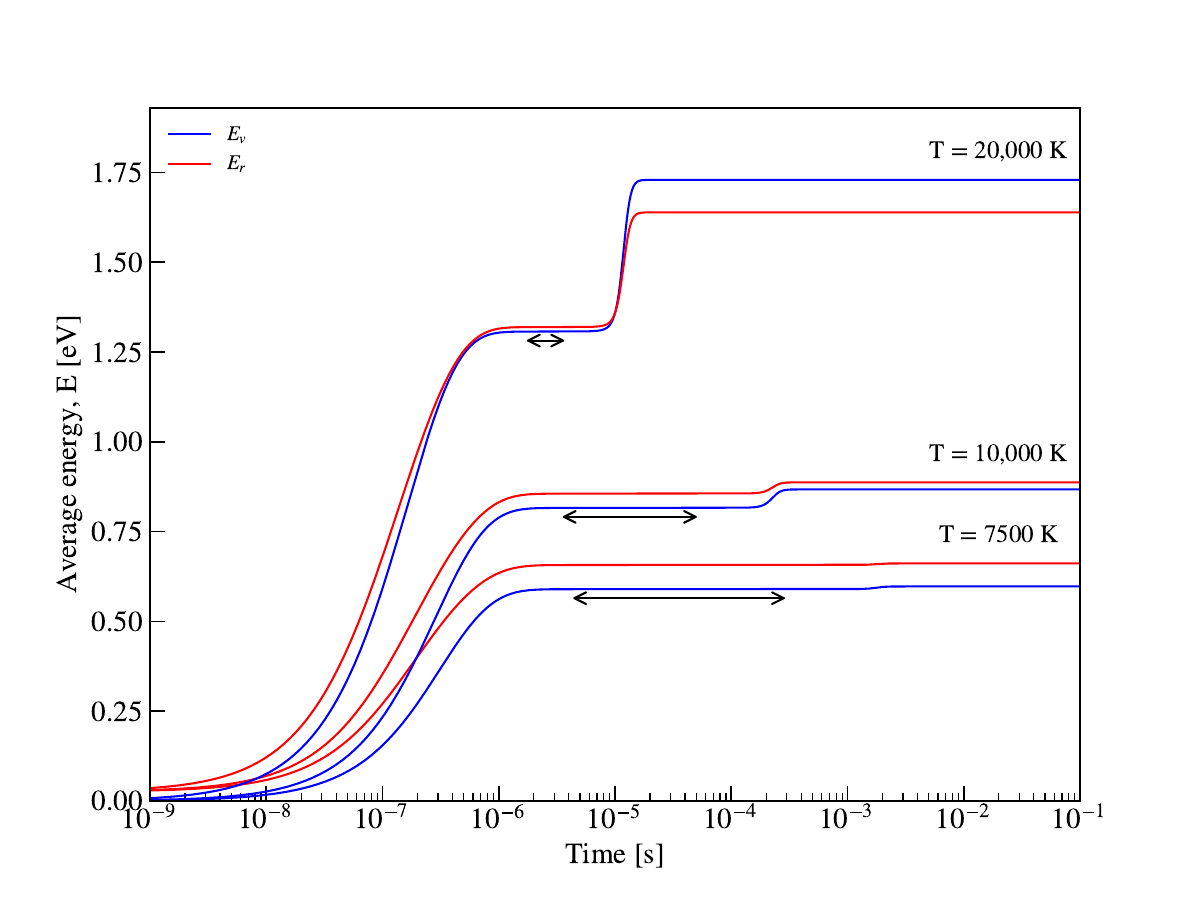}
     \caption{Average rotational and vibrational energies.}
\end{subfigure}
\begin{subfigure}[t]{0.49\textwidth}
     \centering
     \includegraphics[width=\textwidth]{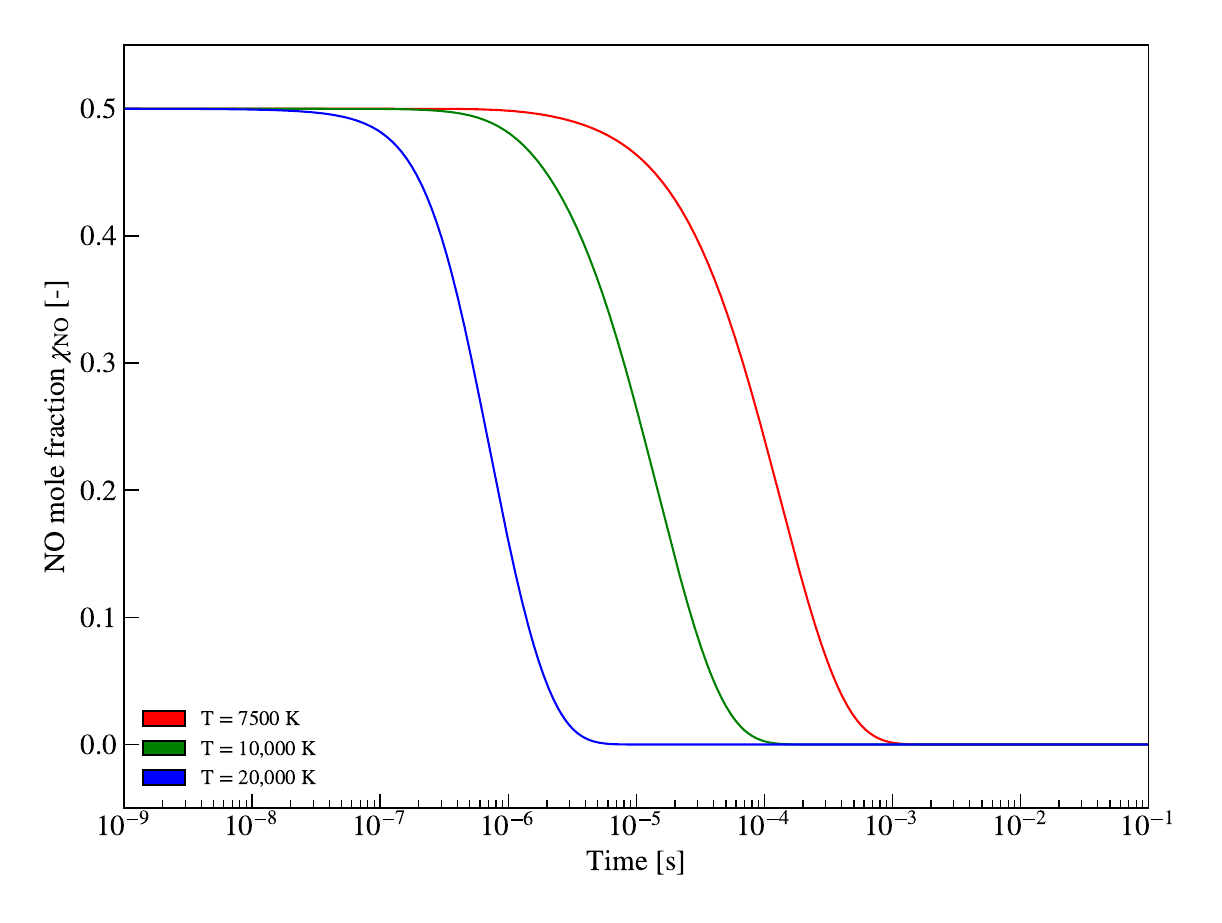}
     \caption{Mole fraction.}
\end{subfigure}
\caption{Temporal evolution of the average rotational and vibrational energies of \ce{NO(X^2\Pi)} and of its mole fraction, under the action of the total dissociation, at $T = 7500$, 10,000 and 20,000~K (heterogeneous exchange reactions excluded).}
\label{DissEnergyMolar_NOC}
\end{figure*}

\begin{figure*}[!ht]
\centering
\begin{subfigure}[t]{0.49\textwidth}
     \centering
     \includegraphics[width=\textwidth]{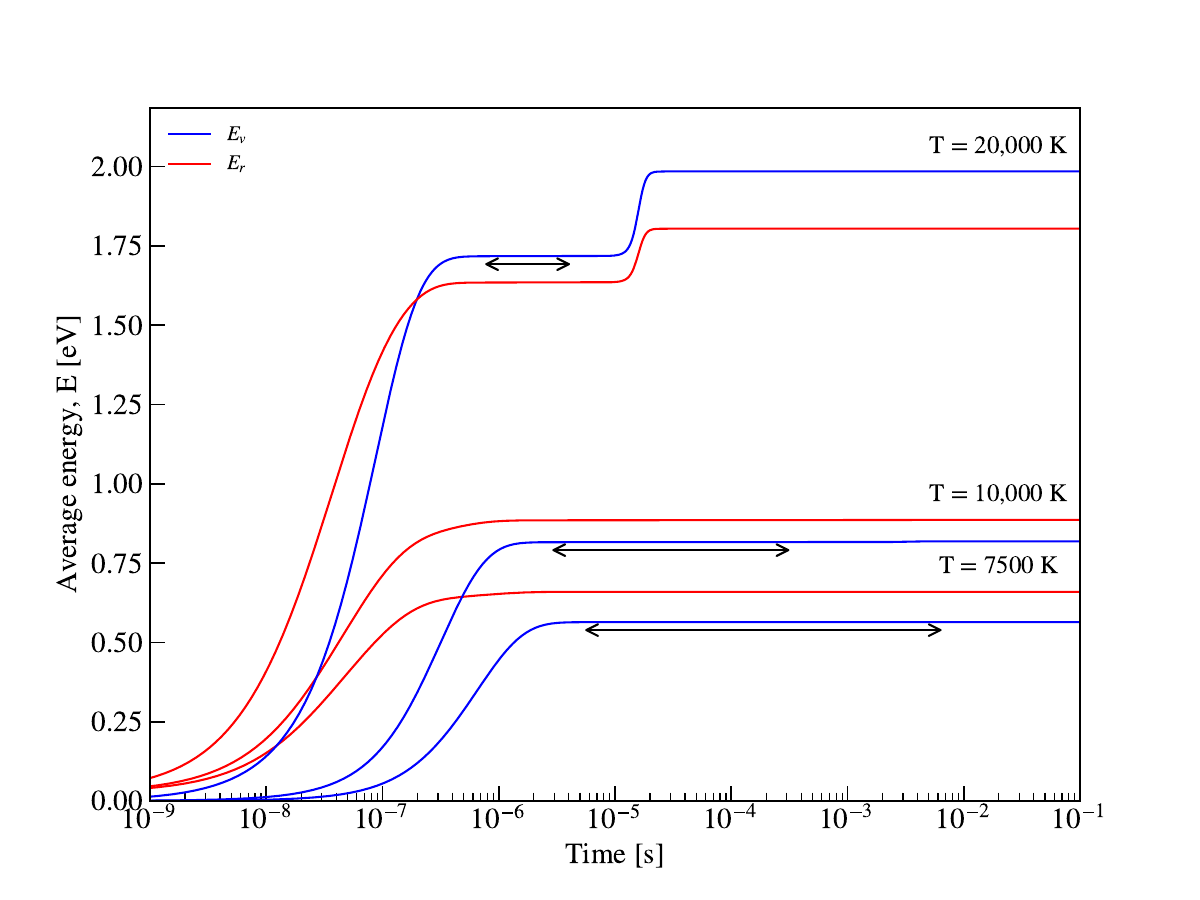}
     \caption{Average rotational and vibrational energies.}
\end{subfigure}
\begin{subfigure}[t]{0.49\textwidth}
     \centering
     \includegraphics[width=\textwidth]{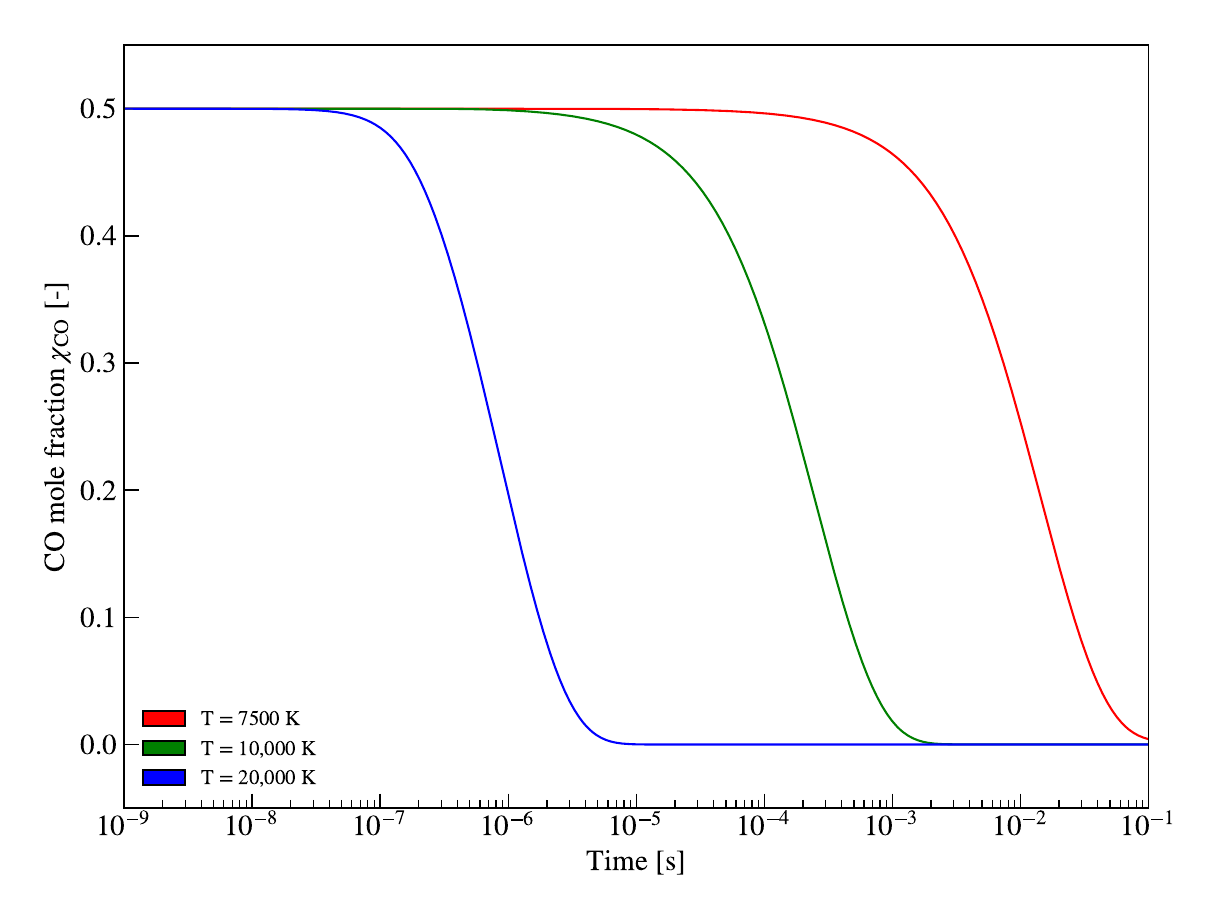}
     \caption{Mole fraction.}
\end{subfigure}
\caption{Temporal evolution of the average rotational and vibrational energies of \ce{CO(X^1 \Sigma^{+})} when colliding with \ce{N(^4S)} and of its mole fraction, under the action of the total dissociation, at $T = 7500$, 10,000 and 20,000~K (heterogeneous exchange reactions excluded).}
\label{DissEnergyMolar_CON4S}
\end{figure*}

\begin{figure*}[!ht]
\centering
\begin{subfigure}[t]{0.49\textwidth}
     \centering
     \includegraphics[width=\textwidth]{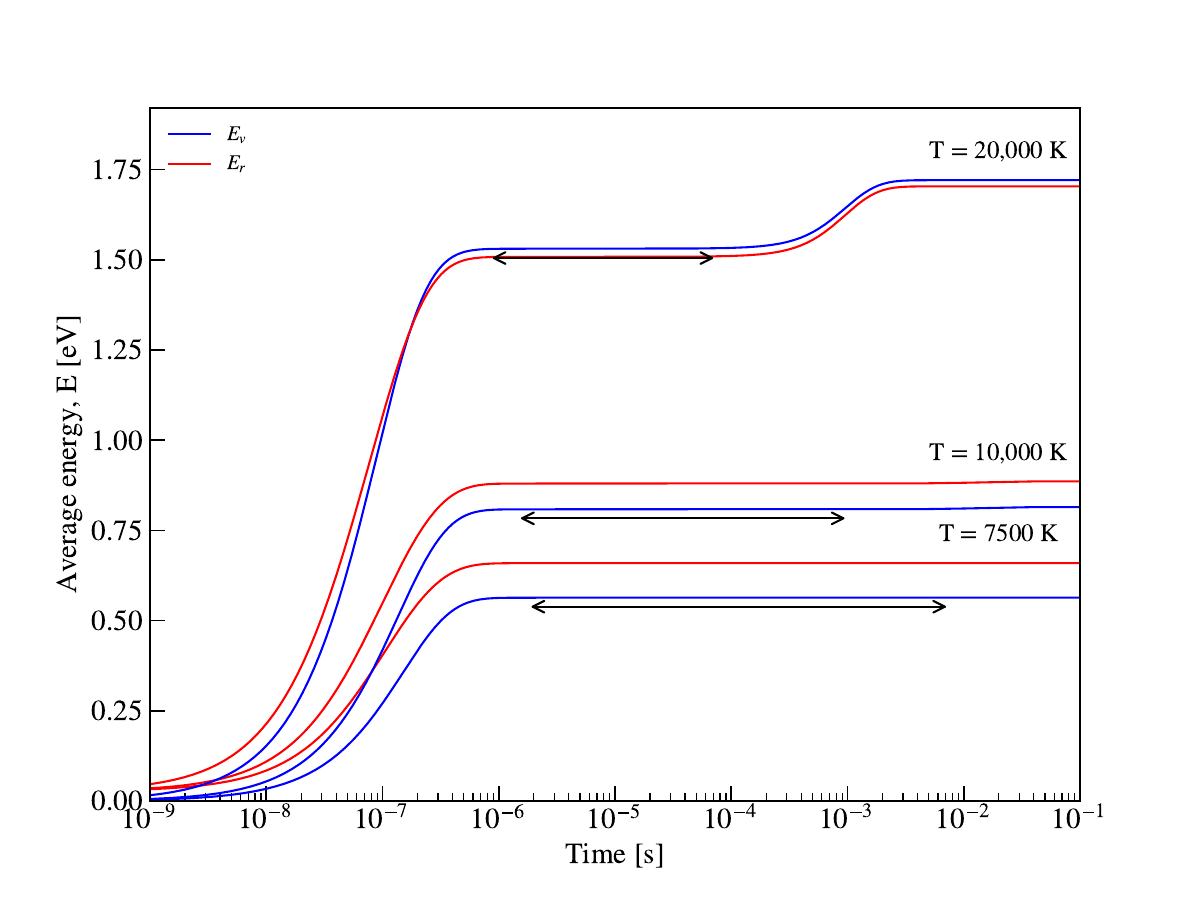}
     \caption{Average rotational and vibrational energies.}
\end{subfigure}
\begin{subfigure}[t]{0.49\textwidth}
     \centering
     \includegraphics[width=\textwidth]{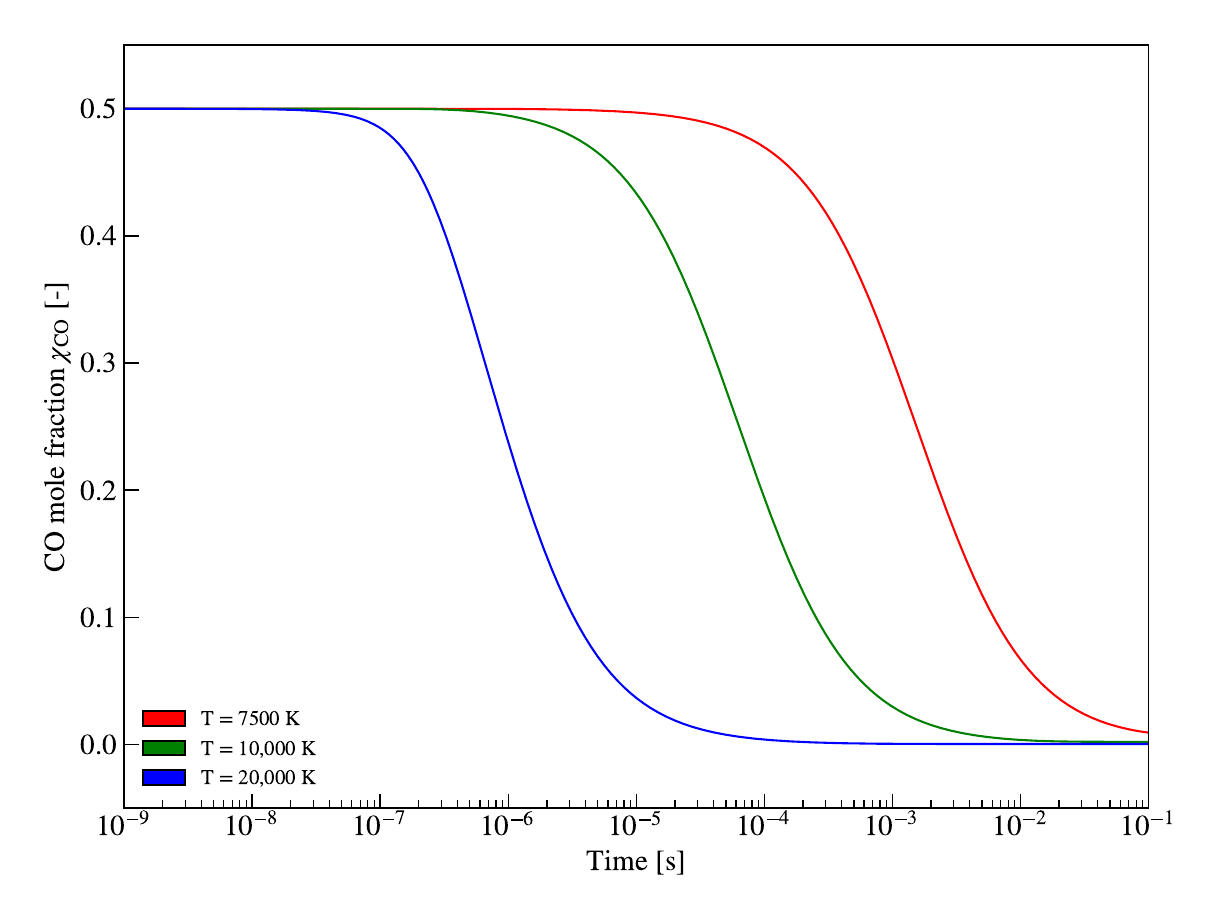}
     \caption{Mole fraction.}
\end{subfigure}
\caption{Temporal evolution of the average rotational and vibrational energies of \ce{CO(X^1 \Sigma^{+})} when colliding with \ce{N(^2D)} and of its mole fraction, under the action of the total dissociation, at $T = 7500$, 10,000 and 20,000~K (heterogeneous exchange reactions excluded).}
\label{DissEnergyMolar_CON2D}
\end{figure*}

\begin{figure*}
\centering
    \begin{subfigure}[t]{0.49\textwidth}
    \centering
    \includegraphics[width=\textwidth]{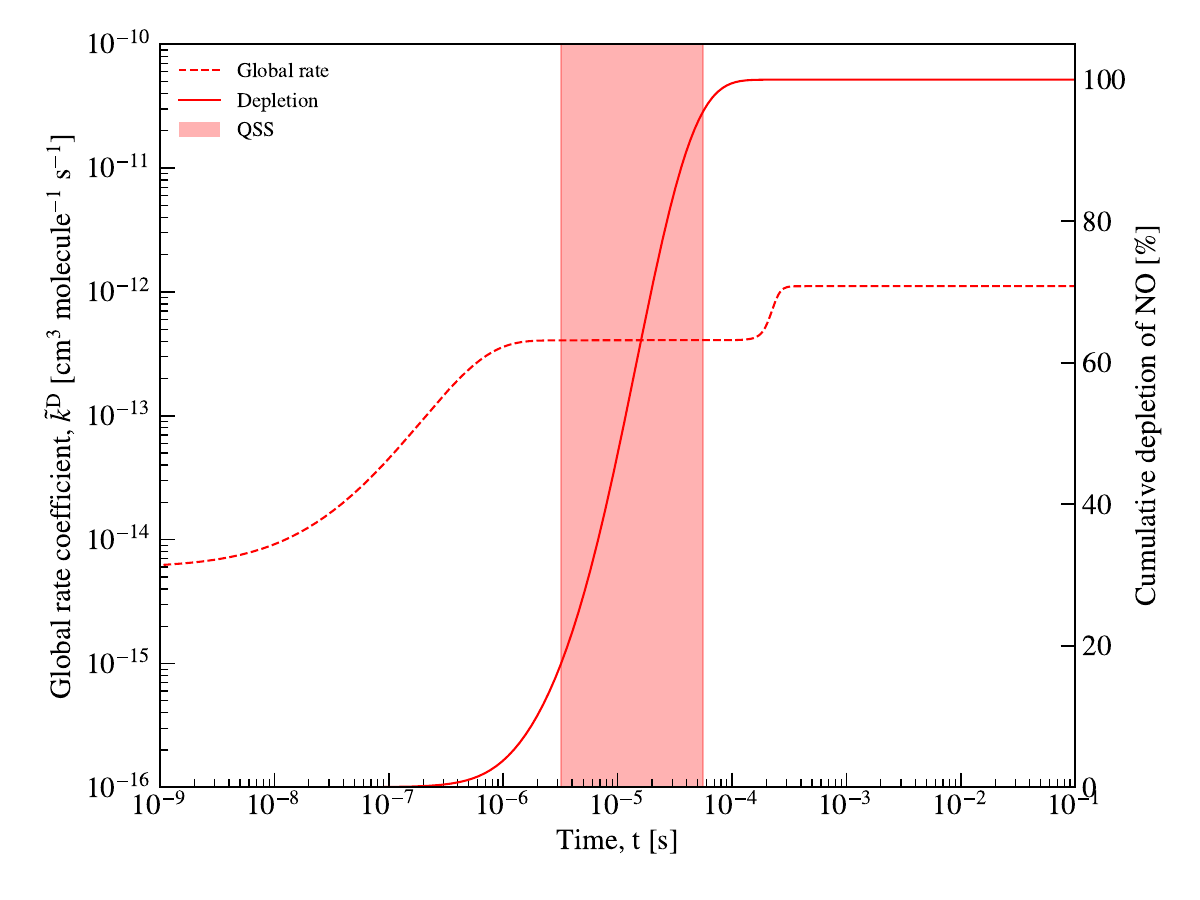}
    \caption{\ce{NO(X^2\Pi) + C(^3P)}.}
    \end{subfigure}
    \begin{subfigure}[t]{0.49\textwidth}
    \centering
    \includegraphics[width=\textwidth]{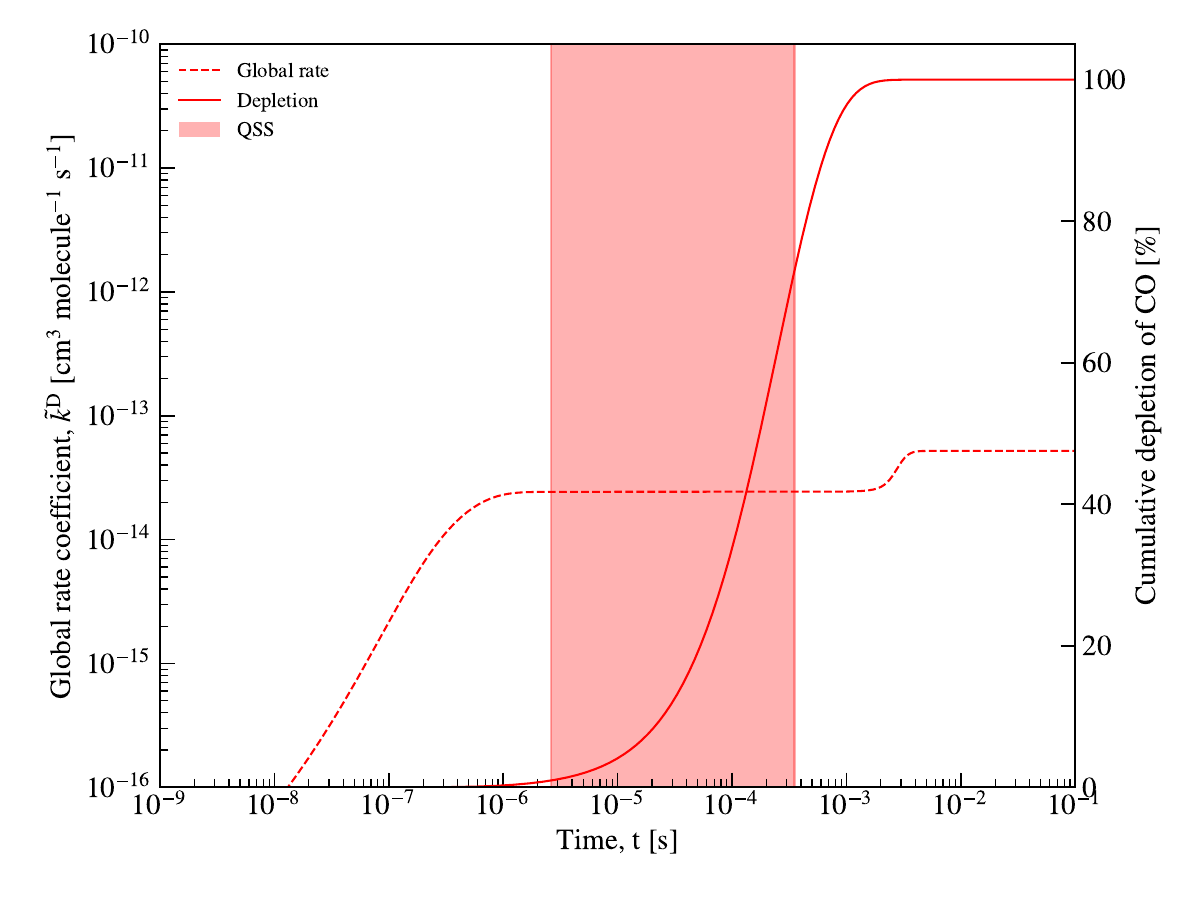}
    \caption{\ce{CO(X^1 \Sigma^{+}) + N(^4S)}.}
    \end{subfigure}
    \begin{subfigure}[t]{0.49\textwidth}
    \centering
    \includegraphics[width=\textwidth]{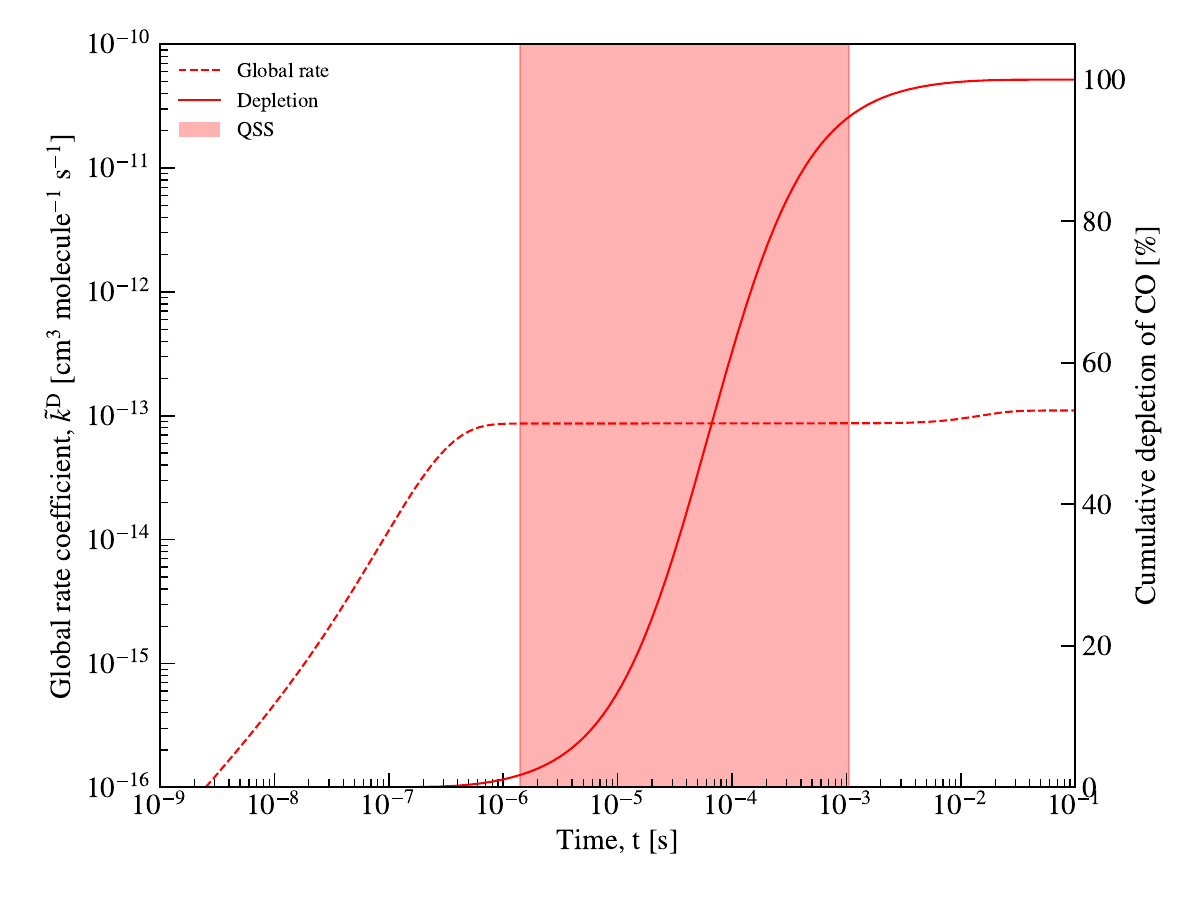}
    \caption{\ce{CO(X^1 \Sigma^{+}) + N(^2D)}.}
    \end{subfigure}
    \caption{Time-cumulative depletion of the diatomic species and global dissociation rate coefficient of the \ce{NO(X^2\Pi) + C(^3P)}, \ce{CO(X^1 \Sigma^{+}) + N(^4S)} and \ce{CO(X^1 \Sigma^{+}) + N(^2D)} collisional systems at $T = 10{,}000$~K (heterogeneous exchange reactions excluded).}
    \label{CumulativeProduction}
\end{figure*}

\begin{figure*}[!ht]
\centering
    \begin{subfigure}[t]{\textwidth}
    \centering
    \includegraphics[width=\textwidth]{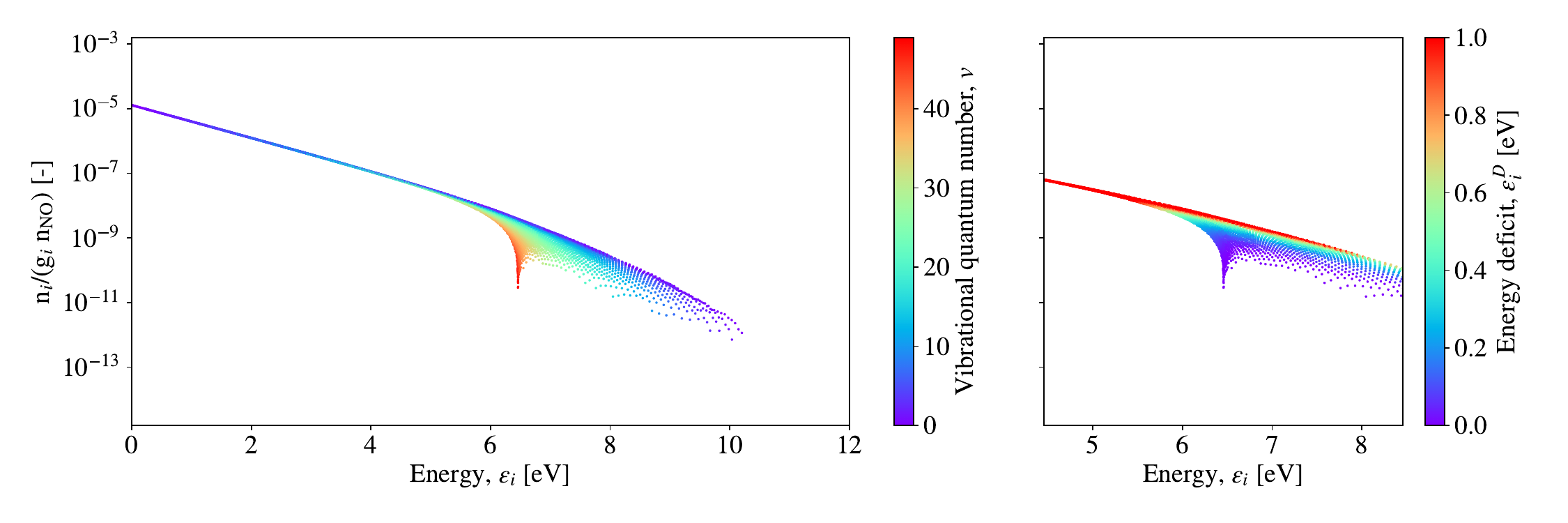}
    \caption{\ce{NO(X^2\Pi) + C(^3P)}.}
    \end{subfigure}
    \begin{subfigure}[t]{\textwidth}
    \centering
    \includegraphics[width=\textwidth]{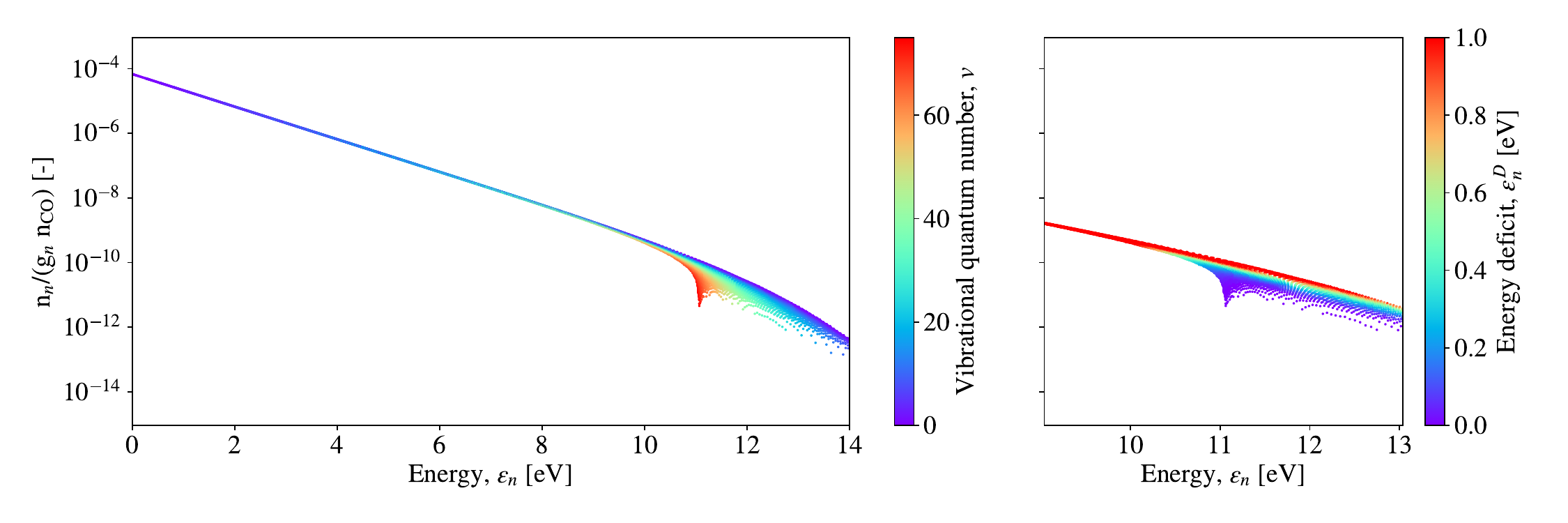}
    \caption{\ce{CO(X^1 \Sigma^{+}) + N(^4S)}.}
    \end{subfigure}
    \begin{subfigure}[t]{\textwidth}
    \centering
    \includegraphics[width=\textwidth]{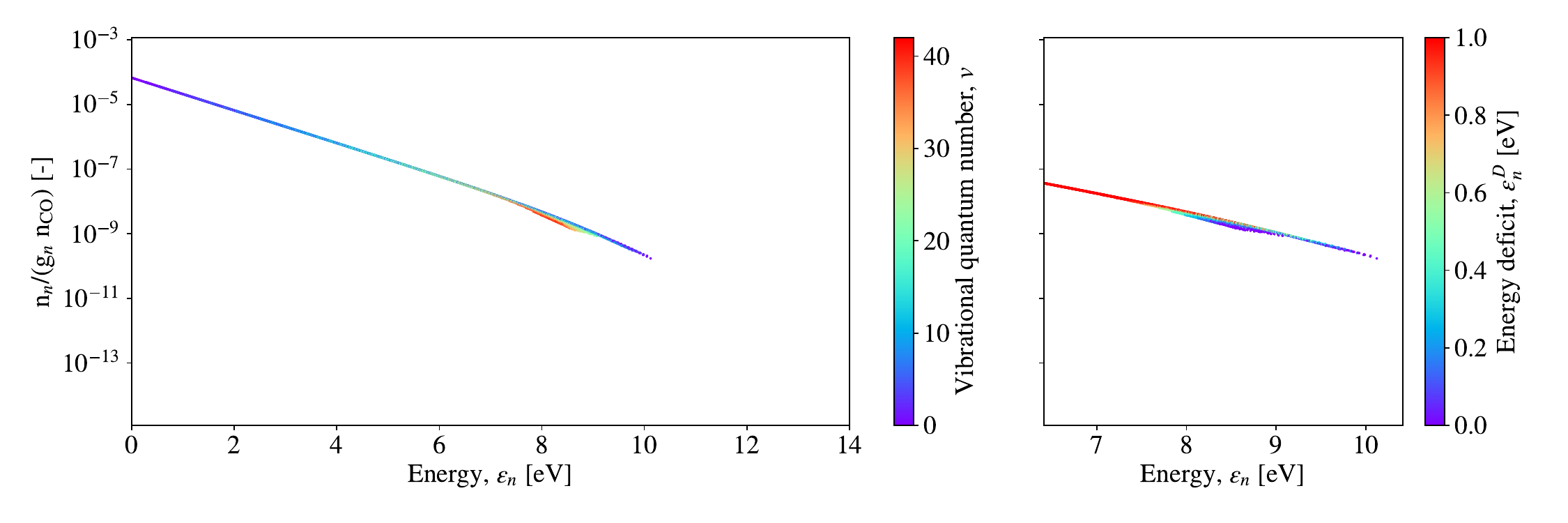}
    \caption{\ce{CO(X^1 \Sigma^{+}) + N(^2D)}.}
    \end{subfigure}
    \caption{Rovibrational distributions of the \ce{NO(X^2\Pi) + C(^3P)}, \ce{CO(X^1 \Sigma^{+}) + N(^4S)} and \ce{CO(X^1 \Sigma^{+}) + N(^2D)} collisional systems at the midpoint of the QSS regime at $T = 10{,}000$~K (heterogeneous exchange processes excluded). The left panels present the population distributions coloured by the vibrational quantum number, and the right panels a zoom of the rovibrational states near the rotationless dissociation limit, coloured by the state-specific energy deficit.}
    \label{MidQSS_Others}
\end{figure*}

The rotational and vibrational energy loss coefficients due to dissociation are curve-fitted onto a continuous functional form defined as, 
\begin{equation}
    C^{\text{D}} = \text{exp}\left( \frac{K_1}{T} + K_2 + K_3 \text{ln}T \right),
\end{equation}
where the fitting parameters of the rotational and vibrational energy loss coefficients are tabulated in Tabs.~\ref{RotEnergyLoss} and~\ref{VibEnergyLoss}, respectively.

\begin{table*}[!ht]
    \centering
    \caption{Fitting parameters for the rotational energy loss coefficient for the \ce{NO(X^2 \Pi) + C(^3P)}, \ce{CN(X^2 \Sigma^{+}) + O(^3P)} and \ce{CO(X^1 \Sigma^{+}) + N(^4S,^2D)} collisional systems.}
    \label{RotEnergyLoss}
    \begin{tabular*}{\textwidth}{@{\extracolsep{\fill}}cccc@{}}
        Collision & $K_1$ & $K_2$ & $K_3$ \\
        \hline
        \ce{NO(X^2 \Pi) + C(^3P)} & $-2019$ & $-1.538$ & $0.04643$ \\
        \ce{CN(X^2 \Sigma^{+}) + O(^3P)} & $-1803$ & $-1.946$ & $0.09949$ \\
        \ce{CO(X^1 \Sigma^{+}) + N(^4S)} & $-2980$ & $-1.382$ & $0.02396$ \\
        \ce{CO(X^1 \Sigma^{+}) + N(^2D)} & $-2094$ & $0.3877$ & $-0.1416$ \\
    \end{tabular*}
\end{table*}

\begin{table*}[!ht]
    \centering
    \caption{Fitting parameters for the vibrational energy loss coefficient for the \ce{NO(X^2 \Pi) + C(^3P)}, \ce{CN(X^2 \Sigma^{+}) + O(^3P)} and \ce{CO(X^1 \Sigma^{+}) + N(^4S,^2D)} collisional systems.}
    \label{VibEnergyLoss}
    \begin{tabular*}{\textwidth}{@{\extracolsep{\fill}}cccc@{}}
        Collision & $K_1$ & $K_2$ & $K_3$ \\
        \hline
        \ce{NO(X^2 \Pi) + C(^3P)} & $260.3$ & $1.481$ & $-0.2332$ \\
        \ce{CN(X^2 \Sigma^{+}) + O(^3P)} & $-223.6$ & $2.277$ & $-0.3068$ \\
        \ce{CO(X^1 \Sigma^{+}) + N(^4S)} & $-473.7$ & $2.619$ & $-0.3273$ \\
        \ce{CO(X^1 \Sigma^{+}) + N(^2D)} & $263.2$ & $1.344$ & $-0.2228$ \\
    \end{tabular*}
\end{table*}

\subsubsection{Heterogeneous Exchange Processes}

\begin{figure*}[!ht]
\centering
\begin{subfigure}[t]{0.49\textwidth}
     \centering
     \includegraphics[width=\textwidth, trim=1.5cm 0cm 2cm 2cm, clip]{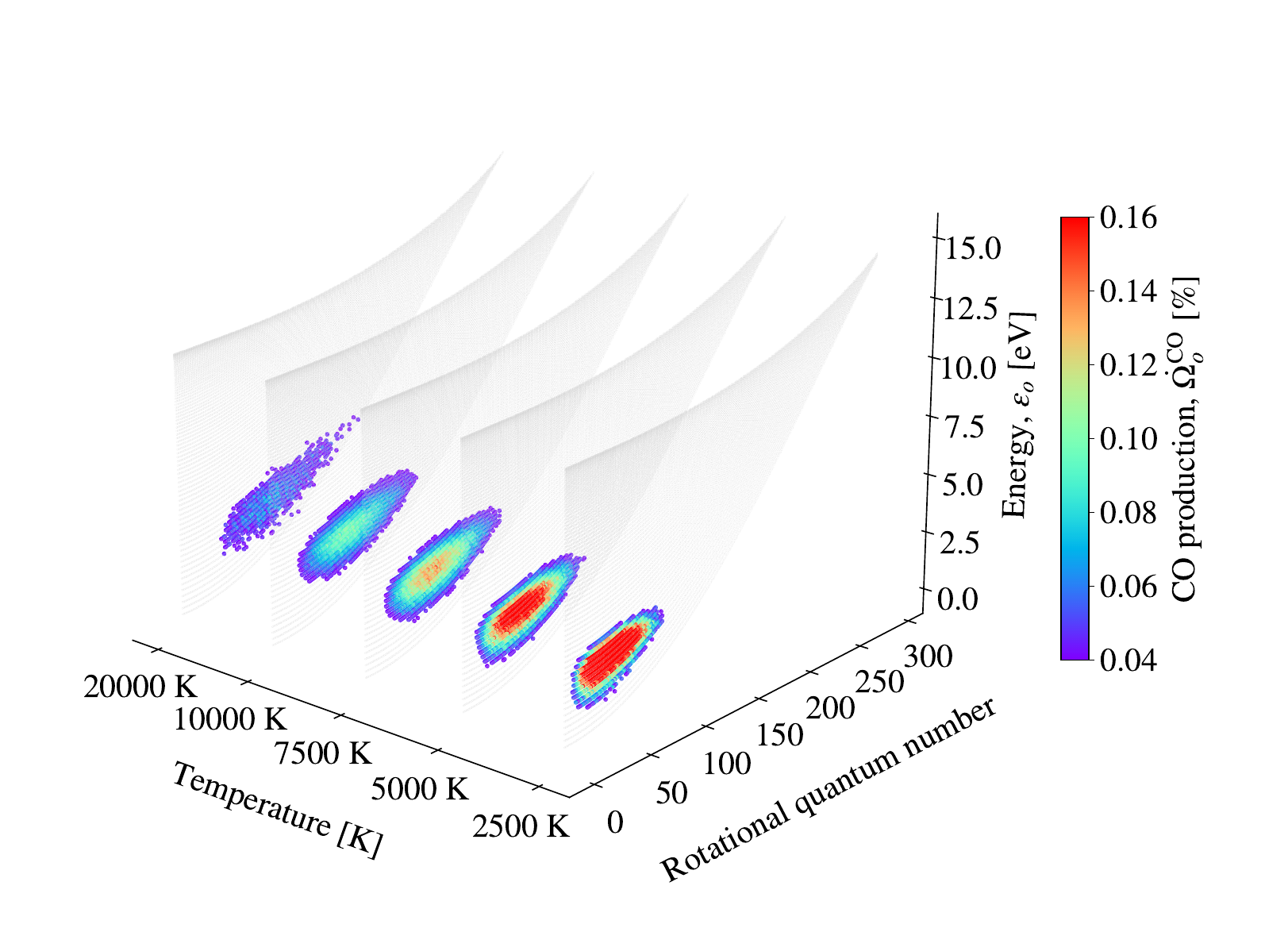}
     \caption{\ce{CO(X^1\Sigma^{+})} production.}
\end{subfigure}
\begin{subfigure}[t]{0.49\textwidth}
     \centering
     \includegraphics[width=\textwidth, trim=1.5cm 0cm 2cm 2cm, clip]{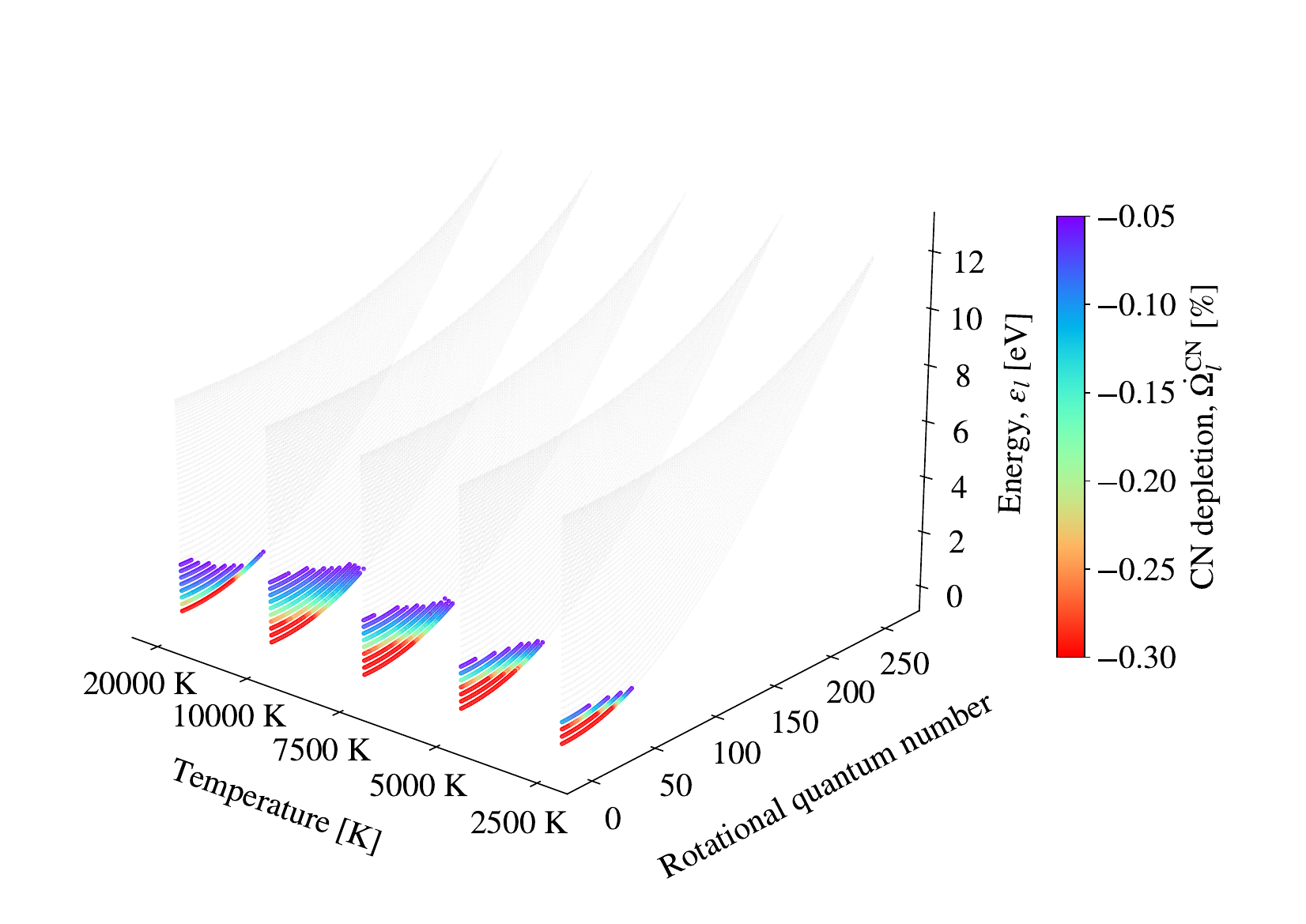}
     \caption{\ce{CN(X^2\Sigma^{+})} depletion.}
\end{subfigure}
\caption{Distribution of the exchange production and depletion rates for \ce{CN(X^2\Sigma^{+}) + O(^3P) <=> CO(X^1\Sigma^{+}) + N(^4S)} at peak production time for five temperatures. The grey dots indicate that the production and depletion rates are below or above a threshold.}
\label{ProdDep_CNO_CON4S}
\end{figure*}

\begin{figure*}[!ht]
\centering

\begin{subfigure}[t]{0.49\textwidth}
     \centering
     \includegraphics[width=\textwidth, trim=1.5cm 0cm 2cm 2cm, clip]{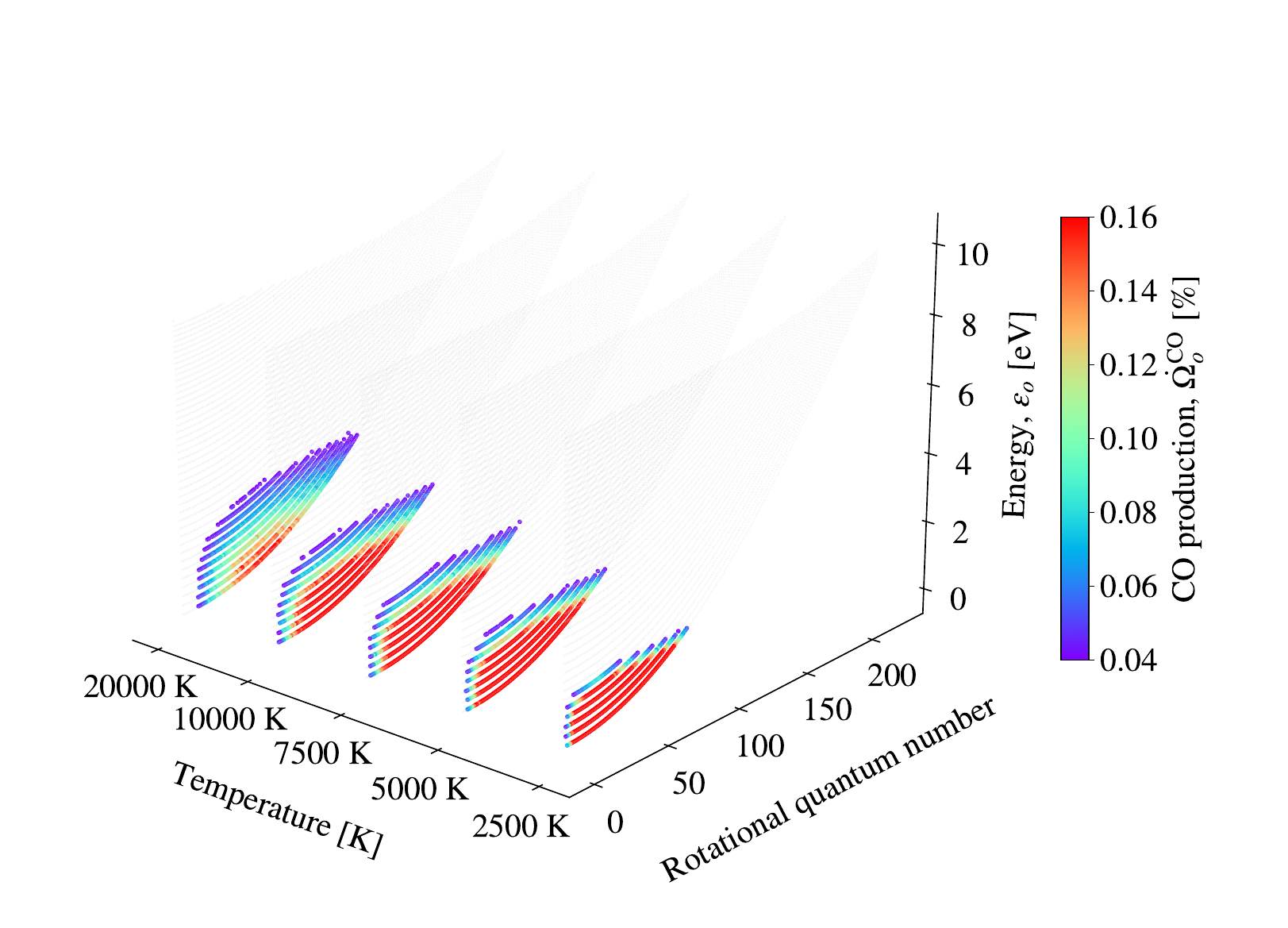}
     \caption{\ce{CO(X^1\Sigma^{+})} production.}
\end{subfigure}
\begin{subfigure}[t]{0.49\textwidth}
     \centering
     \includegraphics[width=\textwidth, trim=1.5cm 0cm 2cm 2cm, clip]{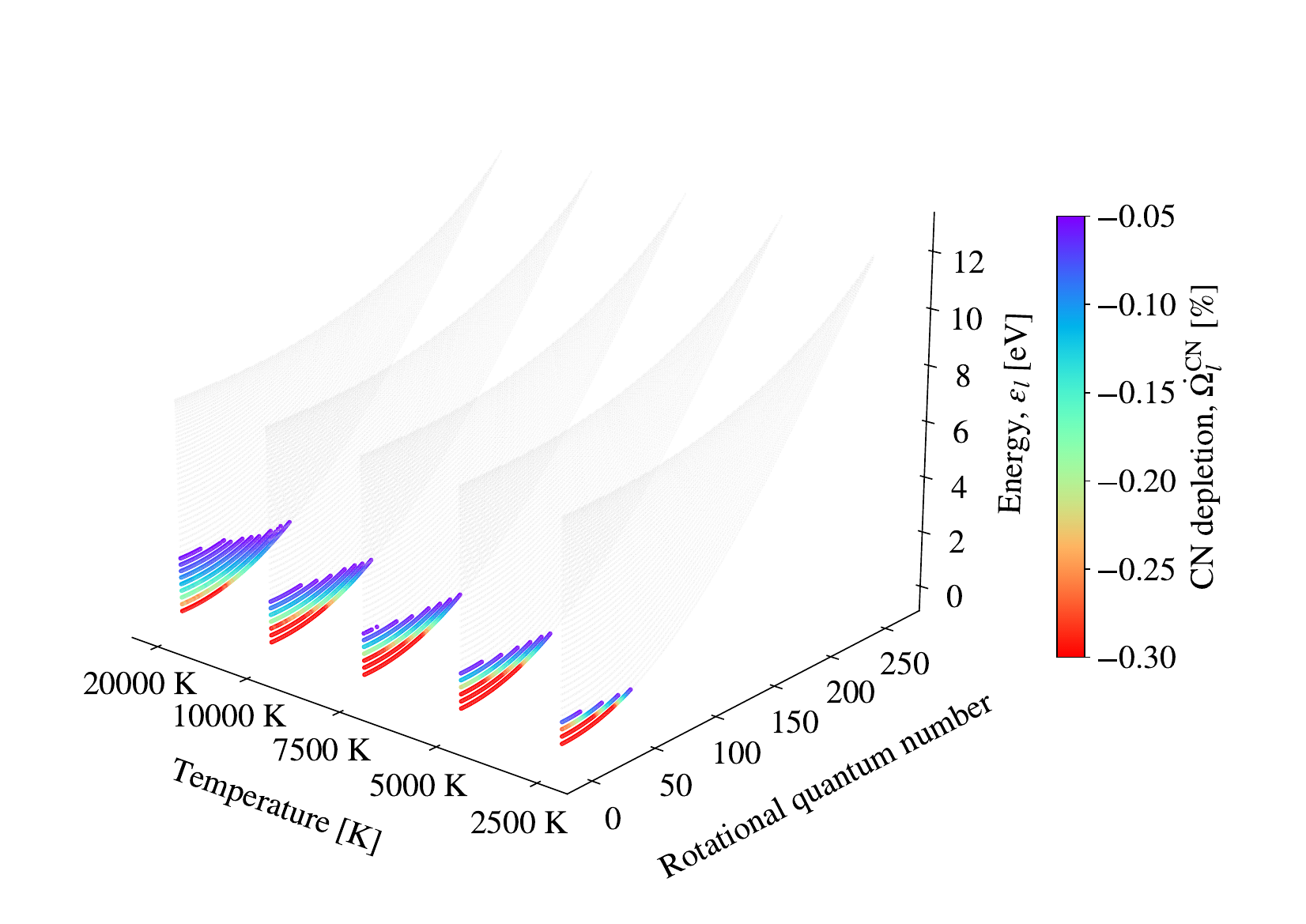}
     \caption{\ce{CN(X^2\Sigma^{+})} depletion.}
\end{subfigure}

\caption{Distribution of the exchange production and depletion rates for \ce{CN(X^2\Sigma^{+}) + O(^3P) <=> CO(X^1\Sigma^{+}) + N(^2D)} at peak production time for five temperatures. The grey dots indicate that the production and depletion rates are below or above a threshold.}
\label{ProdDep_CNO_CON2D}
\end{figure*}

\begin{figure*}[!ht]
\centering
\begin{subfigure}[t]{0.49\textwidth}
     \centering
     \includegraphics[width=\textwidth, trim=1.5cm 0cm 2cm 2cm, clip]{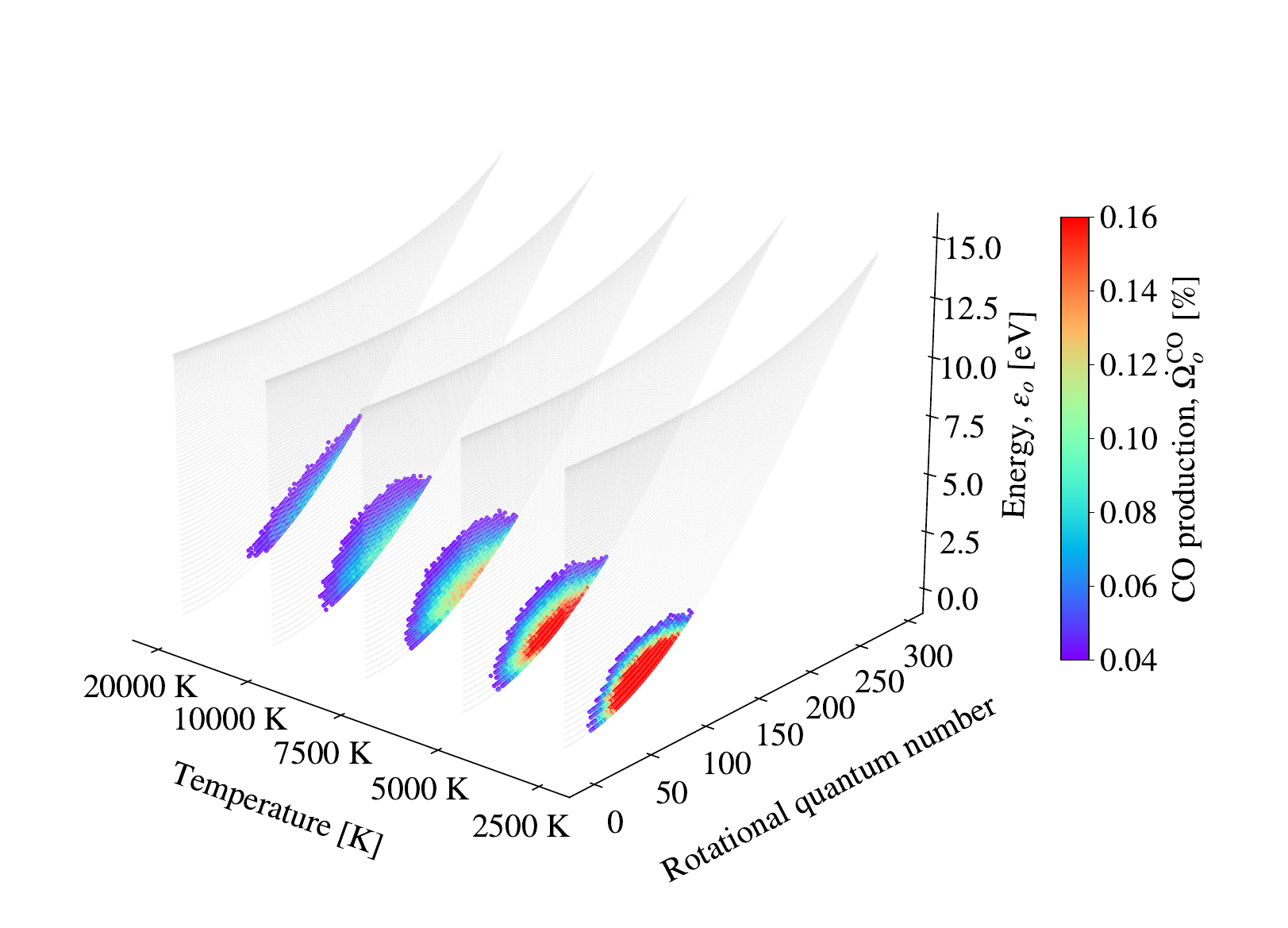}
     \caption{\ce{CO(X^1\Sigma^{+})} production.}
\end{subfigure}
\begin{subfigure}[t]{0.49\textwidth}
     \centering
     \includegraphics[width=\textwidth, trim=1.5cm 0cm 2cm 2cm, clip]{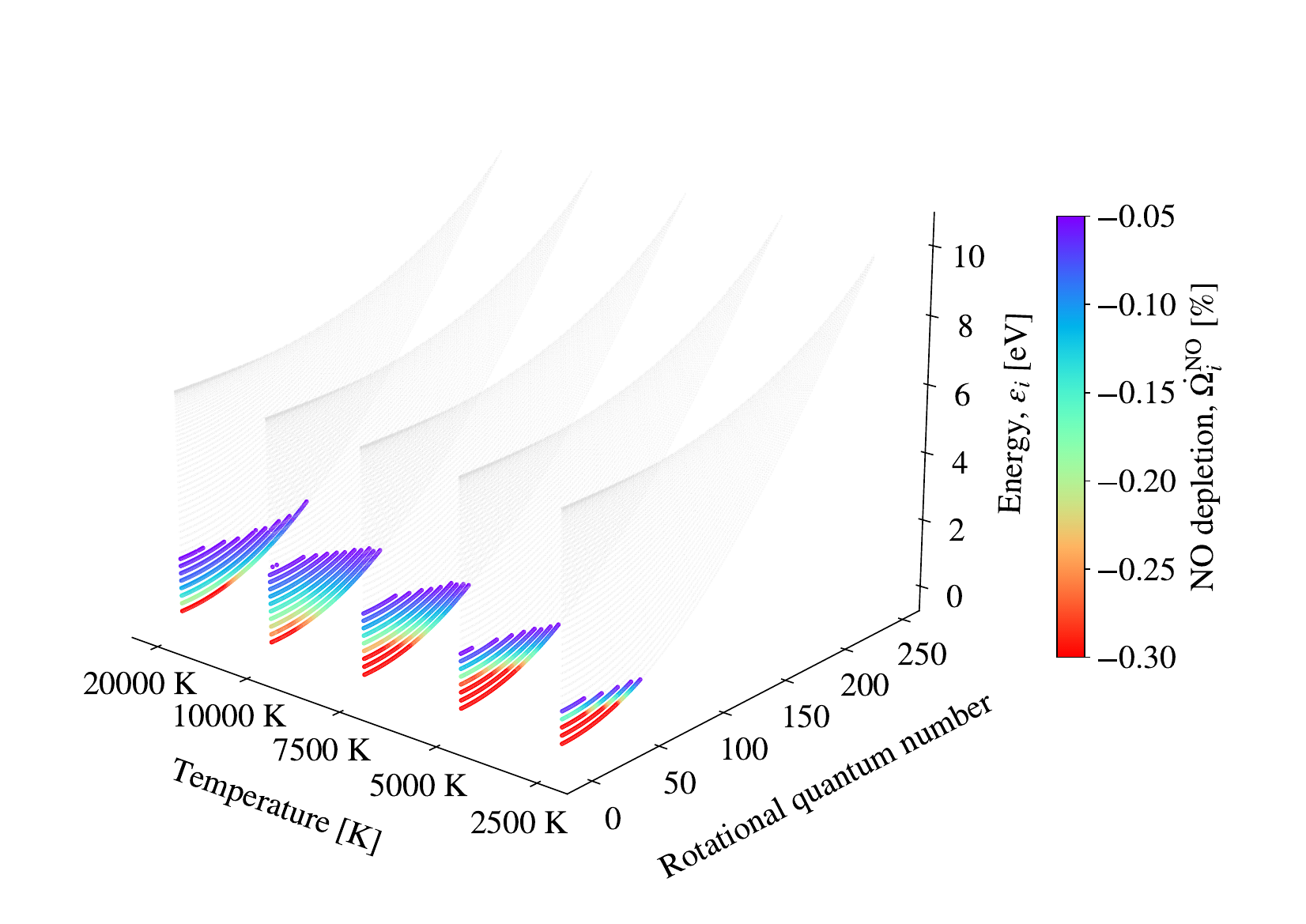}
     \caption{\ce{NO(X^2\Pi)} depletion.}
\end{subfigure}

\caption{Distribution of the exchange production and depletion rates for \ce{NO(X^2\Pi) + C(^3P) <=> CO(X^1\Sigma^{+}) + N(^4S)} at peak production time for five temperatures. The grey dots indicate that the production and depletion rates are below or above a threshold.}
\label{ProdDep_NOC_CON4S}
\end{figure*}

\begin{figure*}[!ht]
\centering

\begin{subfigure}[t]{0.49\textwidth}
     \centering
     \includegraphics[width=\textwidth, trim=1.5cm 0cm 2cm 2cm, clip]{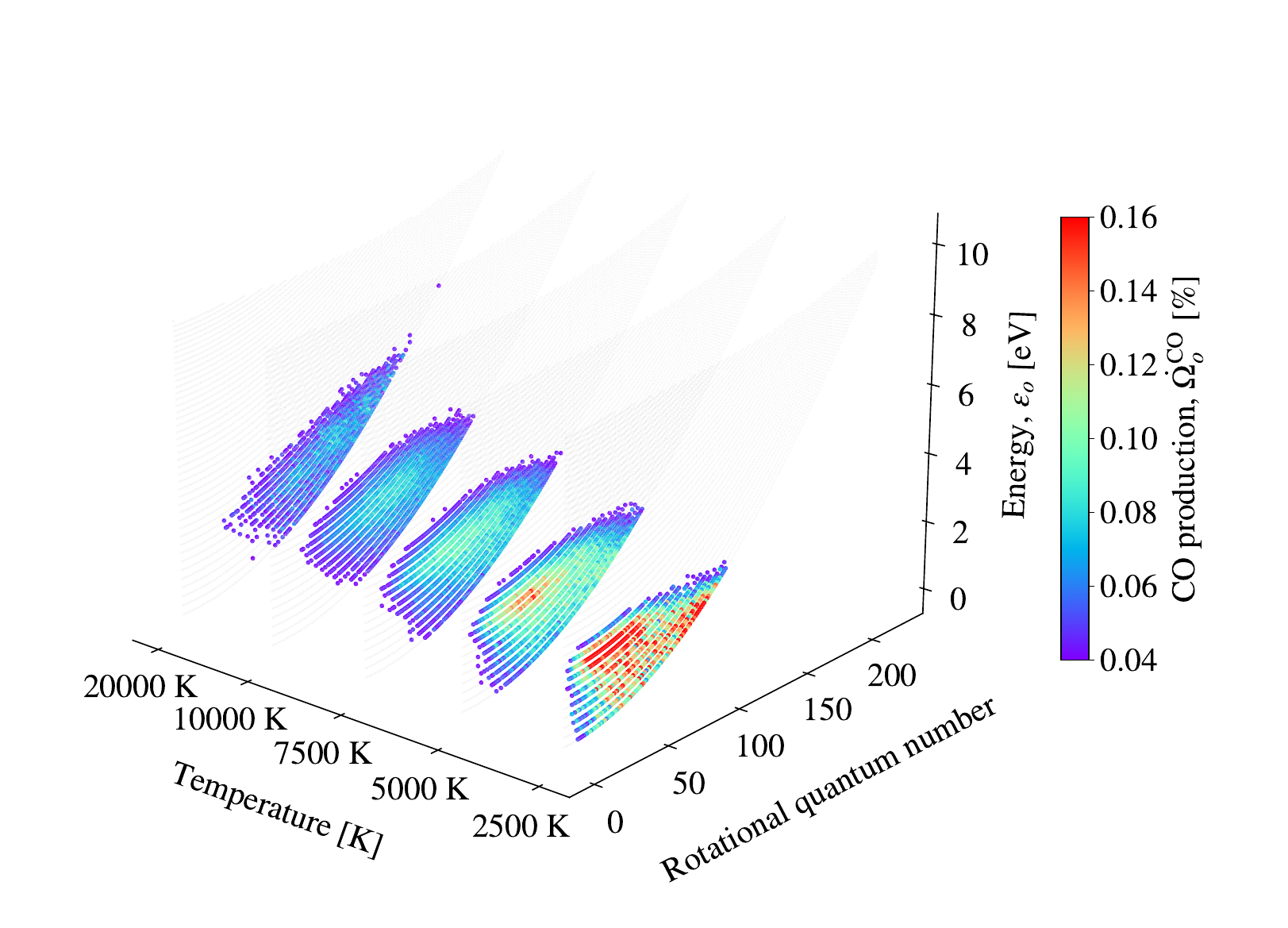}
     \caption{\ce{CO(X^1\Sigma^{+})} production.}
\end{subfigure}
\begin{subfigure}[t]{0.49\textwidth}
     \centering
     \includegraphics[width=\textwidth, trim=1.5cm 0cm 2cm 2cm, clip]{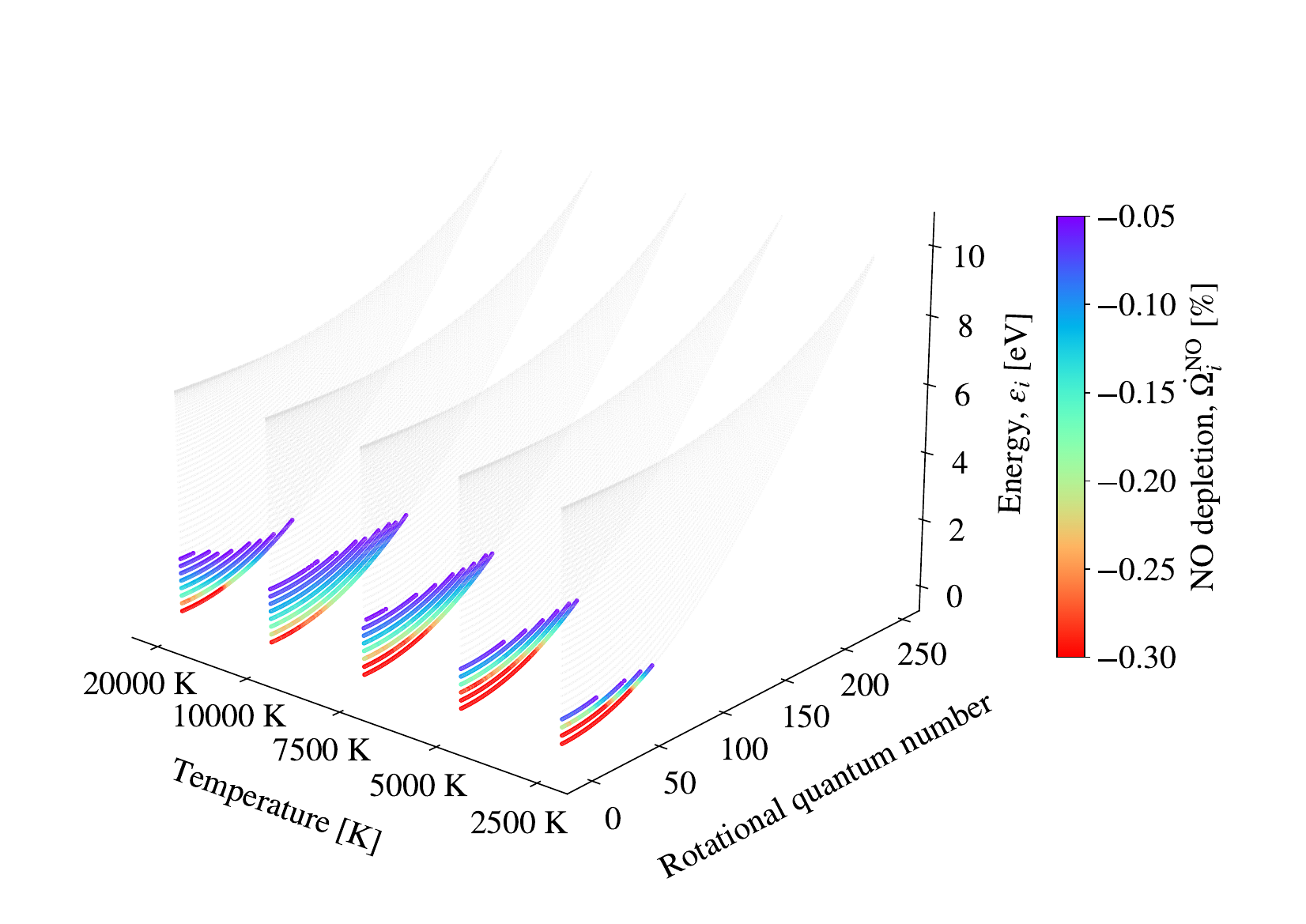}
     \caption{\ce{NO(X^2\Pi)} depletion.}
\end{subfigure}

\caption{Distribution of the exchange production and depletion rates for \ce{NO(X^2\Pi) + C(^3P) <=> CO(X^1\Sigma^{+}) + N(^2D)} at peak production time for five temperatures. The grey dots indicate that the production and depletion rates are below or above a threshold.}
\label{ProdDep_NOC_CON2D}
\end{figure*}


\begin{figure*}[!ht]
\centering
\begin{subfigure}[t]{1\textwidth}
     \centering
     \includegraphics[width=\textwidth]{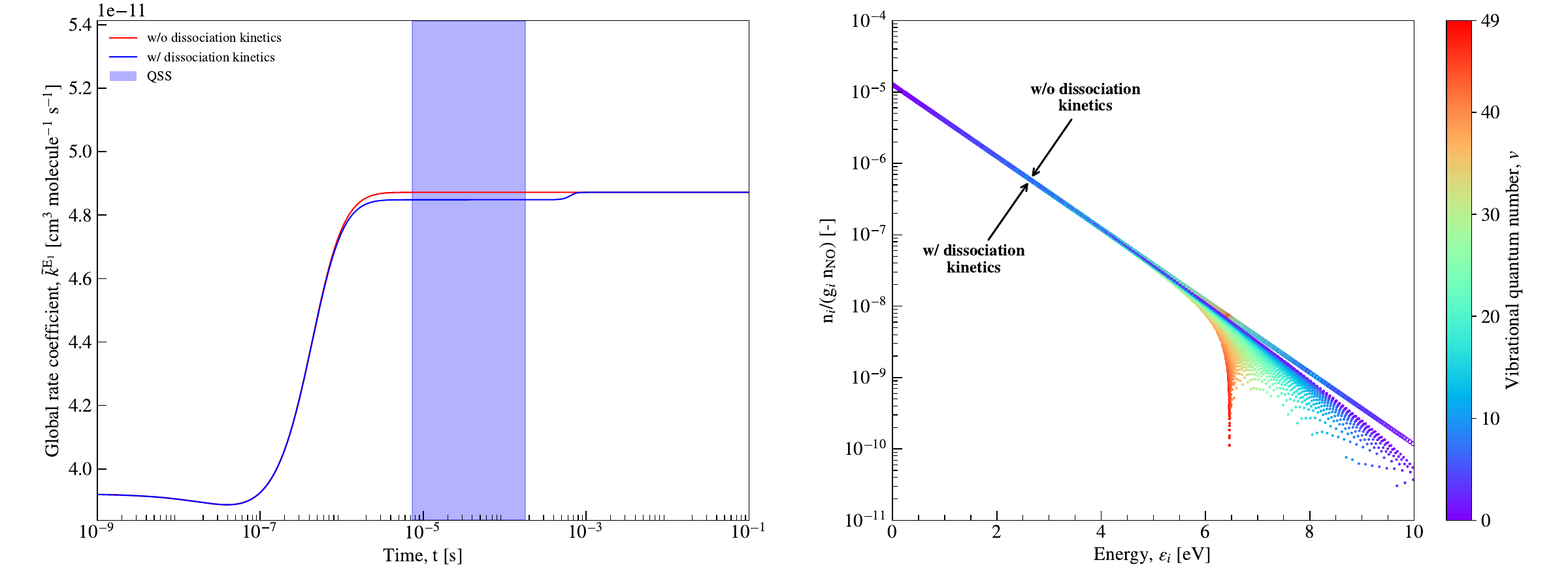}
     \caption{\ce{NO(X^2\Pi) + C(^3P) -> CN(X^2\Sigma^{+}) + O(^3P)}.}
\end{subfigure}
\begin{subfigure}[t]{1\textwidth}
     \centering
     \includegraphics[width=\textwidth]{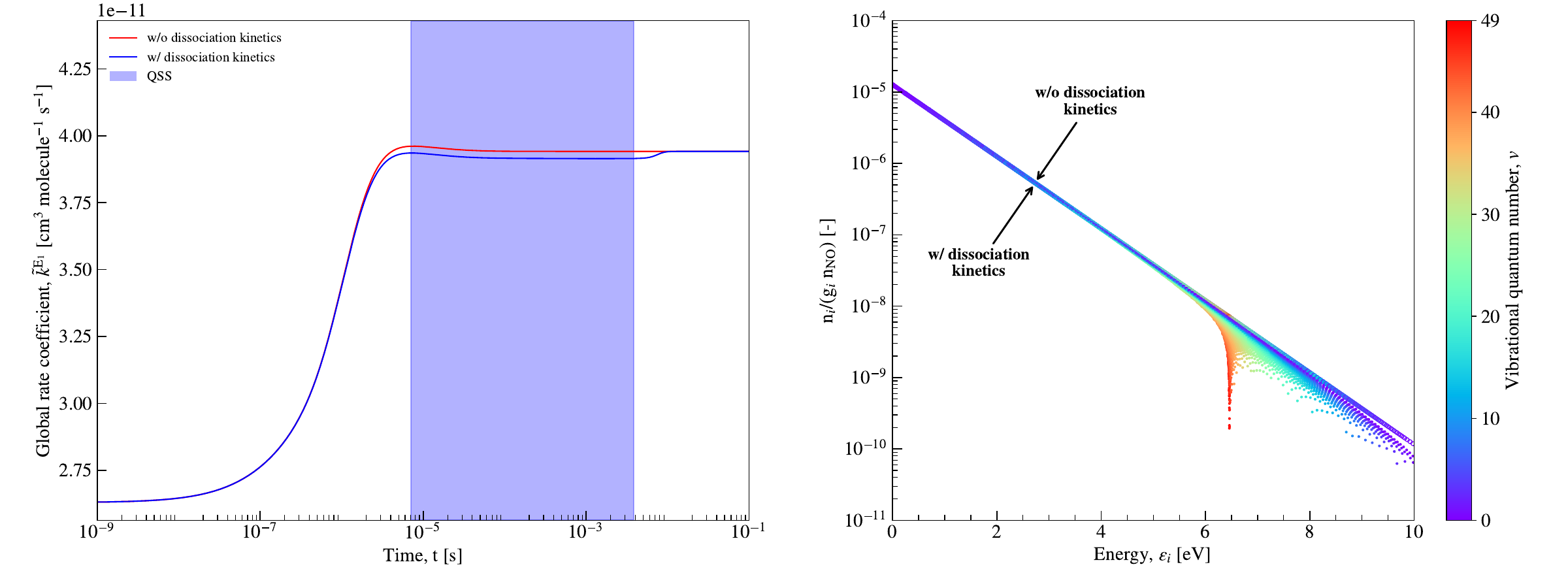}
     \caption{\ce{NO(X^2\Pi) + C(^3P) -> CO(X^1\Sigma^{+}) + N(^4S)}.}
\end{subfigure}
\begin{subfigure}[t]{1\textwidth}
     \centering
     \includegraphics[width=\textwidth]{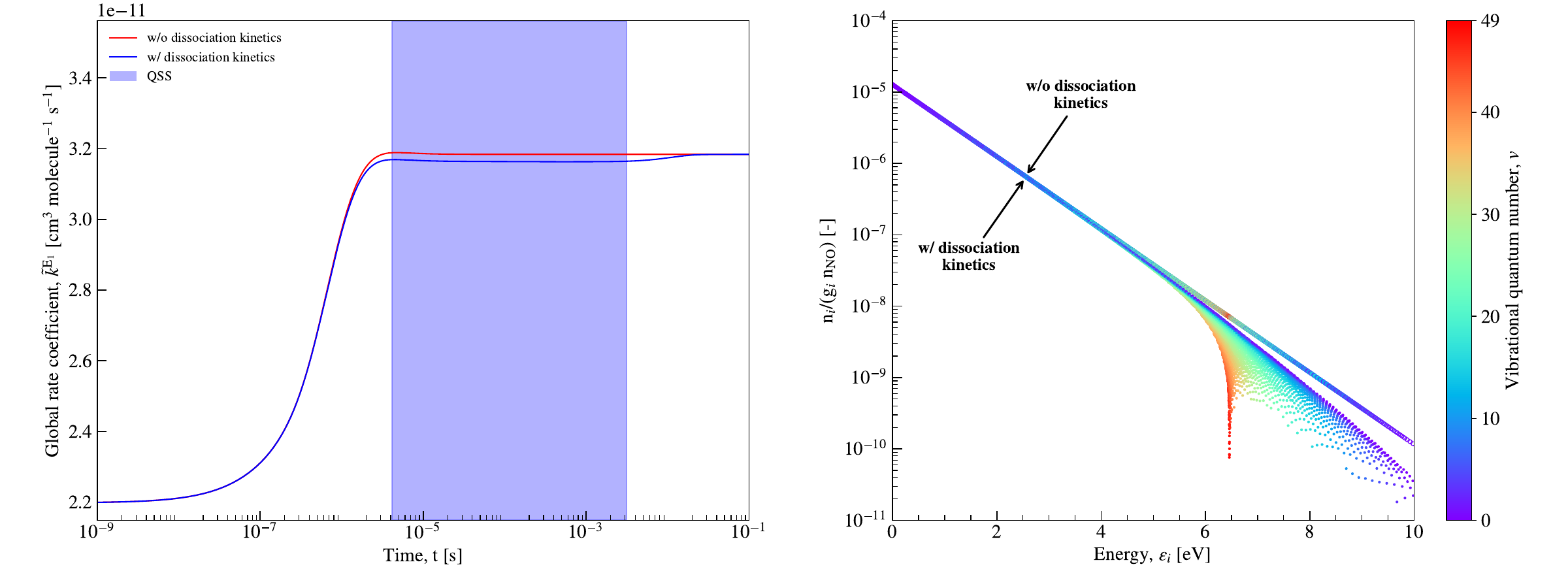}
     \caption{\ce{NO(X^2\Pi) + C(^3P) -> CO(X^1\Sigma^{+}) + N(^2D)}.}
\end{subfigure}
\caption{Comparison of the global rate coefficients of the heterogeneous exchange reactions and their corresponding rovibrational distributions of \ce{NO(X^2\Pi)} with and without dissociation kinetics at the midpoint of the QSS regime at $T = 10{,}000$~K. The initial mole fraction was $\chi_{\ce{NO}} = \chi_{\ce{C}} = 0.5$.}
\label{GlobalRateExchange_NOC}
\end{figure*}

\begin{figure*}[!ht]
\centering
\begin{subfigure}[t]{1\textwidth}
     \centering
     \includegraphics[width=\textwidth]{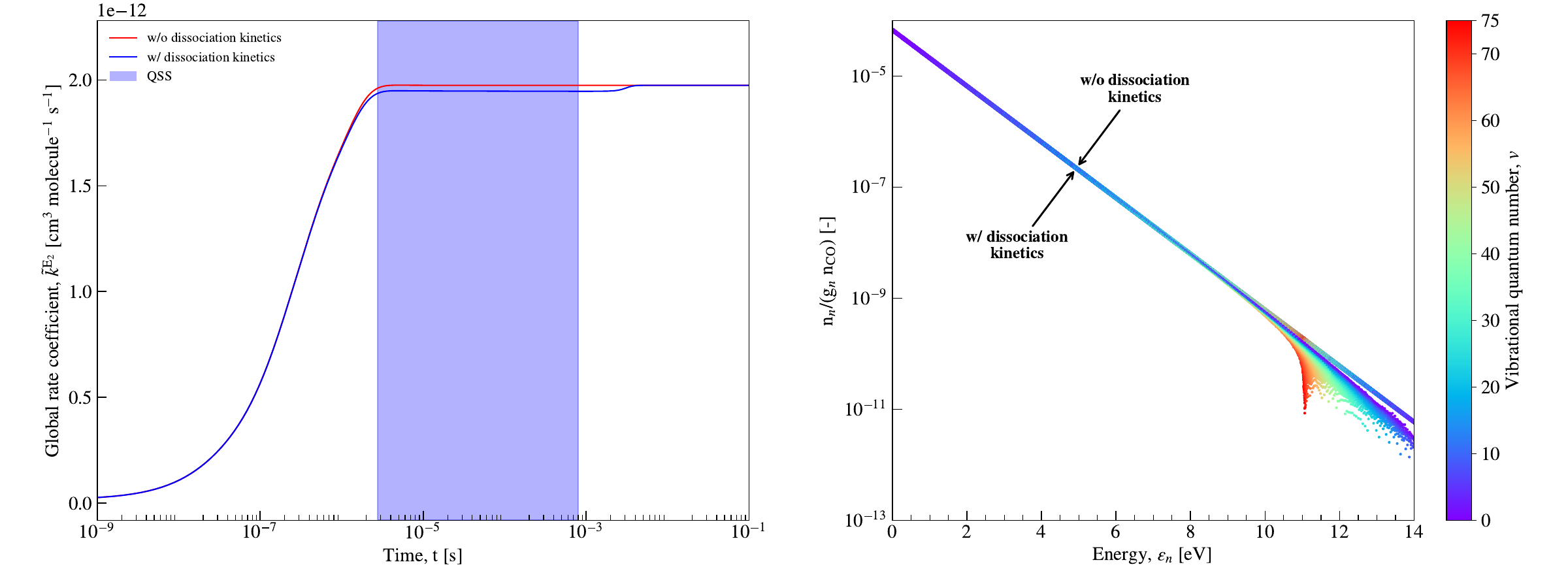}
     \caption{\ce{CO(X^1\Sigma^{+}) + N(^4S) -> NO(X^2\Pi) + C(^3P)}.}
\end{subfigure}
\begin{subfigure}[t]{1\textwidth}
     \centering
     \includegraphics[width=\textwidth]{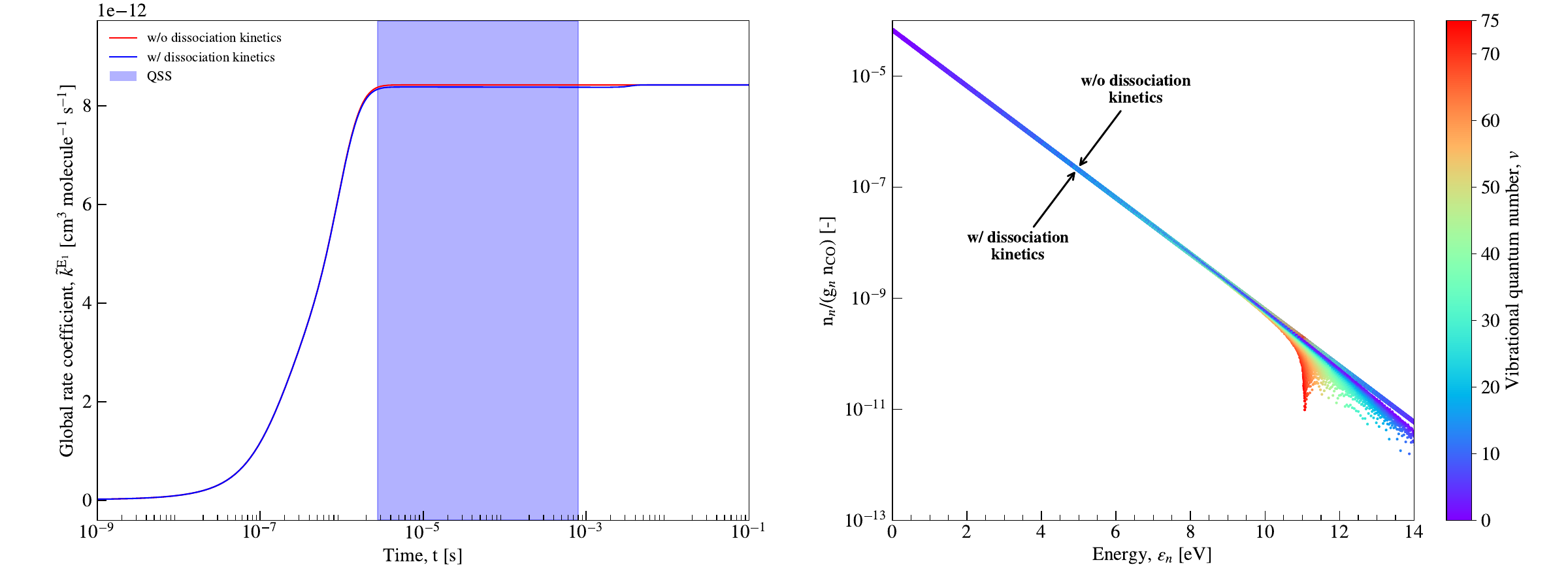}
     \caption{\ce{CO(X^1\Sigma^{+}) + N(^4S) ->  CN(X^2\Sigma^{+}) + O(^3P)}.}
\end{subfigure}
\caption{Comparison of the global rate coefficients of the heterogeneous exchange reactions and their corresponding rovibrational distributions of \ce{CO(X^1\Sigma^{+})} with and without dissociation kinetics at the midpoint of the QSS regime at $T = 10{,}000$~K. The initial mole fraction was $\chi_{\ce{CO}} = \chi_{\ce{N(^4S)}} = 0.5$.}
\label{GlobalRateExchange_CON4S}
\end{figure*}

\begin{figure*}[!ht]
\centering
\begin{subfigure}[t]{1\textwidth}
     \centering
     \includegraphics[width=\textwidth]{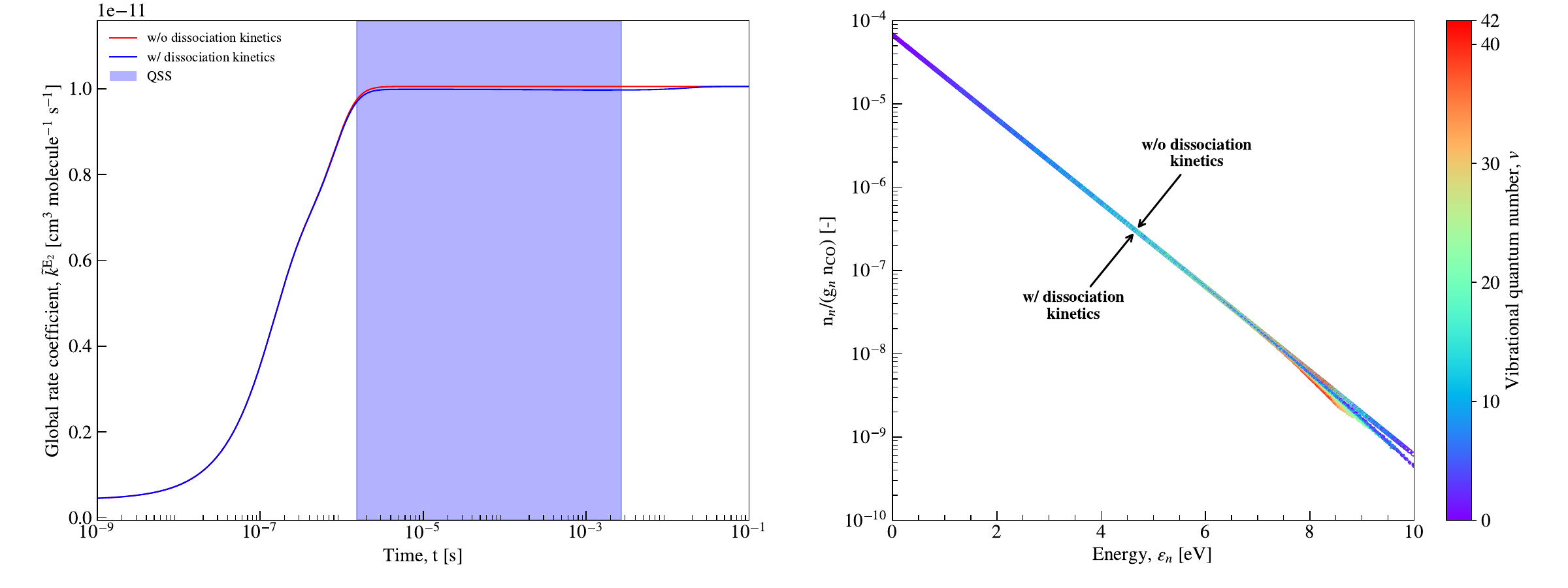}
     \caption{\ce{CO(X^1\Sigma^{+}) + N(^2D) -> NO(X^2\Pi) + C(^3P)}.}
\end{subfigure}
\begin{subfigure}[t]{1\textwidth}
     \centering
     \includegraphics[width=\textwidth]{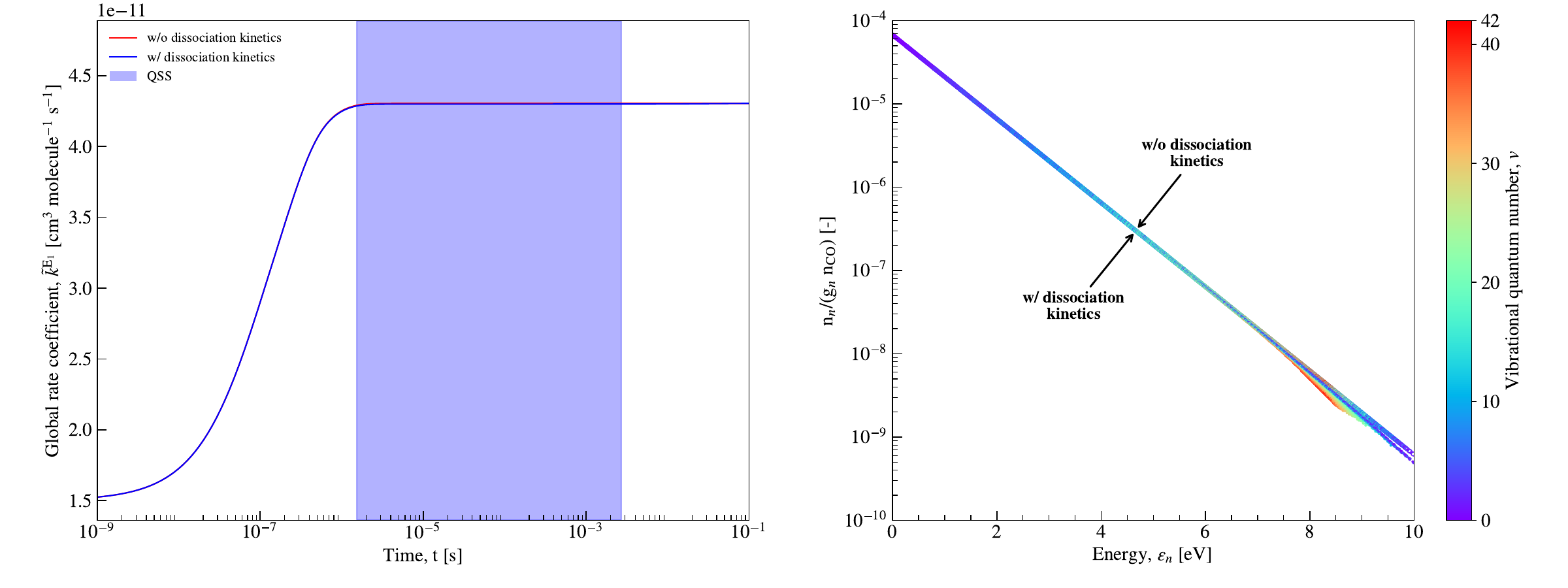}
     \caption{\ce{CO(X^1\Sigma^{+}) + N(^2D) -> CN(X^2\Sigma^{+}) + O(^3P)}.}
\end{subfigure}
\caption{Comparison of the global rate coefficients of the heterogeneous exchange reactions and their corresponding rovibrational distributions of \ce{CO(X^1\Sigma^{+})} with and without dissociation kinetics at the midpoint of the QSS regime at $T = 10{,}000$~K. The initial mole fraction was $\chi_{\ce{CO}} = \chi_{\ce{N(^2D)}} = 0.5$.}
\label{GlobalRateExchange_CON2D}
\end{figure*}

\subsection{Rate Coefficients}

\begin{table*}[!ht]
    \centering
    \caption{Recommended modified Arrhenius fitting parameters for the \ce{NO(X^2 \Pi) + C(^3P)}, \ce{CN(X^2 \Sigma^{+}) + O(^3P)} and \ce{CO(X^1 \Sigma^{+}) + N(^4S,^2D)} collisional systems.}
    \label{ArrheniusCoefficientFit}
    \begin{tabular*}{\textwidth}{@{\extracolsep{\fill}}cccc@{}}
        Reaction & $A$ [m$^3$~molecule$^{-1}$~s$^{-1}$] & $\eta$ [-] & $\theta$ [K] \\
        \hline
        \ce{NO(X^2 \Pi) + C(^3P) -> CN(X^2 \Sigma^{+}) + O(^3P)} & $4.156\times 10^{-20}$ & 0.739 & $-2283$ \\
        \ce{NO(X^2 \Pi) + C(^3P) -> CO(X^1 \Sigma^{+}) + N(^4S)} & $4.123\times 10^{-20}$ & 0.747 & 519 \\
        \ce{NO(X^2 \Pi) + C(^3P) -> CO(X^1 \Sigma^{+}) + N(^2D)} & $4.148\times 10^{-19}$ & 0.495 & 2666 \\
        \ce{NO(X^2 \Pi) + C(^3P) -> N(^4S) + O(^3P) + C(^3P)} (QSS) & $3.054\times 10^{-11}$ & $-1.130$ & $76861$ \\
        \ce{NO(X^2 \Pi) + C(^3P) -> N(^4S) + O(^3P) + C(^3P)} (equilibrium) & $2.150\times 10^{-11}$ & $-0.988$ & $76744$ \\
        & & & \\
        \ce{CN(X^2 \Sigma^{+}) + O(^3P) -> NO(X^2 \Pi) + C(^3P)} & $6.694\times 10^{-20}$ & 0.761 & $11970$ \\
        \ce{CN(X^2 \Sigma^{+}) + O(^3P) -> CO(X^1 \Sigma^{+}) + N(^4S)} & $5.107\times 10^{-19}$ & 0.558 & $598$ \\
        \ce{CN(X^2 \Sigma^{+}) + O(^3P) -> CO(X^1 \Sigma^{+}) + N(^2D)} & $2.681\times 10^{-18}$ & 0.331 & $-1611$ \\
        \ce{CN(X^2 \Sigma^{+}) + O(^3P) -> C(^3P) + N(^4S) + O(^3P)} (QSS) & $1.232\times 10^{-10}$ & $-1.166$ & $92262$ \\
        \ce{CN(X^2 \Sigma^{+}) + O(^3P) -> C(^3P) + N(^4S) + O(^3P)} (equilibrium) & $5.692\times 10^{-12}$ & $-0.774$ & $91409$ \\
        & & & \\
        \ce{CO(X^1 \Sigma^{+}) + N(^4S) -> NO(X^2 \Pi) + C(^3P)} & $4.565\times 10^{-19}$ & 0.743 & $54398$ \\
        \ce{CO(X^1 \Sigma^{+}) + N(^4S) -> CN(X^2 \Sigma^{+}) + O(^3P)} & $3.511\times 10^{-18}$ & 0.533 & $40224$ \\
        \ce{CO(X^1 \Sigma^{+}) + N(^4S) -> C(^3P) + O(^3P) + N(^4S)} (QSS) & $1.571\times 10^{-7}$ &$-1.724$& $136300$ \\
        \ce{CO(X^1 \Sigma^{+}) + N(^4S) -> C(^3P) + O(^3P) + N(^4S)} (equilibrium) & $4.803\times 10^{-11}$ & $-0.825$ & $130808$ \\
        & & & \\
        \ce{CO(X^1 \Sigma^{+}) + N(^2D) -> NO(X^2 \Pi) + C(^3P)} & $1.837\times 10^{-18}$ & 0.491 & $28880$ \\
        \ce{CO(X^1 \Sigma^{+}) + N(^2D) -> CN(X^2 \Sigma^{+}) + O(^3P)} & $7.373\times 10^{-18}$ & 0.306 & $10351$ \\
        \ce{CO(X^1 \Sigma^{+}) + N(^2D) -> C(^3P) + O(^3P) + N(^4S)} (QSS) & $2.180\times 10^{-9}$ &$-1.429$&$107911$ \\
        \ce{CO(X^1 \Sigma^{+}) + N(^2D) -> C(^3P) + O(^3P) + N(^4S)} (equilibrium) & $9.946\times 10^{-12}$ & $-0.853$ & $104865$ \\
    \end{tabular*}
\end{table*}

\bibliography{reference}